%% file: 2020_LiteBIRD_PTEP.tex
\documentclass{ptephy_v1}
\pdfoutput=1

\input{authors_affiliations_LB_2020.01_key_ptepv2}

\usepackage{lineno}

\usepackage{glossaries}
\usepackage{caption}
\makenoidxglossaries 
\input{acronyms}
\input{journal-macros}
\usepackage{pdfpages}
\usepackage{graphicx}
\usepackage{siunitx}
\usepackage{textcomp}
\usepackage{multirow} 
\usepackage{soul}
\usepackage{url} 
\usepackage{array} 

\def\gtorder{\mathrel{\raise.3ex\hbox{$>$}\mkern-14mu
\lower0.6ex\hbox{$\sim$}}}
\def\ltorder{\mathrel{\raise.3ex\hbox{$<$}\mkern-14mu
\lower0.6ex\hbox{$\sim$}}}
\def\pixelred{\color[RGB]{127, 9, 9}}
\def\pixelyellow{\color[RGB]{103, 102, 1}}
\def\pixelgreen{\color[RGB]{1, 102, 1}}
\def\pixelblue{\color[RGB]{3, 4, 120}}

\newcommand{\LB}{\textit{LiteBIRD}}
\newcommand{\lb}{\LB}
\newcommand{\LiteBIRD}{\LB}
\newcommand{\liteBIRD}{\LB}

\newcommand{\litebird}{\LB}
\newcommand{\Planck}{\textit{Planck}}
\newcommand{\planck}{\Planck}
\newcommand{\WMAP}{\textit{WMAP}}

\def\arraynet{2.2}

\begin{document}

\title{Probing Cosmic Inflation with the \LiteBIRD\ Cosmic Microwave Background Polarization Survey}







\begin{abstract}%
\lb\, the Lite (Light) satellite for the study of $B$-mode polarization and Inflation from cosmic background Radiation Detection, is a space mission for primordial cosmology and fundamental physics. The Japan Aerospace Exploration Agency (JAXA) selected \lb\ in May 2019 as a strategic large-class (L-class) mission, with an expected launch in the late 2020s using JAXA's H3 rocket.
\lb\ is planned to orbit the Sun-Earth Lagrangian point L2, 
where it will map the cosmic microwave background (CMB) polarization over the entire sky for three years, with three telescopes in 15 frequency bands between 34 and 448\,GHz, to achieve an unprecedented total sensitivity of $\arraynet$\,$\mu$K-arcmin, with a typical angular resolution of 0.5$^\circ$ at 100\,GHz. 
The primary scientific objective of \lb\ is to search for the signal from cosmic inflation, either making a discovery or ruling out well-motivated inflationary models. The measurements of \lb\ will also provide us with insight into the quantum nature of gravity and other new physics beyond the standard models of particle physics and cosmology. We provide an overview of the \lb\ project, including scientific objectives, mission and system requirements, operation concept, spacecraft and payload module design, expected scientific outcomes, potential design extensions and synergies with other projects.
\end{abstract}

\subjectindex{\lb\, cosmic inflation, cosmic microwave background, $B$-mode polarization, primordial gravitational waves, quantum gravity, space telescope}

\maketitle

\tableofcontents


\clearpage\newpage

\section{Introduction} 
\label{s:introduction}

\subsection{CMB Polarization as the New Frontier and the \LiteBIRD\ Satellite}
\label{ss:intro_cmb_pol}
Observations of \gls{cmb} temperature fluctuations have played a pivotal role in establishing the standard cosmological model, called the \gls{lcdm} model~\cite{Weinberg:2008zzc}, and provide insights into the origin of structure, the density of baryons, dark matter, dark energy, the number of neutrino species, and the global properties of spacetime~\cite{Peebles:2009zz,Sugiyama:2014lga,ScottSmoot,1804633}. Observations have reached a point at which most of the information about the early Universe available in temperature fluctuations has been exhausted \cite{komatsu/bennett:2014,Planck2018I}. However, precise measurements of the fainter \gls{cmb} polarization anisotropies hold the key to answering many remaining questions about the Universe.
Observations have so far only begun to scratch the surface~\cite{Aiola:2020azj,Adachi:2020knf,Sayre:2019dic,SPT-3G:2021eoc,BK2021,SPIDER:2021ncy}.

Perhaps the biggest remaining question is what mechanism created the small primordial fluctuations that seeded the observed \gls{cmb} anisotropies and eventually grew into stars and galaxies. The most widely studied idea is ``cosmic inflation''~\cite{starobinsky:1980,sato:1981,guth:1981,albrecht/steinhardt:1982,linde:1982,linde:1983}. According to this idea the primordial fluctuations originated as quantum fluctuations during a period of nearly exponential expansion of the very early Universe~\cite{mukhanov/chibisov:1981,starobinsky:1982,hawking:1982,guth/pi:1982,bardeen/turner/steinhardt:1983}. Eventually this period ended and the Universe became filled with a hot and dense plasma that subsequently cooled and led to the Universe we see around us. As a consequence of the nearly exponential expansion that stretched microscopic regions of spacetimes to macroscopic scales, the plasma is homogeneous and isotropic except for the minute quantum fluctuations that were also stretched to macroscopic scales.   

Cosmic inflation predicts primordial density fluctuations that are consistent with the observed temperature fluctuations~\cite{mukhanov/chibisov:1981,starobinsky:1982,hawking:1982,guth/pi:1982,bardeen/turner/steinhardt:1983}. In addition, inflation predicts quantum fluctuations in the fabric of spacetime itself~\cite{grishchuk:1975,starobinsky:1979}. These primordial gravitational waves lead to a characteristic imprint in \gls{cmb} polarization, commonly referred to as ``$B$-mode'' polarization~\cite{seljak/zaldarriaga:1997,kamionkowski/kosowsky/stebbins:1997,HuWhite1997Total}, and many of the best-motivated models predict a signal that is large enough to be detected with \lb\ 
(the Lite (Light) satellite for the study of $B$-mode polarization and Inflation from cosmic background Radiation Detection)~\cite{kamionkowski/kovetz:2016}.

A detection of this signal would open an unexplored frontier of physics, shedding light on fundamental processes at energies far beyond the reach of CERN's Large Hadron Collider, revolutionizing our understanding of physics and the early Universe \cite{Lyth:1996im}.
A detection of primordial gravitational waves with \lb\ has important implications for many aspects of fundamental physics. A detection would, for example, indicate that inflation occurred near the energy scale associated with grand unified theories, providing additional evidence in favor of the idea of the unification of forces. Knowledge of the energy scale of inflation also has important implications for several other aspects of fundamental physics, such as axions and, in the context of string theory, the fields that control the shapes and sizes of the compact dimensions.

To search for the imprint of gravitational waves, \lb\ will conduct a survey of the entire sky that is 30 times more sensitive than previous full-sky experiments, corresponding to a raw sensitivity of nearly 1000 \Planck\ missions.\footnote{This is based on a comparison between the inverse-variance weighted combination of the white noise for CMB polarization of all channels for~\lb\ versus \Planck.}   
\lb\ will be the natural next step in the series of \gls{cmb} space missions, following
the \gls{nasa}['s] \glsentryshort{cobe}~\cite{Smoot1990} and \glsentryshort{wmap}~\cite{Bennett2003a}, and the \gls{esa}['s]  \Planck~\cite{Tauber2010a}, each of which has made its own landmark scientific discoveries. 
See Ref.~\cite{Peebles:2009zz} for a comprehensive list of CMB experiments and their pioneering contributions. 

The CMB polarization anisotropies can be decomposed according to their transformation properties under parity transformations into ``$E$ modes'' and ``$B$ modes.'' The $E$-mode polarization is predominantly caused by acoustic waves present at recombination, and the signal is strongest on angular scales of a few to tens of arcminutes (corresponding to multipoles of $\ell\sim 1000$). 
The $B$-mode polarization pattern imprinted by gravitational waves peaks on degree angular scales (corresponding to multipoles of $\ell\simeq 80$) and on very large angular scales (corresponding to multipoles of $\ell\lesssim 10$)~\cite{seljak/zaldarriaga:1997,Zaldarriaga:1996xe}. The ``recombination peak'' near $\ell\simeq 80$ is imprinted during the epoch when electrons and protons combine to form hydrogen and the Universe becomes neutral, while the ``reionization bump'' below $\ell\simeq 10$ is imprinted around the time when the first stars reionize the Universe~\cite{zaldarriaga:1997}.  At linear order, the density perturbations do not generate $B$-mode polarization, which makes the $B$-mode power spectrum the most natural observable to search for primordial gravitational waves. However, CMB photons are deflected by the gravitational potentials associated with the matter along the line of sight. This is referred to as weak gravitational lensing and converts some of the ``$E$-mode'' polarization generated by density perturbations into $B$ modes \cite{zaldarriaga/seljak:1998}. Like the $E$ modes, this effect peaks on much smaller scales of a few arcminutes (corresponding to multipoles of $\ell\simeq 1000$) but must be taken into account. This contribution is well-understood theoretically, and \lb\ targets any excess over the lensing signal caused by the imprint of gravitational waves. 
While ground-based experiments only target the recombination peak, \lb\ can see both peaks in the $B$-mode power spectrum.

In addition to the $B$ modes caused by weak gravitational lensing of $E$ modes, there are additional ``foreground'' sources of $B$-mode polarization at microwave frequencies. Thermal emission by interstellar dust grains that are aligned with the Galactic magnetic field and synchrotron emission from electrons spiraling in the Galactic magnetic field provide the dominant contributions. Fortunately, the frequency dependence of the primordial signal and foreground emission differ significantly so that multi-frequency observations allow us to disentangle the primordial and foreground contributions \cite{Leach2008,CORE:2017yri,Planck2018IV}.

To separate these primordial and foreground components, \lb\ will survey the full sky in 15 frequency bands from 34 to 448\,GHz, with effective polarization sensitivity of $2\,\mu$K-arcmin and angular resolution of $31$\,arcmin (at 140\,GHz). Rapid polarization modulation, a densely linked observation strategy, and the stable environment of an orbit around L2 (the second Lagrangian point for the Sun-Earth system), provide unprecedented ability to control systematic errors, especially on the largest angular scales below $\ell \simeq 10$. Taken together, the control of foregrounds and systematic errors gives \lb\ the ability to detect both the reionization and recombination bumps in the $B$-mode power spectrum, giving much higher confidence that a primordial signal has been uncovered. Importantly, if a hint of the recombination peak is seen by a ground-based or balloon-borne experiment, \lb\ will make a definitive statement on the detection of the signal and greatly improve the quantitative constraints on the physics of inflation. The forecast for \lb's ability to measure the primordial $B$-mode power spectrum is shown in Fig.~\ref{fig:cl}, together with currently available measurements.

Even in the absence of gravitational waves, on the largest angular scales, scattering of photons during the reionization epoch at $z\simeq 6$--10 generates $E$-mode polarization~\cite{zaldarriaga:1997}. \lb\ will measure this signal with high precision and will make a definitive determination of the optical depth to the surface of last scattering. The optical depth contains key information about the nature of the epoch of reionization and will, for example, constrain models of the first stars. The optical depth is currently the least well-constrained parameter of the standard cosmological model and currently limits any constraints that rely on comparisons of the amplitude of \gls{cmb} anisotropies and clustering of the matter distribution, such as the measurement of the sum of neutrino masses~\cite{DiValentino:2015sam,allison/etal:2015,Giusarma:2016phn,Archidiacono:2016lnv,boyle/komatsu:2018}. \lb\ will provide a cosmic-variance-limited measurement\footnote{A measurement is cosmic-variance-limited if the error bar is limited only by the fraction of sky available for the cosmological analysis and can no longer be decreased by improving the instrument~\cite{abbott/wise:1984:2,White:1993jr,ScottSrednickiWhite:1994,Knox:1995}.} of $E$-modes at low multipoles. This will complement measurements by high-resolution ground-based \gls{cmb} experiments such as the \gls{spo} \cite{10.1117/12.2561995}, \gls{so} \cite{Ade:2018sbj}, and \gls{cmbs4}~\cite{Abazajian:2016yjj} and will significantly improve cosmological measurements of the sum of neutrino masses. 

Finally, the \lb\ all-sky polarized maps in 15 frequency bands will be a rich legacy data set for understanding the large-scale magnetic field structure in the Milky Way, having five times greater sensitivity to Galactic magnetic fields than ESA's \planck\ mission~\cite{Planck2018XI,Planck2018XII}.


\subsection{Outline of this Review}
\label{ss:intro_outline}

This paper is arranged as follows.
Section~\ref{s:cmb_polarization} introduces CMB $B$-mode tests of cosmic inflation, including constraints expected by \lb\ and an argument for the necessity for CMB $B$-mode measurements from space.  Section~\ref{ss:litebird_overview} gives a broad overview of the \lb\ mission, including the science requirements, a description of the instrument, and a description of flight operations.  Section~\ref{s:payload} describes the \lb\ instrument, including the telescope designs, the bolometric detector arrays, the readout, the cryogenics, and calibration strategy.  Section~\ref{s:cosmological_forecasts} gives a detailed analysis of the statistical and systematic uncertainties in the tensor-to-scalar ratio measurement; this section also includes an analysis of the impact of foregrounds and instrumental uncertainties.  Section~\ref{s:scientific_outcomes} describes the scientific outcomes of \lb\ beyond the detection of primoridial gravitational waves, including measurement of the optical depth to reionization, determination of neutrino masses, a search for cosmic birefringence, mapping hot gas in the Universe, a search for anisotropic CMB spectral distortions, a probe of primordial magnetic fields, and measurements to elucidate the astrophysics of the Milky Way.  Section~\ref{s:discussion} describes possible extensions to the \lb\ mission design, including extending the frequency range, as well as synergy with other cosmology and astrophysics projects.  Section~\ref{s:conclusions} concludes this review. 




\section{CMB \textit{B} Modes as Tests of Cosmic Inflation}\label{s:cmb_polarization}

\subsection{CMB Polarization Power Spectra}

\lb\ will provide maps of the temperature and polarization anisotropies in 15 frequency bands from 34 to 448\,GHz. Fundamental theory does not predict the detailed structure of the maps, only their statistical properties, like the expected correlations between temperature and polarization anisotropies between different points on the sky. Earlier measurements by \WMAP\ and \Planck\ imply that the anisotropies are nearly Gaussian
so that their statistical properties are predominantly characterized by the 2-point correlation functions~\cite{komatsu2003,Planck2013XXIV,Planck2015XVII,Planck2018IX}. 
The observations by \WMAP\ and \Planck\ also tightly constrain departures from statistical isotropy~\cite{Bennett2010,Planck2013XXI,Planck2015XVI,Planck2018VII}. Under the assumption that the underlying probability distribution is isotropic, the correlations between anisotropies at different points in the sky only depend on the angle between them. 

Given these properties, it is natural to consider angular correlation functions that measure the correlations between different points in the sky as a function of this angular separation. While it would be possible to work with maps and their angular correlation functions, in practice it is more convenient to expand the temperature $T$ maps in terms of spherical harmonics,
\begin{equation}
\Delta T(\hat{n})=\sum\limits_{\ell,m}a^T_{\ell m} Y_{\ell m}(\hat{n})\,,
\end{equation}
and work with the coefficients of this expansion $a^T_{\ell m}$, referred to as multipole coefficients. 

\begin{figure}[htbp!]
\centering
\includegraphics[width=\textwidth]{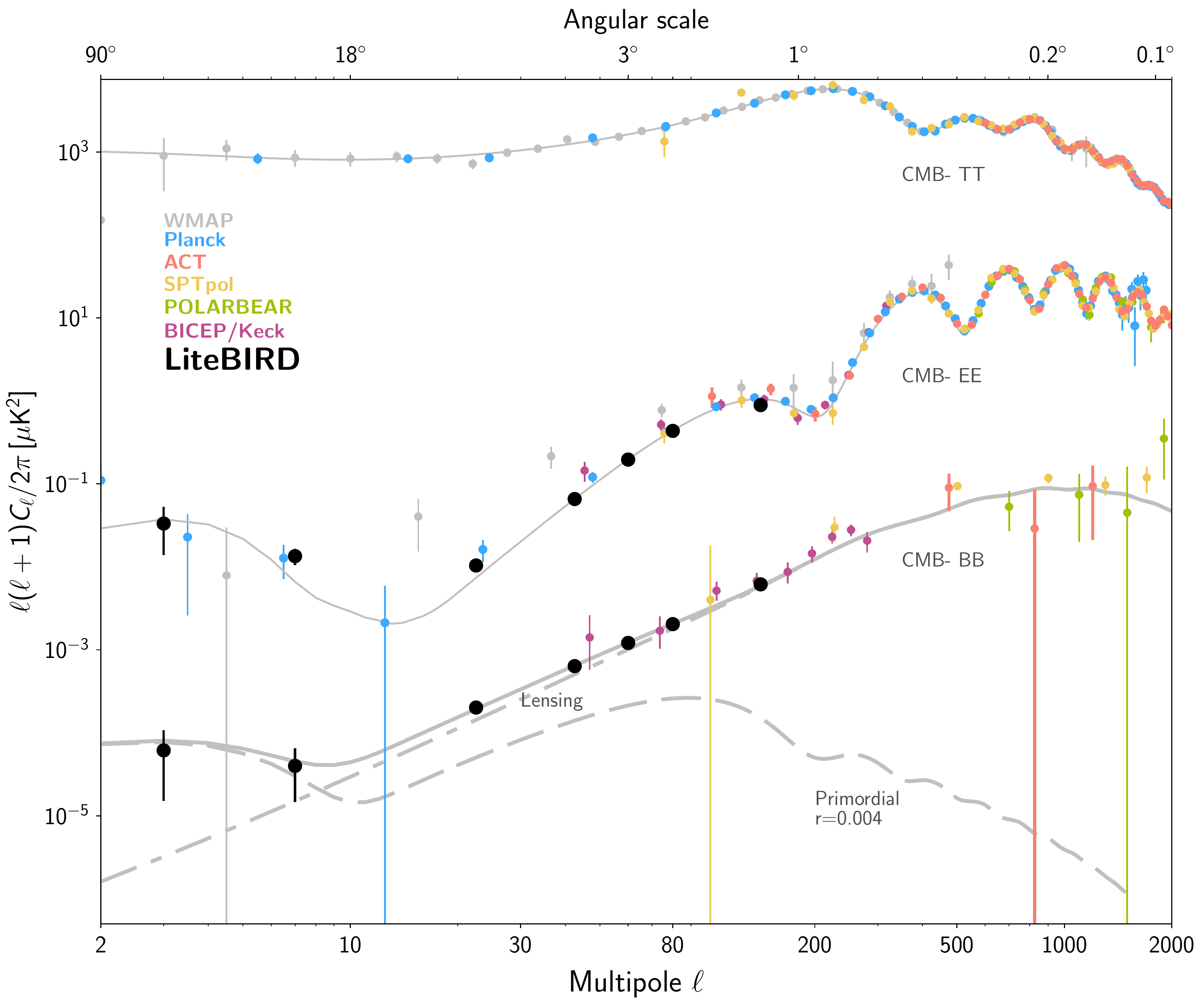}
\caption{CMB power spectra of the temperature anisotropy (top), $E$-mode polarization (middle), and $B$-mode polarization (bottom). 
The solid lines show the angular power spectra for the best-fit \gls{lcdm} model in the presence of a scale-invariant tensor (gravitational wave) perturbation with a tensor-to-scalar ratio parameter of $r=0.004$. The thin dashed line shows the contribution to the $B$-mode spectrum from scale-invariant tensor perturbation with $r=0.004$. A summary of present measurements of CMB power spectra (colored points)~\cite{hinshaw/etal:2013,Bennett2012,Ade:2017uvt,newKeckArrayBicep2,Henning:2017nuy,Planck2018V,Sayre:2019dic,Aiola:2020azj,Choi:2020ccd,Adachi:2020knf,BK2021} and the expected polarization sensitivity of \lb\ (black points) are also shown.}
\label{fig:cl}
\end{figure}

Similarly, it is convenient to expand the maps of the Stokes $Q$ and $U$ parameters, which characterize the linear polarization, in terms of spin-weighted spherical harmonics, ${}_{2} Y_{\ell m}$,
\begin{equation}
Q(\hat{n})+iU(\hat{n})=-\sum\limits_{\ell,m}(a^E_{\ell m}+i a^B_{\ell m})\, {}_{2} Y_{\ell m}(\hat{n})\,,
\end{equation}
and work with the expansion coefficients $a^E_{\ell m}$ and $a^B_{\ell m}$. This decomposition into $E$- and $B$-mode polarization is convenient because $E$- and $B$-mode patterns have distinct parity, i.e., they transform differently under the inversion of spherical coordinates, $\hat n\to -\hat n$. Specifically, the spherical harmonics coefficients transform as $a^E_{\ell m}\to (-1)^\ell a^E_{\ell m}$ and $a^B_{\ell m}\to (-1)^{\ell+1} a^B_{\ell m}$. As a result, when forming the angular power spectra\footnote{The use of angular power spectra rather than correlation functions is convenient because it leads to a nearly diagonal covariance matrix. For a non-expert review of the physics of the CMB and CMB observables see, for example, Ref.~\cite{Samtleben:2007zz} and for a more technical overview of CMB polarization see Ref.~\cite{HuWhite1997Primer}. The expressions given here are idealized. In practice, foreground emission near the Galactic plane is too bright and must be masked, the instrument has finite resolution, the maps are pixelized, and so on, all of which lead to (known) corrections to these expressions.}
\begin{equation}\label{eq:Cl}
C^{XY}_{\ell}=\frac{1}{2\ell+1}\sum_m
a^X_{\ell m}a^{Y*}_{\ell m}\,,
\end{equation}
where $X$ and $Y$ are either $T$, $E$, or $B$, there are ``parity-even'' combinations such as $C^{TT}_{\ell}$, $C^{TE}_{\ell}$, $C^{EE}_{\ell}$, and $C^{BB}_{\ell}$ that do not change sign under the inversion of spherical coordinates, as well as ``parity-odd'' combinations such as $C^{TB}_{\ell}$ and $C^{EB}_{\ell}$ that do change sign. All of the parity-even combinations from the density fluctuations (scalar perturbation) have been measured already, as seen in Fig.~\ref{fig:cl}, whereas $C^{BB}_{\ell}$ from the primordial gravitational waves (tensor perturbation), the target of the \lb\ mission, have not been detected yet~\cite{newKeckArrayBicep2,Sayre:2019dic,Tristram:2020wbi,SPIDER:2021ncy}. The parity-odd combinations of the CMB polarization can be used to probe new physics which violates parity symmetry~\cite{lue/wang/kamionkowski:1999,feng/etal:2005,liu/lee/ng:2006,saito/ichiki/taruya:2007,contaldi/magueijo/smolin:2008}. While no significant evidence for parity-odd power spectra of the CMB has been found (see~\cite{PlanckIntXLIX,kaufman/keating/johnson:2016,minami/komatsu:2020} for summaries), $C^{TB}_{\ell}$ from the polarized dust emission in our Galaxy has been found ~\cite{PlanckIntXLIX,Planck2018XI}.

As we briefly mentioned in the introduction, the decomposition into $E$ and $B$ modes is convenient also because, to linear order, the well-measured density perturbations only generate temperature and $E$-mode anisotropies, whereas gravitational waves lead to $B$ modes in addition to temperature anisotropies and $E$ modes. Somewhat heuristically, this can be understood from the fact that $E$ modes behave much like the gradient component of a vector field, whereas the $B$ modes behave like the curl-like component. At linear order, one can construct a gradient component from density perturbations, but it is impossible to construct a curl-component. For more details, we refer the interested reader to Refs.~\cite{kamionkowski/kovetz:2016, Baumann:2009ds}.
So $B$ modes provide the cleanest way for CMB experiments to search for primordial gravitational waves. 

To see more explicitly how the information about the very early Universe is encoded, note that the contributions of primordial density perturbations to the angular power spectra of temperature or $E$-mode anisotropies are schematically given by
\begin{equation}
C_{({\rm s}),\ell}^{XX}=\int \frac{dk}{k}\Delta^2_\zeta(k)\left|\int\limits_0^{\tau_0} d\tau\, S^X_{({\rm s})}(k,\tau)\,j_\ell\left[k(\tau_0-\tau)\right]\right|^2\,,
\label{eq:clscalar}
\end{equation}
where $k$ is the wave number of a Fourier mode and $\tau$ is the so-called conformal time, which is related to the physical time $t$ as $d\tau=dt/a(t)$ with $a(t)$ being the scale factor for the homogeneous and isotropic expansion of space. The subscript ``0'' indicates the present-day epoch.
The integrand factorizes into three pieces: 
\begin{itemize}
    \item The primordial power spectrum of density perturbations as a function of $k$ (or equivalently radians per distance), $\Delta^2_\zeta(k)$, which contains information about the very early Universe.
    \item The source functions, $S^X_{({\rm s})}(k,\tau)$, which contain information about the physics of the medium, largely from recombination to the present.
    \item The spherical Bessel functions, $j_\ell(x)$, for a spatially flat universe.
\end{itemize}

Similarly, the contributions of primordial gravitational waves to the angular power spectra of temperature, $E$-mode, and $B$-mode anisotropies are given by
\begin{equation}
C_{({\rm t}),\ell}^{XX}=\int \frac{dk}{k}\Delta^2_h(k)\left|\int\limits_0^{\tau_0} d\tau\, S^X_{({\rm t})}(k,\tau)\,\chi^X_\ell\left[k(\tau_0-\tau)\right]\right|^2\,,
\label{eq:cltensor}
\end{equation}
where $\Delta^2_h(k)$ is now the primordial power spectrum of gravitational waves, $S^X_{({\rm t})}(k,\tau)$ are source functions for tensor perturbations in the medium, and $\chi^X_\ell(x)$ is a function of the spherical Bessel functions and their derivatives appropriate for $X=T$, $E$, or $B$~\cite{Seljak:1996is,Zaldarriaga:1996xe,Kamionkowski:1996ks,HuWhite1997Total}.

A wealth of information about the Universe is contained in the dependence of the source functions and the argument of the spherical Bessel function on cosmological parameters, like the matter density, the number of relativistic degrees of freedom, and so on. Here we are most interested in the information contained in the primordial power spectra, $\Delta^2_\zeta(k)$ and $\Delta^2_h(k)$, which are conventionally parameterized as~\cite{Lyth:1998xn}
\begin{equation}
\Delta^2_\zeta(k)=\Delta^2_\zeta \left(\frac{k}{k_\ast}\right)^{n_{\rm s}(k)-1}\quad\text{and}\quad \Delta^2_h(k)=\Delta^2_h \left(\frac{k}{k_\ast}\right)^{n_{\rm t}(k)}\,,
\end{equation}
where $k_\ast$ is a pivot scale that will be taken to be $k_\ast=0.05\,{\rm Mpc}^{-1}$ throughout this document, and $n_{\rm s}$ and $n_{\rm t}$ are referred to as the scalar and tensor spectral indices, respectively. In general $n_{\rm s}$ and $n_{\rm t}$ are functions of $k$. However, the scale dependence is expected to be weak, and the default analyses typically report $n_{\rm s}=n_{\rm s}(k_\star)$. So-called ``scale-invariant'' spectra correspond to $n_{\rm s}=1$ and $n_{\rm t}=0$.

Unfortunately,\footnote{for the prospect of detecting primordial B-modes} $B$ modes are not only generated by primordial gravitational waves, but also by weak lensing of the CMB by matter along the line of sight, which converts $E$-modes into $B$-modes~\cite{zaldarriaga/seljak:1998}, and by polarized Galactic emission from interstellar dust grains and relativistic electrons~\cite{Leach2008,CORE:2017yri,Planck2018IV}. As a consequence, in order to detect the $B$ modes from primordial gravitational waves, both lensing and foreground contributions must be carefully accounted for. As we will discuss in more detail, \lb\ employs 15 frequency bands to characterize and remove the foreground emission. The weak lensing signal has the same frequency dependence as the gravitational wave signal, but its angular dependence is theoretically well-understood. Furthermore, because the weak lensing is caused by large-scale structure along the line of sight, some of the weak lensing signal can be removed by combining \lb\ with other data sets~\cite{smith/etal:2012,simard/hanson/holder:2015,sherwin/marcel:2015,namikawa/nagata:2015,larsen/etal:2016}.

The theoretical predictions and current measurements of the angular power spectra are shown in Fig.~\ref{fig:cl}. The \lb\ error bars, which are used for the constraints presented in the next sections, include foreground residuals as detailed in Sect.~\ref{s:cosmological_forecasts}.

\subsection{Cosmic Inflation}
\label{ss:cmbpol_inflation}

The remarkable insight gained from analysing cosmological data is that all cosmic structures, such as galaxies, stars, planets, and eventually us, appear to have originated from tiny quantum fluctuations in the early Universe. Within this inflationary picture, there was a very early period of nearly exponential expansion that generated the seed fluctuations for today's structure. 

According to general relativity, spacetime expands exponentially if the energy budget is dominated by vacuum energy. However, from our existence, we know that this early period of cosmological inflation must have ended. This requires a clock, or more formally a scalar field, that keeps track of time and eventually causes inflation to end. Within quantum mechanics, this scalar field will experience quantum fluctuations, and according to cosmic inflation these initially microscopic quantum fluctuations were stretched to macroscopic scales by the nearly exponential expansion, serving as the seeds of structure formation \cite{mukhanov/chibisov:1981,starobinsky:1982,hawking:1982,guth/pi:1982,bardeen/turner/steinhardt:1983}. 

In this scenario, the matter sector must include a scalar field, the ``inflaton'', $\phi$. For the simplest models of inflation, the action contains
\begin{equation}
\int d^4x\sqrt{-g}\left[-\frac12g^{\mu\nu}\partial_\mu\phi\partial_\nu\phi-V(\phi)\right]\,,
\end{equation}
and is characterized by the potential $V(\phi)$. As usual, we denote the expansion rate of the universe (called the ``Hubble rate'') by $H=\dot{a}/a$, where $a(t)$ is the scale factor appearing in the \gls{flrw} line element. For a flat \gls{flrw} universe this line element is \mbox{$ds^2=-dt^2+a^2(t)d\mathbf{x}^2$}, and the dynamics of the scale factor is governed by the Friedman equation
\begin{equation}
H^2=\frac{8\pi G}{3}\rho\,,
\end{equation} 
where $\rho={1\over2}\dot{\phi}^2+V(\phi)$ is the energy density in the inflaton.

The slow-roll parameter $\epsilon$, the fractional rate of change of the expansion rate in one Hubble time ($1/H$), is given by~\cite{Lyth:1998xn}
\begin{equation}
\epsilon\equiv -\frac{\dot{H}}{H^2}=\frac{3\dot{\phi}^2}{\dot\phi^2+2V(\phi)}\,.
\end{equation}
If the energy density of the scalar field is dominated by the potential energy density $\dot{\phi}^2\ll V(\phi)$, the slow-roll parameter is small, $\epsilon\ll1$. In this case the scale factor grows nearly exponentially.
To be phenomenologically viable, inflation must last sufficiently long to solve the horizon and flatness problems~\cite{guth:1981}. The simplest way to satisfy this requirement is to have a potential that is flat enough so that the fractional rate of change of the inflaton velocity per Hubble time is small. Such models of inflation, based on a single slowly rolling scalar field, predict statistically homogeneous and isotropic, adiabatic, and nearly Gaussian primordial density perturbations with a spectrum of primordial density perturbations given by~\cite{Lyth:1998xn}
\begin{equation}\label{eq:Dz2}
\Delta^2_\zeta(k)=\frac{1}{2\epsilon M_{\rm P}^2}\left(\frac{H}{2\pi}\right)^2\,,
\end{equation}
where the slow-roll parameter $\epsilon$ and the Hubble  rate $H$ are to be evaluated at a time when $k=aH$, and $M_{\rm P}$ is the (reduced) Planck mass. Since both $\epsilon$ and $H$ are slowly varying functions of time, the spectrum is expected to be nearly (but not exactly) scale invariant $n_{\rm s}\simeq 1$. Furthermore, as inflation proceeds, the Hubble rate decreases. The slow-roll parameter $\epsilon$ is small during inflation, approaches unity as inflation ends, and in the simplest models it increases monotonically. Decreasing $H$ and increasing $\epsilon$ implies that the simplest models predict a red spectrum, i.e. an amplitude of the power spectrum that decreases with increasing wave number, corresponding to $n_{\rm s}<1$. All these predictions, including the deviation from an exactly scale invariant spectrum, have been confirmed by CMB data from \WMAP~\cite{Komatsu2009,komatsu/etal:2011,hinshaw/etal:2013}, the \Planck\ satellite~\cite{Planck2013XXII,Planck2015XX,Planck2018X}, and various ground-based observations \cite{story/etal:2013,sievers/etal:2013,aylor/etal:2017,louis/etal:2017}. 

So far we have discussed the period during which the Universe expands nearly exponentially. From the Cosmos today, we know that eventually this period must have ended, and the energy density in the inflaton must have been converted to a plasma of standard-model particles. This process is referred to as ``reheating''~\cite{Kofman:1994rk,Kofman:1997yn,Lozanov:2019jxc}. The details of reheating are unknown, but rather remarkably, the observational predictions only mildly depend on these details, at least for the single-field models discussed here~\cite{Bardeen:1980kt,Weinberg:2003sw,Weinberg:2004kr}. The main effect on observables arises from the amount by which the Universe expands during reheating. The amount of expansion during this period affects how physical scales today are related to physical scales during inflation, or more quantitatively how long has elapsed 
before the end of inflation $k_*=aH$.

\subsection{Primordial Gravitational Waves from Cosmic Inflation}
\label{ss:cmbpol_pgw}

Constraints on the primordial spectrum of density perturbations from observations of temperature and $E$-mode anisotropies provide strong evidence for the quantum mechanical origin of cosmic structure, and to many they already suggest that the early Universe underwent a period of inflation. However, extraordinary claims require extraordinary evidence. 

Like the scalar field, the spacetime metric also fluctuates, and just like the fluctuations in the scalar field, the microscopic fluctuations in the spacetime metric were also stretched to macroscopic scales by the inflationary expansion. So inflation predicts a statistically homogeneous and isotropic, nearly Gaussian background of primordial gravitational waves~\cite{grishchuk:1975,starobinsky:1979,abbott/wise:1984}. For models based on a single slowly rolling scalar field the power spectrum is given by~\cite{Lyth:1998xn}
\begin{equation}\label{eq:Dh2}
\Delta^2_h(k)=\frac{8}{ M_{\rm P}^2}\left(\frac{H}{2\pi}\right)^2\,,
\end{equation}
where $H$ is again to be evaluated when $k=aH$. Since $H$ is a slowly decreasing function of time, the primordial gravitational wave spectrum is expected to be nearly scale invariant $(n_{\rm t}\simeq 0)$ and red $(n_{\rm t}<0)$. According to equation~(\ref{eq:Dh2}), in the context of inflation a detection of a primordial gravitational wave signal would allow a determination of the expansion rate of the Universe during inflation. In single-field slow-roll models, the expansion rate is directly related to the energy scale of inflation, $V\simeq 3 H^2 M_{\rm P}^2$.

These gravitational waves are a remarkable prediction of inflation, and their detection would provide strong independent evidence for inflation, arguably providing definitive confirmation. A detection of this signal would also be the first observation of quantum fluctuations of spacetime itself, and have other important implications to be discussed below.

\subsection{Implications of LiteBIRD Power Spectrum Measurements for Inflation}
\label{ss:cmbpol_r}

To discuss the implications of \lb's $B$-mode power spectrum measurements for inflation, it is convenient to introduce the ratio of the power in primordial gravitational waves, given in Eq.~(\ref{eq:Dh2}), to the power in primordial density perturbations, defined in Eq.~(\ref{eq:Dz2}), referred to as the ``tensor-to-scalar ratio'',
\begin{equation}
r=\frac{\Delta_h^2(k)}{\Delta_\zeta^2(k)}\,.
\end{equation}
Key quantities like the energy scale of inflation and the range traveled by the scalar field are closely related to this parameter, and different classes of models of inflation make different predictions for $r$.

In single-field slow-roll models, the amplitude of the primordial density perturbations inferred from measurements of temperature and $E$-mode perturbations, together with the Friedmann equation, allows us to express the energy scale of inflation in terms of $r$ through
\begin{equation}
V^{1/4}=1.04 \times 10^{16} {\rm GeV}\left(\frac{r}{0.01}\right)^{1/4}\,.
\end{equation}
Thus, a detection achievable by \lb\ would imply that the inflationary energy scale is close to that associated with grand unified theories, and would provide additional evidence for the idea of grand unification~\cite{Lyth:1996im}.

Under the same assumptions, the tensor-to-scalar ratio not only constrains the energy scale of inflation, but also the distance traveled by the inflaton~\cite{Lyth:1996im},
\begin{equation}
\frac{\Delta\phi}{M_{\rm P}}\gtrsim \left(\frac{r}{8}\right)^{1/2}N_\ast\,,
\end{equation} 
where $N_\ast$ represents the number of ``$e$-folds'', the natural logarithm of the change in linear scale of the Universe, between the time when $k_\ast=a H$ and the end of inflation. As briefly discussed earlier, the exact time when $k_\ast=a H$, and hence the value of $N_\ast$, depends on the details of reheating, the process that converts the energy density in the inflaton into a hot plasma of standard model particles. This process is not well constrained, but taking $N_\ast=30$ as a conservative lower limit, we see that a detection of gravitational waves above $r=0.01$ would imply an excursion in field space that exceeds $M_{\rm P}$. Such a detection would significantly constrain theories of quantum gravity, such as superstring theories (see for example, Ref.~\cite{Obied:2018sgi} and references therein).

In the absence of a detection, \lb\  will set an upper limit of $r < 0.002$ at 95\,\% C.L. (accounting for both statistical and systematic uncertainties). Since both the energy scale and the field range vary slowly with $r$, an upper limit does not immediately translate into stringent constraints on either the energy scale or the distance traveled by the inflaton. To explain the implications of an upper limit and to understand the motivation for the \lb\ design sensitivity we will require an additional concept, that of the characteristic scale of the potential~\cite{Abazajian:2016yjj,Linde:2016hbb}. To introduce this quantity and highlight its importance, we will begin with an argument that does not involve the microscopic details of a particular model of inflation. 

Provided the fractional rate of change of the expansion rate is small compared to the expansion rate, $\epsilon\ll1$, the tensor-to-scalar ratio $r$ obeys a simple differential equation in terms of the number of $e$-folds $N$ until the end of inflation~\cite{Mukhanov:2013tua,Roest:2013fha,Creminelli:2014nqa}:
\begin{equation}
\frac{d\ln r}{dN}=\left[n_{\rm s}(N)-1\right]+\frac{r}{8}\,.
\end{equation}

The cosmic microwave background allows us to observe a window of a few $e$-folds around $N_*$, which we typically expect to be between 50 and 60. 
The observed departure of the primordial power spectrum from scale invariance is numerically close to $(p+1)/N_*$, where $p$ is some number of order unity. In the simplest models of inflation, we expect additional small or large numbers beyond $N_*$ to be absent, which means that we expect $n_{\rm s}(N)-1=-(p+1)/N$. In this case, we can solve the differential equation and find the general solution up to an integration constant $N_{\rm eq}$. If we continue with the assumption that there are no additional large or small numbers, the solution is well-described by one of two limiting behaviors, 
\begin{equation}
r(N)=\frac{8p}{N}\qquad\text{and}\qquad r(N)=\frac{8p}{N}\left(\frac{N_{\rm eq}}{N}\right)^p\,,
\label{eq:r_vs_N}
\end{equation}
where $p$ is constrained to be positive, consistent with the observed red spectrum $n_{\rm s}<1$~\cite{hinshaw/etal:2013,Planck2013XXII}, and by assumption $N_{\rm eq}$ is expected to be of order unity.

We previously saw that the simplest single-field models are completely characterized by a potential. It is then natural to ask which potentials give rise to these solutions. It can be shown that the first solution in Eq.~(\ref{eq:r_vs_N}) corresponds to potentials that at least during inflation are well approximated by a monomial $V(\phi)\simeq\mu^{4-2p}\phi^{2p}$. For $p$ of order unity, we see that this class of models predicts $r\gtrsim 0.01$, which is easily within reach of \lb . 

For the second solution, the qualitative behavior depends on the value of $p$. For $p>1$ the potential corresponds to so-called ``hilltop'' inflation models~\cite{Boubekeur:2005zm} for which the potential near the origin in field space approaches a constant from below like a power of the field set by $p$. Inflation occurs as the field rolls off the hill toward a minimum at larger field values. For $p<1$ the potentials correspond to so-called ``plateau'' models, for which the potential approaches a constant from below at large field values, again with a power set by $p$. In this case inflation occurs as the field rolls off the plateau toward a minimum near the origin. 

Rather intriguingly, the current measurement of $n_{\rm s}$ favors $p\simeq1$, which is a special case. It corresponds to plateau models in which the plateau is approached exponentially,
\begin{equation}
V(\phi)\simeq V_0\left(1-e^{-\phi/{\cal M}}\right)\,.
\end{equation}
This behavior occurs in many models of inflation, including  Starobinsky $R^2$ model for inflation~\cite{starobinsky:1980} (discussed in more detail below), models in which inflation is driven by the Higgs boson~\cite{Bezrukov:2007ep,Ballesteros:2016xej}, or more generally models with a non-minimally coupled inflaton~\cite{Komatsu:1999mt}, fibre inflation~\cite{Cicoli:2008gp}, Poincar\'e disk models~\cite{Ferrara:2016fwe,Kallosh:2017ced}, $\alpha$-attractors more generally~\cite{kallosh/linde/roest:2013,Carrasco:2015pla,Planck2015XX,Kallosh:2014rga}, or the Goncharov-Linde model~\cite{Goncharov:1983mw,Goncharov:1985yu}, just to name a few. 

The ``characteristic scale of the potential,'' ${\cal M}$ 
is related to the integration constant $N_{\rm eq}$ according to ${\cal M}=\sqrt{N_{\rm eq}}M_{\rm P}$. This allows us to express the tensor-to-scalar ratio in this class of models in terms of the characteristic scale as

\begin{equation}
r \simeq 0.0025\left(\frac{57}{N_\ast}\right)^2\left(\frac{{\cal M }}{M_{\rm P}}\right)^2\,.
\end{equation}

Instantaneous reheating corresponds to $N_\ast\simeq 57$.
Any delay in reheating will decrease $N_\ast$, and hence will increase the expected tensor-to-scalar ratio for a given characteristic scale. 
As a consequence, for ${\cal M}\gtrsim M_{\rm P}$ we expect $r\gtrsim 0.0025$, so that an upper limit from \lb\ with $r<0.002$ at 95\,\% C.L. (accounting for both statistical and systematic uncertainties) would disfavor any of the simplest models of inflation with a characteristic scale of the potential larger than the Planck scale.

\begin{figure}[htbp!]
\centering\includegraphics[width=\textwidth]{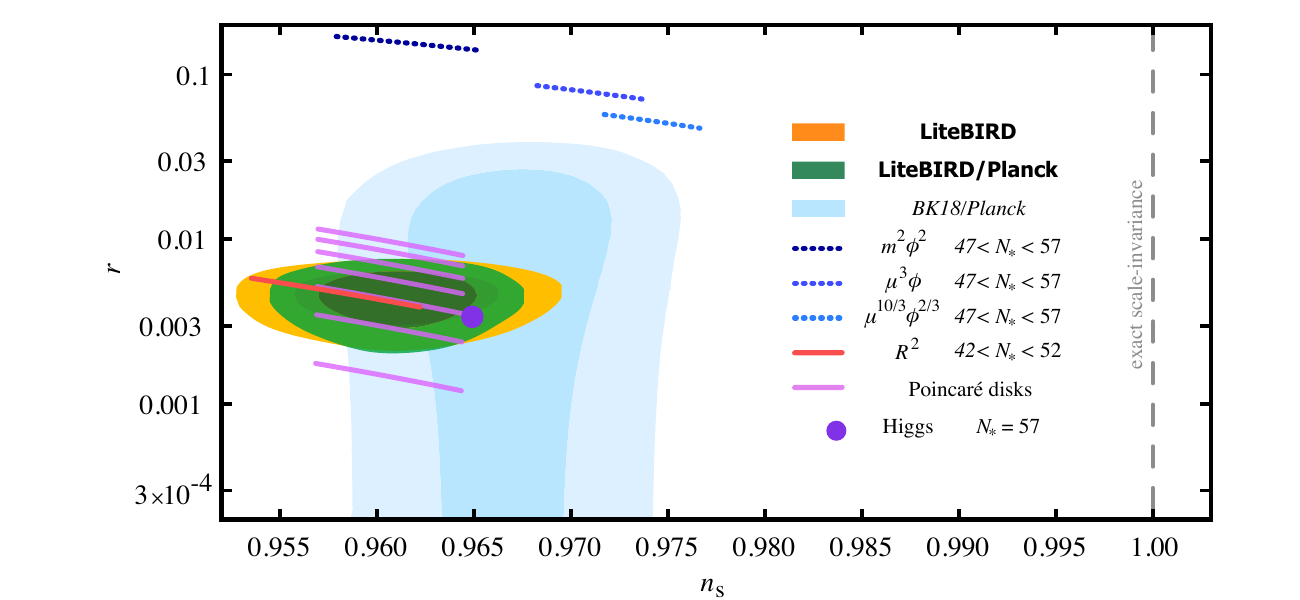}
 \caption{
\lb\ constraints on the tensor-to-scalar ratio $r$ and the scalar spectral index $n_{\rm s}$ assuming Starobinsky's $R^2$ model for inflation~\cite{starobinsky:1980} with $N_\ast=51$ (specifically the analytic prediction described in the text) as the fiducial model. The lighter and darker green regions show 68\,\% and 95\,\% confidence level limits achievable with \lb\ and \Planck. The lighter and darker orange regions (partly hidden behind the green regions) show 68\,\% and 95\,\% confidence level limits achievable with \lb\ alone. The current limits are shown in light blue.  The dotted blue lines show representative cases of the first class of models described in the text, monomial models. The red line and the dark purple dot show the predictions of the Starobinsky model~\cite{starobinsky:1980} (labeled as $R^2$) and models that invoke the Higgs field as the inflaton~\cite{Bezrukov:2007ep,Ballesteros:2016xej}, respectively. The light purple lines shows the prediction for Poincar\'e disk models~\cite{Ferrara:2016fwe,Kallosh:2017ced}.
}
\label{fig:nsr-detection}
\end{figure}

In models such as the Starobinsky model, the Planck scale does not occur by accident, but appears because the characteristic scale and the Planck scale are set by the same dimensionful coefficient in the action, the coefficient of the Einstein-Hilbert term. This makes models with ${\cal M}\gtrsim M_{\rm P}$ a natural target for \lb. In Fig.~\ref{fig:nsr-detection} we take the Starobinsky model as our fiducial model to showcase what a detection of primordial gravitational waves with \lb\ would look like in the $n_{\rm s}$--$r$ plane.

For a given reheating history, a model makes a definitive prediction, corresponding to a point in the $n_{\rm s}$--$r$ plane. However, since the reheating history is uncertain, we represent the predictions of models by bars corresponding to $47<N_\ast<57$. One exception to this general rule is the Starobinsky model. Unlike for most models, the underlying idea of this example is that inflation is a consequence of a short-distance modification of the theory of gravity rather than a consequence of the matter sector. Even though the Starobinsky model can be written as a scalar-tensor theory, like any generic $f(R)$ theory~\cite{Sotiriou:2008rp,DeFelice:2010aj}, this idea naturally predicts that the couplings to matter fields responsible for reheating are gravitational couplings in the $f(R)$ frame. In this case, reheating is somewhat delayed. More detailed studies suggest that the delay corresponds to a change in $N_\ast$ of about 5. Thus, we take $N_\ast=51$ for our fiducial model. In our simple analytic approximation, this leads to $r\simeq0.0046$ and $n_{\rm s}\simeq0.961$. Since reheating is expected to be slower, but the details are uncertain in the Starobinsky model as well, we show the model prediction with $42<N_\ast<52$. The second exception to the rule are models in which the inflaton is identified with the Higgs field. In this case we know the couplings to the matter fields, and the reheating history is calculable. To reflect this, we represent these models by a single point in the $n_{\rm s}$--$r$ plane at $N_\ast=57$, even though some uncertainty exists here as well. 

\begin{figure}[htbp!]
\includegraphics[width=\textwidth]{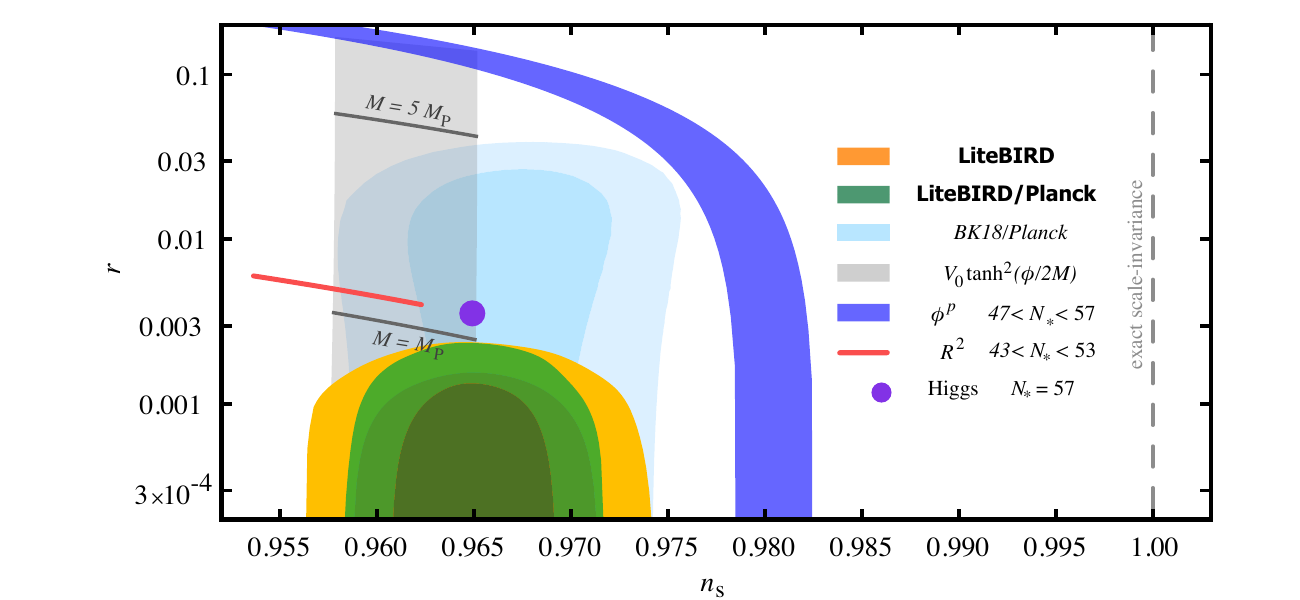}
\caption{
\lb\ constraints for a fiducial model with $r = 0$. The lighter and darker green regions show 68\,\% and 95\,\% confidence level upper limits achievable with \lb\ and \Planck. The lighter and darker orange regions (partly hidden behind the green regions) show 68\,\% and 95\,\% confidence level upper limits achievable with \lb\ alone. The blue band shows the first class of models mentioned in the text, monomial models. The gray band shows a concrete representative second class of plateau models with $p=1$, $\alpha$-attractors~\cite{kallosh/linde/roest:2013}. As discussed in the text, the second class of models depends on the characteristic scale of the potential ${\cal M}$. The darker gray lines show $\alpha$-attractors with ${\cal M}=M_{\rm P}$ and ${\cal M}=5  M_{\rm P}$. In the absence of a detection, \lb\ will exclude the first class of models at high significance, and will exclude models in the second class with a super-Planckian characteristic scale, which includes the Starobinsky model~\cite{starobinsky:1980} and models that invoke the Higgs field as the inflaton~\cite{Bezrukov:2007ep,Ballesteros:2016xej}, shown as the red line and the purple dot, respectively. }
\label{fig:nsr-upper}
\end{figure}

Among the examples given above, the Goncharov-Linde model~\cite{Goncharov:1983mw,Goncharov:1985yu} predicts a sub-Planckian characteristic scale, and $\alpha$-attractors~\cite{kallosh/linde/roest:2013,Carrasco:2015pla,Planck2015XX,Kallosh:2014rga} with a sub-Planckian characteristic scale also exist. So a detection of primordial gravitational waves with \lb\ is by no means guaranteed. We thus also showcase what an upper limit would look like in Fig.~\ref{fig:nsr-upper}. Let us note that in addition to being simpler in the sense that they do not contain a large hierarchy of scales, models with ${\cal M}\gtrsim M_{\rm P}$ are also simpler in a different sense. One may ask whether inflation will begin for general initial conditions for a given model, and it has recently become possible to investigate this question in numerical general relativity, assuming that the description in terms of a single scalar field is already appropriate at that time~\cite{East:2015ggf,Clough:2016ymm,Clough:2017efm,Aurrekoetxea:2019fhr}. The simulations show that models with ${\cal M}\gtrsim M_{\rm P}$ are significantly more robust to inhomogeneities and than those with ${\cal M}<M_{\rm P}$~\cite{Aurrekoetxea:2019fhr}. This does not imply that inflation cannot begin in models with a sub-Planckian characteristic scale, but it does suggest that additional dynamics (which could simply be in the form of another field) is needed to set up initial conditions that are appropriate for inflation to begin in such models. So an upper limit from \lb\ would disfavor the simplest models of inflation that naturally predict the observed value of $n_{\rm s}$ and would also be a milestone for early Universe cosmology that provides key information about the inner workings of the earliest moments of the Cosmos.

\subsection{Beyond the \textit{B}-mode Power Spectrum}
\label{ss:cmbpol_beyondpower}

Single-field slow-roll inflation predicts a stochastic background of gravitational waves that originated from quantum vacuum fluctuations in spacetime and is nearly scale invariant, nearly Gaussian, and parity conserving~\cite{grishchuk:1975,starobinsky:1979,abbott/wise:1984}. The detection of a violation of any of these properties would point to new physics beyond the simplest models of inflation~\cite{Komatsu:2022nvu}. The first condition can be tested by reconstructing the power spectrum $\Delta_h^2(k)$ from the observed $B$-mode power spectrum \cite{hiramatsu/etal:2018,campeti/poletti/baccigalupi:2019}, the second property can be tested through measurements of the three-point function (bispectrum)~\cite{shiraishi/etal:2016,agrawal/fujita/komatsu:2018,agrawal/fujita/komatsu:2018b,dimastrogiovanni/etal:2018}, and the third property can be tested by parity-violating correlation functions such as the cross-correlation between the temperature and the $B$-mode polarization, between the $E$- and $B$-mode polarizations \cite{lue/wang/kamionkowski:1999,saito/ichiki/taruya:2007,contaldi/magueijo/smolin:2008,sorbo:2011,thorne/etal:2018}, or parity-violating contributions to the  three-point function~\cite{shiraishi:2019}.

These conditions can be violated when non-minimal couplings of the inflaton are present~\cite{lue/wang/kamionkowski:1999,Bartolo:2017szm,Bartolo:2018elp}, or when other fields are present during inflation and source gravitational waves. The energy density in these fields must be sub-dominant compared to the energy density in the inflaton. However, their energy density may still be sufficient to produce gravitational waves with an amplitude within reach of \lb.

The additional sources could be scalar fields \cite{cook/sorbo:2012,carney/etal:2012,biagetti/fasiello/riotto:2013,senatore/silverstein/zaldarriaga:2014}, a U(1) gauge field \cite{sorbo:2011,anber/sorbo:2012,barnaby/peloso:2011,barnaby/etal:2012,peloso/sorbo/unal:2016,Campeti:2022acx}, or an  SU(2) gauge field \cite{maleknejad/sheikh-jabbari:2011,maleknejad/sheikh-jabbari:2013,dimastrogiovanni/peloso:2012,adshead/etal:2013,adshead/martinec/wyman:2013,maleknejad:2016,obata/soda:2016,dimastrogiovanni/fasiello/fujita:2016}. All these sources can produce strongly scale-dependent gravitational waves (and, in general, density perturbations) that are highly non-Gaussian. The latter two types of source can produce parity-violating gravitational waves. Hence a stochastic gravitational wave background generated during inflation need not satisfy any of the conditions predicted by single-field slow-roll inflation.  If the gravitational waves sourced by the matter fields dominate over the vacuum fluctuations in the metric, then detecting $B$-mode polarization from primordial gravitational waves no longer generally implies the discovery of the quantum nature of space (although the matter that created them was still produced quantum-mechanically). Nevertheless, such a discovery would still provide definitive evidence for inflation because we need inflation to stretch the wavelengths of gravitational waves to billions of light years.

An example of a U(1) gauge field is the primordial magnetic field, which can source tensor perturbations that are non-scale-invariant, non-Gaussian, and parity-violating (see Ref.~\cite{shiraishi:2019} and references therein). Magnetic fields can also induce a spatially-dependent rotation of polarization angles of the CMB by means of Faraday rotation. The effect can be detected using multi-frequency data because the Faraday rotation angle is inversely proportional to the square of the frequency. We discuss this possibility further in Sect.~\ref{ss:pmf}.

The axion-SU(2) model described in  Ref.~\cite{dimastrogiovanni/fasiello/fujita:2016} provides an example that illustrates the benefits of a satellite mission with full sky coverage and access to the reionization bump. This model contains the inflaton, an axion, and SU(2) gauge fields. The energy density is always dominated by the inflaton field.  The axion field $\chi$ has a potential of the form $V(\chi)\propto 1+\cos(\chi/f)$, where $f$ is the axion decay constant, and the axion is coupled to the gauge fields through a Chern-Simons term, $\chi F\tilde{F}$. For time-dependent $\chi$, one of the helicities of the gauge field is amplified. 
This produces chiral gravitational waves~\cite{dimastrogiovanni/peloso:2012,adshead/martinec/wyman:2013,maleknejad/sheikh-jabbari/soda:2013} with parity-violating correlations in the CMB power spectra and circular polarization for laser interferometers \cite{thorne/etal:2018}. The shape of the tensor power spectrum is determined by the evolution of $\chi$ during inflation, and hence by the shape of $V(\chi)$. For the cosine potential, the axion velocity increases initially, reaches the maximum at the inflection point, $\chi(t_*)=\pi f/2$, and then decreases. The resulting tensor power spectrum is approximately log-normal \cite{thorne/etal:2018,fujita/sfakianakis/shiraishi:2019},
\begin{equation}
\label{eq:su2power}
\Delta^2_{h\,{\rm L,\,sourced}}(k)
= r_{*,{\rm sourced}}\Delta_\zeta^2(k)\exp\left(-\frac1{2\sigma^2}\left[\ln\left(\frac{k}{k_p}\right)\right]^2\right)\,,
\end{equation}
where $r_{*,{\rm sourced}}$ (hereafter $r_*$) and $k_p$ are the tensor-to-scalar ratio and the wavenumber at the maximum, and $\sigma^2$ is the width of the power spectrum. The subscript ``L'' (for ``left'') stands for one of the polarization states (determined by the sign of $\dot\chi$), while the other polarization state is not amplified and is negligible. These quantities are determined by the parameters of the model~\cite{thorne/etal:2018,fujita/sfakianakis/shiraishi:2019,Campeti:2020xwn}. This sourced contribution is added to the vacuum contribution characterized by the vacuum tensor-to-scalar ratio, $r_{\rm vac}$. The self-interaction of the gauge fields leads to non-Gaussian gravitational waves~\cite{agrawal/fujita/komatsu:2018,agrawal/fujita/komatsu:2018b,dimastrogiovanni/etal:2018}.

\begin{figure}[htbp!]
\centering
\includegraphics[width = 0.75\textwidth]{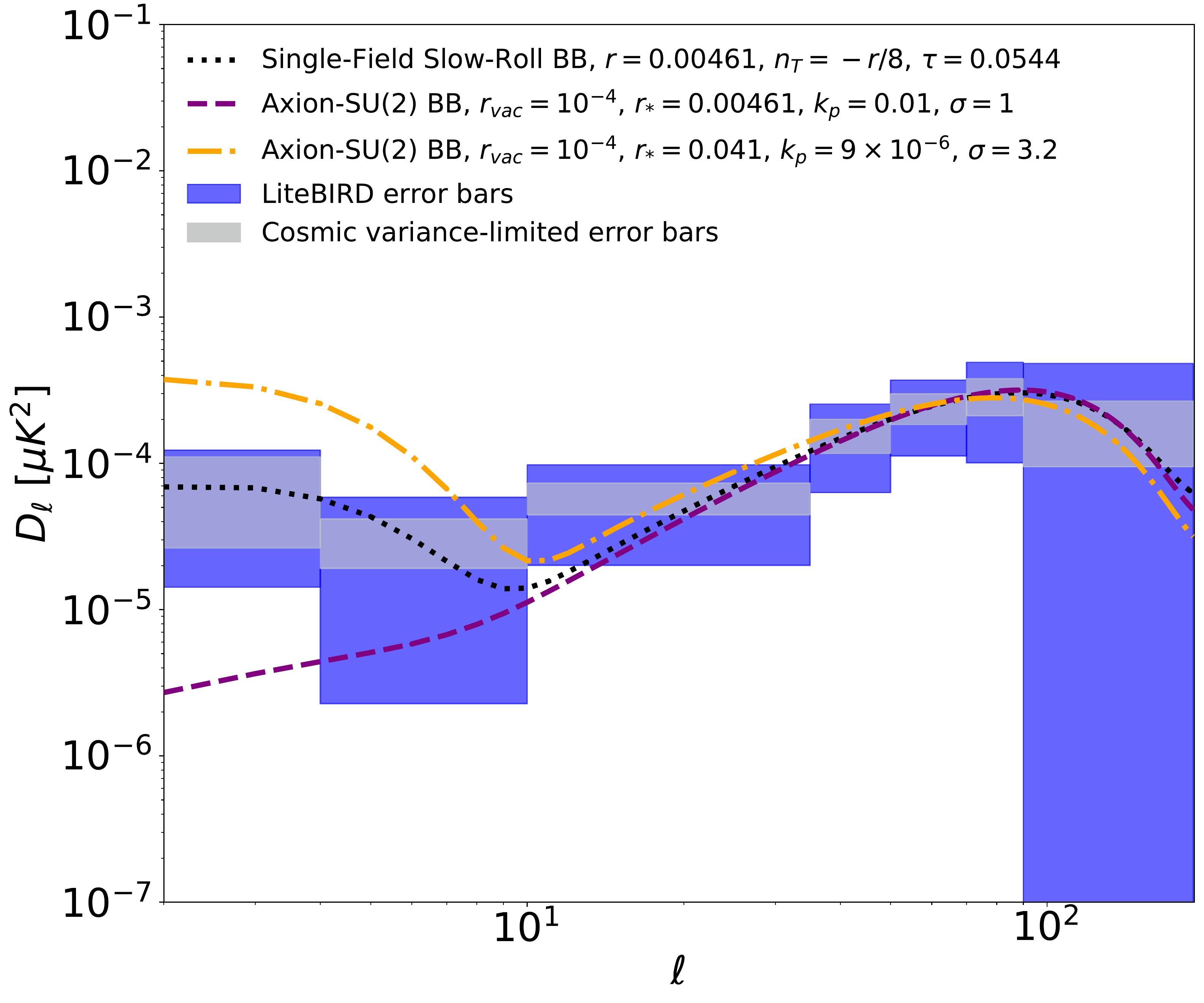}
\caption{$B$-mode power spectra, $D_\ell=\ell(\ell+1)C^{BB}_{\ell}/2\pi$, for the Starobinsky model with $r=0.00461$ and $n_{\rm t}=-r/8$ (black dotted line) and for axion-SU(2) inflation with two sets of parameters (see Eq.~(\ref{eq:su2power}) for the definition): one with $r_{*}=0.00461$, $k_{p}=0.01\,{\rm Mpc}^{-1}$ and $\sigma=1$ (purple dashed line) and another with $r_\ast=0.041$, $k_{p}=9 \times 10^{-6}~{\rm Mpc}^{-1}$ and $\sigma=3.2$ (orange dash-dotted line). The tensor-to-scalar ratio of the vacuum fluctuations is chosen to be $r_{\rm vac}=10^{-4}$.
The cosmic-variance-only (including primordial and lensing $B$-mode variance) and total \lb\ $\pm1\,\sigma$ error bars (including foreground residuals) are shown as the gray and blue regions, respectively.}
\label{fig:su2_axion}
\end{figure}

In Fig.~\ref{fig:su2_axion}, we compare example $B$-mode power spectra of this model (dot-dashed and dashed lines) and that of the Starobinsky model (dotted line). The parameters are chosen such that they all have indistinguishable recombination bumps ($\ell\simeq 80$), whereas they are very different in their reionization bumps ($\ell\simeq 4$). Therefore, a full-sky survey enabled by a space mission such as \lb\ is necessary for establishing the origin of the primordial gravitational waves: tensor vacuum fluctuation versus sourced gravitational waves.

We conclude that, in the case of a detection, it will be important to confirm the detailed predictions of single-field slow-roll models using the \lb\ data before claiming discovery of the quantum nature of spacetime. Applying the established methodology for the CMB bispectrum estimation to the \lb\ $B$-mode data will allow us to improve constraints on tensor non-Gaussianities by several orders of magnitude \cite{shiraishi:2019}. A violation of any of the basic predictions of single-field slow-roll models of inflation for gravitational waves (i.e., nearly scale invariant, nearly Gaussian, parity conserving) would have profound implications for our understanding of the dynamics of inflation.

\subsection{The Need for Measurements from Space}
\label{ss:cmbpol_space}

The \textit{COBE}, \WMAP, and \planck\ data sets are recognized as the reference experiments for their respective \gls{cmb} science goals.   The success of these experiments depended on the advantages of the space environment for making high-fidelity observations of the CMB. These advantages include the following.
\begin{itemize}

\item{All frequencies are accessible, unlike on the ground where water
and oxygen lines block access and reduce the ability to build a detailed model of the foreground emission.  In particular, space observations can measure frequencies far into the  Wien tail of the CMB to distinguish 
CMB fluctuations from Galactic dust emission.}

\item{Detector sensitivity is higher in space than on the ground 
due to the absence of the atmospheric loading
and the disparity increases rapidly with frequency, giving much better per-detector leverage on Galactic 
dust measurements from space. As a rule of thumb, one detector in space is equivalent to 100 detectors (of the same quality) on the ground.}

\item{The absence of atmospheric emission and its large brightness fluctuations in space-based measurements give high-fidelity maps on large angular scales corresponding to $2 \le \ell \lesssim 30$.}

\item{Bright sources such as the Earth and Sun are kept far from the  boresight of the telescope by a large angle, giving very low systematic errors due to pickup
of those sources in the telescope sidelobes.}
\end{itemize}

 \lb's ability to measure the entire sky at the largest angular scales with 15 frequency bands is complementary to that of ground-based experiments, which will focus on deep observations of low-foreground sky to search for an inflationary signal.  \lb\ observations have the potential to detect {\it both\/} the recombination peak at $\ell\,{\simeq}\,80$
{\it and\/} the reionization peak at $\ell\,{\simeq}\,4$.  As highlighted earlier, high significance detections of both peaks would provide firm evidence that we have detected the signature of inflation.  A primary science requirement for \lb\ is to detect both peaks with greater than $5\,\sigma$ significance for a relatively high value of $r\,{=}\,0.01$.  Detection of both peaks is necessary to distinguish between cosmological models with a similar recombination peak, as discussed in the previous section. For all detectable values of $r$, the all-sky data from \lb\ can be tested for isotropy, which is a critical feature of a true cosmological $B$-mode signal. 

Finally, \lb\ can provide valuable foreground information for ground-based experiments. Ground-based experiments can improve \lb's observations with high-resolution lensing data.  The \lb\ data set will be timely, since it will be available at the same time as ground-based CMB data from Chile and the South Pole, as well as other powerful cosmological data sets such as those from the Vera C. Rubin Observatory, \textit{Euclid}, and the \textit{Nancy Grace Roman Space Telescope}.

\subsection{Comparison with Other Probes}
\label{ss:cmbpol_other_probes}

Several methods have been proposed for observing primordial gravitational waves other than through CMB polarization measurements.
These include future projects for a gravitational wave interferometer~\cite{Moore:2014lga}, a technique using pulsar timing arrays~\cite{NANOGRAV:2018hou,NANOGrav:2020bcs}, and a method using the 21-cm line~\cite{Bharadwaj2009}.
The sensitivity of the CMB to primordial gravitational waves is much better than those of the other probes, assuming the spectrum expected from standard cosmology; for example, the sensitivity of \LiteBIRD\ is about 10 million times greater than that of {\it LISA}~\cite{Campeti:2020xwn}.
Therefore, the discovery of primordial gravitational waves seems dramatically more likely to come from CMB observations in this case. 
Once the primordial gravitational waves are discovered with CMB polarization observations, it will give us a concrete target for future projects using other methods. 
In some non-standard models, primordial gravitational waves can be enhanced at shorter wavelengths. 
Their observability by various probes including that of \LiteBIRD\ is discussed elsewhere~\cite{Campeti:2020xwn} with the conclusion that \lb\ is competitive even in those non-standard models.



\section{LiteBIRD Overview} 
\label{ss:litebird_overview}

\subsection{Project Overview}
\label{ss:overview_project}

After some initial conceptual studies~\cite{Hazumi2008AIP,Hazumi2011PTPS,Hazumi2012SPIE,Matsumura2014JLTP,Matsumura2014SPIE} that started in 2008,
we proposed \LiteBIRD\ in 2015 as JAXA's large-class (L-class) mission candidate.
JAXA's L-class is for flagship science missions with a 30\,Billion Yen cost cap.
There will be three L-class missions in about ten years, launched using JAXA's H3 rocket.
\LiteBIRD\ passed an initial down-selection and completed a two-year Pre-Phase-A2 concept development phase in 2018.
JAXA selected \LiteBIRD\ in May 2019 as the second L-class mission after
\textit{MMX}, the Martian Moons Exploration, which will be launched in the mid-2020s.

The \LiteBIRD\ Collaboration has more than 300 researchers as of January 2022, based in Japan, North America, and Europe, with experience in CMB experiments, X-ray satellite missions, and other large projects in high-energy physics and astronomy.
In particular, a large number of researchers who worked on the \Planck\ satellite are members of \LiteBIRD.
We thus consider \LiteBIRD\ to be the successor to the \Planck\ satellite.

\LiteBIRD\ will survey the polarization of the CMB radiation over the full sky with unprecedented precision. 
The full success criterion of \LiteBIRD\ is to achieve $\delta r < 0.001$ for a fiducial model with $r=0$, where $\delta r$ is the total error on the tensor-to-scalar ratio.

Specifically, we define this as the value covering the 68\,\%
area of the posterior probability function for $r$:
\begin{equation}
\frac{\int_0^{\delta r}L(r)dr}{\int_0^{\infty} L(r)dr} = 0.68\,.
\end{equation}
The posterior, $L(r)$, including both statistical and systematic components, will be described in Sect.~\ref{s:cosmological_forecasts}.

This section gives a concise overview of \LiteBIRD.
In Sect.~\ref{ss:overview_mission}, we describe our Level-1 mission requirements, or scientific requirements, and the rationale behind them.
In Sect.~\ref{ss:overview_system}, we introduce our measurement requirements and their flow down to system requirements.
After describing the launch vehicle (Sect.~\ref{ss:overview_launch_vehicle}), we introduce the spacecraft and the payload module (Sect.~\ref{ss:overview_spacecraft}), the service module (Sect.~\ref{ss:overview_svm}), and the operation concept (Sect.~\ref{ss:overview_operation}.)

\subsection{Science Requirements}
\label{ss:overview_mission}

In Fig.~\ref{fig:cl} (Sect.~\ref{s:cmb_polarization} above), we summarized the present measurements of the CMB power spectra, including $B$ modes, with
the expected polarization sensitivities of \lb\ displayed.
The $B$-mode power is proportional to the tensor-to-scalar ratio, $r$, which has been observationally constrained by BICEP/Keck
to be $r < 0.036$ (95\,\% C.L.)~\cite{BK2021},
with a recent update folding in a re-analysis of \Planck\ data yielding $r < 0.032$ (95\,\% C.L.)~\cite{Tristram:2020wbi,Tristram:2021}
and $r < 0.037$ (95\,\% C.L.)~\cite{Campeti:2022vom}.
The next generation of CMB polarization experiments on the ground have the potential to observe the signal around $\ell \simeq 80$,
coming from the recombination epoch.
However, if $r$ is less than approximately 0.03, the $B$ modes due to gravitational lensing become dominant.
Removing contamination of the lensing $B$ modes, often called ``delensing,'' is needed in this case.
In contrast, the other primordial signal, at $\ell < 10$, which is due to reionization, is larger than
the lensing $B$ modes, even for $r=0.001$.
In order to access the reionization peak, one needs to survey the full sky, where
the advantage of observing in space is clear.

The critical question is: to what precision should $r$ be measured?
Here we introduce the total uncertainty on $r$, $\delta r$, 
which consists of five components: (instrumental) statistical uncertainties; systematic uncertainties; uncertainties due to contamination of foreground components; uncertainties due to gravitational lensing; and uncertainties due to observer biases.
There are many different inflationary models under active discussion, which predict different values of $r$.
Among them, there are well-motivated inflationary models that predict $r > 0.01$~\cite{kamionkowski/kovetz:2016}.
If our requirement is $\delta r < 0.001$, we can provide more than $10\,\sigma$ detection significance for such models. On the other hand,
if \LiteBIRD\ finds no primordial $B$ modes and obtains an upper limit on $r$, then
this limit will be stringent enough to set severe constraints on the physics of inflation.
As discussed in Sect.~\ref{ss:cmbpol_r},
if we obtain an upper limit at $r\sim 0.002$, we can completely rule out
one important category of models, namely any single-field model in which
the characteristic field-variation scale of the inflaton potential is greater than the reduced Planck mass.

\begin{table}[htbp!]
\centering
\caption{The two basic science requirements for \LiteBIRD, also called Level~1 (Lv1) mission requirements.}
\label{tbl:lv1}
\begin{tabular}{|lp{9.5em}|p{24em}|} 
\hline
ID	& Title & Requirement description\\
\hline\hline
Lv1.01 & Tensor-to-scalar ratio $r$ measurement sensitivity & 
The mission shall measure $r$ with a total uncertainty of $\delta r\,{<}\,1 \times 10^{-3}$.
This value shall include contributions from instrumental statistical noise fluctuations, instrumental systematics, 
residual foregrounds, lensing $B$ modes, and observer bias, and shall not rely on future external data sets.\\
\hline 
Lv1.02 & Polarization angular power spectrum measurement capability &
The mission shall obtain full-sky CMB linear polarization maps for achieving ${>}\,5\,\sigma$ significance 
using $2\,{\leq}\,\ell\,{\leq}\,10$ and $11\,{\leq}\,\ell\,{\leq}\,200$ separately, assuming $r\,{=}\,0.01$. 
We adopt a fiducial optical depth of $\tau=0.05$ for this calculation.\\
\hline
\end{tabular}
\end{table}
Based on all the considerations described above,
we decided to impose the requirements described in Table~\ref{tbl:lv1}.
The first, Lv1.01, shall be achieved {\it without\/} delensing using external data;
if external data are available, we may further reduce $\delta r$~\cite{Diego-Palazuelos2020}.
The second requirement, Lv1.02, is essential to cover the case where $r$ turns out to be large.
If there are already indications of the primordial $B$ modes before the observations by \LiteBIRD, 
that would imply a relatively large value of $r$. In this case,
data from \LiteBIRD\ will allow us to measure the $B$-mode signals from reionization and recombination
simultaneously.
If the spectral shape is consistent with expectations
from the standard cosmology, that will narrow down the list of possible inflationary scenarios, and
provide a much deeper insight into the correct model. 
If we observe an unexpected power spectrum, beyond the standard model prediction,
that will lead to a revolution in our picture of the physics of the early Universe.      
Lv1.02 also sets the angular resolution requirement for \LiteBIRD.

\subsection{Measurement Requirements and System Requirements}
\label{ss:overview_system}

To satisfy the science requirements described in the previous section, we use the requirements flow-down framework shown in Table~\ref{tab:reqflow}. 
To derive Lv2 measurement requirements from Lv1 science requirements,
we also consider program-level constraints, such as the cost cap, which are not controlled by the \LiteBIRD\ team.
We use agreed-upon assumptions between the \LiteBIRD\ team and other parties or within the \LiteBIRD\ team; examples include assumptions on the complexity of the astronomical foreground components, the cooling-chain lifetime, and basic system redundancy guidelines.
There are in total 11 Lv2 measurement requirements on the statistical uncertainty (Lv2.01), the systematic uncertainty (Lv2.02), the scan strategy (Lv2.03), the angular resolution (Lv2.04), calibration measurements (Lv2.05), error budget allocation (Lv2.06), systematic error budget allocation (Lv2.07), the duration of the normal observation phase (Lv2.08), the orbit (Lv2.09), observer bias (Lv2.10), and noise-covariance knowledge (Lv2.11).
Our error budget (Lv2.06) is defined such that an equal amount is given to the total statistical error after foreground separation $\sigma_{\rm stat}$, and the total systematic error $\sigma_{\rm syst}$. 
The requirements we chose are thus $\sigma_{\rm stat} < 0.6\times 10^{-3}$ on the statistical uncertainty (Lv2.01) and $\sigma_{\rm syst} < 0.6\times 10^{-3}$ on the systematic uncertainty (Lv2.02)\footnote{The requirement allows us to keep a sufficient margin to absorb additional noise penalty due to debiasing as described in Sect.~\ref{ss:forecasts_delta_r}.}.
Since we assume no delensing using external data, $\sigma_{\rm stat}$ includes uncertainties from the lensing $B$-mode component. 
Uncertainties due to foreground separation are also in $\sigma_{\rm stat}$.
The observer bias (Lv2.10) shall be much smaller than $\sigma_{\rm syst}$.
The requirement on the statistical uncertainty (Lv2.01) has six sub-requirements on: (1) CMB sensitivity; (2) dust emission; (3) synchrotron emission; 
(4) separation of CO lines; (5) the number of observing bands; and (6) the observing frequency range.
These are determined through detailed simulation.
We require full-sky surveys (Lv2.03) to obtain the $B$ modes to the lowest multipole of $\ell = 2$.
The angular resolution (Lv2.04) shall be better than 80\,arcmin \gls{fwhm} in the lowest frequency band
in order to perform precision measurements at $\ell\,{=}\,200$.
The regular observation phase (Lv2.08) shall be three years, considering the total cost cap and cooling-chain lifetime.
The lifetime is determined by the degradation of working gas and moving parts of the mechanical coolers. The degradation is suppressed through our technology development to assure the required lifetime.
The satellite shall be in a Lissajous orbit (Lv2.09) around the Sun-Earth L2 point to avoid the influence of radiation from the Sun, Moon, and Earth (discussed further in Sect.~\ref{ss:overview_operation}).
Requirements on calibration measurements (Lv2.05, Lv2.11) and systematic error budget allocation (Lv2.07) will be explained in Sect.~\ref{s:payload}.

Lv1 and Lv2 requirements are collectively called ``mission requirements.''
In general, several possible designs meet mission requirements, and so
we performed implementation trade-off studies to choose the best design.
We also considered program-level constraints and assumptions that we used to set Lv2 requirements.

Lv3 integrated system requirements constitute top-level system requirements.
An essential distinction between Lv2 and Lv3 is that Lv3 requirements are for the system chosen from trade-off studies, while Lv2 measurement requirements do not assume a specific system in principle. 
Lv3 requirements include general system requirements not only for mission instruments but also for the bus system,\footnote{Also called the ``service module,'' or ``SVM'' for short.} ground segments, and ground-support equipment.
There are too many Lv3 requirements to list here.
The requirement flow's tree structure is also too detailed to show, since some Lv3 requirements derive from more than one Lv2 requirement; however, we will explain some essential Lv3 requirements in Sect.~\ref{s:payload}.
%
\begin{table}[htbp!]
\centering
\caption{Definitions of five requirement levels used in \LiteBIRD's requirements flow-down.
We split the requirements into five levels, from the top-level science requirements (Lv1) to (instrumental) unit requirements (Lv5).
Each level is allowed to have a sub-structure; for example, a Level~2 requirement Lv2.01 has six sub-requirements (Lv2.01.01, Lv2.01.02, \dots).}
\label{tab:reqflow}
	\begin{tabular}{|p{0.15\textwidth}|p{0.18\textwidth}|p{0.57\textwidth}|}
	\hline
	   \multicolumn{1}{|c|}{Class}
	& \multicolumn{1}{c|}{Symbol}
	& \multicolumn{1}{c|}{Description}\\ \hline\hline
	\multicolumn{3}{|c|}{Mission requirements}\\
	\hline
	Level~1 (Lv1)     & Lv1.XX                 & Top-level quantitative science requirements that are\\ 
	science           & (e.g., Lv1.01)        & directly connected to the full success of the mission.\\
	requirements   &                            & \\ 
	\hline
	Level~2 (Lv2)     & Lv2.XX(.YY) & Measurement requirements to achieve Lv1.\\
	measurement  & (e.g., Lv2.01, & No assumption is made on an instrument.\\
	requirements   &  Lv2.01.01)  & \\ 	
	\hline
	%
	\multicolumn{3}{c}{$\downdownarrows$}\\
	\multicolumn{3}{c}{Implementation trade-off studies}\\
	\multicolumn{3}{c}{$\downdownarrows$}\\
	\hline
	\multicolumn{3}{|c|}{System requirements}\\
	\hline
	Level~3 (Lv3) & Lv3.XX(.YY)    & Top-level implementation requirements for a chosen\\
	integrated    & (e.g., Lv3.01, & instrument to achieve Lv2. Between Lv2 and Lv3 are\\
	system        & Lv3.01.01)     & tradeoff studies for instrument selection.\\
	requirements  & & \\
	\hline   
	Level~4 (Lv4) & Lv4.XX(.YY)    & Instrument requirements to achieve Lv3.\\
	instrument    & (e.g., Lv4.01, & \\
	requirements  &   Lv4.01.01)   & \\
	\hline   
	Level~5 (Lv5) & Lv5.XX(.YY)    & Requirements on units composing each instrument\\
	Unit          & (e.g., Lv5.01, & to achieve Lv4.\\
	requirements  & Lv5.01.01)     & \\
	\hline   
  	\end{tabular}
\end{table}

\subsection{Launch Vehicle}
\label{ss:overview_launch_vehicle}

\LiteBIRD\ will be launched on an H3~\cite{JAXA_H3}, Japan's new flagship rocket.
It will achieve greater flexibility, reliability, and performance at a lower cost than the currently used H-IIA rocket.
The H3 rocket is under development through its prime contractor, Mitsubishi Heavy Industries, with a maiden flight scheduled in 2022.
The first stage of the H3 rocket will adopt the newly-developed liquid engine, LE-9, which achieves a thrust 1.4 times larger than the LE-7A engine currently in use.
Its second-stage engine, LE-5B-3, and the solid rocket booster, SRB-3, will also be improved.
The launch capability of the H3 rocket to a geostationary transfer orbit will be the highest ever among JAXA's launch vehicles,
exceeding that of the existing H-IIA and H-IIB launch vehicles.
The launch facility at Tanegashima Space Center will also be upgraded following the development of H3.

The design of the H3 rocket allows for several different configurations.
The rocket type is defined by the combination of the number of first-stage engines (2 or 3), the number of
solid rocket boosters (0, 2, or 4), and the length of the fairing (short or long).
These setups make it possible to cope with various payload sizes and orbits.
Considering the size, weight, and orbit of \LiteBIRD, we plan to adopt the H3-22L configuration,
which means two first-stage engines, two boosters, and the long fairing.
The estimated launch capability
with this configuration is larger than 3.5\,t. We thus set a provisional requirement on the total weight of \LiteBIRD\ as ${<}\,3.5$\,t. This requirement may be updated after the first flight of the H3 rocket.

In most cases, the launch environment of H3 is expected to be similar or more moderate than that of H-IIA.
Details of the launch environment may depend on the rocket configuration, especially on the number of solid rocket boosters,
the satellite mass, and the flight path.
We conservatively assume the launch environment of H-IIA in general for the design of \LiteBIRD.
However, when the launch environment is critical in the design, such as for the mechanical requirement on the fundamental frequency
of the satellite, we adopt requirements based on the current best estimation of the performance of H3.

\subsection{Spacecraft Overview}
\label{ss:overview_spacecraft}

\begin{figure}[htb!]
\centering
\includegraphics[width=1\textwidth]{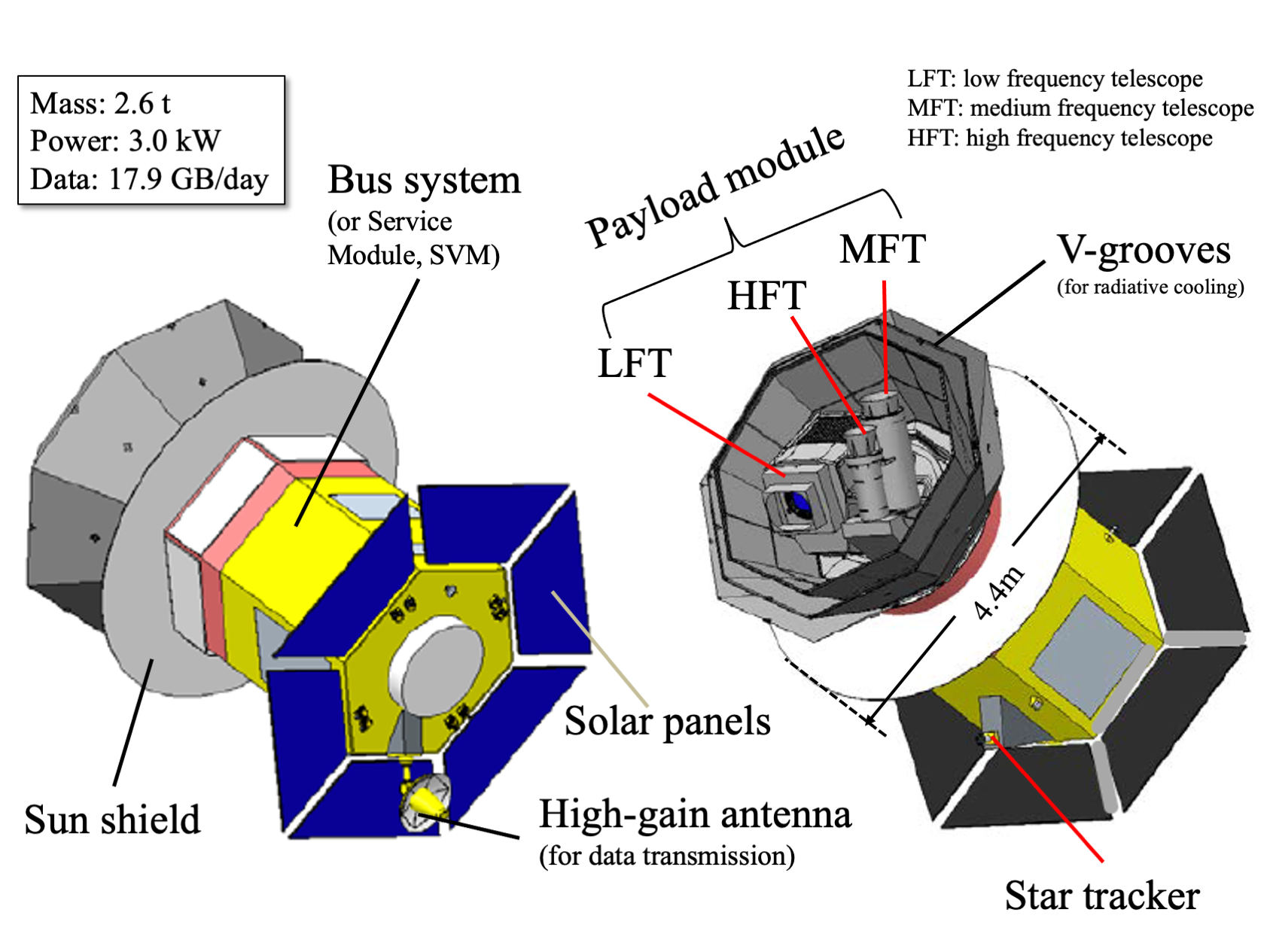}
\caption{Conceptual design of the \LiteBIRD\ spacecraft. The \glsentryfull{plm} houses the \gls{lft},
the \gls{mft}, and the \gls{hft}.}
\label{fig:satellite_overview} 
\end{figure}

\noindent
The overall structure of the spacecraft for \LiteBIRD\ is determined directly from the mission requirements.
The axisymmetric shape of the spacecraft is selected for reducing the moment of inertia to make the spin easier.
We chose to place the \gls{plm}, including the telescopes, at the top of the spacecraft and the
solar panels at its bottom, perpendicular to the spin axis. 
The high-gain antenna should be placed on the bottom side of the satellite, i.e., opposite the mission instruments, to point to the Earth and reduce interference with the telescopes. 
Based on these considerations, we show the basic structure of the spacecraft in Fig.~\ref{fig:satellite_overview}.

In this configuration, the whole spacecraft spins, and the possibility of using a slip-ring to rotate only the \gls{plm} is {\it not\/} adopted.
The main reasons for this selection are to handle large heat dissipation in the PLM and to reduce the possibility of a single-point failure.
The PLM is equipped with mechanical coolers, which dissipate a fairly large amount of heat.
A radiator of sufficient size to dissipate the heat can be equipped only in the \gls{svm} and 
it is not easy to transfer heat from the spinning PLM through the slip-ring to a non-spinning \gls{svm}.
The slip-ring introduces a single point whose failure would be critical for the mission.
Furthermore, a slip-ring might produce micro-vibration and could increase the detector noise
significantly.
For these reasons, we decided to rotate the whole spacecraft and
not to adopt the slip-ring.

The spacecraft has a thrust tube at its center, which transfers the \gls{plm} launch load to the rocket. 
We will install the fuel tank inside the thrust tube to utilize the inner space effectively.
The insides of the side panels are used to mount various electric components of both the \gls{svm} and \gls{plm}\@.
PLM components are preferentially placed on the upper parts of the side panels, whereas SVM components are on the lower parts of the side panels.
The outer sides of the upper parts of the side panels are used to mount radiators, which radiate the heat dissipated
in the \gls{plm}, such as from the mechanical coolers and electronics boxes.

\begin{figure}[htb!]
\begin{center}
\includegraphics[width=1.0\linewidth]{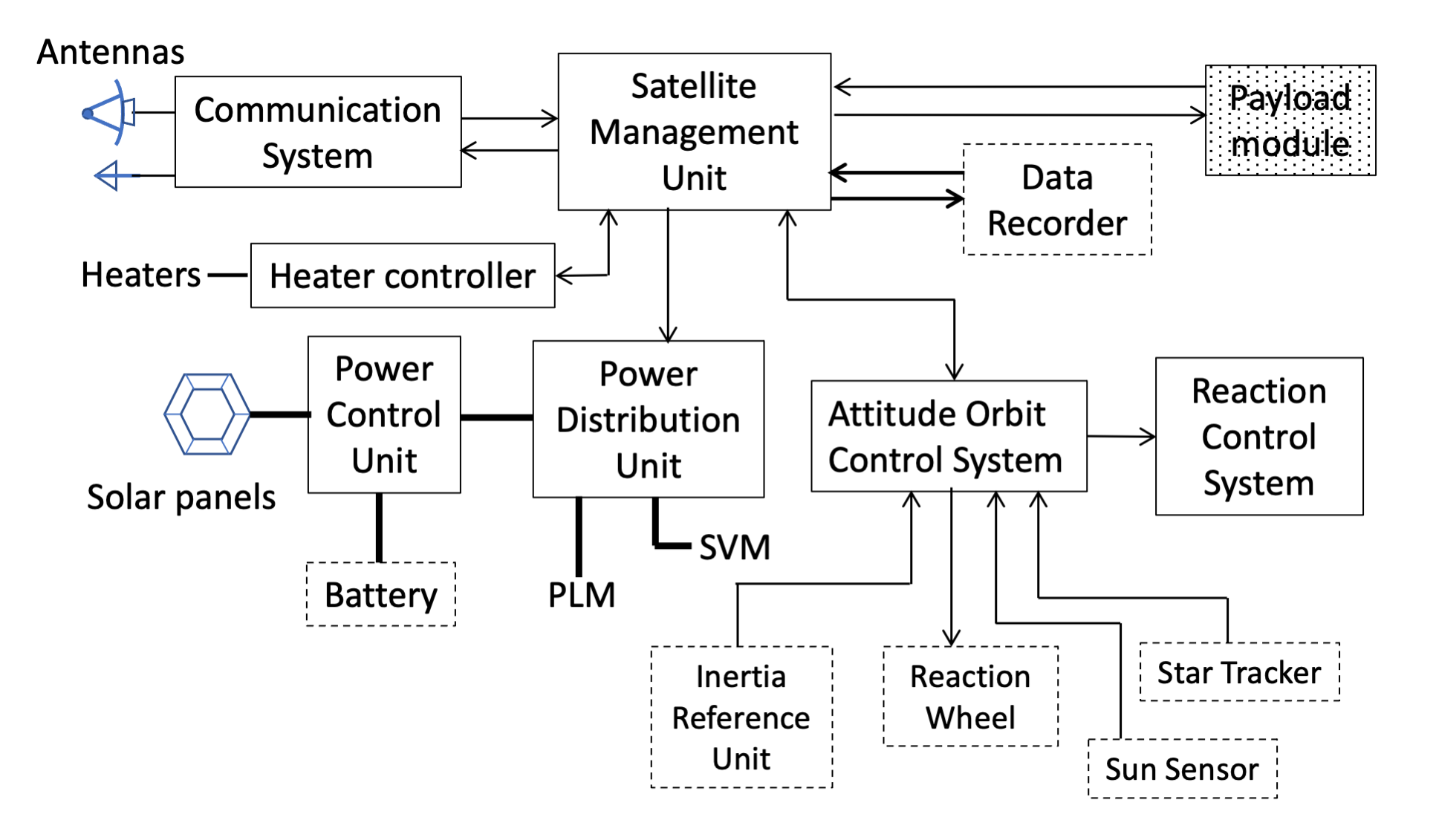}
\caption{Block diagram of the spacecraft for \LiteBIRD. Boxes with broken lines represent electric equipment, while those with solid lines are subsystems composed of multiple equipment types.  Lines and arrows connecting boxes are only representative.}
\label{fig_spacecraftBlockDaigram}
\end{center}
\end{figure}
We show the block diagram of the spacecraft in Fig.~\ref{fig_spacecraftBlockDaigram}.
The \LiteBIRD\ spacecraft uses a typical satellite configuration. 
Observation of the entire sky is conducted through the scan strategy that is detailed in Sect.~\ref{section:sensitivity}.
The slow spin rate of 0.05\,rpm 
makes it possible to adopt three-axis attitude control that satisfies the \LiteBIRD\ attitude accuracy requirements, even if the spacecraft spins.
The spacecraft will have a total weight of 2.6\,t, including the fuel of approximately 400\,kg, and a total height of 5.3\,m.
Thus the current weight has a large margin compared to the rocket's capability.
We estimate the total power of the spacecraft to be 3.0\,kW\@.
The downlink rate will be 10 Mbps in X-band and will transfer a total of 17.9\,GB of scientific data every day.
All these parameters are subject to change as the conceptual design of the satellite continues to be developed.

\subsection{Service Module}
\label{ss:overview_svm}

The \glsentryfull{svm} of \LiteBIRD\ includes an \gls{aocs}, thermal control system,
communication system, data handling system, power system, and other subsystems.
The SVM of \LiteBIRD\ utilizes existing technology as much as possible to reduce the development cost and risks.
In what follows, we briefly describe some characteristics of the SVM\@.

\LiteBIRD\ is a zero-momentum, 3-axis stabilized spacecraft designed to realize the required pattern of an all-sky survey,
i.e., the combination of spin and precession.
\Glspl{mw} are used to control the attitude of the spacecraft, and the \gls{rcs}
is used to unload the MWs and to control the orbit.
Because the spacecraft is operated with zero momentum, relatively large MWs are required to cancel
the angular momentum due to the spin.
The MWs also cancel the spin-axis component of the angular momentum due to the polarization modulators.
Other wheels are used to control the precession of the spacecraft.
The spacecraft receives a small amount of external torque even at L2, mostly from solar radiation.
This causes a steady increase or decrease of the rotational frequencies of the MWs, and thus the RCS is used to unload the MWs regularly.
The RCS also provides the required $\Delta V$ for the initial correction of the orbit and for the orbit insertion at L2.
In addition, we use the RCS once every few months to correct orbit errors, since L2 is a gravitational saddle point and any orbit around it is intrinsically unstable.

The \gls{aocs} uses \glspl{stt} and \glspl{iru} to determine the spacecraft's attitude
because a good attitude solution is required for \LiteBIRD\@.
The spin rate of 0.05\,rpm corresponds to 0.3\,deg\,${\rm sec}^{-1}$.
This spin speed is easily handled by the currently available STTs and the degradation of the attitude
solution is negligible.
The situation is the same for the currently available IRUs.
We expect that the STTs can track stars continuously, but the IRUs will be used to interpolate the attitude
in case the STTs temporarily fail to track stars.

A thermal control system keeps the temperature of the on-board components in the required range.
This is not easy when some of the components have large heat dissipation or require tight temperature stability.
Components with large heat dissipation are the mechanical coolers, their drivers, and the signal processing units.
Adiabatic demagnetization refrigerator controllers and squid controller units require tight temperature stability.
A total heat dissipation of the payload module is about 1.4\,kW\@.
Because there is no room in the payload module for sufficient radiators for this amount of heat dissipation, radiators are 
placed on the upper parts of the side panels of the spacecraft. 
The total area of the radiators may be estimated to be approximately 10\,${\rm m}^2$.
Heat pipes may be used to transfer heat from the components to the radiators.
When accurate temperature control is required, heaters are used in combination with the radiators.

\LiteBIRD\ uses X-band for telemetry and command (and ranging), and also for the downlink of the mission data.
The main reason to use X-band for telemetry and command is that the primary \gls{great} supports only
X-band, not S-band, for uplink.
We will also use the 34-m antenna at \gls{usc} as a secondary station.  
Because GREAT is the only deep space station in Japan, 
many interplanetary satellites will use the station.
The secondary station is useful when GREAT is unavailable due to tracking other satellites.
However, the USC 34-m antenna is smaller than GREAT (54-m antenna), and so the downlink rate at USC 34m will be only 1/3 of that of GREAT\@.
The required downlink rate for the mission data, 10\,Mbps, will be achieved only with GREAT\@.
To achieve this rate, we will use a parabolic antenna with a diameter of 0.5 m mounted on a 2-dimensional gimbal with 20\,W of output power.
The total amount of mission data is estimated to be 17.9\,GB\,${\rm day}^{-1}$.
This means that approximately 4.5\,hours are needed for the downlink of the mission data every day.

The power system and data-handling system of \LiteBIRD\ use the heritage of past JAXA science missions, such as ASTRO-H,
as much as possible.  
Thus we adopt Space Wire for the on-board data-handling system.  
This makes the interface checks of the electronic components easier.  
We adopt a 50\,V unregulated bus for the power system.
Because of the scan strategy, the \glspl{sap} receive solar radiation at an incident angle of $45^{\circ}$.
This may reduce the efficiency of the SAPs by $1/\sqrt{2}$, but we selected fixed panels to make the system simple.
We choose the \glspl{sap} to be close to disk-shaped
to avoid diffraction of microwaves from the Sun interfering with the telescopes.

\subsection{Operation Concepts}
\label{ss:overview_operation}

\subsubsection{Basic Principles of Operation}
Because \LiteBIRD\ needs to observe the whole sky as much as possible, we adopt the following principles of operation:
(1) \LiteBIRD\ does not make real-time observations or \gls{too} observations, but continuously makes an
all-sky survey; (2) mission data will be downlinked to the primary tracking station, GREAT, once a day for a duration of 
about 4.5\,hours; (3) observations may be interrupted 
during 
the unloading of the MWs
and orbit-keeping maneuvers; and (4) precession of the spacecraft may be stopped and observations may be interrupted 
in the case of an emergency in the SVM, such as due to a hardware failure.

In what follows, we describe the outline of mission operation for each operations phase.

\subsubsection{Outline of Operation}
\LiteBIRD\ will be launched from the Tanegashima Space Center with the H3 rocket.
It will be directly inserted into an orbit that approaches the L2 point.
Soon after separation from the rocket, the spacecraft establishes 3-axis controlled attitude and
deploys solar panels to obtain enough power.
Then, the spacecraft starts to spin, but not to precess, in order to achieve uniform temperature distribution around the spin axis,
which is followed by an initial checkout of the SVM\@.
This initial operations phase and checkout will take approximately a week.

After the initial checkout of the SVM, we will start the health check of the PLM\@.
The launch-locks of the PMUs are released after the health check.
The half-wave plate will be supported by the holding mechanism.
We then start cooling the telescopes with a combination of radiative cooling and the shield cooler.
The \gls{JT} coolers and the \glspl{adr} are turned on at the appropriate timing.
This initial cooling takes a relatively long time, approximately 70 or 80 days.
When the nominal operation temperature of the focal-plane detectors are reached, their
function and performance are checked.
Thus most of the cruising phase out to L2, which may last approximately 100 days, will be
spent for the initial cooling.

When the spacecraft arrives at L2, an insertion maneuver into the Lissajous orbit will be carried out.
Then, test observations are conducted in the all-sky scanning mode, i.e., combination of precession and spin.
Various functions and performance of the mission instruments are verified, and operational parameters are optimized.

When the test observations are completed, regular observations begin and continue for 3 years.
In this phase, all-sky survey observations are conducted steadily and we obtain as much data as possible.
Our scan strategy is described in Sect.~\ref{section:sensitivity} and Fig.~\ref{fig:PLM-param-Imo}. 
As for calibration of the instruments, our baseline plan is to rely on data from the regular observations, not to pause them for special calibration data taking.
Our in-flight calibration plan is discussed in more detail in Sect.~\ref{ss:plm_calib}.
After these regular observations, we may extend the mission, if approved by the \gls{jaxa} review.

\subsection{Data Processing and Analyses}
\label{ss:overview_analysis}

CMB missions typically scan the sky continuously with many detectors at multiple frequencies, measuring temperature and linear polarization, while simultaneously recording the orientation of the telescope. The \gls{tod} then consist of the time stamp, observation direction
determined by satellite attitude monitor, HWP position, thermal and electric monitor for the instruments,
and the measured signal for each detector sample. The measured signal includes not just the CMB, but also emission from astrophysical foregrounds, together with other systematic effects, as well as instrumental noise. 

\begin{figure}[htbp!]
\centering
\includegraphics[width=0.7\textwidth]{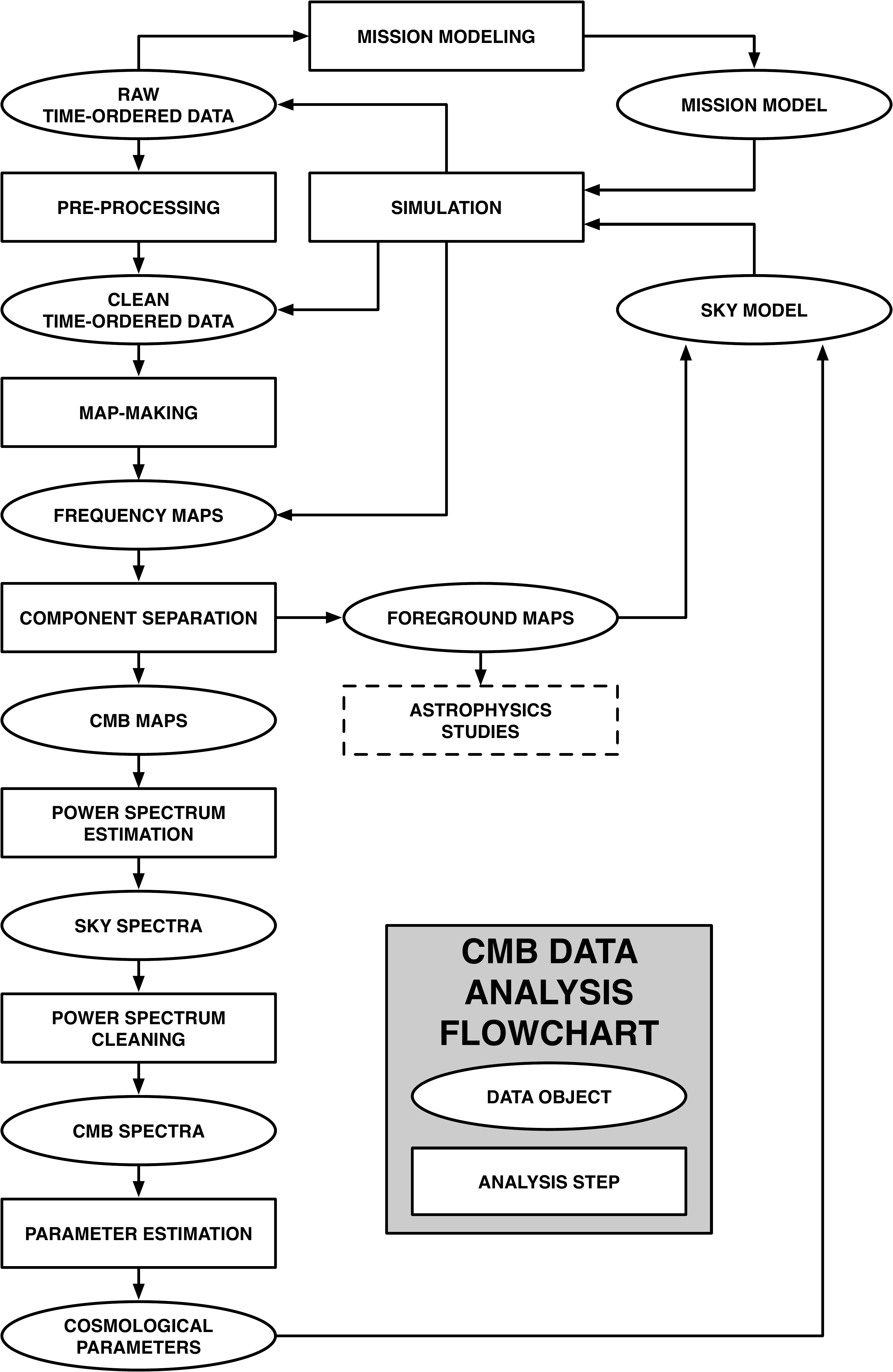}
\caption{Flowchart indicating the sequence of steps leading to the
determination of cosmological parameters from mission TOD via several
intermediate products that can also be scientifically exploited. Note
the iterative nature of the processing.}
\label{fig:DA_Flowchart} 
\end{figure}

The processing of CMB experimental data can be described as a series of steps that move from the TOD to sky maps at each frequency, which are then separated into different physical components. The statistics of the CMB map are finally compared to the cosmological prediction from theoretical models. These data processing steps, indicated in Fig.~\ref{fig:DA_Flowchart}, reduce the statistical and systematic uncertainties in the data, first reducing the systematic contamination by appropriate mitigation in a given data domain, and then reducing the dimensionality of the data set by exploiting its redundancies to project onto a lower-dimensional domain. 

Systematic mitigations include removing glitches (e.g., from cosmic-ray hits) in the time domain, separating the CMB from the astrophysical foreground components in the map domain, and quantifying contamination from unresolved emission sources in the spectral domain. Throughout the analysis process we also have to account for non-idealities in the instrument, both optical (e.g., from asymmetric and mismatched beams, including side-lobes, differences in the bandpass for each detector) and electronic (e.g., gain drift, non-linearity and crosstalk).

Finally, we need an accurate description of the uncertainties in the products and their correlations. The data volumes we need to amass in order to detect the tiny CMB signals preclude exact analyses, and so we typically use Monte Carlo methods for debiasing and forward propagation of uncertainties. We therefore need to be able to generate and reduce large numbers of very accurate simulated data sets, whose input mission and sky models are themselves informed by our analyses of the satellite data.

With 4508 detectors sampling at 19.1\,Hz for 3 years, the \LiteBIRD\ mission will gather $8\times 10^{12}$ detector samples. Manipulating this data volume while capturing all of the correlations in the CMB signals, foregrounds, and instrumental noise and systematics, with sufficient precision to yield reliable, unbiased results, is a computationally challenging task, requiring the use of state-of-the-art high-performance computing systems. Similarly, tracking the provenance of all of the data products, including the myriad data cuts used to check the robustness of the analysis, requires dedicated databases accessible to all collaboration members.
The biggest data challenges for \LiteBIRD\ will be removing astrophysical foreground contamination using the sky maps at the 15 observing frequency bands (Sect.~\ref{ss:forecasts_fg_cleaning}), and mitigating systematic effects and precise characterization of their residuals (Sect.~\ref{ss:forecasts_syst}).





\section{Payload Module of LiteBIRD} 
\label{s:payload}

\subsection{Overview} 
\label{ss:plm_overview}

\begin{figure}[htbp!]
\centering
\includegraphics[width = 0.7\textwidth]{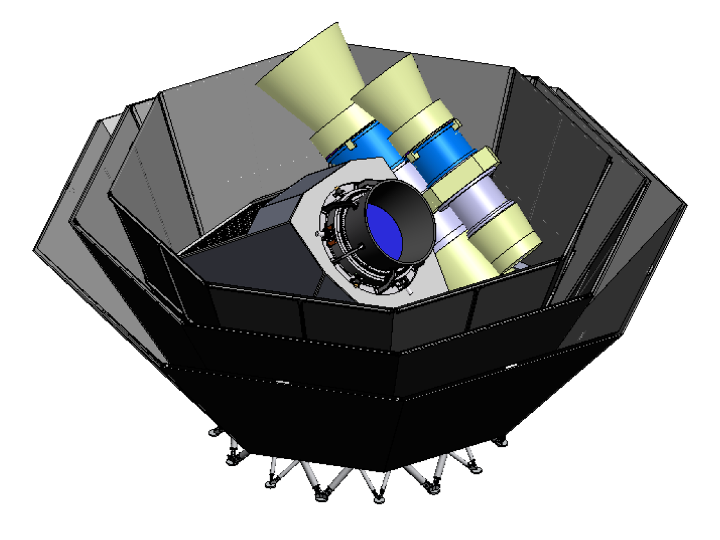}
\caption{Payload module overview  showing the surrounding V-grooves acting as passive coolers up to 30\,K, with the 4.8-K enclosure of the low-frequency crossed-Dragone telescope in the front, and the two 4.8-K tubes of the mid- and high-frequency on-axis telescopes in the back.}  
\label{fig:plm}
\end{figure}

The \litebird\ PLM consists of three telescopes -- at low, medium, and high frequencies -- with their respective cryo-structure and focal planes cooled down to 0.1\,K. The PLM also includes the global cooling chain from 300\,K to 4.8\,K, and room-temperature elements, such as drivers and warm readout electronics of the detectors. Requirements for the PLM have been derived from the top-level requirement of achieving a tensor-to-scalar ratio error of $\delta r < 0.001$ (see Sect.~\ref{ss:overview_mission}). Table~\ref{tbl:plm:sensitivities} gives information on the frequency bands, beam sizes, and NETs for each of the 15 frequency channels.  A discussion of the NET calculations is given in Sect.~\ref{section:sensitivity}.  

The \litebird\ requirements imply technical challenges for the PLM, in terms of sensitivity, optical properties, stability, or even compactness, over a wide range of frequencies, from 34 to 448\,GHz. In order to achieve such a challenging set of scientific requirements, an important feature of \liteBIRD\ is its observing strategy, focusing on the largest scales over the sky to maximize the signal expected from the reionization and recombination peaks of the $B$-mode power spectrum, which requires an unprecedented sensitivity over multipoles $2 \leq \ell \leq 200$. This demands only reasonably low resolution for the telescopes ($<80$\,arcmin), but associated with a strong control of the systematics in order to minimize the $1/f$ noise.

\begin{table}[htbp!]
\setlength{\tabcolsep}{4pt}
\centering
\caption{\liteBIRD\ sensitivities. We show the values related to the sensitivity in 15 frequency bands. From left to right the columns are: the telescope covering the band; the band identification number; the band center frequency in GHz; the bandwidth in GHz and its ratio; the main beam FWHM in arcmin; the detector pixel size in mm; the number of bolometers used; the NET value of a single detector in $\mu$K\,$\sqrt{\rm s}$; and the NET value of the detector array in $\mu$K\,$\sqrt{\rm s}$.  
}
\label{tbl:plm:sensitivities}
\begin{tabular}{|c|c|c|c|c|c|c|c|c|}
\hline

 Tel. & ID&  $\nu$   & $\delta\nu$ ($\delta\nu/\nu$) & Beam size & Det. pixel  & No.\ of & NET detector & NET array   \\ 
                 &              &  [GHz]    &    [GHz]    & [arcmin]     & size [mm]              &   bolo                        & [$\mu$K\,$\sqrt{\rm s}$\,] & [$\mu$K\,$\sqrt{\rm s}\,]$ \\ \hline\hline
 LFT &	1 &	40 &  12 (0.30) & 70.5 &	32 & 48  &  114.63	       &  18.50  \\  \hline
 LFT &	2 &	50 &  15 (0.30) & 58.5 &	32 & 24  &	72.48          &  16.54  \\  \hline
 LFT &	3 &	60 &  14 (0.23) & 51.1 &	32 & 48  &	65.28          &  10.54  \\  \hline
 LFT &	4 &	68 &  16 (0.23) & 41.6 &	16 & 144 &	105.64         &  9.84   \\ 
 &   &	       &  16 (0.23) & 47.1 &	32 & 24  &	68.81          &  15.70  \\ 
 &   &         &            &      &       &     &                 &  8.34 (comb.) \\  \hline
 LFT &	5 &	78 &  18 (0.23) & 36.9 &	16 & 144 &  82.51          &  7.69  \\ 
     &    &    &            & 43.8 &	32 & 48  &  58.61          &  9.46  \\ 
     &    &    &            &      &       &     &                 &  5.97 (comb.) \\  \hline
 LFT &	6 &	89 &  20 (0.23) & 33.0 &	16 & 144 &  65.18          &  6.07 \\  
     &    &    &            & 41.5 &	32 & 24  &  62.33          &  14.22 \\  
     &    &    &            &      &       &     &                 &  5.58 (comb.)\\ \hline
 LFT &	7 &	100 & 23 (0.23) &	30.2 &  16  & 144 &	54.88          &  5.11 \\  
 MFT &    &     &           &   37.8 & 11.6 & 366 & 71.70          &  4.19 \\
     &    &     &           &        &      &       &              &  3.24 (comb.) \\ \hline
 LFT &  8 & 119 & 36 (0.30) &	26.3 &	16   &	144 & 40.78        &  3.80   \\ 
 MFT &    &     &           &	33.6 &	11.6 &	488 & 55.65        &  2.82   \\ 
     &    &     &           &        &       &      &              &  2.26 (comb.)  \\ \hline
 LFT &	9 &	140 & 42 (0.30) &	23.7 &	16   &	144 & 38.44        &  3.58  \\  
 MFT &    &     &           &   30.8 &  11.6 &	366 & 54.00        &  3.16  \\  
     &    &     &           &        &       &      &              &  2.37 (comb.)  \\ \hline
 MFT & 10 &	166 & 50 (0.30) &	28.9 &	11.6 & 488  & 54.37        &  2.75  \\ \hline
 MFT & 11 &	195 & 59 (0.30) &   28.0 &  11.6 & 366  & 59.61        &  3.48  \\ 
 HFT &    &     &           &   28.6 &   6.6 & 254  & 73.96        &  5.19  \\ 
     &    &     &           &        &       &      &              &  2.89 (comb.)  \\ \hline
 HFT & 12 &	235 & 71 (0.30) &	24.7 &	6.6  &	254  & 76.06       &  5.34 \\  \hline
 HFT & 13 &	280 & 84 (0.30) &	22.5 &	6.6  &	254  & 97.26       & 6.82 \\  \hline
 HFT & 14 &	337 & 101 (0.30) &	20.9 &	6.6  &	254  & 154.64     & 10.85 \\  \hline
 HFT & 15 &	402 & 92 (0.23)  &	17.9 &	5.7  &	338  & 385.69      & 23.45 \\   \hline
 Tot. &    &     &            &       &       &  4508 &             &   \\  \hline
\end{tabular}
\end{table}

In this context, a critical technical choice made for \LiteBIRD\ was to use as the first optical element a continuously-rotating \gls{hwp}. This allows us to distinguish between the instrumental polarized signal and the sky signal, which is modulated at $4f_{\rm HWP}$. While, without the HWP, data from a pair of detectors mutually orthogonal in their polarization orientations are usually combined, causing leakage from temperature to polarization if there are any differences in the beams, gains, or band-passes between the two detectors; this can be removed through the use of HWPs that enable us to measure the polarization using a single detector. Lastly, the presence of the continuously-rotating HWP performs an effective suppression of the $1/f$ noise. A detailed trade-off analysis, including the polarization effects induced by the HWP itself, has been carried out between the two cases, i.e., with and without the HWP, demonstrating that the performance without HWP are expected to have potential large systematic effects compared with those with HWP, as described in Sect.~\ref{sss:syst_intro}.
Hence, in order to guarantee appropriate thermal performance in terms of stability and minimal heat load, the three telescopes will be equipped with \glspl{pmu} continuously rotating at a few Hz around a stable temperature below 20\,K, using a magnetic levitating mechanism with superconducting bearing, as described in more detail in Sects.~\ref{sss:plm_lft_pmu} and \ref{sss:plm_mhft_subsystems}. 

The distribution and the number of bands over the wide range of frequencies, from 34\,GHz to 448\,GHz, have been optimized to deal with the following constraints: the spectral resolution has to ensure the appropriate characterization of the expected complexity of the spectral energy distribution of the synchrotron and dust Galactic foregrounds, leading to 15 broad and partially overlapping bands; the limited frequency range of HWP materials (sapphire and metal-mesh) and associated anti-reflection coating required us to split into three telescopes; the spectral mapping of the CO lines has to be optimized by rejecting such molecular lines from some of the bands and including them in others (notice that notch-filters have not been included, since it has been demonstrated that temperature-to-polarization leakage from CO lines is highly reduced by the rotating HWP); and finally an overlap between bands and instruments had to be foreseen to mitigate systematic effects. We ended up with the following distribution: a reflective telescope at low frequency (see Sect.~\ref{ss:plm_lft}), the \gls{lft} (34--161\,GHz), and two refractive telescopes at middle and high frequencies (see Sect.~\ref{ss:plm_mhft}), the \gls{mft} (89--225\,GHz) and \gls{hft} (166--448\,GHz), as illustrated in Fig.~\ref{fig:plm}. The MFT and HFT telescopes are mounted on the same mechanical structure, and point in the opposite direction compared to the LFT, but cover the same circle over the sky when spinning. 

The focal planes of the three telescopes with large field of view ($18^{\circ}\times9^{\circ}$ for LFT, and $28^{\circ}$ diameter for MFT and HFT) are populated with multichroic polarized \gls{tes} detectors (one to three bands per pixel). This multichroic technology allows for a very compact design with sufficient flexibility on the optimization of the sensitivity per band that is needed to improve the performance of the component-separation techniques. Two detector technologies have been used, lenslet-coupled detectors for the LFT and MFT, and horn-coupled detectors for the HFT, for a total of 4508 detectors cooled down to 100\,mK, as detailed in Sect.~\ref{ss:plm_detection}. The readout electronics takes advantage of the frequency multiplexing scheme to accommodate this large set of detectors without loss of information and minimal power dissipation on the focal planes. 

The temperature stability of the instruments is another crucial point for CMB $B$-mode polarization probes because of the following aspects: the temperature fluctuation of the optical components contributes to noise stability and $1/f$ noise; and temperature variation of the mechanical structures has a direct impact on pointing stability.
Hence the three \LiteBIRD\ telescopes are fully cooled down to 4.8\,K, minimizing the heat load on the focal planes. The proposed 300-K to 4.8-K cryogenic chain for \LiteBIRD\ is based on the architecture developed as part of the \textit{SPICA}-SAFARI mission. It combines radiative cooling (V-grooves) down to 30\,K combined with mechanical cryo-coolers to provide cooling to temperatures down to about 4.8\,K. 
In its current definition, a 15-K pulse-tube cooler associated with three V-groove radiators, respectively at 160\,K, 90\,K, and 30\,K, intercept part of the thermal loads. Then, one helium \gls{JT} loop (4-K JT, 4He), pre-cooled by two 2-stage Stirling coolers (100\,K / 20\,K).
All telescopes have intermediate cold stages at 1.75\,K and 0.35\,K between their mechanical enclosure at 4.8\,K and the detectors at 0.1\,K. The 1.75-K cooler is based on 2-K Joule Thompson cooler, to provide a continuous cooling at 1.75\,K. The Sub-kelvin cooler is made of two \gls{adr} stages in series to provide stable and continuous cooling at 0.35\,K, combined with two other ADR stages in parallel for the 0.1-K stage. Again the design of this cryo-chain has been optimized to ensure maximum stability of the temperature of the focal planes and the optical elements of the telescopes. 

In the following sections, we provide more details on the instrumental setup, following the natural path of the scientific signal, i.e., starting with the optical and mechanical descriptions of the LFT (Sect.~\ref{ss:plm_lft}) and MFT and HFT (Sect.~\ref{ss:plm_mhft}) telescopes. Since the detection chain of the three telescopes follows a common architecture, it is globally described in Sect.~\ref{ss:plm_detection} for LFT, MFT, and HFT, while the global electrical architecture of the PLM is detailed in Sect.~\ref{ss:plm_electronics}. A description of the whole cooling chain, from 300\,K down to 100\,mK is provided in Sect.~\ref{ss:plm_cryo}. Finally the ground and in-flight calibration plan of the whole instrumental setup is discussed in Sect.~\ref{ss:plm_calib}.

\subsection{Low-Frequency Telescope} 
\label{ss:plm_lft}

\subsubsection{Overview}

\begin{figure}[htbp!]
\begin{center}
   \includegraphics[width =120mm]{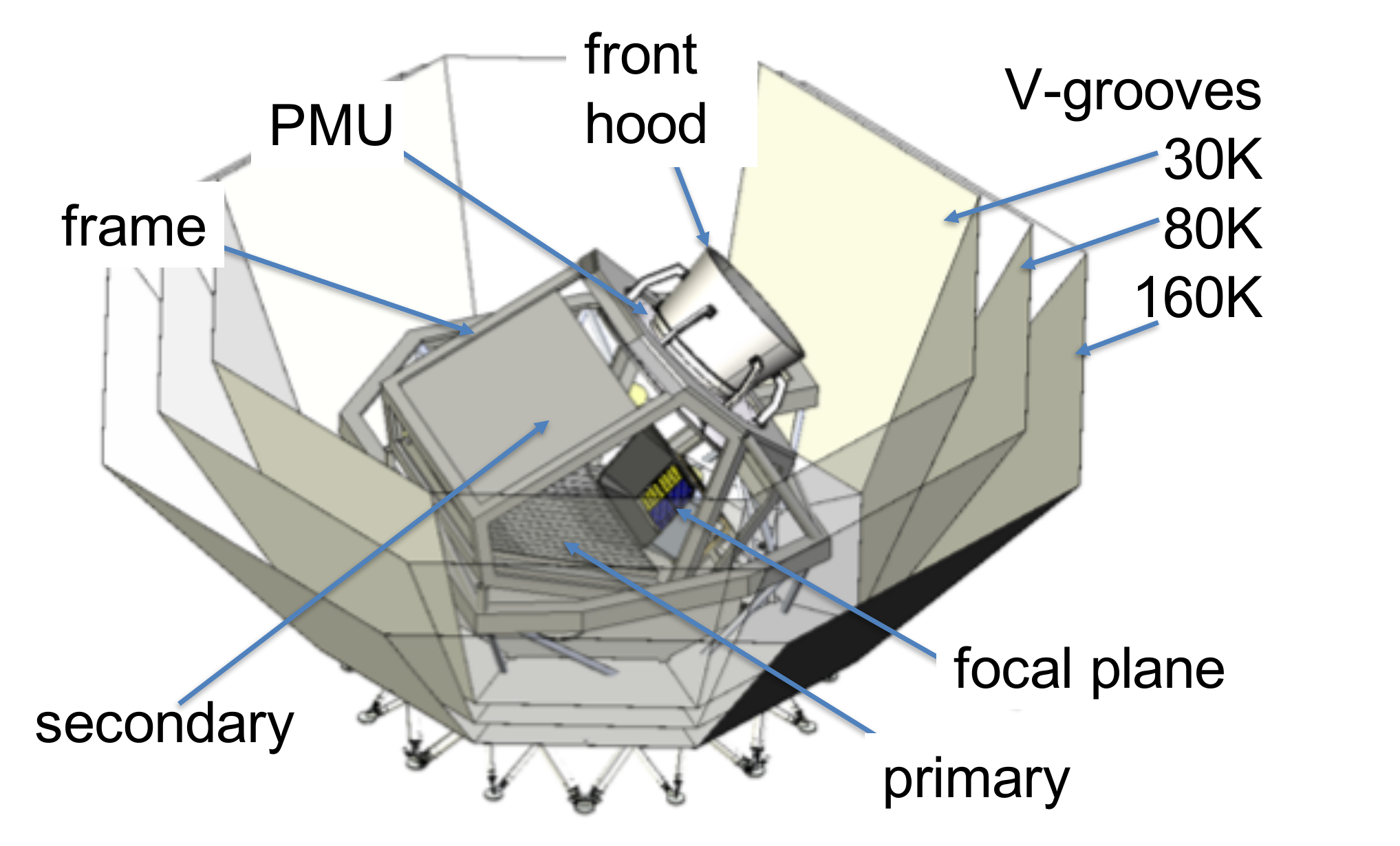}
 \end{center}
 \caption{Overview of the Low-Frequency Telescope (LFT). The Mid- and High-Frequency telescopes and side panels are not shown for clarity.}
 \label{fig:LFT-overview}
\end{figure}

With an aperture diameter of 400\,mm and an angular resolution ranging from 24 to 71\,arcminutes, LFT consists of nine broad frequency bands spanning from 34 to 161\,GHz, in order to cover the spectral domains of both CMB and synchrotron radiation emission. 
It is operated at a cryogenic temperature of 4.8\,K to reduce the optical loading, and surrounded by radiators called V-grooves, acting as passive coolers. 
The LFT optical design follows a crossed-Dragone configuration, with an antenna made of aluminum. As the first optical component, a \glsentryfull{pmu}, consisting of a continuously rotating transmissive half-wave plate, is mounted in front of an aperture stop, allowing us to minimize straylight contamination.
The LFT focal plane is based on multi-chroic TES detectors cooled down to 100\,mK, as described in Sect.~\ref{sss:detection_fpu}.
A frame structure at 5\,K supports all components: the PMU; focal plane; primary and secondary reflectors; and absorbers.
An overview of the LFT is presented in Fig.~\ref{fig:LFT-overview}, introducing the various components listed above.  

\begin{figure}[htbp!]
\centering
   \includegraphics[width =1\textwidth]{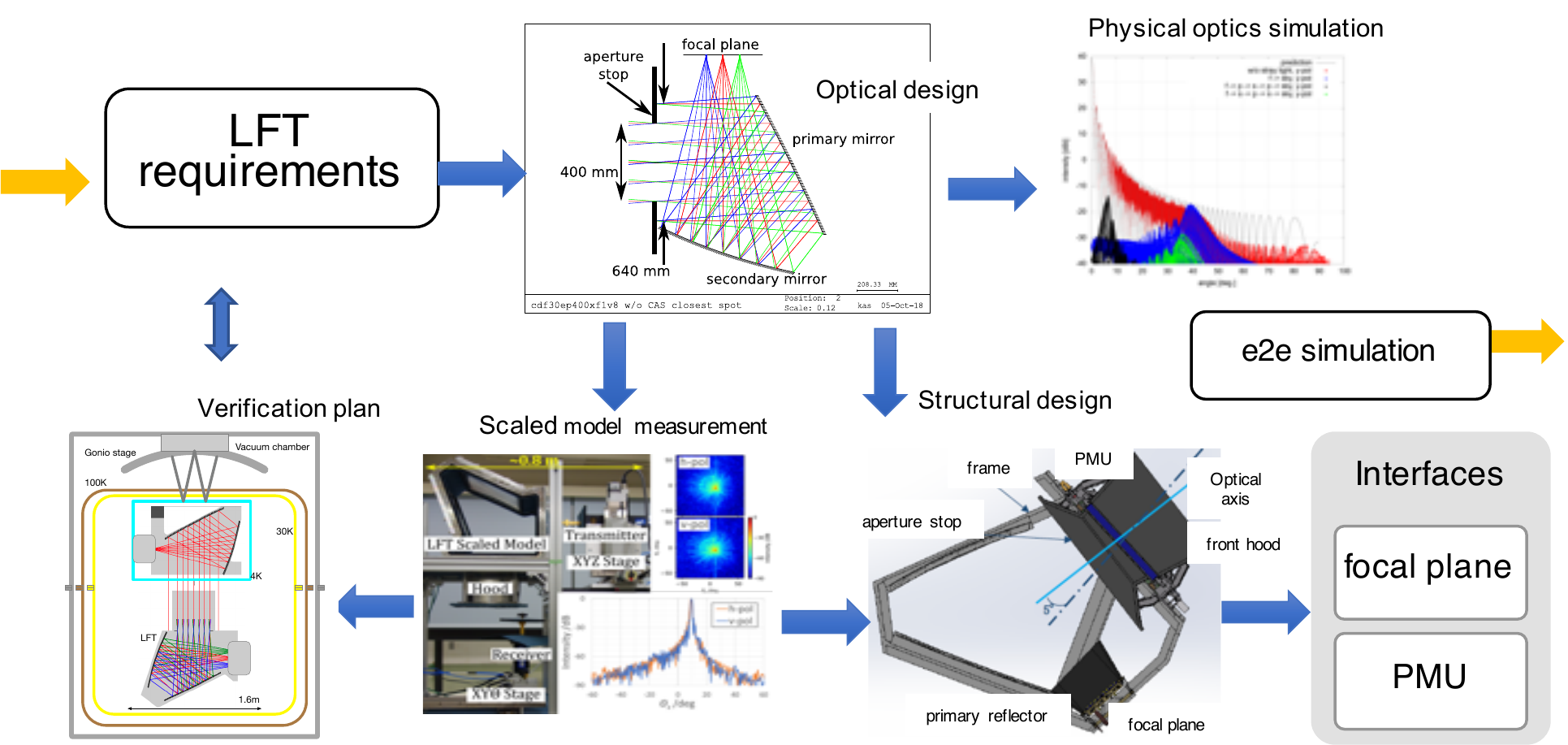}
 \caption{Design flow of LFT. From the instrument requirements, the optical design, physical optics simulations, structural design, interface requirements, a scaled version of LFT, and verification plan are designed and developed either sequentially or in parallel.}
 \label{fig:wide-fov-design-flow}
\end{figure}

Performance requirements of the LFT are described in Ref.~\cite{Sekimoto2020} and a design flow is shown in Fig.~\ref{fig:wide-fov-design-flow}.
Starting from an optical design satisfying the instrument requirements, we proceeded with physical optics simulations and by completing the structural design to correctly match the interface requirements. A scaled version of LFT has allowed us to validate the models based on the scaled model measurements~\cite{Takakura2019IEEE}.

\subsubsection{Optical Design}
The challenges of \LiteBIRD\ include its wide \gls{fov} and broadband capabilities for millimeter-wave polarization measurements, 
which are derived from the sensitivity requirements.
The wide FoV corresponds to a large focal plane area, so that a detector pixel has different spill-over or edge-taper at reflectors depending on the pixel position on the focal plane.
Possible paths of straylight increase with a wider FoV.
After various trade-off studies of various optical configurations, including a front-fed Dragone \cite{Bernacki2012}, we concluded that the crossed-Dragone antenna is the best option for LFT,
because it has good beam and polarization performance over the required FoV.

\begin{table} [htbp!]
\centering
\caption{Optical specifications of LFT~\cite{Sekimoto2020}.
The \gls{psf} flattening is defined as $(\sigma_{\mathrm{maj}} - \sigma_{\mathrm{min}}) / (\sigma_{\mathrm{maj}} + \sigma_{\mathrm{min}})$ where $\sigma_{\mathrm{maj}}$ and $\sigma_{\mathrm{min}}$ are for the major and minor axes, respectively.}
\begin{tabular}{|l|c|}
\hline
Aperture diameter & 400\,mm \\
Field of view & $18^\circ \times 9^\circ$ \\
Strehl ratio &  $> 0.95$ at 161\,GHz\\
Focal plane telecentricity &  $< 1.0^\circ$ \\
F-number, f$/N$  & $2.9 < N < 3.1$ \\
PSF flattening & $< 5$ \% \\
Cross polarization  & $< -30$\,dB \\
Rotation of polarization angle across FoV & $<\pm1.5^\circ$ \\
\hline
\end{tabular}
\label{tbl:lft-specification}
\end{table}

\begin{figure}[htbp!]
\begin{minipage}{0.5\hsize}
  \begin{center}
  \vspace{0.5mm}
\includegraphics[width = 0.8\textwidth]{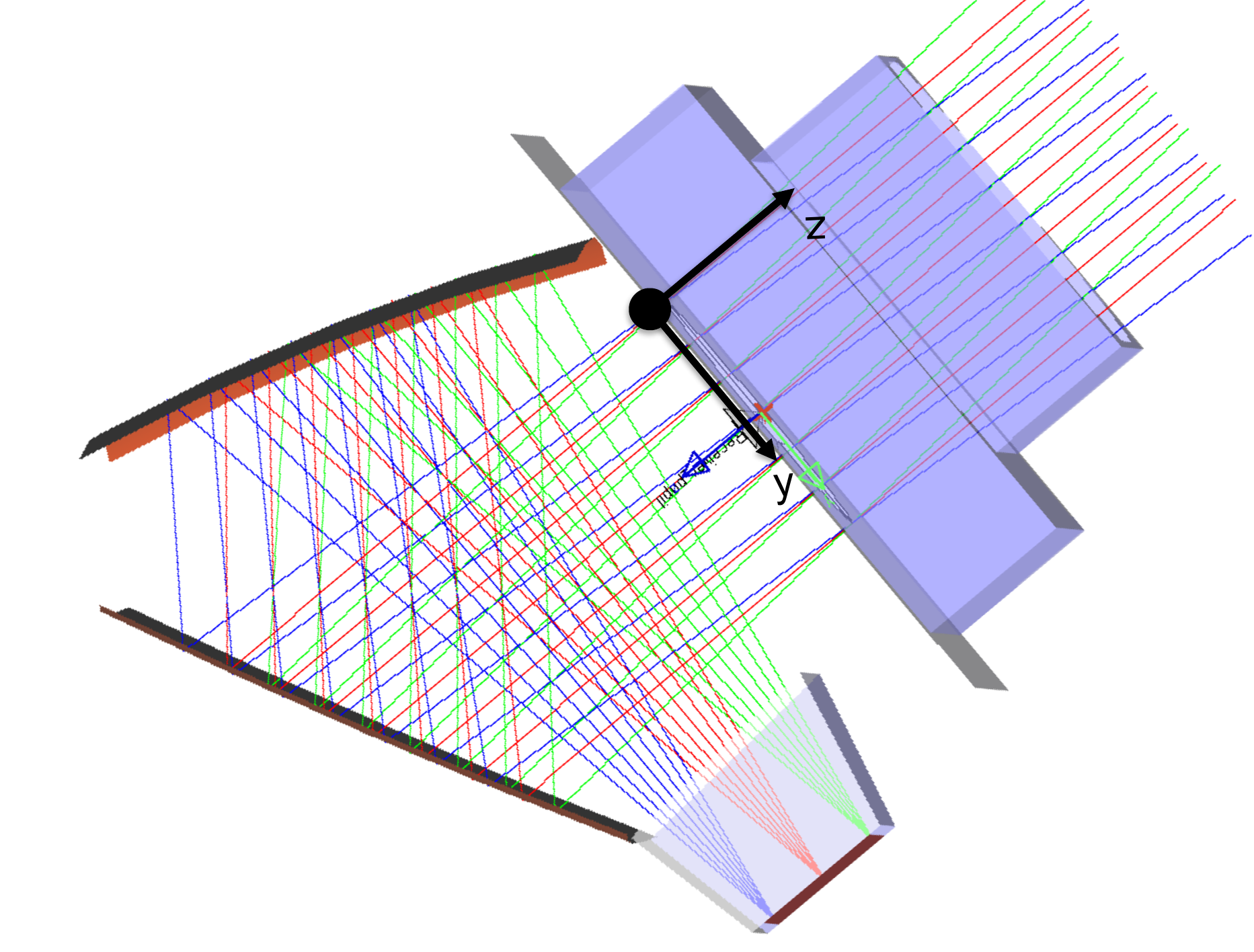}
\caption{Ray tracing diagram of LFT.}
\label{fig:lft-ray-trace}
\end{center}
 \end{minipage}
 \begin{minipage}{0.5\hsize}
  \begin{center}
  \includegraphics[width = 0.8\textwidth]{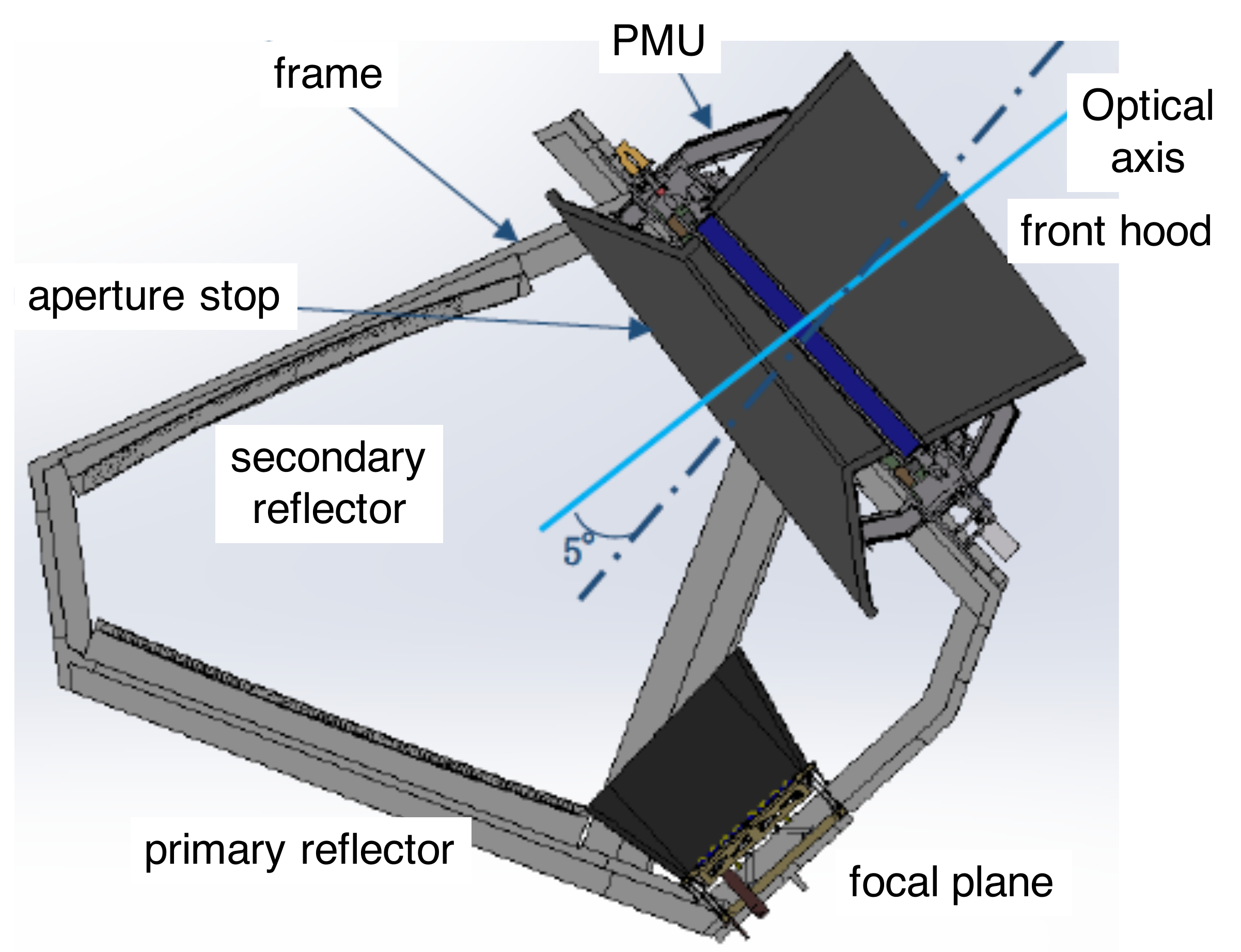}
\caption{LFT structure.}
\label{fig:lft}
  \end{center}
 \end{minipage}
\end{figure}

The crossed-Dragone antenna of LFT has been designed with anamorphic aspherical surfaces \cite{Kashima:2017yub} to achieve the specifications listed in Table \ref{tbl:lft-specification}.
A ray diagram of LFT is shown in Fig.~\ref{fig:lft-ray-trace}, which has an FoV of $18^{\circ}\times9^{\circ}$.
The f/3.0 ratio and crossing angle of $90^\circ$ between the aperture to primary and the secondary to focus axes have been chosen after an extensive straylight study.
While the requirements on far sidelobe knowledge (about $-56$\,dB) is one of the most challenging requirements, 
the optical design has been optimized to achieve far-sidelobe levels as low as possible, ideally below the knowledge requirements threshold, in order to be able to mitigate the need of far-sidelobe high-accuracy measurements. Indeed, the expected beam calibration sequence plans to go through room temperature measurements of far sidelobes, cryogenic measurements of near sidelobes, and in-flight calibration of near sidelobes, where we define the near sidelobes to be the region out to $3^\circ$ with respect to the main lobe peak direction, and the remaining part of the beam is defined as the far sidelobes.

\paragraph{Optical Components}
An aperture stop at 4.8\,K, with an inner diameter of 400\,mm, is made of millimeter absorber, TK-RAM \cite{TK-RAM,Saily2004} on aluminum plate.
 This works to create the desired beam shape and to reduce the photon noise for the configuration with an edge taper of about $3$\,dB.
Primary and secondary reflectors made of aluminum (A6061) have a rectangular shape of $835\,\mathrm{mm}\times795\,\mathrm{mm}$ and $872\,\mathrm{mm}\times739\,\mathrm{mm}$, respectively, 
with serrations to reduce diffraction patterns from the edges of mirrors.

Millimeter absorbers (to reduce reflections) are attached on the inside surface of the 5-K frame, which plays the role of a cavity.
TK-RAM, Simons Observatory meta-material microwave absorber \cite{Xu2021}, and 3D-printed absorber \cite{Petroff2019} are candidates for such an absorber as described in Section \ref{sss:plm_mhft_subsystems}. 
Eccosorb AN72 and HR10 are also candidates, however, 
they have large \gls{tml} and \gls{cvcm}. 
According to the NASA outgas database \citep{NASAoutgass}, AN72 washed with ethanol shows reasonable TML and CVCM values.
A front hood whose height is 500\,mm reduces straylight to the far sidelobes. 
The aperture shape of the front hood is described in Ref.~\cite{Takakura2019IEEE}.

\paragraph{Optical Simulations}
Physical optics simulations of LFT with {\tt GRASP10}~\cite{GRASP10} are reported in Ref.~\cite{Imada2018SPIE}, with
the simulated telescope elements including LFT reflectors and the aperture stop.
The feed pattern is assumed to be a Gaussian beam, but the real feed pattern may change the antenna pattern on the sky. 
At this stage, the HWP, which may generate additional sidelobes, 
is not taken into account for the physical optics simulations.

\paragraph{Structural Design}
The frame and reflectors of LFT are made of aluminum (A6061) in order to shrink similarly by 0.4\,\% to 4.8\,K from 300\,K.
The telescope is covered with aluminum and absorbers to reduce straylight from the inner surface of the V-groove.
The mass of LFT, including the focal plane, is estimated to be approximately $200$\,kg.
The telescope is supported by trusses made of aluminum on the 4.8-K interface plate, as illustrated in Fig.~\ref{fig:lft}.

\subsubsection{Scaled Model Demonstration}
\label{sect:LFT-scaled-model}
\begin{figure}[htbp!]
\begin{minipage}{0.6\hsize}
  \begin{center}
\includegraphics[width=80mm]{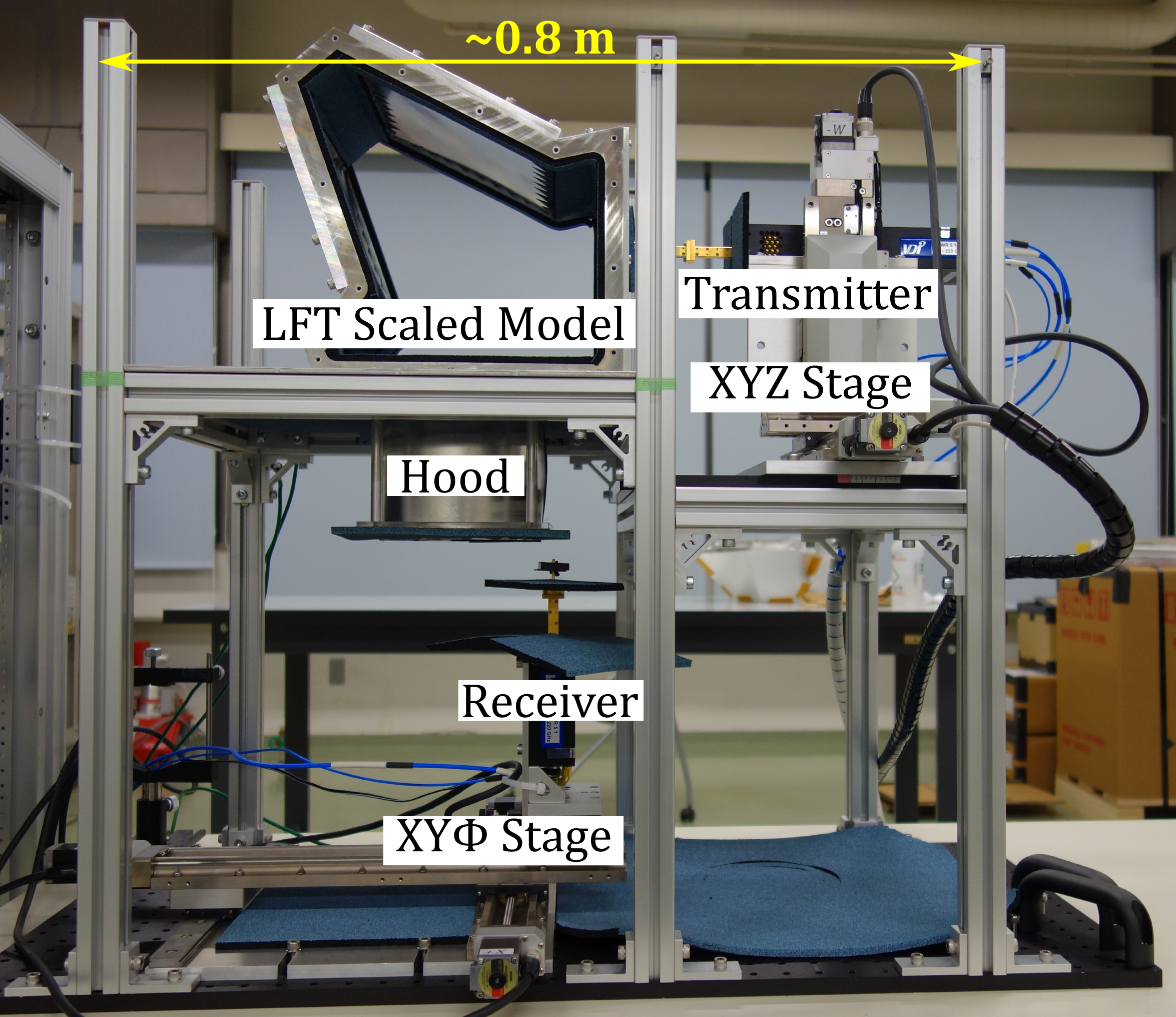}
  \end{center}
 \end{minipage}
 \begin{minipage}{0.4\hsize}
  \begin{center}
   \includegraphics[width=50mm]{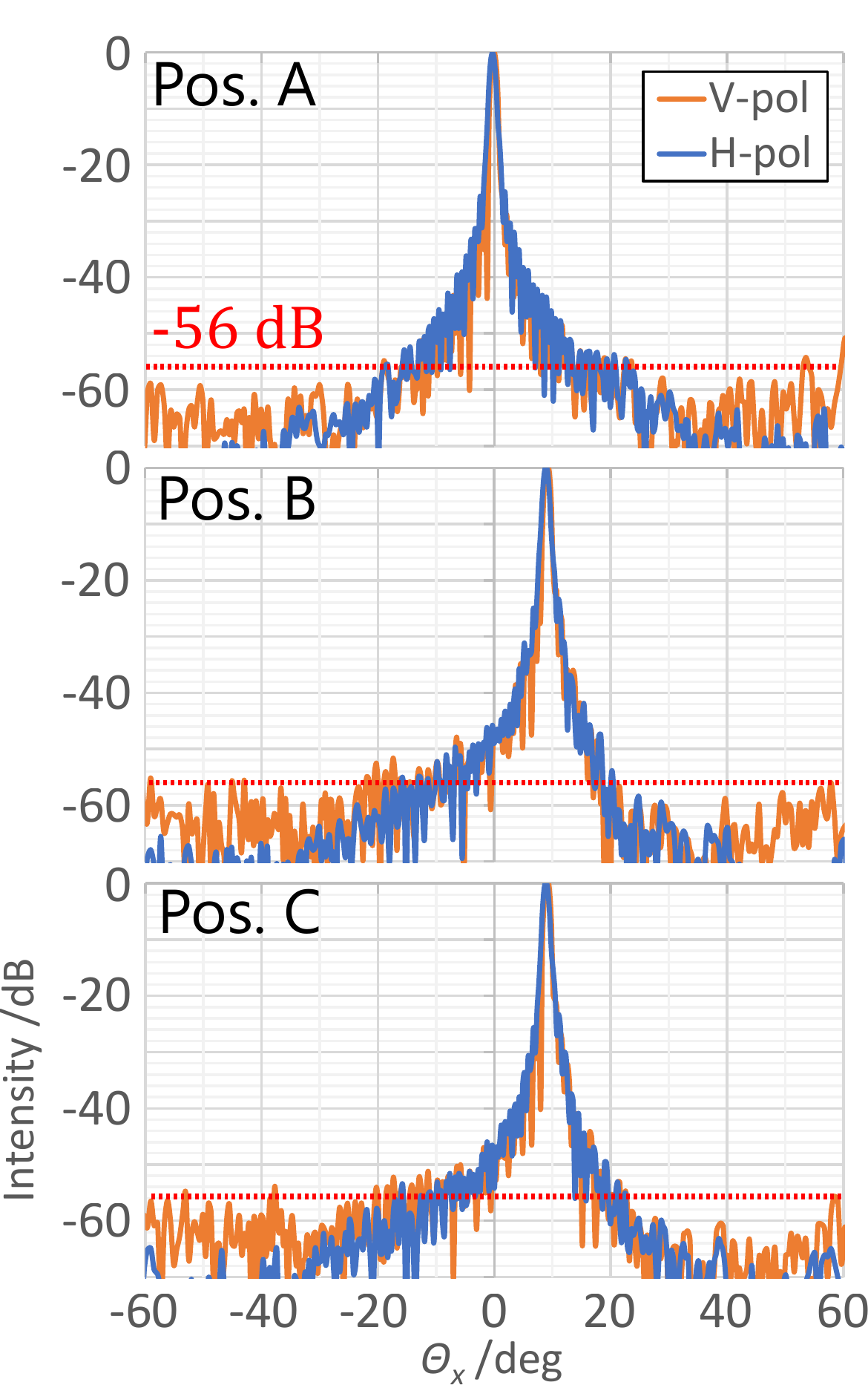}
  \end{center}
 \end{minipage}
 \caption{\textit{Left}: LFT quarter (1/4) scaled model and its near-field measurement system. \textit{Right} Far field patterns of the quarter LFT at the center (Pos. A), the left upper edge (Pos. B), and the left bottom edge (Pos C) of the focal plane measured at 220\,GHz (which corresponds to 55\,GHz for the full-scale LFT~\cite{Takakura2019IEEE}.}
 \label{fig:LFTqscaleMeasSys}
\end{figure}

A quarter (1/4) size scaled model of LFT (see Fig.~\ref{fig:LFTqscaleMeasSys}) has been designed and developed to characterize the antenna pattern in the near field~\cite{Takakura2019IEEE}.
The observed frequencies are also scaled, so that the antenna pattern of the scaled model matches that of the full size one.
The measured antenna patterns are then transformed to far fields, as shown in Fig.~\ref{fig:LFTqscaleMeasSys}. 
Based on these scaled model measurements, we confirmed the wide-field performance of LFT and the suppression of far sidelobes.

\subsubsection{LFT Polarization Modulation Unit}
\label{sss:plm_lft_pmu}

\begin{figure} [htbp!]
\centering
\includegraphics[width=1.0\textwidth]{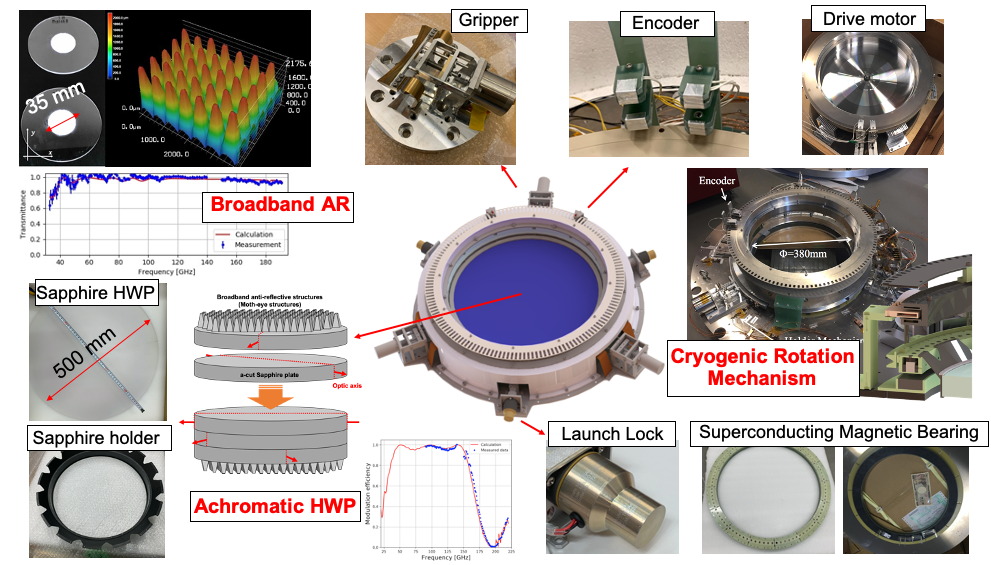}
\caption{\label{fig:pmu_overview} Overview and components of the \LiteBIRD\ LFT PMU BBM. The AHWP is composed of five-layer sapphire plates that are about 500\,mm in diameter with moth-eye sub-wavelength grating structures for anti-reflection on two outer surfaces. The entire AHWP is held in the leaf-like holder, which accounts for the differential thermal contact and yet is strong enough to survive the launch impact and vibration. The rotational mechanism is composed of the cryogenic holder mechanism (called the ``gripper''), the optical encoder for monitoring the rotor position, and the drive motor mechanism to drive the rotor. The rotation is supported by the superconducting bearing, ring magnet, and ring YBCO. The entire rotor is held by the launch lock in order to survive the launch impact.} 
\end{figure}

The baseline configuration of the \LiteBIRD\ LFT PMU consists of a Pancharatnam-type multi-layer sapphire stack as an \gls{ahwp}, which is supported by a cryogenically cooled rotational mechanism.
A similar system was employed by a balloon-borne CMB experiment, \gls{ebex}, and has been under development by multiple ground-based experiments, e.g. POLARBEAR-2 and Simons Observatory \cite{ebex_pmu,pb2b_pmu,so_sat}.
Here we describe a summary of the current development status using the PMU \gls{bbm} during the conceptual design phase. This \gls{bbm} is aimed at demonstrating some of the key technological challenges, as we will highlight later. 
Figure~\ref{fig:pmu_overview} shows the overview and components of the current PMU \gls{bbm}; a more detailed overview can be found in Ref.~\cite{Sakurai2020SPIE}.

The AHWP is continuously rotated by the cryogenic rotation mechanism that consists of the \gls{smb} and the synchronous motor spinning at a rotation speed of 46\,rpm throughout the mission.
The AHWP covers the observational frequency band of LFT.
The broadband anti-reflection coating is achieved using sub-wavelength structures directly machined by a laser.
The small diameter sample demonstration achieved the measured transmittance of 91\,\% and 97\,\% for the 40- and 50-GHz LFT bands, respectively, and above 98\,\% for other frequency bands~\cite{ryota_JAP2020}.
Most recently, we made a sample with a diameter up to 80\,mm \cite{ryota_SPIE2020}. 
The current machining strategy is scalable in diameter, and the demonstration with larger diameter samples (200\,mm and 330\,mm) is in progress.
The high modulation efficiency is realized using multilayered sapphire plates based on a Pancharatnam recipe. 
We plan to employ 5-layer sapphire plates, which achieves an averaged modulation efficiency of about 92\,\% and 96\,\% for the 40- and 140-GHz LFT bands, respectively, and above 98\,\% for other frequency bands, as well as nulling the frequency-dependent fast-axis variation over the LFT bandwidth~\cite{Kunimoto_JATIS_2020,Kunimoto2020SPIEDemo,Kunimoto2020SPIEDesign}.
The multi-layer sapphires will be glued by using a sodium silicate solution.
The first preliminary tests with a polished $4\,\textrm{mm}\times4\,\textrm{mm}$ surface show that the mechanical strength is $>20$\,MPa, which is sufficiently high to mechanically treat the multi-layer sapphire stack as one bulk sample under the launch impact~\cite{Toda2020SPIE}.
We also test if there is any effect to the millimeter-wave transmittance with this bonding scheme using a sapphire sample with a diameter of 100\,mm, and we did not find any effect within the noise level of the measurement, equal to 2\,\% of the signal. We are continuing to address this evaluation with a larger diameter sample and high accuracy millimeter-wave characterization. 
The AHWP is the first optical element of the LFT and it is held to the telescope structure using both the launch lock system and the cryogenic holder mechanism. 
The launch lock will be released soon after launch.
 The cryogenic holder mechanism is used to maintain the rotor in place during launch and cool-down, before the rotor thermalizes and levitates thanks to the high-temperature superconductor bulks. It also functions as a heat path that thermalizes the rotor to the operating temperature of 4.8\,K. 
The combination of the SMB and the synchronous motor achieves fully contact-free rotation to minimize the heat dissipation from physical contact, as well as giving no wear-and-tear during the mission. The required AHWP temperature during science operations has to be less than 20\,K in order to reduce thermal emission from the AHWP itself.
The corresponding maximum heat dissipation that the rotor can accept is ${<}\,1$\,mW.
Despite having no physical contact, a source of heat dissipation can still come from magnetic friction.
The key development items are to achieve an uniform magnetic field of the rotor magnet, along with lightweight mechanical and thermal design, without using metal to minimize the eddy currents.
The status of this development is detailed in Ref.~\cite{Sakurai2019JPCS}.

The absolute and relative angular positions of the AHWP with respect to the instrument frame are monitored by using a cryogenically compatible optical encoder, which consists of an LED and a photo-diode.
We developed an angular encoder readout system, which consists of an optical chopper between the LED and photodiode. 
The encoder from the photodiode is read using an \gls{fpga} with an \gls{adc} and we have demonstrated reconstruction of the position angle to better than 1\,arcmin with the dedicated algorithm. 
This analysis, together with identifying the source of the position angle uncertainties, are reported in Ref.~\cite{Sugiyama2020SPIE}.
The absolute polarization angle referenced to sky coordinates will be calibrated by some other means, e.g., nulling $C_{\ell}^{EB}$ or using polarized astrophysical sources (see Sect.~\ref{ss:plm_calib}). 

Mitigation of systematic effects for low-$\ell$ reconstruction makes the PMU the key to the success of the mission.
Current efforts shall lead the design of the demonstration and engineering models by working with companies that are qualified to procure the flight hardware.

\subsection{Mid- and High-Frequency Telescopes} 
\label{ss:plm_mhft}

\subsubsection{Overview}
\label{sss:MHFT_Overview}
\begin{figure}[htbp!]
\centering
\includegraphics[height=80mm]{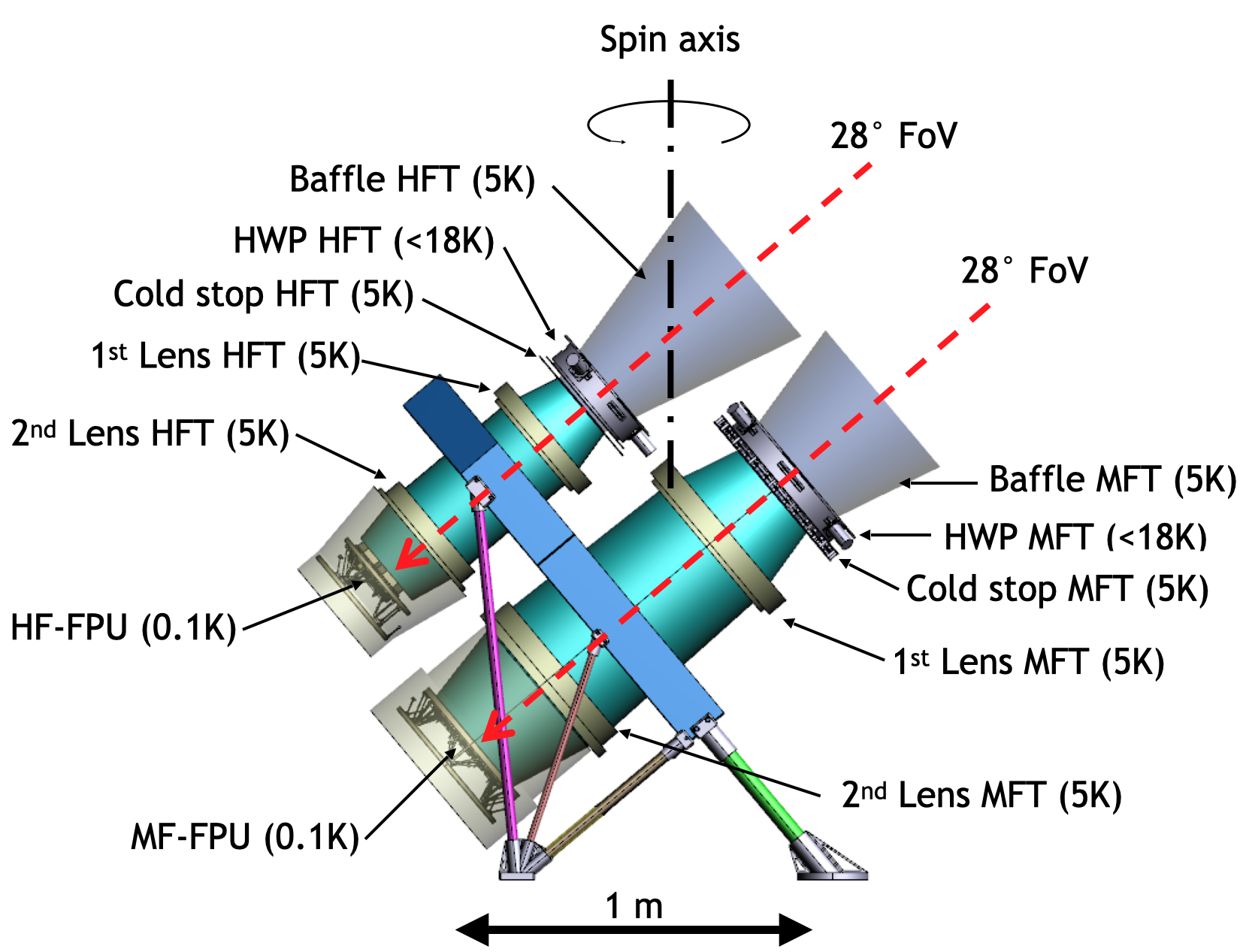}
\caption{Overview of the Mid- and High-Frequency Telescopes.}
\label{fig:mhft}
\end{figure}

The optimization of the instrumental design over the mid- and high-frequency range of \litebird, spanning from 89\,GHz to 448\,GHz, led to a design with two fully refractive telescopes, as shown in Fig.~\ref{fig:mhft}.
The frequency bands of the \gls{mft} range from 89 to 224\,GHz, and from 166 to 448\,GHz for the \gls{hft} (see 
 Ref.~\cite{Montier2020SPIE} for more details).  When considered together, the
 two telescopes are referred to as the \gls{mhft}.

The choice to use two distinct fully refractive telescopes
has been mainly driven by the constraints put on the \gls{hwp} material and the \glspl{arc}, imposed by the broad frequency coverage and the limited mass budget and volume allocation. 
Each telescope features its own polarization modulator, in order to mitigate the bandwidth limitations of the HWP mesh technology. The global design specifications of the \gls{mft} and \gls{hft} are summarized in Table~\ref{tbl:MHFT-specification}.
This design benefits from the broad expertise gained on many current and upcoming sub-orbital CMB experiments, such as BICEP/Keck~\cite{Ade_2015_BKIV}, SPIDER~\cite{Gualtieri2018}, LSPE~\cite{Lamagna2020a}, and Simons Observatory~\cite{Gudmundsson:21}. With these very compact on-axis designs, the \gls{mft} and \gls{hft} match the volume and weight constraints, mitigate straylight issues, and split the entire frequency range into two bands. This choice also offers more flexibility for an optimal design of the separate filtering chains. Dealing with two smaller compact telescopes 
 will considerably simplify the \gls{ait} and \gls{aiv} phases of the project and the ground calibration activities.
On the other hand, careful design and modeling are needed for the \gls{pp} lenses and transmissive meta-materials, see
Sect.~\ref{sss:MHFT_Optical_design}.
Finally \gls{mft} and \gls{hft} assemblies are held on a single mechanical structure cooled down to 4.8\,K, which contains the sub-4.8\,K parts of the cryo-chain.

\begin{table} [htbp!]
\centering
\caption{Global design specifications of MFT and HFT.
}
\begin{tabular}{|l|c|c|}
\hline
                   & MFT    & HFT \\
\hline\hline                   
Frequency coverage       & 89--224\,GHz & 166--448\,GHz\\
                        & 5 observation bands  & 5 observation bands  \\
\hline
Detectors number      & 2074        & 1354 \\
\hline
Aperture diameter & 300\,mm     & 200\,mm \\
\hline
Rotational HWP speed     &    39\,rpm  &  61\,rpm   \\
\hline
Field of view    & \multicolumn{2}{c|}{\O $28^{\circ}$ }\\
\hline
Angular resolution & \multicolumn{2}{c|}{ $<$ 30\,arcmin } \\
\hline
Volume allocation at 5\,K          & \multicolumn{2}{c|}{$1700\,\mathrm{mm}\times1400\,\mathrm{mm}\times 750\,\mathrm{mm}$ } \\
\hline
Mass budget at 5\,K  &\multicolumn{2}{c|}{100\,kg }  \\
\hline
\end{tabular}
\label{tbl:MHFT-specification}
\end{table}

\subsubsection{Optical Design}
\label{sss:MHFT_Optical_design}

\begin{figure}[htbp!]
    \centering
    \begin{tabular}{cc}
    \begin{minipage}{0.47\hsize}
    \includegraphics[width = \hsize ]{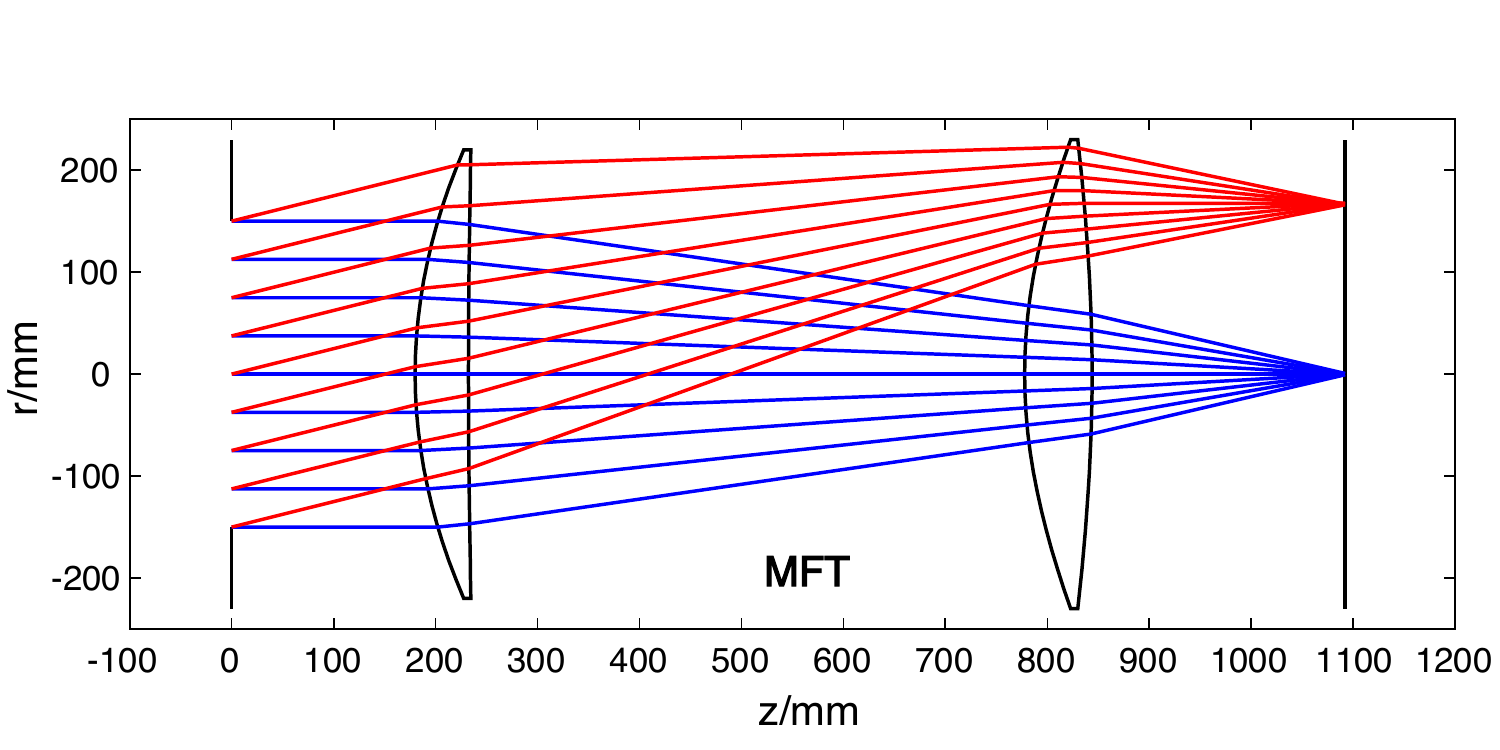}
    \end{minipage} &
    \begin{minipage}{0.47\hsize}
    \includegraphics[width = \hsize ]{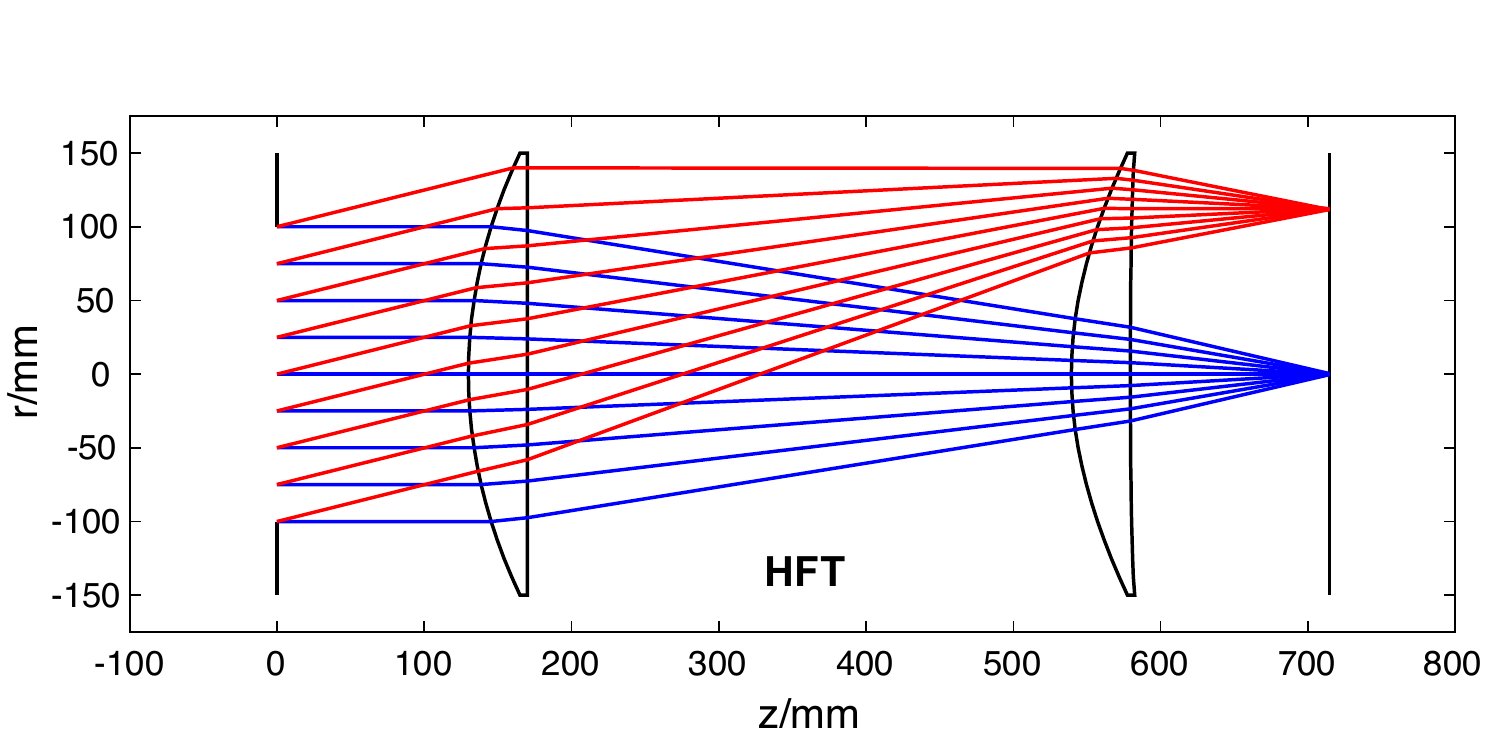}
    \end{minipage}
    \end{tabular}
    \caption{Ray-tracing diagrams for MFT (left) and HFT (right). The on-axis and off-axis fields ($14^\circ$) are the blue and red rays, respectively. The telescope aperture is located at $ z = 0 $, with a diameter of 300\,mm and 200\,mm for MFT and HFT, respectively. The lens material is polypropylene.}
    \label{fig:opt_config}
\end{figure}

The optical configurations of \gls{mft} and \gls{hft} are shown in Fig.~\ref{fig:opt_config}.
Both telescopes have f/2.2, with an aperture stop (300\,mm and 200\,mm in diameter, for \gls{mft} and \gls{hft}, respectively) located skywards of the two lenses. The stop also serves as a convenient location for the transmissive HWP, 
which is located very close to (but slightly sky-side of) the stop.

Both systems are diffraction-limited to the upper edge of their respective bands, across a $28^\circ$ diameter field of view, and have telecentric focal planes.
Initial physical optics simulations for the \gls{mft} also indicate uniform beam shapes, with low cross-polarization and ellipticity smaller than 0.03, across the field of view. The \gls{hft} optical performance is very similar. 
A more detailed review of the performance analysis of the \gls{mft} and the \gls{hft} systems is discussed in Ref.~\cite{Lamagna2020SPIE}. 

An axisymmetrical optical design for the \gls{mft} and \gls{hft} has been adopted to remain simple with excellent performance, while being relatively relaxed in terms of optical tolerances. 
However, this creates optical modeling challenges due to refractive surfaces and metamaterials.
Some performance validation is needed to address both the issues related to intrinsic frequency-, polarization-, and temperature-dependent properties of the refracting materials. 
It will be essential to
understand the higher-order optical coupling effects arising from a broad set of non-idealities, like imperfect dielectric-to-vacuum matching at the optical interfaces, imperfectly-absorbing tube surfaces and focal planes, scattering off internal surfaces, and thermal radiation pickup from the payload environment. 

\paragraph{Half-wave Plate} While the electromagnetic behavior of the mesh HWP is very well modeled and validated at the subsystem level, the presence of a mesh HWP along the optical path to the focal plane makes it necessary to model the whole system with accuracies beyond what is achievable with currently available simulation software. The need to capture the impact of imperfect and frequency-dependent phase shifts, and the effect of non-normal incidence on the final performance of the device creates an extremely complex modeling effort. The HWP imperfections might modulate the background and lead to a strong synchronous signal peaked at harmonics of the HWP rotation frequency. Such an \gls{hwpss} has been observed using Maxima~\cite{Johnson_2007}, EBEX~\cite{chapman2014}, POLARBEAR~\cite{POLARBEAR2016} and NIKA~\cite{ritacco2017},
each of which used a 
continuously rotating HWP, similarly to the \litebird\ case. 
The parasitic HWPSS can be modeled and subtracted in the data analysis process~\cite{ritacco2017}. However any residual could act as an additional noise in the polarization data and increase the total error budget. In order to reach the required accuracy it will be crucial to characterize and model this at the system level as a result of beam propagation through the whole optical chain, including the HWP, quasi-optical filters, optical elements and focal-plane beam formers.
This effort is in progress. 

\begin{figure}[htbp!]
    \centering
    \begin{tabular}{cc}
    \begin{minipage}{0.34\hsize}
    \includegraphics[width=\hsize]{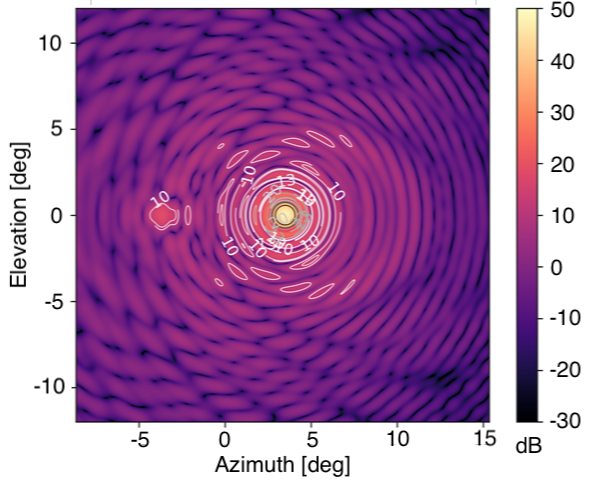}
    \end{minipage}&
    \begin{minipage}{0.61\hsize}
    \includegraphics[width=\hsize]{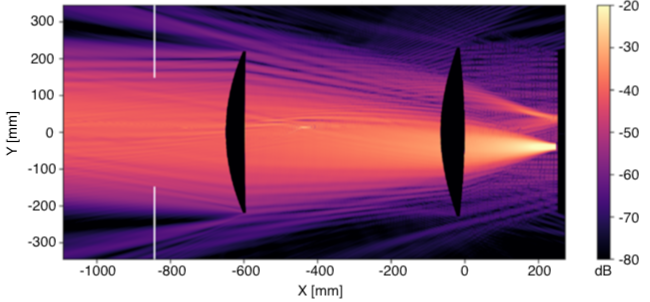}
    \end{minipage}
    \end{tabular}
    \caption{\textit{Left}: Example of optical ghosting at 90\,GHz, from the combined effect of a non-ideal ARC, (in this case a single-layer ARC optimized for 150\,GHz,) on \gls{mft} lenses and imperfect absorption on the focal plane, as seen in the telescope far field. A secondary beam-like feature can be shown out of the main beam pattern in this example. \textit{Right}: Corresponding power distribution in the plane of the telescope, as predicted by these simulations.}    
    \label{fig:ghost_examples}
\end{figure}

\paragraph{Optical Ghosts}
Ghosts are detectable effects arising from a variety of non-idealities in the optical system, mostly due to multiple internal reflections that partially re-focus the light away from the intended location on the focal plane.
Image and polarization artifacts due to this effect must be clearly identified, modeled and compared with the actual behavior of the optical system. A systematic study on these effects needs to be carried out to evaluate their impact on the final performance of the MHFT instruments.
As reported in Ref.~\cite{Lamagna2020SPIE} a preliminary study was performed at 90\,GHz on the \gls{mft} to characterize
multiple reflections off optical surfaces. 
This analysis showed a diffuse contamination from ghosting across the focal plane at the level of $-40$\,dB of the peak level of the nominal focused image, with resulting artifacts in the reconstructed image at scales of a few degrees on the sky (see Fig.~\ref{fig:ghost_examples}).

\subsubsection{Sub-systems}
\label{sss:plm_mhft_subsystems}
We detail below the various sub-systems of the two telescopes, starting from the first optical component encountered, the rotating HWP, down to the mechanical structure.   

\begin{figure}[ht]
\centering
\includegraphics[width=55mm,height=45mm]{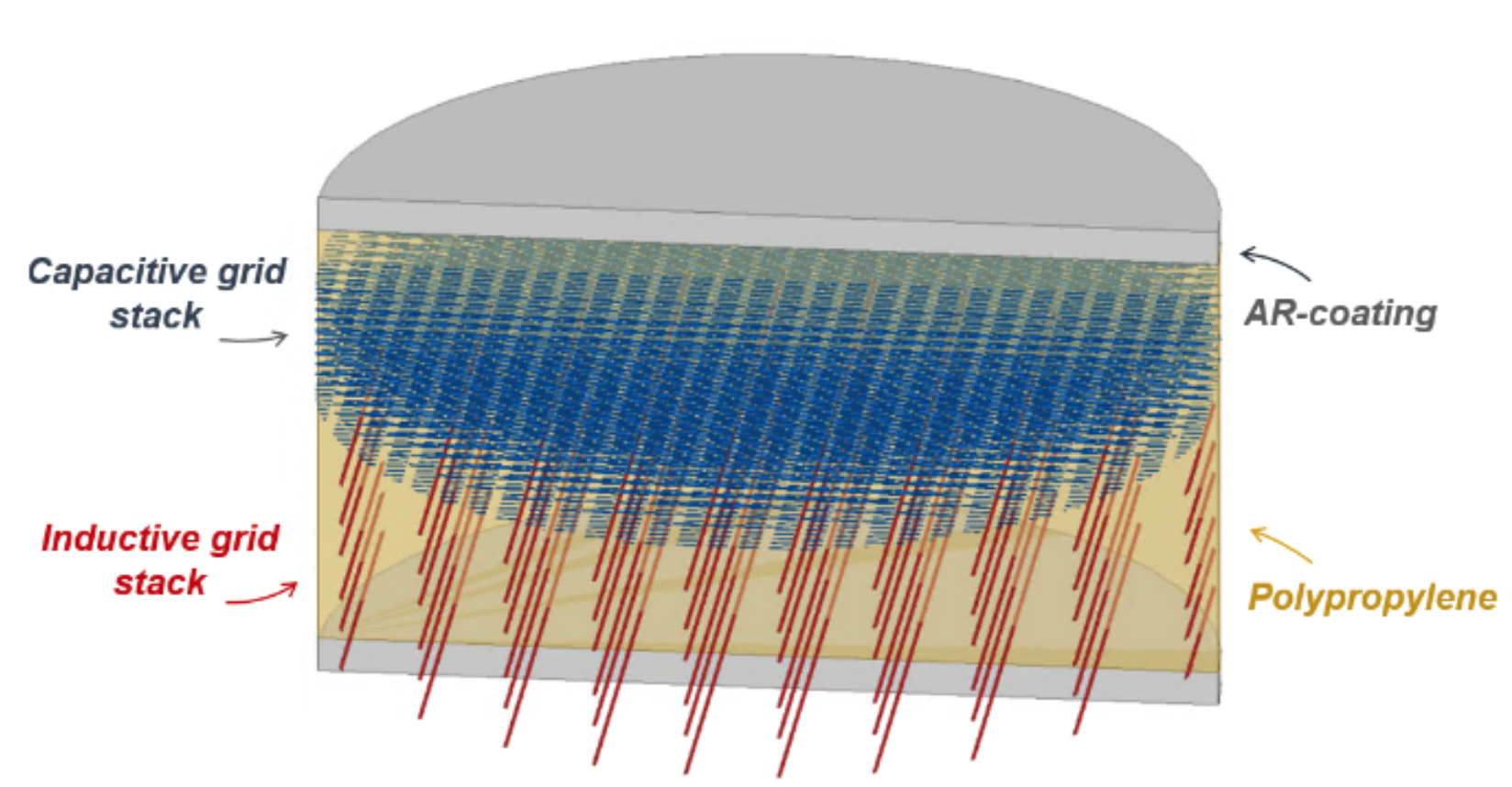}
\includegraphics[width=65mm,height=40mm]{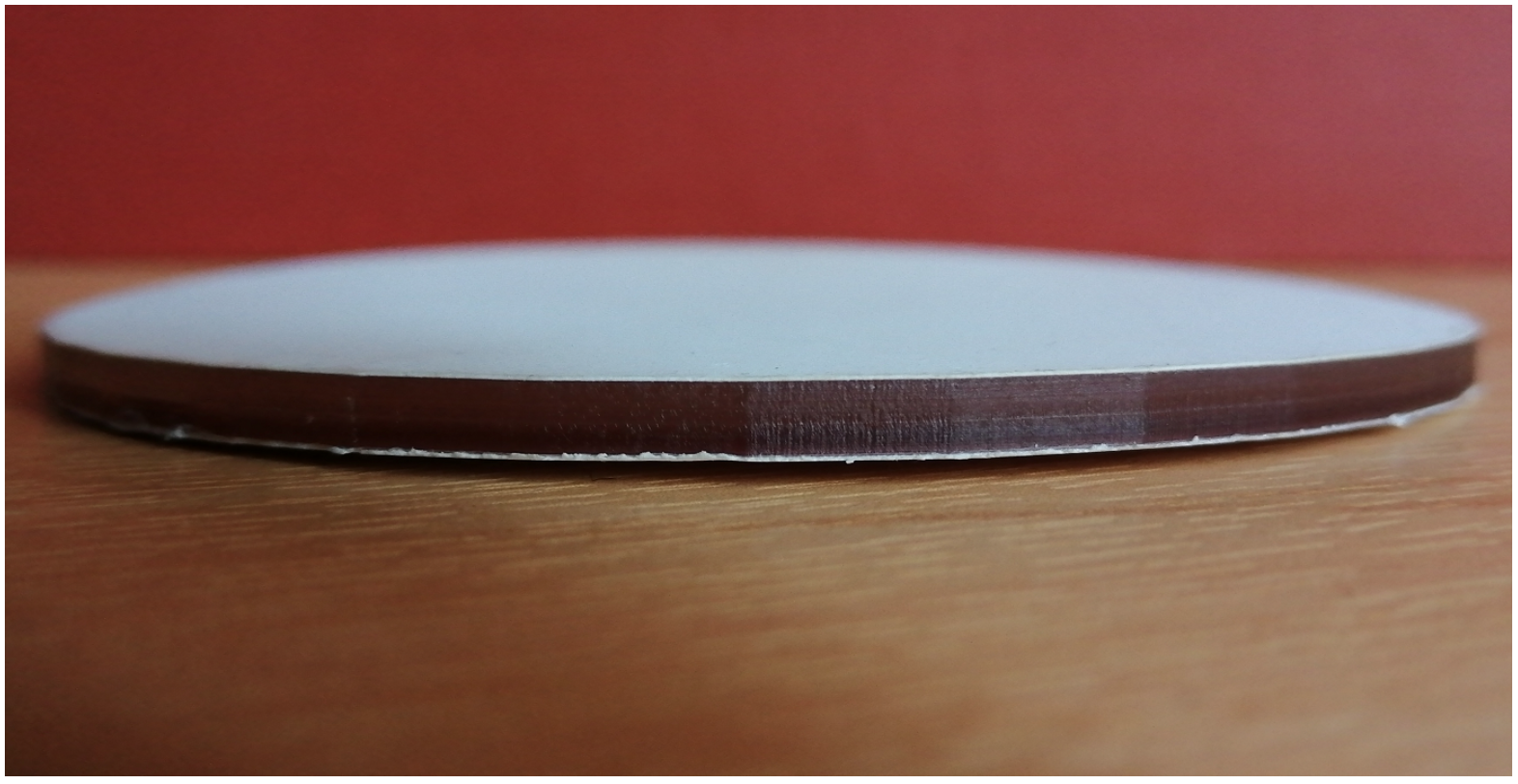}
\caption{
\textit{Left}: Model showing a small portion of a mesh-HWP with the embedded stacks of capacitive and inductive grids.  \textit{Right}: Mesh HWP prototype working between \SI{100}{} and \SI{300}{\giga\hertz} and being around $\SI{3}{\milli\meter}$ thick.}
\label{fig:HWP}
\end{figure}

\paragraph{PMU design and development}
The polarization modulator units (PMUs) for the \gls{mft} and \gls{hft} are based on spinning HWPs. 
The mesh-filter technology~\cite{Pisano1} provides ultra-light transmissive HWPs  and has been adopted for both telescopes. While this technology offers a large gain in mass compared to sapphire HWPs, the current mesh-HWP bandwidth would not cover the whole LFT bandwidth, whereas it can more easily accomodate the \gls{mft} and \gls{hft} bandwidths. Moreover, facilities already in place allow us to manufacture large diameter devices (up to 60\,cm diameter). The mesh HWPs consist of polypropylene-embedded anisotropic metal grids, AR-coated on both sides (see Fig.~\ref{fig:HWP}). These devices emulate the behavior of birefringent materials by means of orthogonally oriented stacks of capacitive and inductive grids~\cite{Pisano2,Pisano3}. The expected transmission coefficients and the polarization modulation efficiencies across the MHFT bands are, respectively, approximately 95\,\% and greater than 95\,\%,  for \gls{mft} and \gls{hft}, respectively. The estimated weights of the \gls{hft} and \gls{mft} waveplates are approximately $\SI{100}{\gram}$ and $\SI{400}{\gram}$, respectively. A detailed description of the mesh-HWP design and manufacture can be found in Ref.~\cite{Pisano4}.

The rotation mechanisms have the same design for both \gls{mft} and \gls{hft}, and shall meet several stringent requirements, in terms of mass ($<\SI{20}{\kilogram}$), dimension ($<\SI{200}{\milli\meter}$ and $<\SI{300}{\milli\meter}$), stiffness, power dissipation ($<\SI{4}{\milli\watt}$), HWP temperature ($<\SI{20}{\kelvin}$~\cite{Columbro2019}), and \gls{trl}, for the levitation, driving, and gripping mechanisms, as well as position encoder.

\begin{figure}[htbp!]
\centering
\includegraphics[width=55mm,height=50mm]{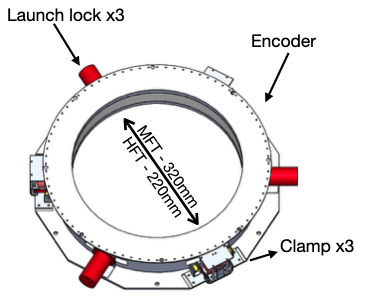}
\includegraphics[width=65mm,height=50mm]{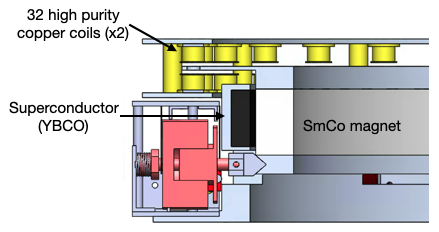}
\caption{
\textit{Left}: Overview of the PMU design. The coil rings are removed to show the encoder on the edge of the rotor.  \textit{Right}: PMU cross-section.
}
\label{fig:PMU_design}
\end{figure}

A very low friction superconducting magnetic bearing~\cite{Johnson2017} based on the magnetic levitation between a YBCO superconductor ring (stator) and an SmCo (Samarium-Cobalt) permanent magnet ring (rotor) hosts the HWPs (Fig.~\ref{fig:PMU_design}).

The rotation is driven by an electromagnetic motor, based on the interaction between eight small cylindrical SmCo magnets placed on the edge of the rotor and 64 high purity copper coils placed on the stator.
The motor magnets and the hysteresis in the main magnet assembly are the main contributions to the power load on the 5-K stage and must be minimized as much as possible. In addition, the eddy currents induced by the motor coils heat up the rotor, which can be cooled down only by radiation, a process that is not very efficient at cryogenic temperatures. The total heat load expected {\it from both mechanisms\/} is 2.8\,mW, which is within the requirement, with a 40\,\% margin.

The rotor temperature is monitored by capacitive sensors~\cite{PdB_levitation_measurement2020} (3\,\% accuracy), which are also used to monitor the levitation height and the wobble of the rotor (\SI{10}{\micro\meter} resolution). 
A thermal model built by Comsol Multiphysics\footnote{\url{https://www.comsol.com}} shows that after a slow rise the equilibrium temperature should be approximately $\SI{16}{\kelvin}$ for both modulators, assuming HWP emissivities of 0.02 and 0.03 for \gls{mft} and \gls{hft}, respectively, while the assumed emissivity of aluminum is 0.5,  achievable with a blackened surface~\cite{Columbro_SPIE}.

The position readout system uses an optical encoder, with 64 precision slits in the rotor ring periphery. The light from a warm temperature modulated LED is transferred to the rotor via an optical fiber, and the light transmitted through the slits is transferred back to a  warm temperature detector via another optical fiber. This configuration minimizes the heat load on the different stages of the cryogenic system~\cite{2011AA...528A.138S}. The expected angular encoding accuracy is better than 0.1\,arcmin and can be improved through a Kalman filter using the high rotation stability ($<\SI{1}{\milli\hertz}$) and the inertia of the system.
A simple and reliable clamp/release system~\cite{Columbro2018} (with zero power dissipation while holding the rotor, and zero power dissipation when the rotor is released) has been developed. It is expected to be used only once at the beginning of the flight, nevertheless the system was designed to be able to clamp and release the rotor as many times as needed.

\paragraph{Lens components}
The baseline design employs lenses manufactured from polypropylene, and anti-reflection coating with a matching layer of porous \gls{ptfe}. This coated lens technology has been well characterized and validated through an extensive ESA-funded Technical Research Programme. 
Lenses of similar sizes have been previously manufactured by members of the project team, and successfully deployed on suborbital CMB polarimetry experiments. 
In addition, porous PTFE coated lenses have also been deployed in space on the {\it Herschel}-SPIRE satellite instrument.
As such, the coated lenses demonstrate a high level of technical maturity, operating in a cryogenic vacuum environment. However, over the coming months, we will additionally undertake a significant program of precision characterization of these materials at cryogenic temperature (losses, stress-induced birefringence, etc.).

\paragraph{Absorbers}
While far-infrared filters will mainly be used to prevent out-of-band radiation from propagating into the telescope, optical absorbers will also be used to prevent or mitigate spurious in-band reflections across the optics tubes. The need for a reliable, light, thermally uniform, space-qualified material has been recognized as a driver for a dedicated study, meant to investigate a set of candidate absorbing materials and/or metamaterial structures. This study will compare traditional absorber candidates such as Thomas-Keating tiles or Eccosorb HR10/AN72 to micromachined, 3D-printed, and injection molded metamaterials \cite{Wollack2014, Wollack2016, Petroff2019, Xu2021}. The study will include detailed laboratory measurements and commensurate electromagnetic simulations of the candidate absorbers spanning the entire MHFT frequency range. The results of this work will allow us to perform a tradeoff study to select the best solution for absorbers.

\paragraph{Filtering scheme}
Although band definition is achieved via on-chip filtering, additional optical filters are required in order to control the out-of-band rejection level, to protect the detectors from straylight and to control the thermal environment.  For MFT we are proposing deployment of a chain of four low-pass filters positioned at the 4.8-K, 2-K, 300-mK, and 100-mK stages.  It is noted that the HWP also acts as a low-pass filter and we have the option of adding a high-pass element at the detectors should this be required.  The average in-band transmission over the frequency range of the MFT (or HFT) for each of these elements can be  targeted to be 95\,\%.

\paragraph{Mechanical structure}
The mechanical structure is composed of two main parts: the two telescope tubes; and an exoskeleton. The telescope tubes are designed to hold the optical elements and the various subsystems, and to ensure optical alignment for each of the two telescopes. 
The exoskeleton holds the two tubes, connects the telescopes to the payload module and provides the thermal link to the cryochain. All the mechanical elements are made of material compliant with the optical, mechanical, and thermal constraints, such as aluminum and \gls{cfrp}.

\begin{figure} [htbp!]
\begin{center}
\begin{tabular}{cc} 
\includegraphics[
 height=5cm]{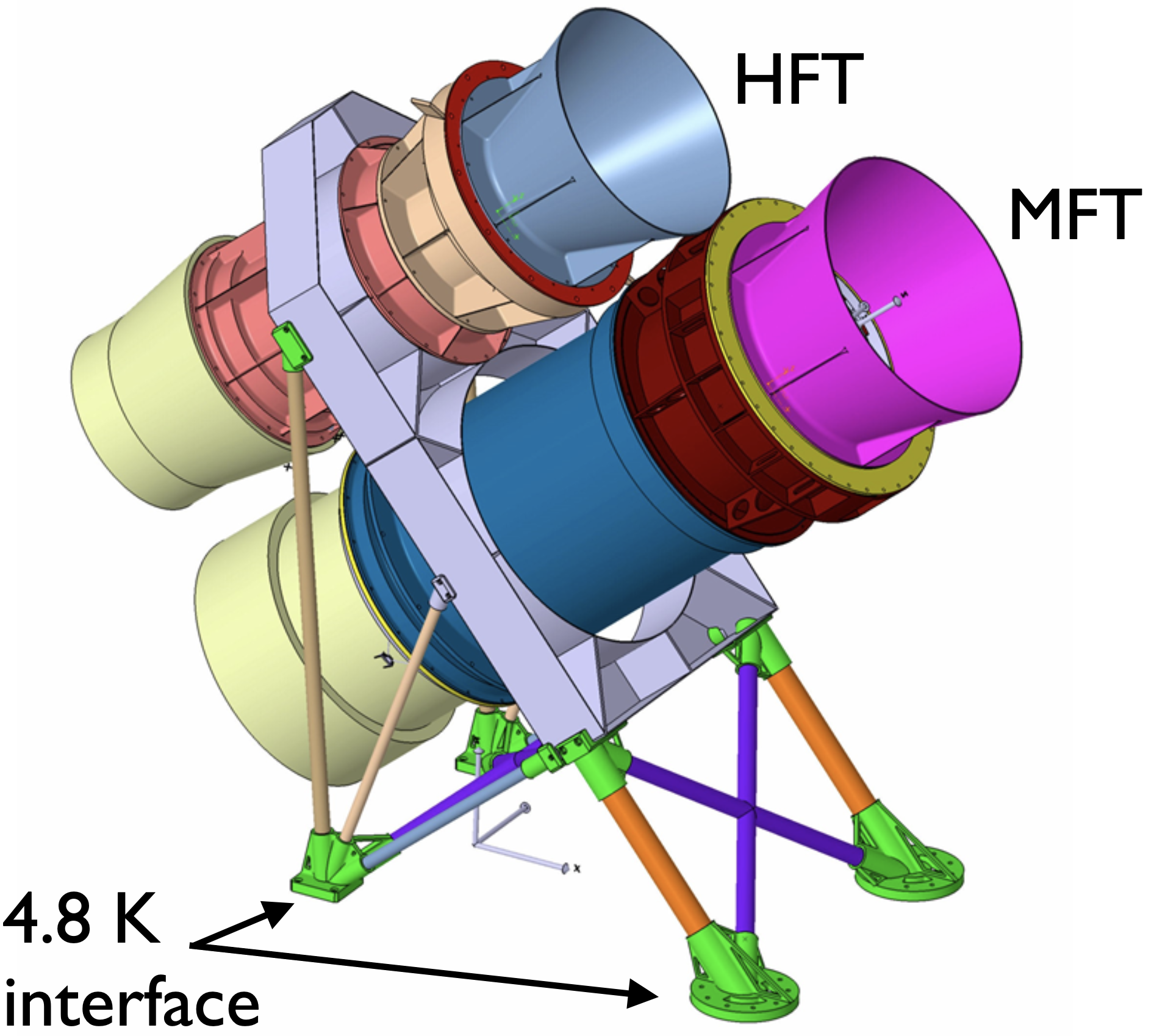}& 
  \includegraphics[
   height=5cm]{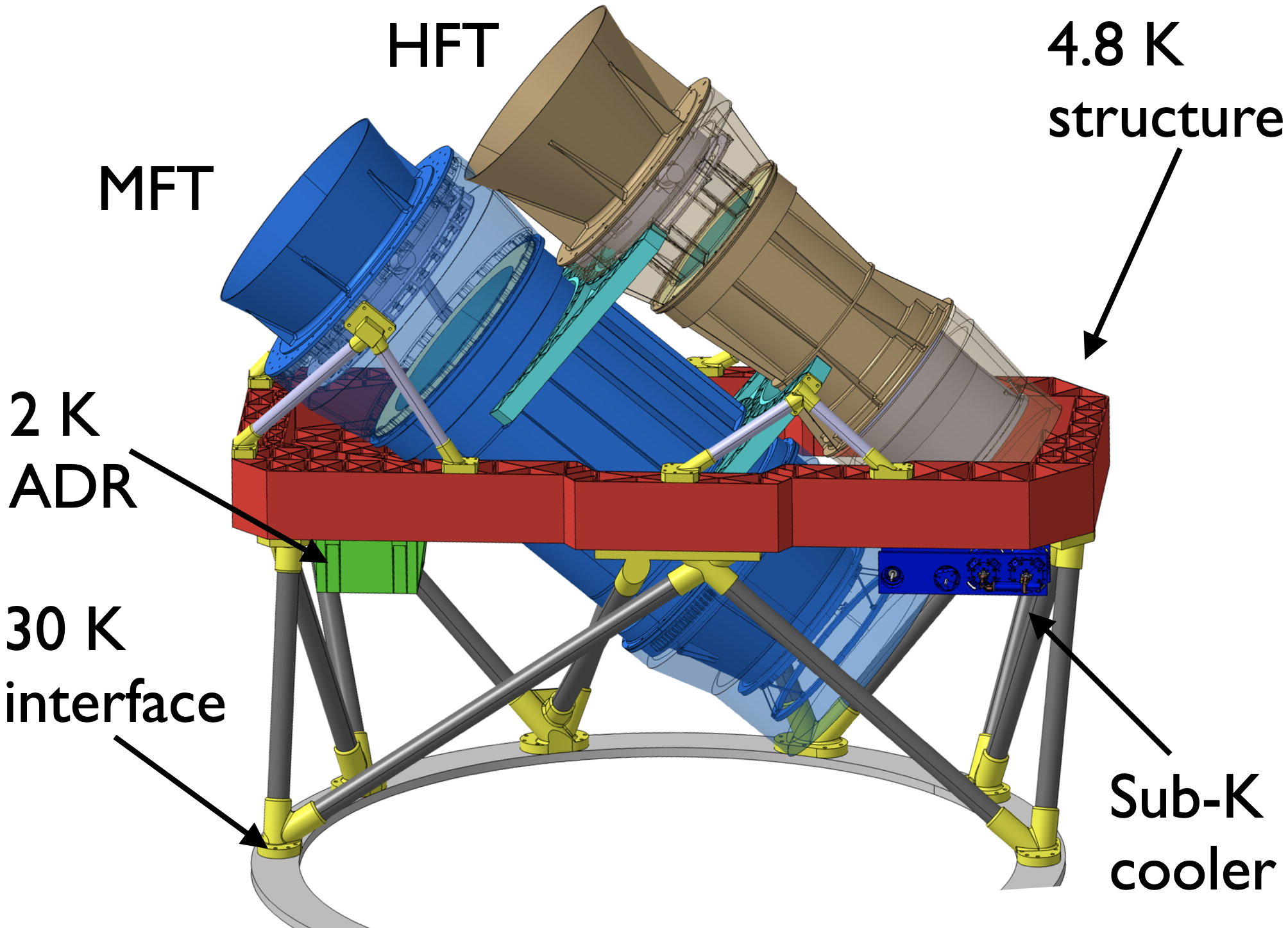}
\end{tabular}
\end{center}
\caption{Two options of the MHFT mechanical structure design based on a mechanical interface with the satellite at the 4.8-K stage (Left), and at the 30-K stage (Right).}
\label{fig:Meca1}
\end{figure}

A first series of iterations on the mechanical design 
has been performed to optimize the mechanical structure, taking into account the various constraints such as minimum eigen-frequencies, total mass, launch load, and thermal conduction (design shown on the left side of Fig.~\ref{fig:Meca1}). 
In the initial design, the mass of the  MHFT is about
118\,kg without margins, which is over the allocated 100\,kg value. This specification on the mass is mostly driven by the parasitic heat load on the 4.8-K stage due to the thermal conductance of the satellite mechanical structure holding both LFT and MHFT. 

To deal with the stringent constraints on the total mass budget at 4.8\,K, we decided to study another option, which consists of moving the mechanical interface with the satellite from the 4.8-K stage to the 30-K stage of the PLM.
The total mass budget has been revised in the framework of this second option. The heavy 4.8-K interface ring, shared by LFT and MHFT in the first option design, has now been replaced by two 30-K to 4.8-K optimized cryomechanical structures made of \gls{cfrp}. 
We designed the 30-K to 4.8-K cryomechanical structure with particular attention to the thermal conductance. The design of the MHFT tubes is similar to the first option (design shown in the right panel of Fig.~\ref{fig:Meca1}).
This solution allows us to drastically reduce the mass of the 4.8-K interface with the satellite, while minimizing the parasitic heat load on the coldest stages.
The study of this mechanical structure design shows that the constraints on the eigen-frequencies and on the parasitic heat-load are satisfied.

\subsection{Detection Chain} 
\label{ss:plm_detection}

\litebird\ will be equipped with 1030 multi-chroic pixels sensitive to polarization, totaling 4508 \gls{tes}s, distributed over the three focal planes of \gls{lft}, \gls{mft}, and \gls{hft}. Detector specifications for each of the \litebird\ bands are show in Tables~\ref{tbl:bolospecsgeneral} and \ref{tbl:bolospecsbands}. 
The detection chains of the three telescopes have been designed using a very similar architecture that is described in this section for the three telescopes. It includes focal planes and associated readout electronics, both cold and warm parts.  More details on the design and fabrication of the detector modules can be found in Ref.~\cite{Westbrook2020SPIE}.

\begin{table}[htbp!]
\centering
\caption{Summary of the common optical and bolometric design goals of the \lb\ detectors.  We expect that there will be little deviation from these goals during the development of the \glspl{fpu} for \lb.}
\label{tbl:bolospecsgeneral}
\begin{tabular}{|c|c|}
\hline
Parameter & Design value \\
\hline \hline
\gls{tes} normal resistance & 1.0 \textbf{}\\
\gls{tes} operating resistance & 0.6--0.8\,$\Omega$ \\
Parasitic series resistance & 0.05--0.2\,$\Omega$ \\
\gls{fpu} ($T_{\rm b}$) & 0.100\,K\\
Transition temperature ($T_{\rm c}$) & 0.171\,K \\
Cross wafer $T_c$ variation & $\leq$ 7\,\% \\
Minimum operating power & $2.5\times$ optical power \\ 
Pixel in-band optical efficiency & $\geq70$\,\% \\
Thermal carrier & Phonon ($n=3$)\\
Intrinsic time constant ($\tau_0$) & 33\,ms \\
Loopgain during operation & $\geq10$ \\
Common 1/$f$ knee & $\leq20$\,mHz\\
\gls{fpu} lifetime & $\geq3$\,years \\
On-sky end-to-end yield & $\geq$ 80\,\%\\
\hline
\end{tabular}
\end{table}

\begin{table}[htbp!]
\centering
\caption{Optical specifications for the \litebird\ detectors. The values listed in this table flow down from the noise specifications listed in Table~\ref{tbl:plm:sensitivities}.}
\label{tbl:bolospecsbands}
\begin{tabular}{|c|c|c|c|c|}
\hline
 \begin{tabular}[c]{@{}c@{}}Pixel\\ ID \end{tabular} &
 \begin{tabular}[c]{@{}c@{}}Frequency\\ {[}GHz{]}\end{tabular} & 
 \begin{tabular}[c]{@{}c@{}}Bolometer \\ count\end{tabular} &
  \begin{tabular}[c]{@{}c@{}}Bandwidth \\ $\Delta_{\nu} / {\nu}$ \end{tabular} &  
  \begin{tabular}[c]{@{}c@{}}$P_{\rm opt}$  \\ {[}pW{]}\end{tabular}
  \\ \hline\hline
  LF1                       & {\pixelred{40/60/78}}     & 48   & 0.30 / 0.23 / 0.23 & 0.358 / 0.300 / 0.303  \\
  LF2                       & {\pixelyellow{50/68/89}}  & 24   & 0.30 / 0.23 / 0.23 & 0.386 / 0.302 / 0.311  \\
  LF3                       & {\pixelgreen{68/89/119}}  & 144  & 0.23 / 0.23 / 0.30 & 0.367 / 0.363 / 0.449  \\
  LF4                       & {\pixelblue{78/100/140}}  & 144  & 0.23 / 0.23 / 0.30 & 0.367 / 0.356 / 0.440  \\
  MF1                       & {\pixelred{100/140/195}}  & 366  & 0.23 / 0.30 / 0.30 & 0.411 / 0.463 / 0.386  \\
  MF2                       & {\pixelyellow{119/166}}   & 488  & 0.30 / 0.30        & 0.496 / 0.416          \\ 
  HF1                       & {\pixelred{195/280}}      & 254  & 0.30 / 0.30        & 0.782 / 0.486          \\
  HF2                       & {\pixelgreen{235/337}}    & 254  & 0.30 / 0.30        & 0.603 / 0.384          \\
  HF3                       & {\pixelblue{402}}         & 338  & 0.23               & 0.290                  \\
  \hline
\end{tabular}
\end{table}

\subsubsection{Focal-Plane Units}
\label{sss:detection_fpu}
\litebird\ will deploy for each of the telescopes an \gls{fpu}, 
which consists of the following components: 
(i) a \gls{fps} providing thermal insulation from the \gls{jt4} and thermal connections to the \gls{jt2} and \gls{skadr}, which provide the cooling power to the intermediate temperature stages, cooled down to 1.75\,K, 350\,mK, and 100\,mK; (ii) and a set of individual arrays of detectors, the \gls{fpm}, operated at 100\,mK. 
Each \gls{fpm} consists of a single detector array, optical coupling hardware (lenslet or horn arrays, backshort wafers, etc.), the \gls{cru}, and mechanical structures that hold these parts together and provide an interface to the \gls{fps}.

\subsubsection{Focal-Plane Modules}
\label{sss:detection_fpm}

\begin{figure}[htbp!]
\centering
\includegraphics[width = 1\textwidth]{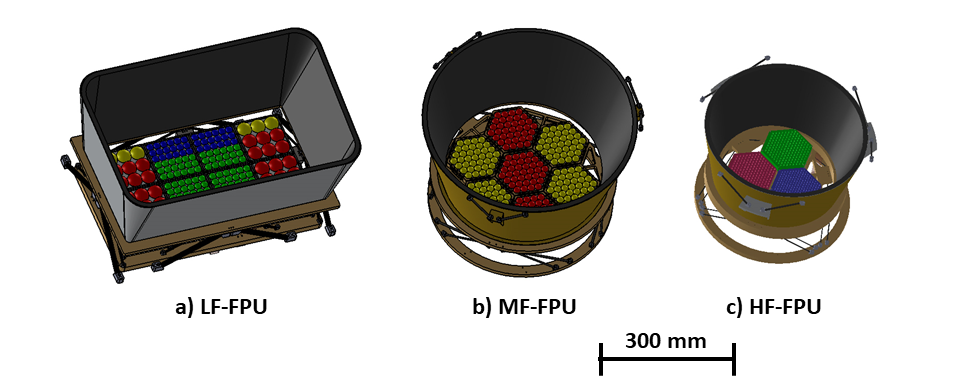}
\caption{Low-, Mid-, and High-Frequency Focal-Plane Units. The LF-FPU rectangular shape is matched to the \gls{lft}'s oblong \gls{fov} using square FPM tiles.   The Mid- and High-Frequency Telescopes both have circular \gls{fov}s and therefore employ hexagonal arrays arranged in a close-packed pattern.   Pixel types are color coded and the details of each can be found in Table~\ref{tbl:focalplanes}.}
\label{fig:focalplanes}
\end{figure}

\begin{table}[htbp!]
\caption{Focal-plane configurations for the \gls{lffpu}, \gls{mffpu}, and \gls{hffpu}. The colors in the frequency column correspond to those in Fig.~\ref{fig:focalplanes}.}
\label{tbl:focalplanes}
\setlength{\tabcolsep}{4pt}
\begin{tabular}{|c|c|c|c|c|c|c|c|}
\hline
Telescope             & \begin{tabular}[c]{@{}c@{}}Detector\\ type\end{tabular}                      & Module                 & \begin{tabular}[c]{@{}c@{}}Frequency\\ {[}GHz{]}\end{tabular} & \begin{tabular}[c]{@{}c@{}}Pixel size \\ {[}mm{]}\end{tabular} & \begin{tabular}[c]{@{}c@{}}Module \\ count\end{tabular} & \begin{tabular}[c]{@{}c@{}}Pixel \\ count\end{tabular} & \begin{tabular}[c]{@{}c@{}}Detector\\ count\end{tabular} \\ \hline\hline
                    &                                                                              &                            & {\pixelred{40/60/78}}                               & 32                                                             &                                                         & 24                                                     & 144                                                      \\
                      &                                                                              &  {\multirow{-2}{*}{LF12}} & {\pixelyellow{50/68/89}}            & 32                                                             & {\multirow{-2}{*}{4}}                                    & 12                                                     & 72                                                       \\
                        &                                                                              &                          & {\pixelgreen{68/89/119}}                       & 16                                                             &                                                         & 72                                                     & 432                                                      \\
{\multirow{-4}{*}{LFT}} & {\multirow{-4}{*}{\begin{tabular}[c]{@{}c@{}}Lenslet/\\ Sinuous\end{tabular}}} &  {\multirow{-2}{*}{LF34}} & {\pixelblue{78/100/140}}                             & 16                                                             & {\multirow{-2}{*}{4}}                                     & 72                                                     & 432                                                      \\
 &  & & & Total LFT & 8 & 180 & 1080\\ \hline
                      &                                                                              & MF1                    & {\pixelred{100/140/195}}                      & 11.6                                                             & 3                                                       & 183                                                    & 1098                                                     \\
 {\multirow{-2}{*}{MFT}} & {\multirow{-2}{*}{\begin{tabular}[c]{@{}c@{}}Lenslet/\\ Sinuous\end{tabular}}} & MF2   & {\pixelyellow{119/166}}                                & 11.6                                                             & 4                                                       & 244                                                    & 976                                                      \\ 
 &  & & & Total MFT & 7 & 607 & 2074\\ \hline
                      &                                                                              & HF1                    & {\pixelred{195/280}}                                & 6.6                                                              & 1                                                       & 127                                                    & 508                                                      \\
                      &                                                                              & HF2                    & {\pixelgreen{235/337}}                                & 6.6                                                              & 1                                                       & 127                                                    & 508                                                      \\
 {\multirow{-3}{*}{HFT}} &  {\multirow{-3}{*}{\begin{tabular}[c]{@{}c@{}}Horn/\\ OMT\end{tabular}}}        & HF3                    & {\pixelblue{402}}                                    & 5.7                                                            & 1                                                       & 169                                                    & 338   \\
  &  & & & Total HFT & 3 & 423 & 1354 \\ \hline
  &  & & & Total mission & 18 & 1210 & 4508\\ \hline
\end{tabular}
\end{table}

\begin{figure}[htbp!]
\centering
\includegraphics[width = 1\textwidth]{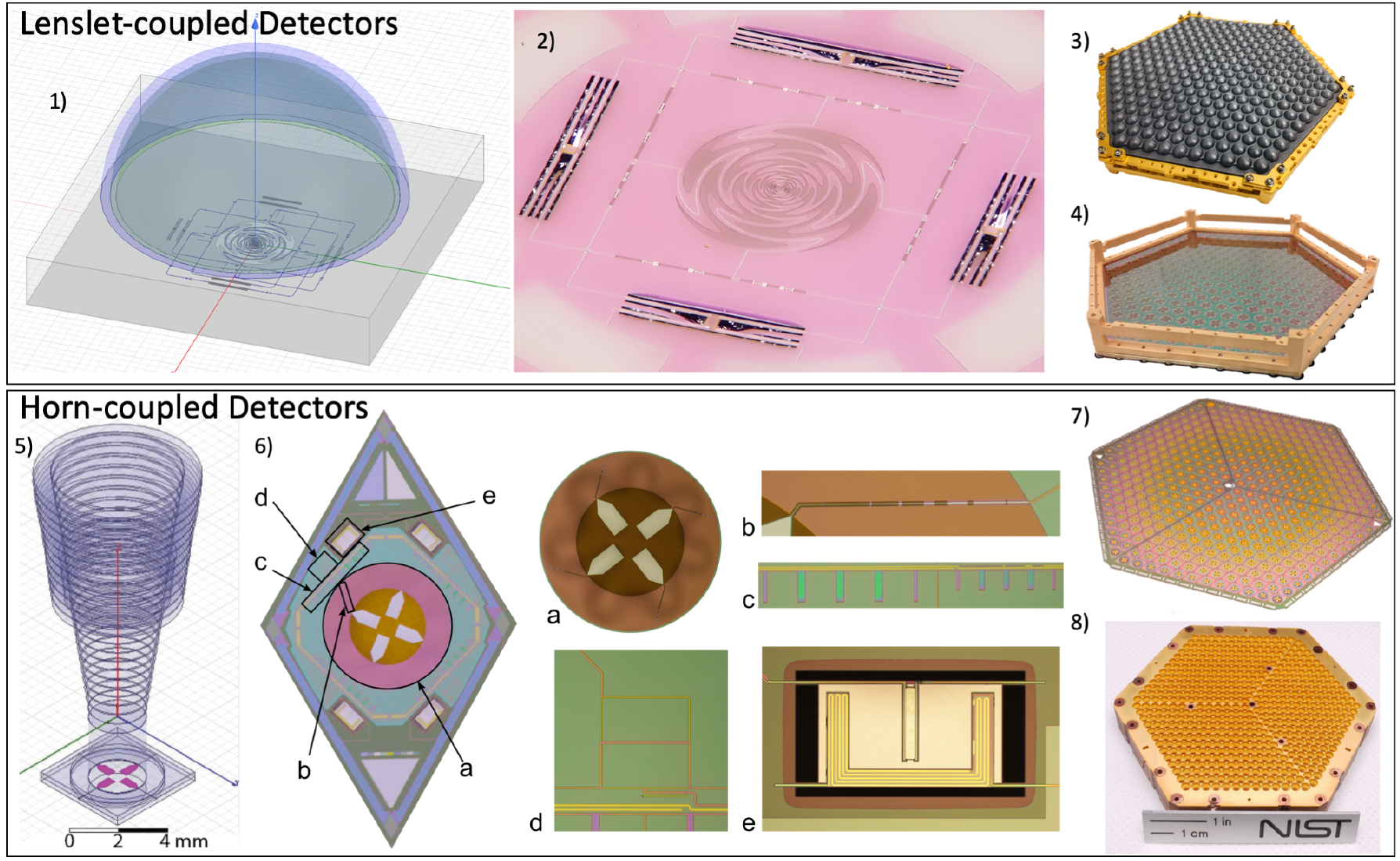}
\caption{\litebird\ detector arrays consist of lenslet-coupled arrays for the LF- and MF-FPUs and horn-coupled detector arrays for the HF-FPU. The individual sub-figures show: (1) a single lenslet-coupled detector; (2) photograph of a microfabricated sinuous antenna-coupled detector; (3) a machined monolithic silicon lenslet array; (4) a microfabricated detector array in a gold-plated detector holder; (5) a single horn-coupled detector; (6) an optical micrograph of a detector with labeled components being (a) planar OMT, (b) coplanar waveguide to microstrip transition, (c) diplexer, (d) 180 hybrid, and (e) TES bolometer; (7) a photograph of a 432-element array of dichroic horn-coupled detectors and mating; and (8) a silicon platelet feedhorn array.}
\label{fig:detector_architecture}
\end{figure}

\begin{figure}[htbp!]
\centering
\includegraphics[width=0.8\textwidth]{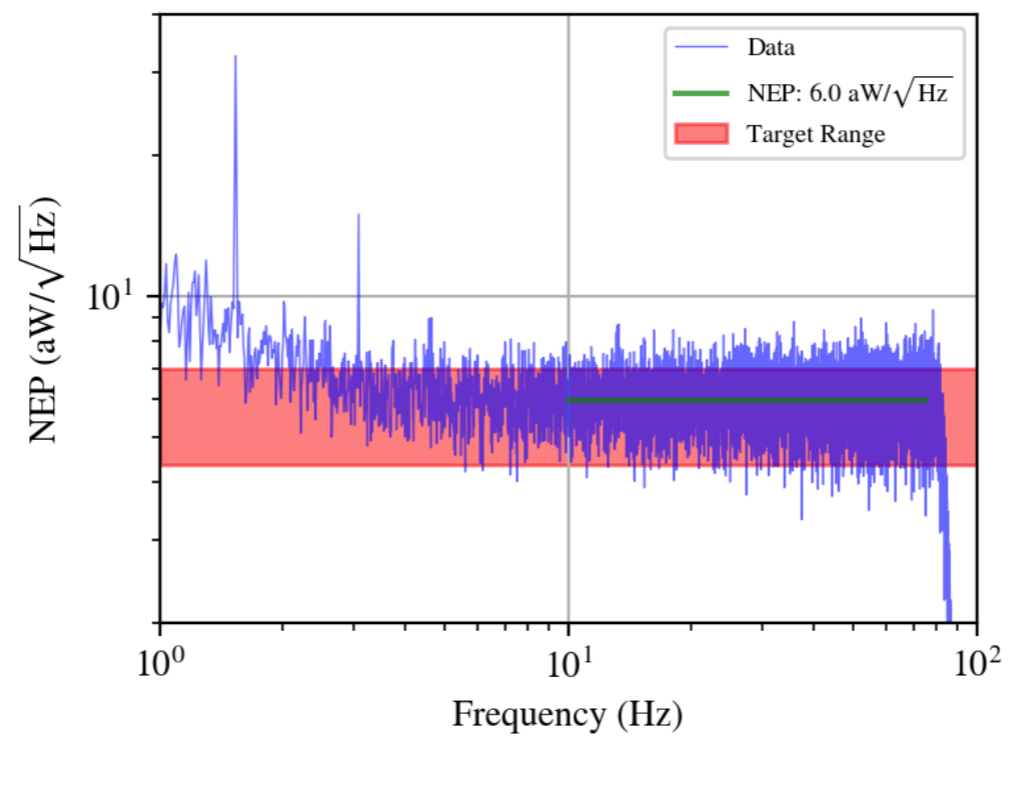}
\caption{Demonstration of a 100-mK \gls{tes} with white noise performance in the \litebird\ target range for readout and phonon noise contributions in the different frequency bands. The 0.5-pW bias power used is consistent with the lowest power band in \litebird.
}
\label{fig:tesnoisedemo}
\end{figure}

The \gls{fpm}s of \litebird\ are filled with arrays of lenslet- or horn-coupled \gls{tes} bolometers fabricated on silicon wafers coupled to a multiplexed readout system.
The telescope designs drive the focal plane layouts; the \gls{lft} \gls{fpu} is rectangular to match the oblong illumination pattern of the crossed-Dragone telescope design, while the \gls{mft} and \gls{hft} \gls{fpu}s are hexagonal arrays that pack the pixels most efficiently into the axisymmetric \gls{mft} and \gls{hft} refractive telescopes, as shown in Fig.~\ref{fig:focalplanes}.
The electromagnetic coupling structures for each focal plane are based on the most mature technology for the specific frequency ranges and required bandwidths.
The \gls{lft} and \gls{mft} \gls{fpm}s share a common architecture of lenslet-coupled sinuous antennas, while the \gls{hft} \gls{fpm}s use horn-coupled orthomode transducers. 
The dual-polarization antennas of each pixel are coupled to on-chip microstrip bandpass filters that split the signal into two or three frequency bands.
The separated signals propagate along superconducting microstrip lines and thermally dissipate at \gls{tes}s on thermally isolated islands for each frequency band and polarization state.
\gls{tes} electrical bias lines carry the induced signal to the edge of the silicon wafer where it is wire-bonded to a flexible circuit leading to the cold readout.
An overview of the two detector architectures is given in Fig.~\ref{fig:detector_architecture}.
The \gls{tes} technology is able to reach the instantaneous sensitivity required by the \litebird\ mission and is a mature design technology for ground-based and sub-orbital \gls{cmb} experiments  heritage~\cite{arnold2010polarbear}.
A demonstration of the  noise equivalent power of a prototype AlMn TES bolometer operated at 100\,mK is shown in Fig.~\ref{fig:tesnoisedemo}.  
The readout used for these measurements utilizes an older \gls{dfmux} system with SQUIDs at 4.8\,K.

Cosmic-ray mitigation is integrated directly into the focal-plane design to limit cosmic-ray impacts on low-$\ell$ systematics and data loss, described in detail in Sect.~\ref{sec:CR-systematics}. 
The transient energy deposition of cosmic rays interacting with the spacecraft can create glitches in the timestream of data sent to the readout system.
It is known from the \Planck\ mission that long timescale glitches are produced when cosmic rays interact with the silicon die and short pulse-like glitches result from interactions close to the thermistor~\cite{Catalano2014a}.
This has yielded a strategy of increasing thermal conductivity between the silicon die and the focal-plane structures to reduce long timescale thermal fluctuations and blocking phonon propagation to the \gls{tes} for the short glitches.
Progress in the lab has been made in blocking the ballistic phonons near the \gls{tes} by removing and adding metal layers and etching silicon. 
Tests with radioactive sources in the lab have shown phonon blocking effects by interrupting the phonon conduction paths to the \gls{tes}~\cite{beckman2018}.  The array will have dark \gls{tes} channels interspersed throughout the array to help monitor and remove common-mode fluctuations in the detector responses across the wafer.

\subsubsection{Focal-Plane Structures}
\label{sss:detection_structure}

Each \gls{fps} contains two intermediate 
temperature stages at 1.8\,K and 300\,mK, between the \gls{fpu}s at 100\,mK and the 4.8-K telescope 
structures.  A free-space low-pass edge filter and a 
\gls{fph} are supported by the 1.8-K stage.  An aluminized thin film 
spans the interstage gaps to block RF and residual warm radiation.  
The requirements include a thermal budget for each stage
and mechanical performance to survive launch loads and 
keep resonances clear of the science band while in operation.
Our current baseline includes aluminum, titanium, and copper
metalic parts and \gls{cfrp} inter-stage
support struts.  A trade study is in progress to determine 
whether struts can be designed to survive launch loads without 
launch locks.

\subsection{Readout} \label{s:readout}
\begin{figure}[htbp!]
\centering
\includegraphics[width = 1\textwidth]{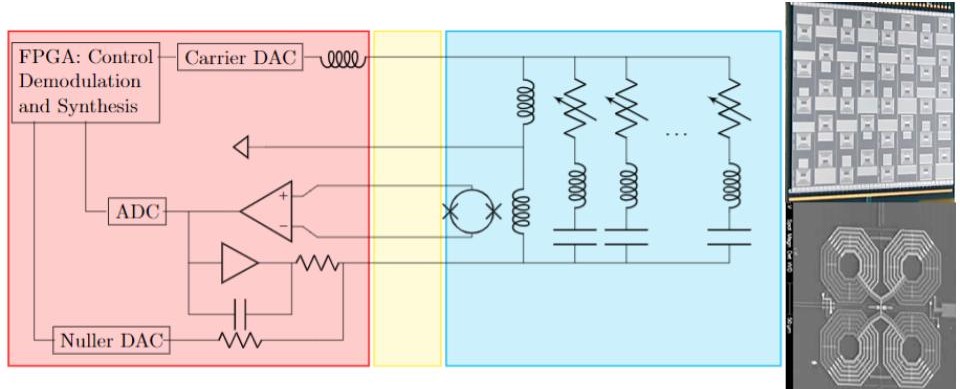}
\caption{
Schematic overview of the cryogenic readout system with digital frequency-domain multiplexing, as described in Sect.~\ref{s:readout} (left) 
and images of a chip of 40 inductor-capacitor resonators (right top) and a single gradiometric SQUID (right bottom).
The red shading indicates the portion of the circuit located at warm temperature, blue is the portion located at the sub-K temperature stages,
and yellow is the twisted-pair wiring harness, which connects them.
The variable resistances correspond to the TES bolometers.
 The nulling line shown here in the left panel uses active feedback to linearize the SQUID amplifier \cite{deHaan2012, Montgomery:PhD}.}
\label{fig:dfmux}
\end{figure}

The \gls{tes} bolometers will be read out using \gls{dfmux} \cite{Dobbs2012}. 
In a \gls{dfmux} system, 
each \gls{tes} is placed in series with an inductor-capacitor (LC) bandpass filter, which separates out the individual biasing sinusoids, allowing the detectors to be operated independently (see Fig.~\ref{fig:dfmux}).
In such a system, the \gls{tes} biasing voltages are provided by sinusoids at MHz frequencies. Up to 68 of these biases are summed together and transmitted to the detector array over a single set of wires, substantially reducing the conductive heat load and enabling large \gls{tes} focal planes.\footnote{
The multiplexing factor of $68\times$ is determined largely by the cryogenic electronics design. The digital electronics and associated firmware can support up to $128\times$ multiplexing.
}
Each \gls{tes} varies in resistance in response to incident radiative power, amplitude-modulating the biasing tones.
Our sky signal is contained in the sidebands, analogous to the way AM radio works.
The resulting current waveforms are amplified using a cryogenic \gls{saa}, before being transmitted to non-cryogenic electronics to be further amplified and digitally demodulated. 
The \gls{tes} bolometers, LC resonators, SAA, and bias elements are all on the 100-mK cryogenic stage and will be discussed in Sect.~\ref{ss:cr}. 

The \litebird\ implementation of \gls{dfmux} has over a decade of development and design, spanning four separate implementations deployed on four ground-based telescopes and one balloon-borne telescope~\cite{Dobbs2008,arnold2010polarbear,Dobbs2012,deHaan2012,Kaneko2020,Bender2014,Carter2018,EBEX2018,montgomery_2015,Montgomery:PhD}.

The success of \gls{dfmux} is the ability to multiplex large numbers of detectors without a substantial increase in system noise, maintaining the photon-noise limited performance of the instrument as a whole. 
The \litebird\ design is based on, and improves upon, the design currently used in the SPT-3G (South-Pole Telescope 3rd Generation) instrument, which has demonstrated  $68\times$ multiplexing with detector-limited noise performance~\cite{Carter2018}.
Section~\ref{ss:we} describes the space-qualified implementation of the non-cryogenic portion of this system.

The overall white noise requirement of the readout system is that it should increase the fundamental statistical noise already in the detection chain by no more than 10\,\% in each band, including yield-reduction due to the readout. 
Other requirements are that the readout-induced crosstalk be lower than 0.3\,\% for every bolometer, and that ambient variable magnetic fields in the spacecraft not significantly increase the noise in the detection chain.

\subsubsection{Cold Readout} \label{ss:cr}
The \litebird\ readout system is improved compared to previously fielded \gls{dfmux} systems by making some modifications to the cold circuitry. 
The previous resistive bias element at 4K has been changed to an inductor on the 100-mK stage and the \gls{saa} has also been moved onto the 100-mK stage\footnote{An alternate location for the SAA under consideration is on the 350 mK stage which has more cooling power than the 100mK stage but is still in close physical proximity to the TES bolometers and LC filters at 100mK.}.
Since the \gls{saa}s, bias elements, and TES bolometer are in close physical proximity, parasitic inducatance between the bias element and the LC filters~\cite{Dobbs2012} is reduced resulting in reduced crosstalk. In addition, the reduction in parasitic inductance increases the dynamic range of the SAA SQUID elements.
The reactive bias element dissipates substantially less power than the resistive bias element, enabling its placement at the 100-mK \gls{adr}-cooled stage.

System resources impose further requirements on the cold circuit: that the thermal power deposition on the 100-mK stage be less than 410\,nW  total; 
that a harness with a length of between 1.8 and 2.5\,m (to be determined more precisely later) be supported by the system; and that the system has a mass less than 2\,kg.

\paragraph{Bias Element and LC Resonator}
Each bolometer is in series with a lithographed planar spiral inductor and interdigitated capacitor, which form a resonance between 1.5\,MHz and 4.5\,MHz and uniquely identifies the bolometer in series with it. 
These are fabricated lithographically and a given die will contain all 68 resonators that will be connected to a single \gls{saa}. 
The optimal frequency schedule balancing anticipated scatter and mechanisms of crosstalk is still under study.  
The LC die must also be shielded by a conductor to ensure consistent inductor values. 
The LC circuit-boards contain the low-impedance bias element and connect to the bolometer wafer with a superconducting stripline.

\paragraph{SQUID Array Amplifier and Wiring Harness}
The \gls{saa} will be used as a low-noise transimpedance amplifier to drive the wiring harness that connects the detectors to the warm \gls{scu}. Requirements on the \gls{saa} to 300K harness, derived from the sensitivity requirement on the readout, are as follows. 
The transimpedance for the \gls{saa}-harness system must be greater than 400\,$\Omega$ across the operating \gls{dfmux} frequencies. 
The noise injected by the \gls{saa} must be such that, when integrated with the warm electronics in the detection chain, the high-level sensitivity requirement is satisfied. 
This requirement must be achieved including the noise-increasing effect of current sharing, which sets an upper bound on the allowable input impedance of the \gls{saa} \cite{SilvaFeaver2018}.
In practice, currently demonstrated noise values of $4\,\mathrm{pA}\,(\mathrm{Hz})^{-1/2}$ meet this requirement. 
Further refinement in the per-component noise injection allocation, given circuit component optimizations, are underway. 
Magnetic shielding of the \gls{saa}s is accomplished, as in other \gls{dfmux} systems, with a hybrid system of niobium foil beneath the \gls{saa} to pin magnetic fields when in the superconducting state, and a $\mu$-metal shield around the \gls{saa} to reduce field amplitude. 

\subsubsection{Warm Readout} \label{ss:we}

\begin{figure}[hbtp!]
  \centering
  \includegraphics[width=0.8\textwidth]{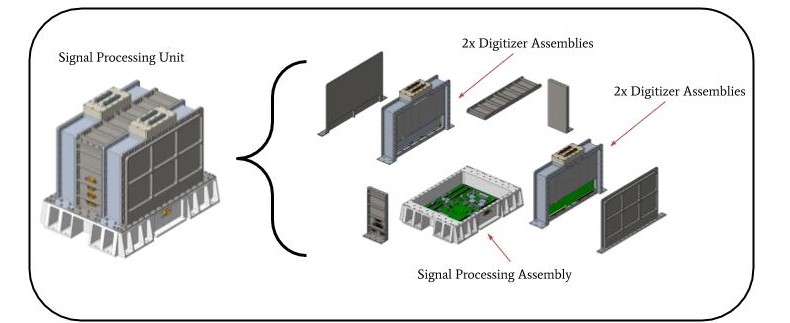}
  \caption{
    Exploded view of the CAD model for the \glsentryfull{spu}. The \gls{spu} contains one \glsentryfull{spa}, on which an FPGA is mounted, and four separate digitizer assemblies. 
    The enclosure is designed to meet thermal dissipation and vibration requirements for launch and flight environments. 
    In total, an \gls{spu} performs the readout for 15 SQUID modules, which means up to 1020 bolometers.}
  \label{fig:spu}
\end{figure}

The \litebird\ \gls{we} implement the non-cryogenic portion of the \gls{dfmux} hardware. This is divided into three separate electronics assemblies, as follows.
\begin{itemize}
    \item The \gls{spa} contains the \glsentryfull{fpga} that performs the digital signal processing and communication. Each SPA supports up to 15 multiplexing modules using four digitizer assemblies. This number of a maximum of 15 SQUID modules per SPA is a design requirement, driven by a projection of the FPGA resources of radiation-hardening techniques such as \gls{tmr}, and \gls{ecc}.
    \item The \gls{da} performs the digital-to-analog conversion to synthesize tones, and analog-to-digital conversion to digitize the resulting waveforms. Each \gls{da} supports four multiplexing modules.
    \item The \gls{sca} houses the SQUID biasing electronics and pre-amplification stages for the SQUID output signals. Each SCA supports four multiplexing modules.
\end{itemize}

The WE redundancy model duplicates SPA and DA electronics with ``cold spares'' that we can switch to in flight.  They are organized in signal-processing units (SPUs) consisting of one SPA and up to four DAs, plus the enclosure (see Fig.~\ref{fig:spu}). Hence there are four SPUs per telescope (two active and two cold redundant).
The SCAs are much simpler, and contain only redundant elements within the electronics, rather than fully redundant spares.
This WE design improves on existing DfMUX readout \cite{montgomery_2015,Montgomery:PhD} in a number of ways: the multiplexing capability per FPGA board is increased from eight multiplexing modules to 15; and the electronics is implemented with a fully radiation-qualified signal chain, including radiation-tolerant components and redundant firmware designed to be robust to the high-radiation environment.
This design meets the requirements for: power consumption (approximately 80\,mW per bolometer including power delivery losses and redundancy); reliability (90\,\% confidence in sufficient yield over the mission lifetime to meet the readout-induced noise requirements); and readout noise \cite{Montgomery:PhD}.

\subsection{Electronics Architecture} 
\label{ss:plm_electronics}

\begin{figure}[htbp!]
\begin{minipage}{1.0\textwidth}
\centering
\includegraphics[width = 1\textwidth]{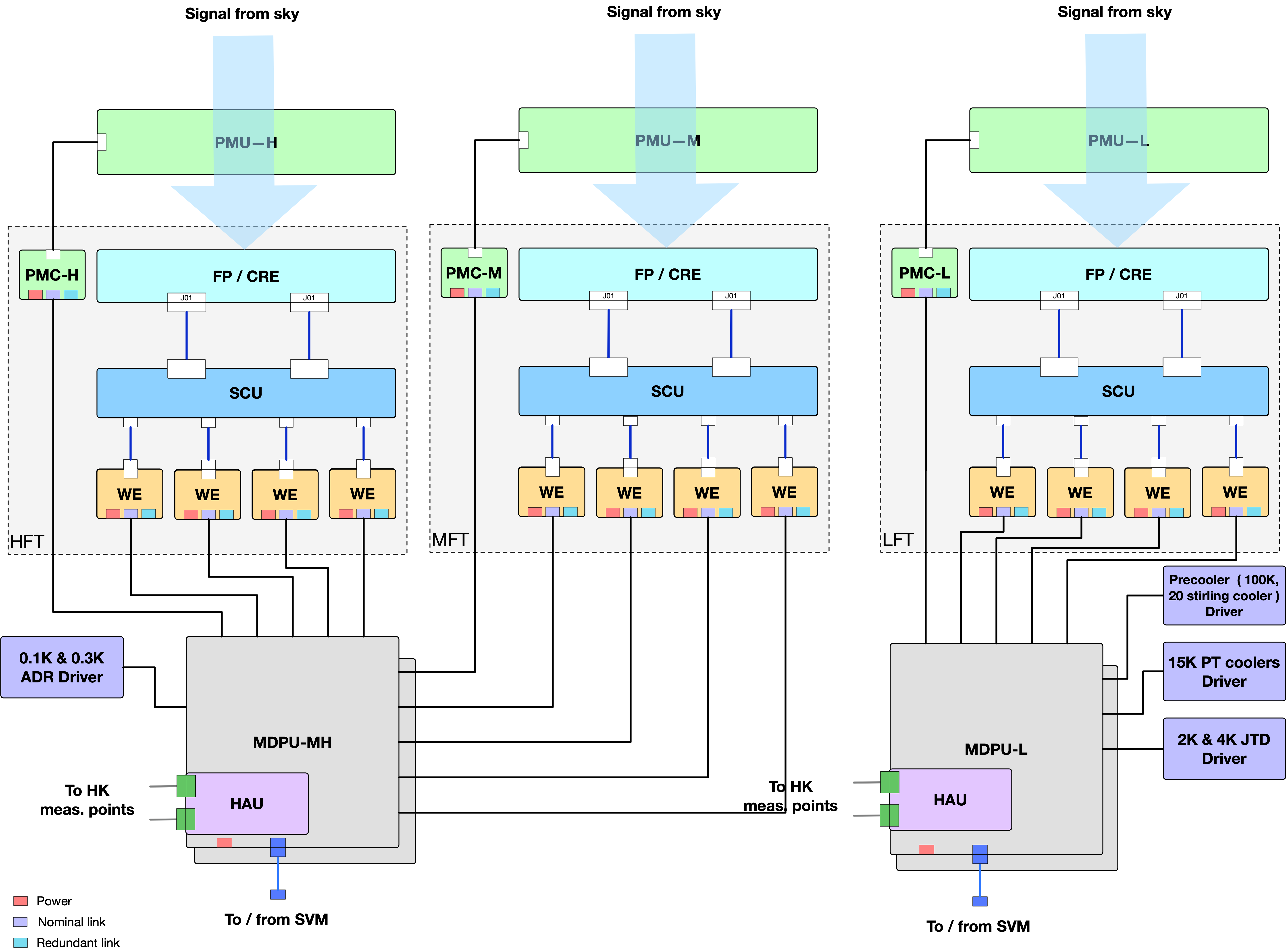}
\caption{Electrical architecture.}
\label{fig:archi_elec}
\end{minipage}
\end{figure}

The electronics architecture is designed to implement the following functions: (i) control and readout of the detectors of the three focal planes; (ii) science signal processing (cosmic-ray glitches capture and scientific signal compression); (iii) control of the cryo-coolers and ADRs; (iv) control of the polarization modulators of the three telescopes; (v) temperature control and monitor of cold stages; (vi) housekeeping acquisition; and (vii) data time-tagging and subsystems synchronization.

To this purpose two mission data-processing units (\glsentryshort{mdpul} and \glsentryshort{mdpuhf}) will centralize these functions for both the LFT and MHFT telescopes, respectively, as shown in Fig.~\ref{fig:archi_elec}. They will address both the scientific signal and cooling chains, as described below. 

On one hand, the scientific signal chain starts with the \gls{cre}, which is part of the focal-plane subsystem of each of the three telescopes. It then consists of the \gls{scu}, which provide bias for detectors and SQUIDs, and the \glsentryfull{we}, which implements the frequency-domain readout electronics. 
Then the \gls{mdpu} provides the last stages of the science signal acquisition; it performs science data preprocessing, compression, detector setup algorithms, focal-plane thermal control (digital part), and sub-systems synchronization. 
Finally, the \gls{hau} performs the acquisition of all housekeeping that cannot be acquired inside the others subsystems. 
Additionally, three polarization modulation controller subsystems (PMC-L, PMC-M, and PMC-H) control the polarization modulation units of the \gls{lft}, \gls{mft}, and \gls{hft}, respectively.

On the other hand, the 0.1-K and 0.35-K ADRs drivers of the cooling chain, responsible for driving the ADRs for the 0.1-K and 0.35-K stages, respectively, are connected to the MDPU-MH. The MDPU-L addresses the higher-temperature stages of the cooling chain, i.e., the 2-K JT and 4-K JT drivers (JTD), the 15-K Pulse Tube cooler driver, and the \gls{pcd}. The two JTDs, responsible for driving the Joule-Thomson cryo-coolers for the 1.75-K and 4.8-K stages, provide the AC power for the 3He JT and 4He JT mechanical coolers, respectively, monitor the temperatures that are provided, and monitor pressures and controlling valves in the JT circuits. The PCD, which is the cooler driver for the Stirling coolers used for the shield and JT pre-cooling, provides AC power for the Stirling mechanical coolers and monitors the temperatures that they are cooling.

\subsection{Cryogenic System} 
\label{ss:plm_cryo}

\begin{figure}[htbp!]
\begin{minipage}{0.5\textwidth}
\centering
\includegraphics[width = 1\textwidth]{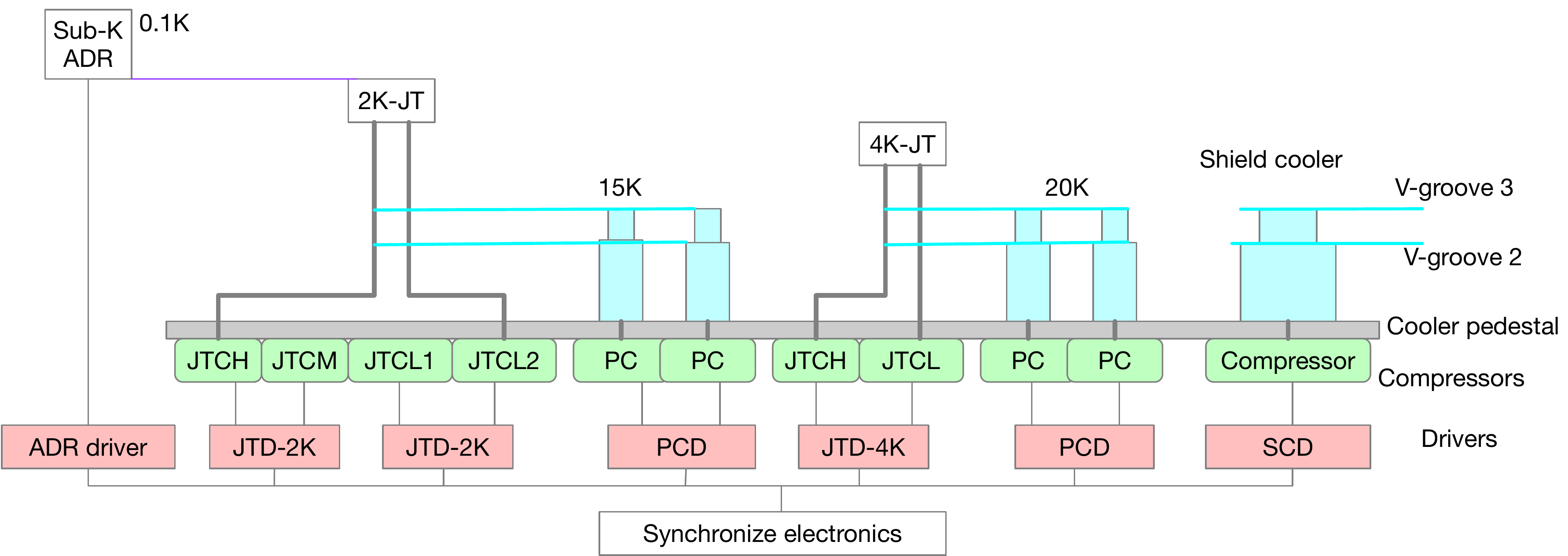}
\caption{Overview of the cryogenic chain.}
\label{fig:cryochain}
\end{minipage}
\begin{minipage}{0.5\textwidth}
\centering
\includegraphics[width = 1\textwidth]{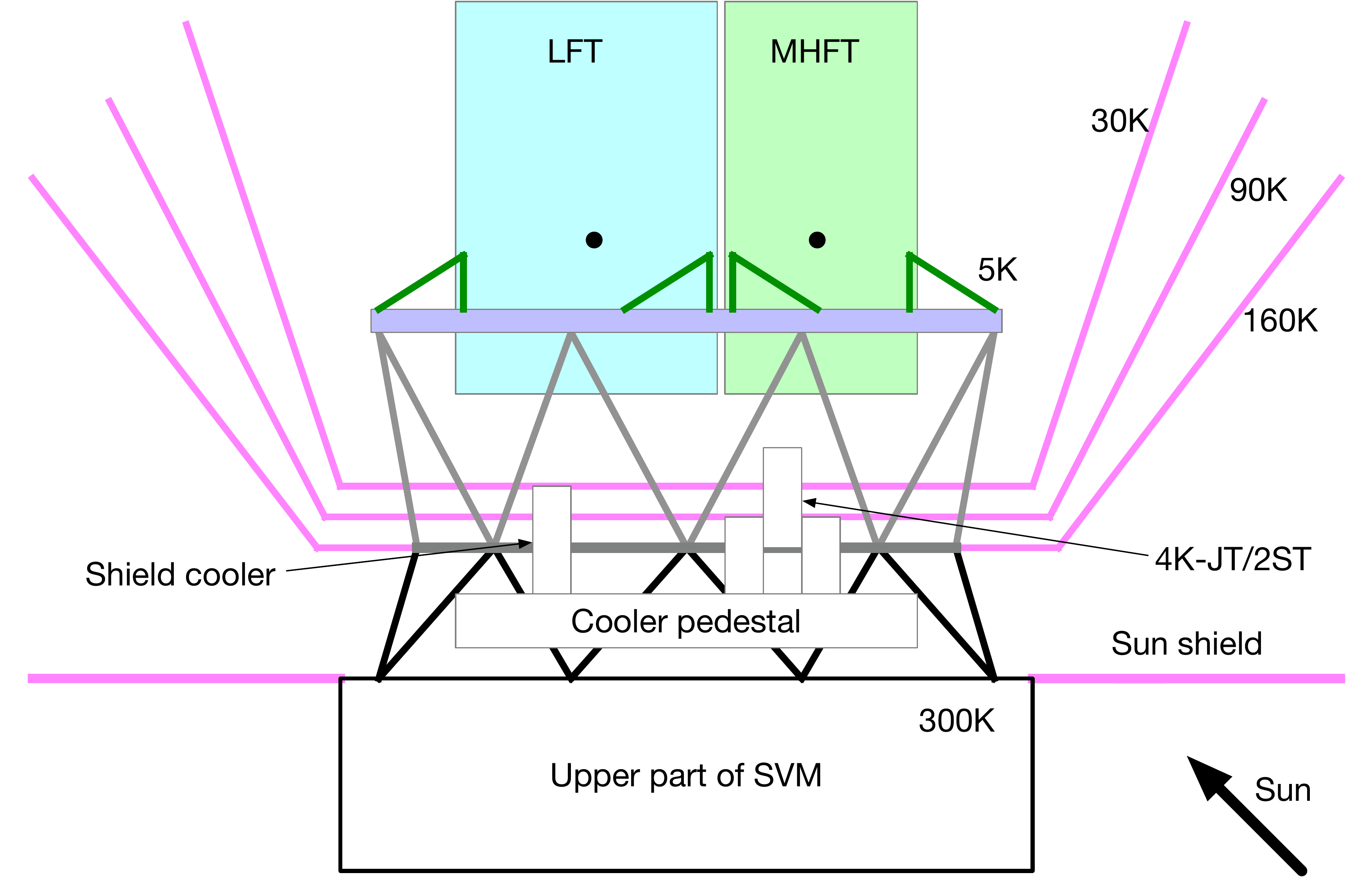}
\caption{Schematic drawing of the cryo-structure and V-grooves.}
\label{fig:cryo-structure5K}
\end{minipage}
\end{figure}

\subsubsection{Down to 5\,K}
The cryogenic telescopes must be cooled to 4.8\,K to achieve the required sensitivity.
The cooling chain between 4.8\,K and  the warm temperature (around 300\,K) is composed of passive radiative  cooling and of active mechanical coolers, as illustrated in Figs.~\ref{fig:cryochain} and \ref{fig:cryo-structure5K}. 
The thermal architecture of \litebird\ has a boundary temperature of 300\,K in between the warm and cryogenic mission instruments. 
The cryostructure, made of CFRP, connects the 5-K interface plate and the  warm temperature SVM/BUS structure, as shown in Fig.~\ref{fig:cryo-structure5K}.  

To reduce the heat loads of the shield cooler and 4K-JT, we employ large thermal shields, which are called ``V-grooves.'' The thermal design of \LiteBIRD\ uses the radiative cooling effect as much as possible. The thermal design of the passive radiative cooler for \litebird\ has been studied in Ref.~\cite{Hasebe2019}. 
The current design has three layers of V-grooves, as shown in Figs.~\ref{fig:satellite_overview} and \ref{fig:cryo-structure5K}. 
The second and third (inner) layers are cooled by the first and second stages of the shield cooler, respectively. 
The first (outer) layer is partially exposed to solar radiation, and therefore the surface is covered by a multi-layer insulation (MLI) blanket.
A Sun shield (Fig.~\ref{fig:cryo-structure5K}) reduces the solar radiation to the outer V-groove.
V-groove passive radiators are cooled down to 160, 90, and 30\,K, thanks to the favorable radiative environment of the L2 orbit. 
Additional cooling capacity for shield cooling is provided by a mechanical cooler. 
A 15-K pulse-tube cooler~\cite{Pennec2016} helps to decrease the temperature of the inner passive radiative enclosure to 30\,K. 

A 4-K JT and two 2ST (two-stage Stirling) pre-coolers keep the telescopes at 4.8\,K. 
The 4-K JT cooler has a cooling capacity of 40\,mW at the cold tip of 4.8\,K at the EOL (End Of Life)~\cite{Sato2010,Sato2012,Sato2014}. 
The 2ST cooler has a cooling capacity of 200\,mW at the 20-K stage at EOL \cite{Sato2010,Sato2012,Sato2014}.
Taking into account 33\,\% margin for the mechanical cooler, the available cooling power at the 4.8-K stage is 30\,mW. 
The 4-K JT cooler and the 2ST cooler have achieved \gls{trl}~8.
Based on the PLM structure, the conductive and radiative loads to the 4.8-K stage are estimated to be 13\,mW. 
The available cooling power for the instruments, including LFT, MHFT and the sub-kelvin coolers at 4.8\,K, is 17\,mW.

\subsubsection{Below 5\,K}

\begin{figure}[htbp!]
\centering
\includegraphics[width = 1\textwidth]{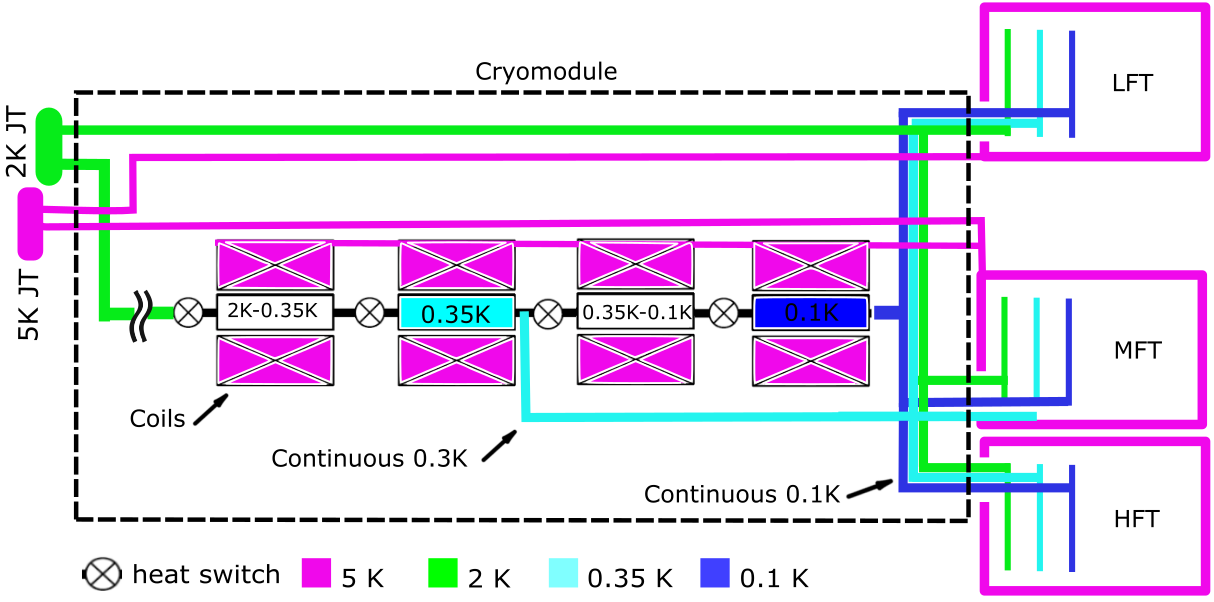}
\caption{Overview of the 5-K to 0.1-K cryogenic chain and thermal links.}
\label{fig:cryomodule}
\end{figure}

\paragraph{Low Temperature Thermal Architecture}
Both MHFT and LFT have intermediate cold stages at 1.75\,K and 0.35\,K
between their outer bodies at 4.8\,K and their detectors. In order to have continuous measurements, these intermediate stages have to
be cooled continuously as well.

A solution combining a 2-K JT cooler and a sub-K (sub-kelvin) cooler with multiple ADRs, has been
selected to cool the instruments below 4.8\,K. The two coolers are used for the two instruments, requiring a web of thermal paths shown in Fig.~\ref{fig:cryomodule}, linking these four components.
As seen in the figure, the thermal links and the sub-K cooler will be integrated on a cryomodule cooled down to 4.8\,K by the 4-K JT cooler.   The cryomodule is going to be located on the 4.8-K mechanical stage of the MHFT. The sub-K cooler side of the cryomodule will be placed close to the focal plane side to minimize the thermal link lengths.

The thermal links have to be designed to minimize parasitic losses. Therefore, heat interception should be used at each of the temperature levels.
Coaxial kevlar cord supports are foreseen to be used to limit heat losses at sub-K temperatures, whereas CFRP supports will be use for both the sub-K thermal links and the 1.75-K ones.
Highly conductive materials, such as copper and high purity aluminum, will be used and superfluid heat pipes will be evaluated at 1.75\,K in comparison with pure aluminum. 

\paragraph{2-K JT cooler}
The 2-K JT cooler is provided by JAXA. It requires pre-cooling stages at 15\,K and 90\,K.  The $^3$He gas is circulated by four compressors. An engineering model has been fabricated for several verification tests and has allowed this technology to achieve TRL 5 \cite{Sato2016Cryo}. A cooling power of 10\,mW at 1.75\,K has been confirmed during these tests \cite{Sato2016Cryo}. Its performance has also been demonstrated in combination with a 50\,mK sub-Kelvin ADR in the case of the Athena cooling chain. \cite{PROUVE2020}. 

\paragraph{Sub-kelvin Cooler}
The sub-K cooler is provided by \gls{cea}. It works from 1.75\,K to 0.1\,K, as schematically shown in Fig.~\ref{fig:cryomodule}
and has been described in Ref.~\cite{Duval:2020ADR}.
The 350-mK continuous cooling stage is obtained thanks to two ADR
stages. Similarly, the 100-mK stage is obtained with two
additional ADR stages. For low temperatures, continuous cooling is achieved
using a series configuration, with one stage being at a
stable temperature, while a second stage is used to extract
heat from the low temperature and dump it to the highest
temperature interface.

\subsection{Calibration} 
\label{ss:plm_calib}

\subsubsection{Overall Status and Strategy Description}

\begin{figure}[htbp!]
\centering
\includegraphics[width = 0.7\textwidth]{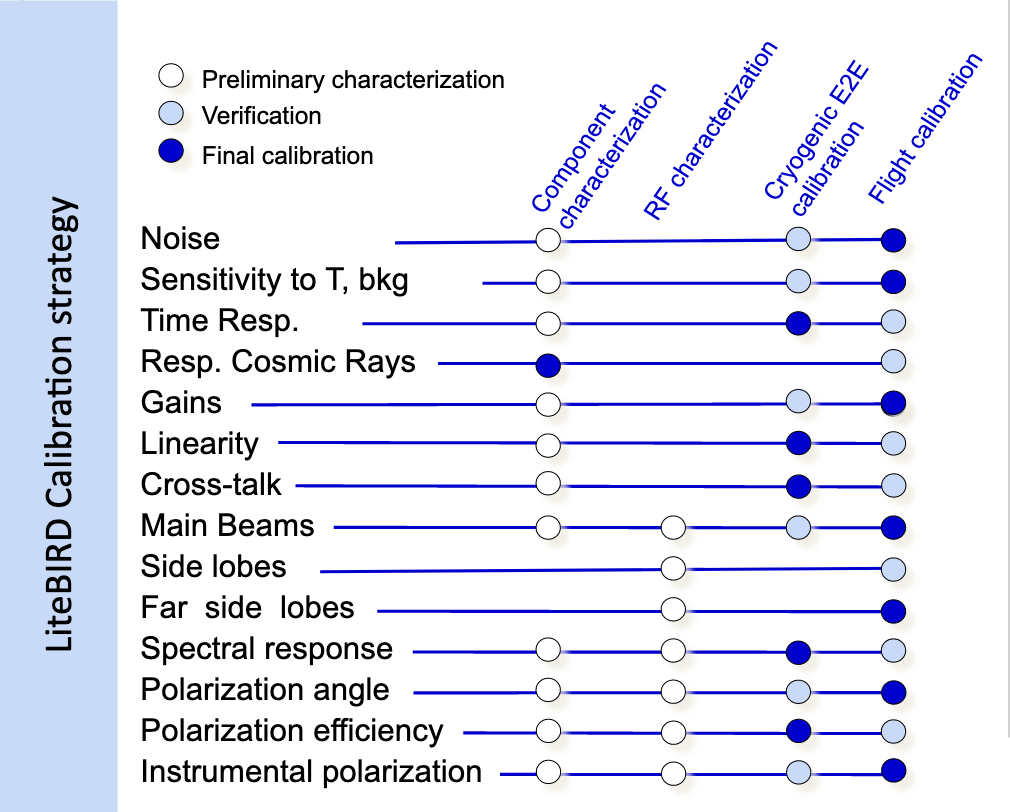}
\caption{Global calibration strategy. The colors of the circles give an indication of the accuracy foreseen for the related calibration operation (white indicates preliminary characterization, light blue is for verification, and dark blue identifies the phase for which the calibration will be ultimately performed). 
The flight calibration phase can be sub-divided into three categories: the calibration and performance verification phase prior to nominal observations; and calibration during and after the nominal observations. These are not detailed here. \label{fig:calib-stra} }
\end{figure}

The strategy of the \litebird\ calibration is to derive a common approach for both instruments, LFT and MHFT, except for some specific items. 
The global picture, as of today, is illustrated in Fig.~\ref{fig:calib-stra}. Four steps are foreseen.
The first is to characterize the performance at the component level.
These charaterizations are part of the deliverables of the sub-systems; they will be based on the \litebird\ specifications and carried out prior to integration at the instrument level.  
The data from these characterizations will be used to build an instrument model and forecast the in-flight 
performance as we develop the system. 
The second step concerns beam calibration. Considering the specific related challenges (linked to the specifications on the beam knowledge and the level of accuracy required for the beam modeling), 
a dedicated set of measurements is foreseen, identified as 
``RF characterization'' (cf.\ Sect.~\ref{beams-calib-section}). 
The next step is the instrument-level calibration, which will be performed for LFT and MHFT independently, in Japan and Europe, respectively,
in a cold flight-like environment. 
The final step of the ground activites is the final verification that will be carried out at the PLM level (system-level testing), 
when the LFT and MHFT will be integrated with the satellite PLM, and the SVM, together with the 
entire \litebird\ cooling system. 
As is highlighted in Fig.~\ref{fig:calib-stra}, it is planned that some instrumental parameters will be ultimately derived from ground calibration operations 
(such as the spectral response, for instance), while, for others, the planned accuracy with flight data should 
allow us to rely on flight data themselves (e.g., main beam and crosstalk using planets). 
It is worth noting that we explore the possibility of mitigating the systematic effects through a post-analysis step
and to solve for the systematic parameters as part of the map-making or component-separation processes. 
Nevertheless, the calibration design philosophy is 
to prioritize the search for a hardware mitigation solution as a starting point.
The calibration plans, the error budget allocation for hardware development, and the post-flight analysis mitigation 
strategies will be refined as the project evolves.

In this section, we highlight the main calibration challenges, by focussing 
on some key parameters, namely the beams (Sect.~\ref{beams-calib-section}), the spectral response (Sect.~\ref{bandpass-calib-section}), and the polarization angle (Sect.~\ref{sss:anglecal}).
For each of them, we detail the requirements, the prospects for ground calibration and, if relevant, the plans for flight calibration. Finally, in Sect.~\ref{section:calib-plan}, we summarize the means that we foresee to characterize the other instrumental parameters.

\subsubsection{Beam-Pattern Characterization}
\label{beams-calib-section}
\begin{figure}[htbp!]
\centering
\includegraphics[width = 1.\textwidth]{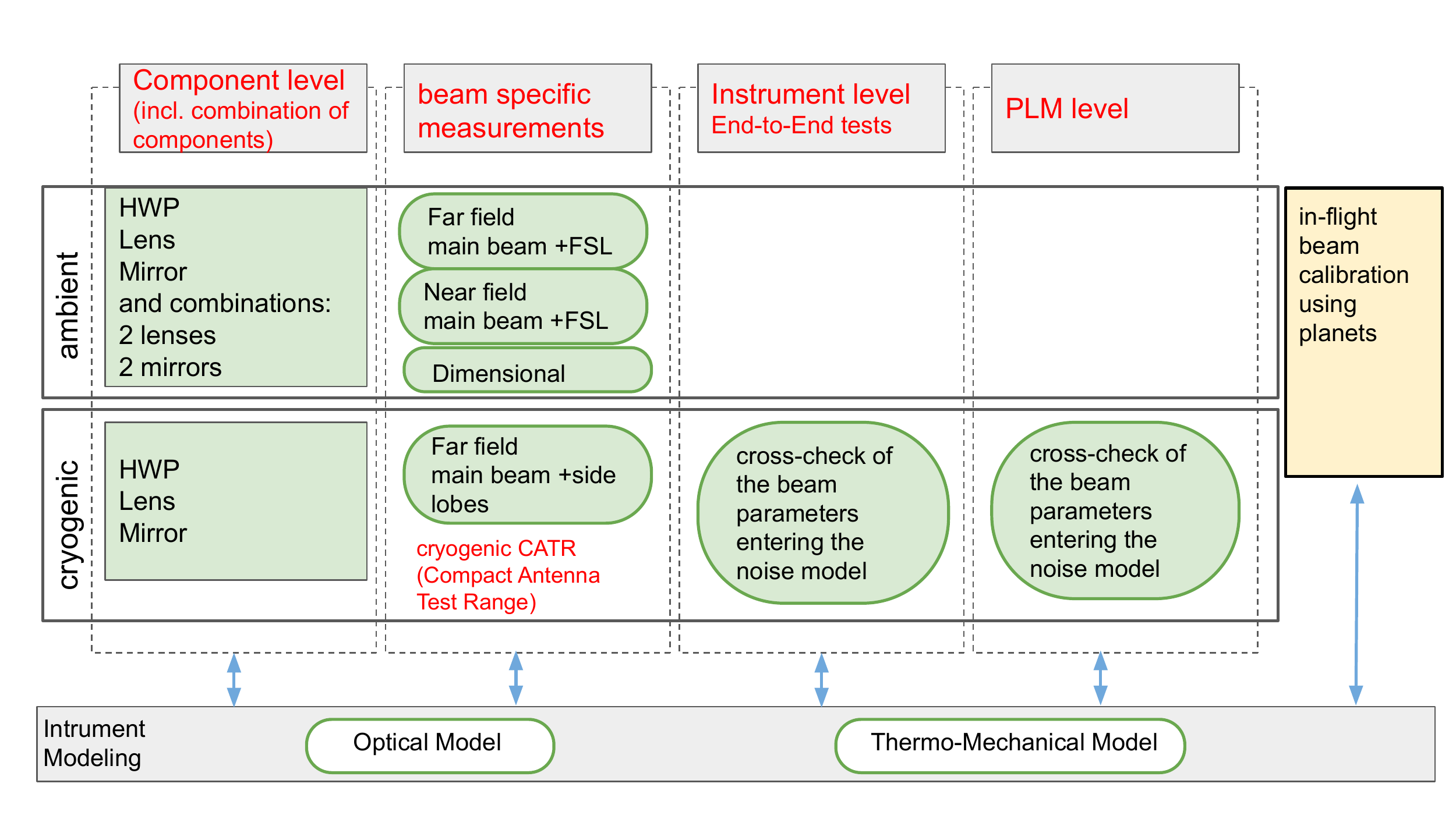}
\caption{Beam-pattern characterization strategy, including ambient and cold calibration operations. The different levels of the operations foreseen are identified by the five columns: at component and sub-system level; at telescope level during dedicated beam-scanning campaigns; at telescope level during integrated end-to-end measurements at PLM level; and finally in-flight using planets. The strong links between measurements and modeling are also shown. \label{fig:beam-calib}}
\end{figure}

This section describes the plans foreseen for the beam-pattern calibration. 
The strategy relies on optical modeling, tuned and consolidated by ground measurements at various levels, from component, subsystems, instrument, and PLM levels and further fed with flight data, as schematically illustrated in Fig.~\ref{fig:beam-calib}.
The cross-polar response and the far sidelobes will be characterized by a combination of ground measurements and optical modeling. The co-polar main and near sidelobe beams are planned to be reconstructed from observations of planets. 

\paragraph{Requirements}

\begin{figure}[htbp!]
    \centering
    \includegraphics[width=1.\textwidth]{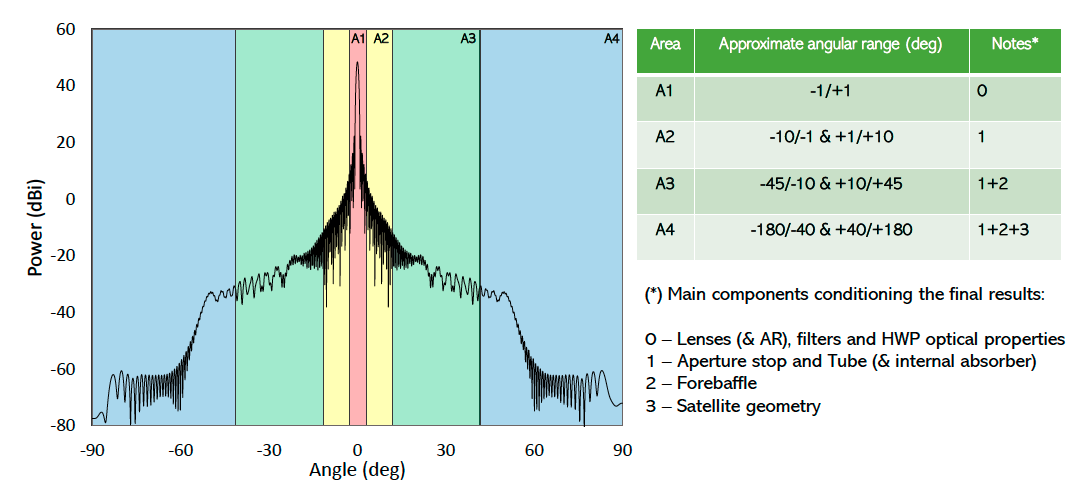}
    \caption{\label{fig:LoK} Radiation pattern of MFT at 100\,GHz for a pixel at the center of the focal plane (including the lenslet-like feed, the two lenses, the stop aperture, and the forebaffle aperture). The sharp dropoff in beam power in the 50--$60^\circ$ range is caused by forebaffle shielding of a Lambertian source placed at the aperture to simulate 1\,\% large-angle scattering. 
    The color scale indicates some qualitative aspects of the beam profile.
    }
\end{figure}

With the example of the MHFT telescope at 100\,GHz, Fig.~\ref{fig:LoK} shows how each telescope element differently affects the angular regions of the co-polar beam shape.
The identified angular ranges and intensity are key ingredients in the estimation of 
the hierarchy driving the calibration efforts. It also gives indications on the level of accuracy needed in the characterization of each individual optical element, as well as at the integrated level. Last (but not least), it gives guidelines to define a suitable way of parameterizing the instrumental effects when feeding them into TOD-based simulations for systematics studies.

The derivation of the beam-calibration requirement is detailed in Sect.~\ref{sec:beam-systematics}.  The 
global requirement reads as follows: the calibration measurement uncertainty, as defined by Eq.~(\ref{sigmacalibbeam}) 
should be smaller than $-57$\,dB, when the optical response is assumed to be normalized to unity at the main peak, for a pixel of $0.5^\circ$ in size.

\paragraph{Prospects for Ground Calibration}
\label{Ground-beams}
As discussed above, Fig.~\ref{fig:beam-calib} illustrates the main steps that are foreseen today for beam characterization. To meet these requirements, the plan is to rely on a very tight connection between measurements performed both at warm and cold temperatures, together with the development of accurate software tools for instrumental modeling.  
\begin{itemize}
\item{} In the first phase, the plans are to go through characterizations of sub-systems (and combinations of components) at warm and cold temperature to consolidate the design of the instrument as well as its optical modeling, allowing us to build the transfer function between warm and cold temperatures.

The MHFT \gls{bbm}
is an example of activities in this phase of the development plan, with comparable activities being planned for LFT.  The purpose of the BBM measurement campaign is to challenge the optical modeling of the MHFT refractive designs, to assess the achievable accuracy both for the modeling and for the \gls{rf} measurements, to prepare the setups that will be used for the cold measurements, and to study various effects such as, for example, misalignment of components (tolerance analysis), standing waves, and ghosting.
    The BBM test plan foresees the following two models.
    \begin{enumerate}[(a)]
        \item BBM-1 (corrugated horn with one lens): simulation and characterization at MFT and HFT frequencies in different labs with a reference common channel as a cross-check. This allows for a verification of the modeling and measurement techniques and facilities.
        \item BBM-2 (corrugated horn with two lenses): increase complexity and possible evaluation of ``combined'' effects due to the coupling of the two lenses. This allows for a verification of the sensitivity to misalignment, AR-coating, etc.
    \end{enumerate}

\item{} The next step is to carry out a 2$\pi$ RF characterization campaign of a representative instrument. This is the case for the 1/4-scaled model for LFT, as described in Sect.~\ref{sect:LFT-scaled-model}. The far sidelobe pattern of the 1/4-model LFT has been measured with 1/4 wavelengths or 4 times higher frequencies~\cite{Takakura2019IEEE}. On the MHFT side, a forthcoming optical RF demonstration model for HFT is being built and will be characterized at warm temperatures. For those measurements, dedicated compact-antenna test ranges (CATRs) are being developed both in Japan and in Europe, using broadband coherent and blackbody sources. The co- and cross-polarization responses are also assessed using a polarized source. From those measurements, the optical modeling of the instrument will be consolidated. 
\item{} A cryogenic characterization will then follow to account for the 5-K nominal working temperature of the telescopes (hence the related thermo-mechanical deformations), as well as for sub-systems whose RF characteristics depend on temperature, such as the index of refraction of MHFT lenses for instance. This will allow us to map (within a limited angular range) the beam response using a 5-K compatible detector with a corresponding beam former. 
Several techniques are under study to measure the beam patterns at cryogenic temperature, e.g., near-field versus far-field using a \gls{catr} with a continuous coherent source or with a blackbody.   

\item{} During the end-to-end calibration campaign for which each instrument will be fully integrated and cryogenically cooled in a chamber for a characterization of in-flight-like conditions, near-field verifications are also foreseen. 
\item{} Finally measurements of diffraction effects due to V-grooves and structures of the MHFT/LFT 
 will be performed at the PLM level at room temperature. Those effects are expected to be small (below $-60\,$dB from the main peak).
\end{itemize}

\paragraph{Prospects for In-flight Calibration} 
For the main beam, as was the case for \planck~\cite{Planck2015IV,Planck2015VII}, we plan to use the signals of outer planets (Mars, Jupiter, etc.) to reconstruct the most important parameters of the radiation pattern for each detector. The idea is that we can assume that a planet, given its limited angular size with respect to the FWHM of \litebird's beams and its considerable brightness,\footnote{Jupiter, the planet with the largest apparent size, is always smaller than 0.5\,arcmin, and the brightness temperature of the outer planets at the frequencies of interest for \litebird\ are typically two orders of magnitude brighter than the CMB.} can be considered as a bright point source that ``samples'' the radiation pattern $\gamma(\mathbf{r})$. With these assumptions, one can estimate the value of $\gamma(\mathbf{r})$ using
\begin{equation}
    \label{eq:beams:inflight:beamFromTemperature}
    \gamma(\mathbf{r}) \propto \frac{\Delta T_\text{pl}(\mathbf{r})}{T_\text{B,pl}} \frac{\Omega_{\rm b}}{\Omega_\text{pl}}.
\end{equation}
Here $\Delta T_\text{pl}$ represents the component of the signal that is caused by the presence of the planet along direction $\mathbf{r}$, $\Omega_{\rm b} = \int_{4\pi} \gamma(\mathbf{r})\,\mathrm{d}\Omega$ is the angular beam size, $\Omega_\text{pl}$ is the solid angle subtended by the planet, and $T_\text{B,pl}$ is the brightness temperature of the planet in the frequency band sampled by the detector.

\begin{figure}[htbp!]
    \centering
    \includegraphics[width=0.48\textwidth]{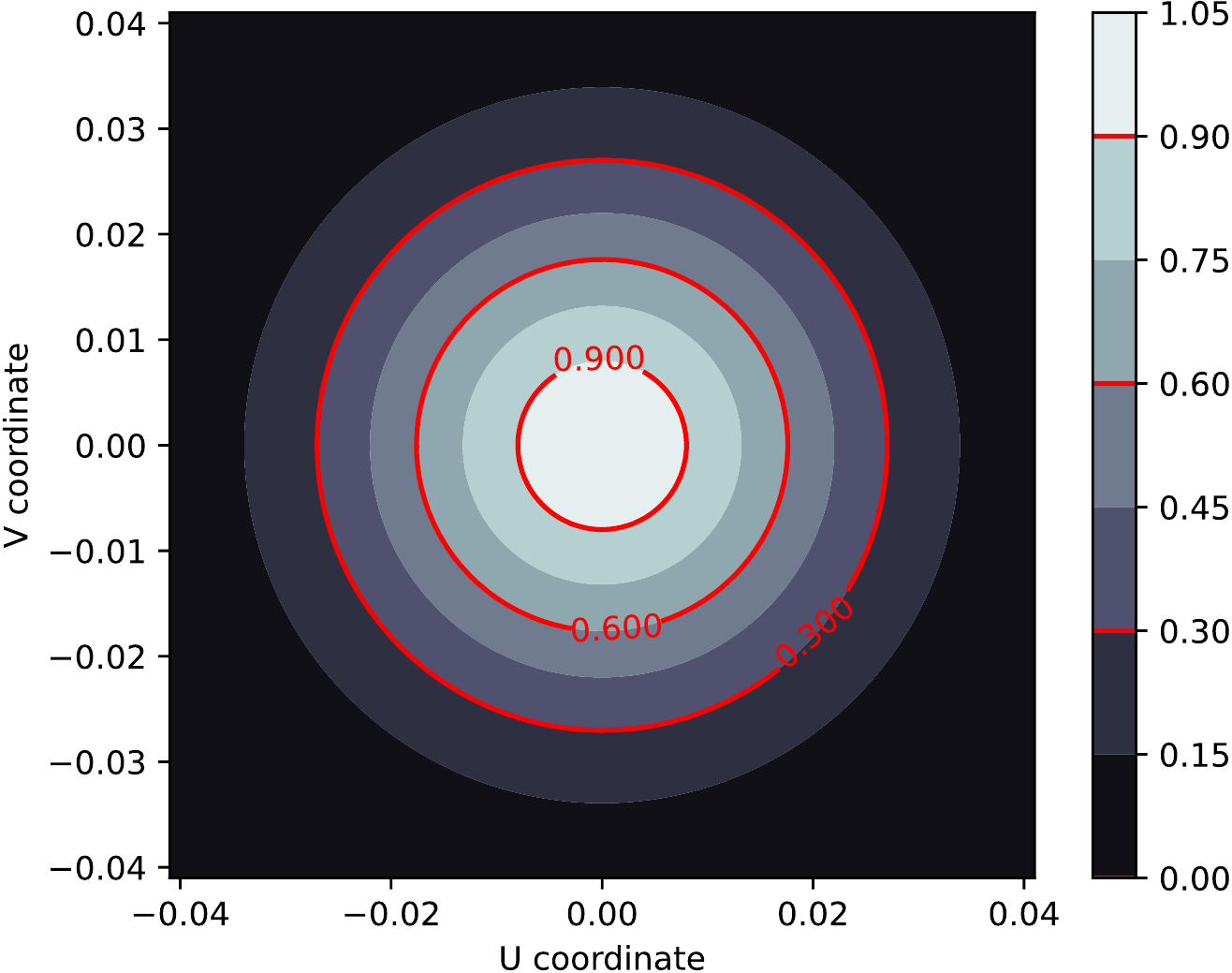}
    \includegraphics[width=0.48\textwidth]{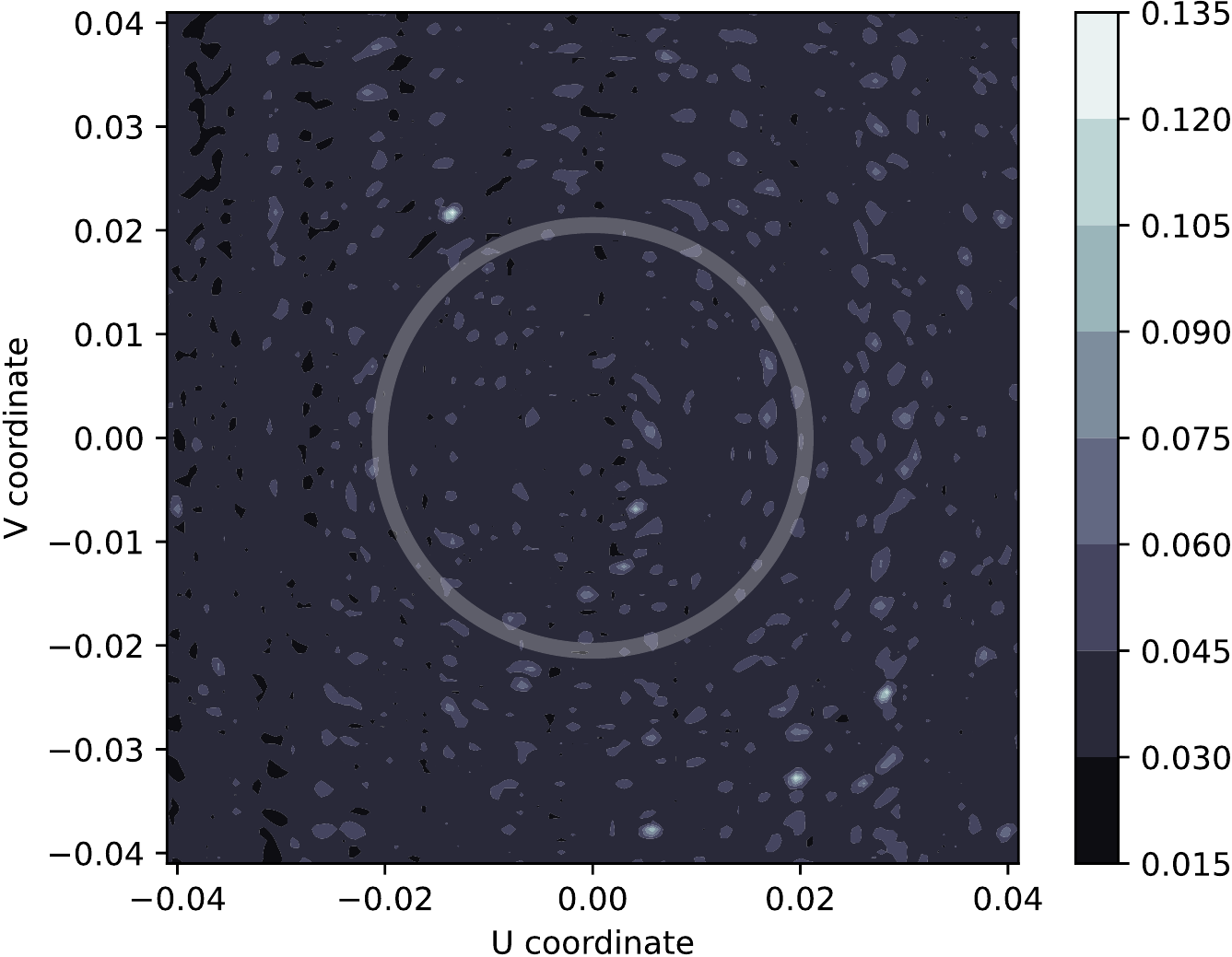}
    \includegraphics[width=0.48\textwidth]{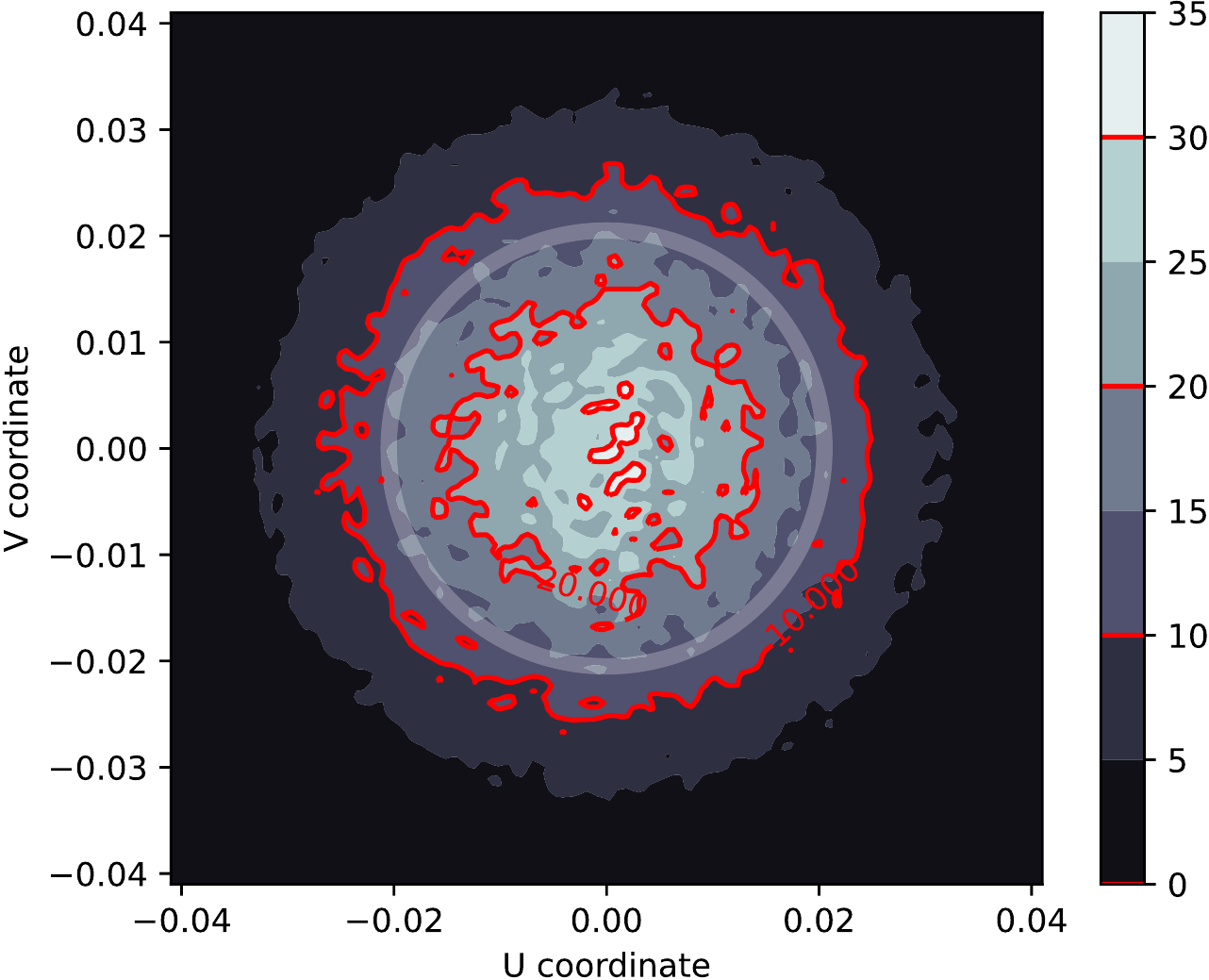}
    \caption{\label{fig:beams:inflight:beammap}\textit{Top left}: Representation of the main beam of the optical response $\gamma(\mathbf{r})$ for a 40-GHz \litebird\ detector, in $(u, v)$ coordinates (in radians), used in the simulations of the in-flight beam calibration. Here $\mathbf{r}=0$ indicates the position of the maximum amplitude of the beam pattern. \textit{Top right}: Representation of the error on the beam reconstruction ($\delta\gamma$), calculated on the timespan of the \litebird\ mission. \textit{Bottom}: The quantity $\gamma / \delta\gamma$, which is the signal-to-noise ratio of the measurement on $\gamma$. The radius of the white ring is equal to the FWHM of the beam and is shown as a reference.}
\end{figure}

We have used Eq.~(\ref{eq:beams:inflight:beamFromTemperature}) to simulate
how accurately we can expect to reconstruct the shape of $\gamma$, assuming that the radiation pattern has a 
perfectly symmetric Gaussian profile.
Specifically, we simulated the scanning strategy of the \litebird\ spacecraft and computed the amount of observation time spent on Jupiter along each direction $\mathbf{r}$ on the $4\pi$ sphere in the reference frame of the detector. 
To estimate $T_\text{B,pl}$, we have used the estimates for Jupiter's spectral energy distribution published in Refs.~\cite{Maris2021} and \cite{PlanckIntLII}. A plot of $\gamma(\mathbf{r})$ and the related expected error for one of the 40-GHz detectors is shown in Fig.~\ref{fig:beams:inflight:beammap}.

We have fitted for the two-dimensional Gaussian FWHM\footnote{Since the term $\Omega_{\rm b} = \int_{4\pi} \gamma(\mathbf{r})\,\mathrm{d}\Omega$ in Eq.~(\ref{eq:beams:inflight:beamFromTemperature}) depends on the unknown quantity $\gamma$, this equation should be typically solved using an iterative algorithm; we chose to neglect this complication and assume that $\Omega_{\rm b}$ is perfectly known in advance.} of the reconstructed beam pattern of 500 Monte Carlo simulations.
The following assumptions have been made: (a) the nominal \litebird\ scanning strategy parameters are used, i.e., we did not study the possibility of dedicated, deep planetary scans; (b) the mission lasts 3 years starting from 1 January 2027; (c) only the part of the $\gamma(\mathbf{r})$ map within 3 times the FWHM from the main beam axis was considered for the fit; and (d) the analysis was run for one detector at 40\,GHz (LFT), one at 166\,GHz (MFT), and one at 402\,GHz (HFT).

The results of our simulations are reported in Table~\ref{tab:beams:inflight:MCMCerrors}. All the errors are well below 1\,\%, with the result at 402\,GHz being the largest in relative terms (0.4\,\%). The variability of the error estimates in our results\footnote{In a real-world scenario, part of this variability could also be due to the dependence of the planetary apparent radius on the observed frequency band, at least for gaseous planets.} depends on the combination of several factors: (a) the brightness temperature within the bandwidth of the detector; (b) the FWHM of the radiation pattern; and (c) the white noise level of the detector.

\begin{table}[htbp!]
    \caption{\label{tab:beams:inflight:MCMCerrors} Estimated error on the reconstruction of the FWHM of the radiation pattern for three \litebird\ detectors.}
    \centering
    \begin{tabular}{|l|c|}
    \hline
    Detector& Error [arcmin]\\
    \hline\hline
    LFT (40\,GHz)& 0.09\\
    MFT (166\,GHz)& 0.03\\
    HFT (402\,GHz)& 0.07\\
    \hline
    \end{tabular}
\end{table}

The estimation of the error does not take into account all sources of errors and systematic effects to be expected from real-world data. Examples include: uncertainty in background subtraction; presence of correlated noise in the timelines variability in the planetary emission; non-Gaussianities of the beams; and non-linearities and saturation of the bolometers. A rough estimation of those effects has been performed while running the same simulation code on \planck\ LFI data. 
The derived error is 2--2.5 times larger than the one quoted in \cite{Planck2015IV}; therefore, we can expect that a similar degradation will be experienced in the error estimates deduced from flight data.

Assuming a $2.5$ degradation factor for the results in Table~\ref{tab:beams:inflight:MCMCerrors} leads to a relative error in the range 0.25--1\,\%, where the lower bound applies to the channels dominated by cosmological signal. We expect that this error is small enough for data analysis, although a full end-to-end simulation is required to propagate the uncertainties down to the scientific products. 
Note that the estimation here only relies on a single source, Jupiter, and thus concatinating other planets in future analyses should give a modest improvement in the statistical error. 
Should the error be too large, we will consider the possibility to run dedicated observations of planets to increase the integration time. 

To complete this picture, a preliminary estimation using data from \planck\ of the additionnel optical loading from head-on observations of Jupiter is approximately 1\,\% at 40\,GHz, 10\,\% at 140\,GHz, and 80\,\% at 337\,GHz of the expected optical loading per band. The global impact of the detector non-linearities and saturation is under study. 

For the sidelobes, cross-checks between in-flight data, simulations, and ground-based characterization are foreseen. For example, using Galactic plane crossings and studying the beam shape as a decomposition of multiple functions.

\paragraph{Beam Effects Coupling with Other Systematics}
Some of the systematic effects appear as a combined effect of the beam together with other instrumental effects. 
Without going into full details, we briefly describe the nature of such effects. 
\begin{itemize}
    \item Frequency-dependent beam shape. This effect can be treated as a photometric gain calibration and will be discussed in a future publication. 
    The detailed study of the photometric gain calibration requirement is described in the next section.
     \item Bolometer transfer function and beam shape degeneracy in the time domain. This effect is related to both the time constant and to the optical crosstalk. For the former, we expect the nominal time constant of the bolometer to be a few tens of milliseconds~\cite{Westbrook2020SPIE}.  
    The two effects can be decoupled by simultaneously fitting them with planet observations, as was done for \planck~\cite{Planck2013VII}. 
    For the latter, any ghosting effect due to the internal reflections within the optical system can result in an extra beam feature that is different from the designed main beam. 
    This can be dealt with in the analysis either as an extra sidelobe feature of the beam or as crosstalk over the focal plane. 
    We will use planetary beam observations to monitor this effect, while electrical crosstalk will be calibrated prior to flight. As such, we foresee the ability to separately assess the electrical crosstalk so that it will not be degenerate with the optically induced crosstalk. 
\end{itemize}

\subsubsection{Bandpasses}
\label{bandpass-calib-section}
\paragraph{Requirements}
Bandpass frequency-resolution requirements have been studied in Ref.~\cite{Ghigna2020}, where an ideal rotating HWP has been assumed. The requirements are driven by the color-correction~\cite{Planck2013IX} uncertainty, that results from the calibration of the foreground signal when the CMB dipole is assumed as the photometric calibrator. With decreasing frequency-resolution of the bandpass response, the color-correction uncertainty increases. This appears clear from the expression for the color-correction factor $\gamma$:
\begin{equation}\label{eq:colorCorrection}
    \gamma_{\rm d,s}=\left(\frac{\int d\nu~G(\nu) \frac{I_{\rm d,s}(\nu)}{I_{\rm d,s}(\nu_0)}}{\int d\nu~G(\nu) \frac{\partial B(\nu, T)}{\partial T}\Big|_{T_0}}\right)\frac{\partial B(\nu_0, T)}{\partial T}\Big|_{T_0},
\end{equation}
where the indices ``d'' and ``s'' refer, respectively, to dust and synchrotron, $G(\nu)$ is the band-pass function, $I_{\rm d}(\nu)$ ($I_{\rm s}(\nu)$) is the dust (synchrotron) spectrum, $\nu_0$ is the effective central frequency of the given band, $B(\nu,T)$ is the blackbody spectrum, and $T_0$ is the CMB temperature.

In Ref.~\cite{Ghigna2020} the color-correction uncertainty resulting from a given finite spectral resolution of the bandpass response has been propagated to a bias on the reconstructed tensor-to-scalar ratio $r$ for all \litebird\ bands. The results show that the high-frequency channels drive the requirements due to the relative brightness of the dust signal.

\begin{figure}[htpb!]
	\centering
	\begin{minipage}{.475\textwidth}
	\centering
	\includegraphics[width=1\textwidth]{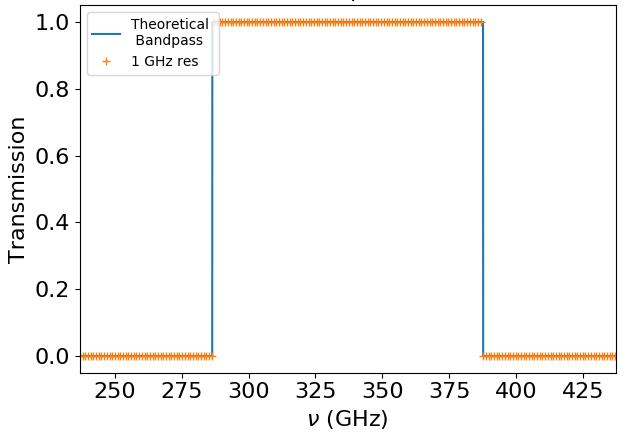}
	\end{minipage}%
	\begin{minipage}{.475\textwidth}
	\centering
	\includegraphics[width=1\textwidth]{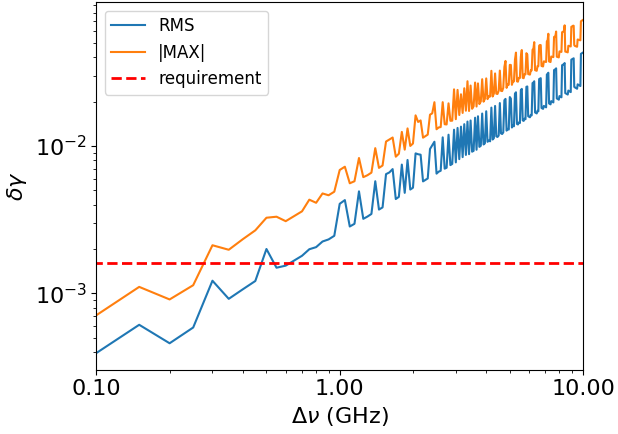}
	\end{minipage}
	\caption{Bandpass and color correction.  \textit{Left:} Top-hat bandpass response (in solid blue) at 337\,GHz with 30\,\% bandwidth. A resampling process with 1-GHz resolution is also shown in orange. \textit{Right:} Calculation of the color correction error for dust ($\delta \gamma_{\rm d}$) as a function of decreasing resolution for the bandpass response on the left. The blue solid line represents the rms value for 100 realizations of the resampling process with a given resolution, while the orange solid line represents the maximum value among 100 realizations. The requirement is shown by the red dashed line. Figure from Ref.~\cite{Ghigna2020}.}
	\label{fig:bandpass_Gamma}
\end{figure}

In Fig.~\ref{fig:bandpass_Gamma} we illustrate the worst case scenario, which corresponds to a top-hat bandpass (with 30\,\% bandwidth) for the 337-GHz channel and show the color-correction uncertainty as a function of the  frequency resolution. The requirement (red dashed line) is found by imposing the resulting tensor-to-scalar ratio bias (after component separation) to be $\leq5.7\times10^{-6}$ and corresponds to $0.2$\,GHz. The complete analysis shows that the spectral resolution should be in the range $\Delta\nu\simeq0.2$--2\,GHz, depending on the mean frequency of the band.

The detailed allocations from the current requirement to each hardware component are yet to be assigned. 
Comprehensive studies of the requirements, including the effects of more complicated foregrounds, such as CO lines, are beyond the scope of this paper.  

\paragraph{Prospects for ground calibration} 
For reference purposes in terms of achievable accuracy for the calibration of a CMB instrument in a space environment, we can consider \planck-HFI. The resolution on the reconstruction of the mean frequency of the bandpass was between 0.56\,GHz at 100\,GHz to 1.57\,GHz at 545\,GHz, while the bandwidth accuracy was between 0.4\,GHz and 3.1\,GHz, and the point-per-point spectral resolution of the measurements was sub-GHz~\cite{Ade2010}.
Still, uncertainties in the signal-to-noise (S/N) ratio achieved in these measurements were set by the limitations of the reference detector used in the calibration facility (see Ref.~\cite{Pajot2010} for a description of the facility). This limitation (especially in the lowest frequency channel, 100\,GHz) placed a tension between binning the spectral measurements to gain S/N against having a higher spectral resolution knowledge 
used in the removal of specific spectral features (such as CO). The strategy for \litebird\ is slightly different. At the component level, the spectral response characterization is planned to be performed using a \gls{vna} with extremely high spectral resolution in order to characterize the potential presence of inherent fringes. 
  At the instrument level, spectral measurements are part of the datacube reconstruction (cf.\ Sect.~\ref{datacube}), which should allow us to achieve the required sub-GHz resolution.
In addition, the out-of-band rejection (particularly important for component separation), which could be characterized down to levels of $10^{-12}$, as was done for the \planck-HFI filter chains~\cite{Ade2010}, will be reconstructed using a combination of component-level and end-to-end cryogenic tests.

\subsubsection{Polarization Angle}
\label{sss:anglecal}
In this section, we address the calibration of the polarization angle for \litebird. 

\paragraph{Requirements}

The derivation of the requirements on the relative and absolute 
polarization angles follows Ref.~\cite{Vielva_2021}. 
It is driven by the bias on $r$ produced by the miscalibration of polarization angles, taking into account the component separation process. 
Assuming a given correlation coefficient matrix $\rho_{\nu_1\nu_2}$ for 
the angle offset (either relative or absolute angles) of the frequency 
detector sets $\nu_1$ and $\nu_2$, 
it can be shown that the bias on $r$ is related to that matrix by the 
following expression: 
\begin{equation}
\label{eq:PVMdeltar2_ave}
\langle \delta_r \rangle \approx c^2 A \left(\sum_{\ell=2}^{\ell_{\mathrm{max}}}{H_\ell}\right)
\sum_{\nu_1=1}^{N_{\rm elem}}\sum_{\nu_2=1}^{N_{\rm elem}}\rho_{\nu_1, \nu_2},
\end{equation}
where 
\begin{eqnarray}
\label{eq:PVMcte}
A & = & 4 \left[ \sum_{\ell = 2}^{\ell_{\mathrm{max}}}{\frac{(B_{\ell}^{\rm fid})^2}{\mathrm{Var}(B_\ell)}} \right]^{-1},\\
H_{\ell} & = & \frac{E_{\ell}B_{\ell}^{\rm fid}}{\mathrm{Var}(B_\ell)} \, ,
\end{eqnarray}
in which $B_{\ell}^{\rm fid}$ and $B_{\ell}$ are the fiducial and observed $B$-mode 
power spectra, respectively, and $E_{\ell}$ is the fiducial $E$-mode power 
spectrum.
Equation~(\ref{eq:PVMdeltar2_ave}) provides the requirements on the polarization angle accuracy $\sigma_{\alpha_\nu}$ assuming a bias of $1\,\%$ of the systematics budget in the $r$ parameter.

\begin{figure}[htbp!]
    \centering
       \includegraphics[width=1.\textwidth]{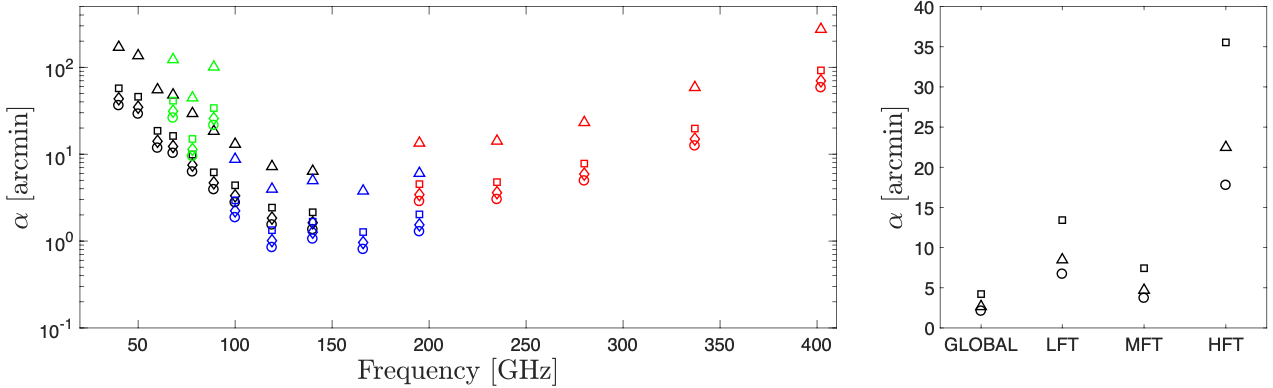}
    \caption{\textit{Left:} Required knowledge of the relative polarization angle for the 22 frequency elements within the LFT (black and green), MFT (blue), HFT (red), and the four correlation cases considered: 0 (square); 1 (circle); 2 (diamond); and 3 (triangle). \textit{Right:} Required knowledge of the absolute polarization angle for the four offsets and the three correlation cases considered: 0 (square); 1 (circle); and 2 (triangle). ``GLOBAL'' stands for the combination of all channels.
    }
    \label{fig:LBangle_requirement}
\end{figure}

The requirements on the relative angles refer to the $N_{\rm elem}=22$ frequency elements that are included in the three focal planes of the LFT, MFT, and HFT
(see Sect.~\ref{section:sensitivity}).
The following cases are considered:
\begin{itemize}
    \item Case 0, all 22 elements are uncorrelated, except those within the same telescope, which are fully correlated;
    \item Case 1, all 22 elements are fully correlated;
    \item Case 2, all 22 elements are partially correlated ($\rho_{\nu_1, \nu_2}=0.5$, for any $\nu_1\neq \nu_2$), except those within the same telescope, which are fully correlated;
    \item Case 3, all 22 elements are uncorrelated (the most ideal case).
\end{itemize}
The 22 elements correspond to the frequency bands per telescope and per detector pixel size, as listed in
Table~\ref{tbl:imo:sensitivities}.
The results of the relative angle requirements for each of those correlation cases are given in Fig.~\ref{fig:LBangle_requirement} (left panel). 

The requirements on the absolute angle consist of a global offset, which accounts for the mismatch between the SVM and the PLM, and three additional ones that account for the mismatch between the PLM and each of the three focal planes. The following cases have been considered:
\begin{itemize}
    \item Case 0, no correlations;
    \item Case 1, the four offsets are fully correlated;
    \item Case 2, the global offset is uncorrelated with any of the three focal plane offsets, and the latter ones are fully correlated.
\end{itemize}
The corresponding requirements are given in 
Fig.~\ref{fig:LBangle_requirement} (right panel). They appear less stringent for the high frequency channels, as opposed to the case of most of the other systematic effects, like, for instance, the beam knowledge requirements, which is more stringent at high frequency mainly because of the effect of the higher resolution in these bands.       

\paragraph{Prospects for In-flight Calibration using $EB$ Cross-correlation} 

For the in-flight absolute angle calibration,
we use the methodology of Refs.~\cite{Minami:2019ruj,Minami:2020xfg,MinamiKomatsu:2020},
which showed that we can calibrate the miscalibration angle, $\alpha$,
using the \textit{observed} power spectra, including both the CMB and the foreground, using full sky maps.
The basic idea is to use the relation between the
observed (``o'') $EB$ cross correlation and the difference of $EE$ and $BB$ power spectra as
\begin{equation}
 \langle C_\ell^{EB,\mathrm{o}}\rangle 
 = \frac{\tan(4\alpha)}{2}\left(\langle C_\ell^{EE,\mathrm{o}}\rangle- \langle C_\ell^{BB,\mathrm{o}} \rangle\right)
 + \langle C_\ell^{EB} \rangle,
\end{equation}
where $\langle C_\ell^{EB} \rangle$ is the ensemble average of the intrinsic $EB$ cross-correlation, which has been assumed to be zero both for the CMB and the foregrounds.
It was shown that we can calibrate the \litebird\ detectors absolute polarization angle with uncertainties of $<2.7$\,arcmin~\cite{MinamiKomatsu:2020}.

As a part of the feasibility demonstrations, a blind analysis was performed using sky maps that were simulated with a polarization angle offset. This study, summarized in Ref.~\cite{racpaper2021}, showed that the offset angles 
could be recovered by imposing the null detection of $C_\ell^{EB}$, using two different analyses.

This way of calibrating angles, called ``self-calibration'' \cite{keating/shimon/yadav:2012}, eliminates \litebird's ability to probe new parity-violating physics using the cosmological $EB$ correlation. The potential presence of the $EB$ correlation intrinsic to the Galactic foreground emission, which is yet to be found \cite{PlanckIntXLIX,Planck2018XI,huffenberger/rotti/collins:2020,clark/etal:2021}, could complicate the analysis further.
The method of Ref.~\cite{Minami:2019ruj} can mitigate these complications, restoring \litebird's ability to probe new physics and account for possible foreground $EB$ signals (see Sect.~\ref{ss:eb}).

\paragraph{Prospects for In-flight Calibration using the Crab Nebula}

The Crab Nebula (Tau~A) is one of the brightest compact sources in the sky in the microwave range, and the brightest one in polarization. For this reason, it has been used to calibrate CMB experiments, and is specifically the main target on the sky used to calibrate the polarization direction (together with the $C_\ell^{EB}$ technique). Interestingly, recent measurements show a discrepancy of $\simeq1^\circ$ between the polarization direction derived through the $C_\ell^{EB}$ method and through the use of the Crab~\cite{2014JCAP...10..007N,2014ApJ...794..171P}). 

Table~\ref{tab:sensitivity_to_crab} shows \litebird's expected sensitivities on the measurement of the Crab polarization angle. To obtain these values we have used: (i) a model for the polarization spectrum $S(\nu)=79.0(\nu/1\,{\rm GHz})^{-0.35}$~\cite{2018A&A...616A..35R}, and a value of the Crab's secular decrease of $-0.218\,\%{\rm year}^{-1}$, derived from \WMAP\ data~\cite{Weiland2010}; and (ii) \litebird\ mission nominal sensitivities, and a simulation of the satellite scanning strategy that gives the expected integration time at the position of the source. As can be seen in Table~\ref{tab:sensitivity_to_crab}, at the best frequencies the polarization angle of Tau~A will be measured with an accuracy of $\simeq 1$\,arcmin. For comparison, the error in the \planck-HFI measurement was $\simeq5$\,arcmin~\cite{Planck2015XXVI}.

\begin{table}[htbp!]
    \centering
    \caption{Expected \litebird\ sensitivity (statistical errors) on the measurement of the Crab Nebula polarization angle ($\Delta\gamma$), per individual bolometer, and when all bolometers in each frequency band are combined.}
    \label{tab:sensitivity_to_crab}
    \begin{tabular}{|c|c|c|}
    \hline
        Frequency & \multicolumn{2}{c|}{$\Delta\gamma$ (arcmin)}\\
        \cline{2-3}
        [GHz] & Per frequency band & Per detector \\
        \hline\hline 
      40     &       2.7	&     21   \\
      50     &       2.2	&     18   \\
      60     &       2.6	&     21   \\
      68     &       1.8	&     21   \\
      78     &       1.8	&     21   \\
      89     &       1.7	&     21   \\
     100     &       1.1	&     14   \\
     119     &       1.0	&     12   \\
     140     &       1.2	&     14   \\
     166     &       2.0	&     45   \\
     195     &       2.5	&     48   \\
     235     &       4.9	&     79   \\
     280     &       5.6	&     89   \\
     337     &       6.1	&     98   \\
     402     &       7.1	&    130   \\
     \hline
    \end{tabular}
\end{table}

Table~\ref{tab:sensitivity_to_crab} shows that \litebird\ will be able to measure the polarization angle of the Crab with a statistical error that is comparable with the requirement (see Fig.~\ref{fig:LBangle_requirement}), after integration of all detectors in each frequency band. The error per detector is, however, far from the requirement, and therefore precise relative calibration between detectors must be achieved by alternative techniques like ground-based calibration (see below). 

An important caveat is that, in this regime of very good statistical accuracy, the error budget will be fully dominated by the uncertainty in the model giving Crab's polarization angle. Currently, the best measurements have statistical errors of $\simeq 0.5^\circ$~\cite{2010A&A...514A..70A,Planck2015XXVI,2018A&A...616A..35R}. However, these observations are dominated by systematic uncertainties, which boost the global error up to to $>1^\circ$~\cite{2020A&A...634A.100A}. Recently Ref.~\cite{2018A&A...616A..35R} presented a new measurement at 150\,GHz obtained with the NIKA camera mounted on the IRAM 30-m telescope. After combination with previous measurements at different frequencies, and assuming no variation of the polarization angle with frequency, they derived a value for the Crab's polarization angle of $-87.7^\circ\pm 0.3^\circ$.  This error of 18\,arcmin is larger than the \lb\ requirement. Ref.~\cite{2020A&A...634A.100A} discussed the prospects for improving the global uncertainty to a level of $\simeq 6$\,arcmin through the addition of new measurements with individual error bars of around $0.2^\circ$ . This could be attempted through independent ground-based observations, using facilities like NIKA2, SCUBA-2, or SRT~\cite{2020A&A...634A.100A,2016SPIE.9914E..03F,doi:10.1142/S2251171715500087}.

Such observations should also allow us to study other possible errors derived from incorrect modeling hypotheses, in particular: (i) the independence of the Crab's polarization angle with frequency; (ii) the time dependence (it is well known that Crab's total-intensity flux density fades with time, but little is known about polarization); and (iii) the impact of the background emission that will affect the \litebird\ measurement due to its coarse beam.
Therefore, ideally these observations must: (i) cover a large frequency range; (ii) be spread in time; and (iii) cover a sufficiently large region around the Crab Nebula ($\simeq 2^\circ$ across).

\begin{figure}[htpb!]
\centering
\includegraphics[width = 0.99\textwidth]{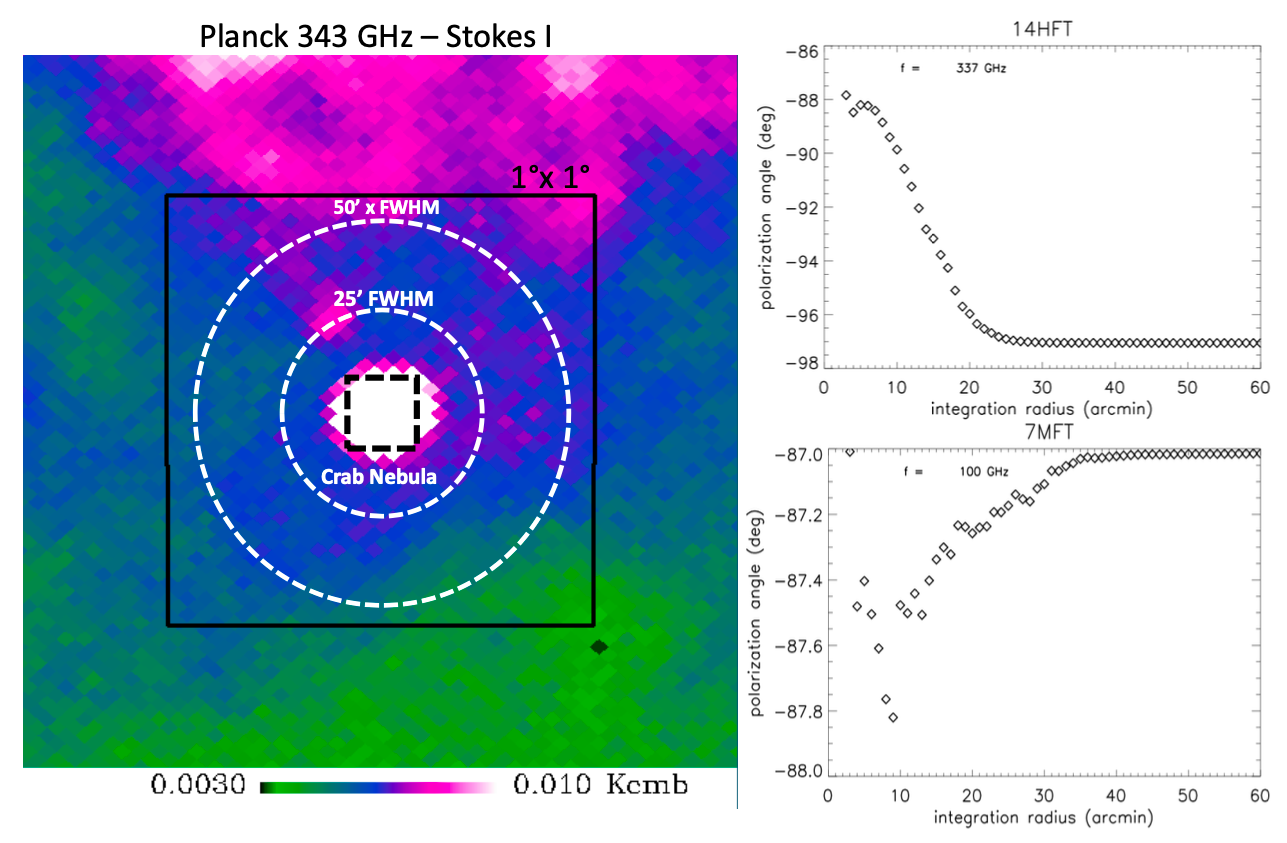}
\caption{Foreground emission surrounding the Crab Nebula. \textit{Left}: 353-GHz \planck\ map (Stokes $I$) centered on the Crab Nebula, displaying foreground emission in its surroundings, compared to the typical beam size of \litebird. \textit{Upper Right}: Polarization angle computed by convolving the \planck\ Stokes parameters maps at 337\,GHz with the \litebird-HFT beam, as a function of the integration radius.   
\textit{Lower Right}: the same at 100\,GHz (MFT). }
\label{fig:crab_reference_angle} 
\end{figure}

The effect of the background contamination is a relevant one, in particular at high frequencies, and we have studied this in detail~\cite{masi2021crab}. This is illustrated in Fig.~\ref{fig:crab_reference_angle} (left panel), where dust emission in the neighbourhood of the Crab Nebula is evident using 353-GHz data from \planck. This emission is partially polarized, and we have analysed its effect using the Stokes parameter maps measured by \planck\ at the relevant frequencies~\cite{Planck2018II,Planck2018III}, convolved with the expected \litebird\ beams. In order to assess the contribution of surrounding diffuse emission we have estimated the measured polarization angle, after convolving the maps with the beam over increasing radii. The resulting measured angle is significantly different from the polarization angle of the nebula itself, and stabilized for integration radii of the order of 1$^\circ$ (see Fig.~\ref{fig:crab_reference_angle}, center and right panels). The depth of the required surveys of the region has been investigated adding white noise to each 1.7\,arcmin pixel in the Stokes $Q$ and $U$ maps, and computing the measured angle obtained by convolution of the $Q$ and $U$ maps with the \litebird\ beams. Repeating the process 10{,}000 times we find that the standard deviation of the distribution of measured angles is smaller than the required 3\,arcmin if the noise per pixel in the reference Stokes parameter maps is below about 30\,$\mu$K rms at 100\,GHz, and 10\,$\mu$K rms at 337\,GHz. 
These stringent requirements motivate us to explore further detailed measurements of the Crab using forthcoming surveys in order to have the reference maps ready at the time of the \lb\ mission.

\paragraph{Prospects for Ground Calibration} 
\label{datacube}
As shown in Fig.~\ref{fig:calib-stra}, the general plan 
is to characterize the relative polarization angles
per focal plane during the end-to-end (E2E) cryogenic operations at the telescope level. Special care will be taken to define a common reference frame or reference transposition from sub-system characteristics to the telescopes, up to the PLM-SVM-integrated instrument.

For reference, the end-to-end ground calibration test campaign for \planck-HFI ultimately achieved an average of $1.5^\circ$ accuracy while the sub-system level (i.e., single cooled pixel) testing had achieved an average accuracy of $0.1^\circ$~\cite{Rosset2010,PlanckIntXLVI}. For \litebird\ since we are using both an HWP and sinuous antenna~\cite{aritokisuzuki-dthesis}, the bandpasses and the polarization angle are therefore highly coupled. It is possible to test for the overall spectral efficiency of a channel with an unpolarized source, and it is possible to accurately measure the maximum efficiency angle of a given detector, but for an end-to-end test of the whole instrument, the two effects are combined in a non-trivial way. We therefore plan to cleanly disentangle them via polarized spectroscopy to obtain the two spectro-polarimetric datacubes~\cite{Savini2009}.
The plan is to test the system with two orthogonal inputs (\gls{fts}, \gls{vna} or other coherent source) and to measure spectra at each small variation of HWP angle position in order to be able to reconstruct the expected modulation curve for any given spectral index source. 

The end-to-end calibration operations are complementary to the RF test campaign (cf.\ Sect.~\ref{Ground-beams}), which will specifically address the impact of the optical distortion and coating inefficiencies. 
A measurement campaign has already been performed on the LFT small-scale model and is reported in Ref.~\cite{Takakura2019IEEE}. 

In addition, various instrumental parameters may also impact the polarization angle and will be characterized at the component level to check their accuracy with respect to the requirements flowdown, in particular on the knowledge of the encoding angle HWP PMU (cf.\ Sects.~\ref{sss:plm_lft_pmu} and \ref{sss:plm_mhft_subsystems}).
In parallel, we are still working on the design and especially on the HWP, since any HWP frequency-dependent fast-axis
within the band can result in a mixing of the $Q$ and $U$ polarization states.  
Nevertheless, for a sapphire Pancharatnam HWP, there is a recipe that nulls the frequency response, and LFT PMU employs such a recipe to minimize this effect~\cite{Kunimoto2020SPIEDesign}, considering the fact that the residual variation may still produce $E$-to-$B$ leakage~\cite{2012ApJ...747...97B}; the study explicitly for \litebird\ is in progress.

\subsubsection{Calibration Development Plan}
\label{section:calib-plan}
The previous sections addressed the most challenging operations, but other instrumental parameters will need to be measured in the end-to-end cold integrated campaign. These are summarized in Table~\ref{tbl:calib}, together with the experimental setup planned to be used to characterize them and the proposed in-flight reconstruction. A dedicated characterization phase between the launch and arrival at L2 is also planned as a performance-verification phase prior to nominal observations; this phase is not detailed in this paper. 

\begin{table}[htbp!]
\caption{Instrumental parameters and calibration strategy to assess them in order to meet the accuracy required to
meet the systematic budget beyond sub-system testing. }
\label{tbl:calib}
\setlength{\tabcolsep}{3pt}
\begin{tabular}{|c|l|l|l|}
\hline
 & Instrumental param. & Instrument-level plans & In-flight plan \\
 &  & in cryogenic facility   & \\
\hline\hline
Beam & Far sidelobes & PLM cross-check & Absolute gains  \\
(co- and cross-polar)  & Near sidelobes &  RF & Planet \\
& Main Beam &  RF & Planet \\
\hline
Polarization & Instrumental polarn. &  E2E/DataCube & $C_\ell^{EB}$  and Crab   \\
& Absolute angle &  E2E/DataCube & $C_\ell^{EB}$ and Crab  \\
& Relative angle &  E2E/DataCube & $C_\ell^{EB}$ and Crab  \\
& Efficiency &  E2E/DataCube & $C_\ell^{EB}$  and Crab \\
\hline
Gain & Relative gain in time &  E2E/$\Delta T$ of the facility & Correlation with HK \\
& Relative and absolute  &  E2E/polar. source &$C_\ell$ norm. + Galaxy \\
& gains &  & + planet + dipole \\
\hline
Cosmic ray & Cosmic-ray glitches &  & Glitches \\
\hline
Spectral response &   &  E2E/FTS/DataCube &  \\
\hline
Transfer function & 
Noise PSD &  E2E & All data \\
& Detector time const.  &  E2E/chopped source & Glitches, planets \\
& Crosstalk &  E2E/polar. source  & Glitches, planets \\
& Linearity  &  E2E/polar. source & Glitches, planets  \\
& Sensitivity to bkg & E2E/$\Delta T$ of the load & Galactic crossings  \\
\hline
\end{tabular}
\end{table}

Finally a viability study (not in the baseline plans of the calibration strategy) has been performed for a calibration satellite (L2-CalSat), flying in formation at L2 and able to emit a reference signal~\cite{Casas_2021}. 
The calibration satellite is based on the CubeSat standard and has been conceived to travel as a piggy-back integrated on the service platform of the main satellite, \lb\ itself. A 6U CubeSat is expected to meet the volume needs of the ancillary satellite. An alternative implementation option would be a micro-satellite capable of reaching L2 autonomously and allowing a major flexibility at the system-design level. However, a major drawback would be the much larger amount of required propellant that would imply a significantly higher cost. The calibration satellite total mass would be 7.2\,kg and the required electrical power would be provided by two steerable solar panels and accumulated in a battery module. The solar panels would continuously generate 30\,W, enough to feed all the subsystems of L2-CalSat and charge the battery to provide the peak power required during calibration. L2-CalSat is required to be at a sufficiently far distance, of about 6\,km, in order to have thermal power levels at least 4 times lower than the saturation power of the detectors (thermal power levels are maximum at the highest frequency bands). The calibration satellite could help to provide strong control over many systematic errors, since those related to gain, non-linearity, spectral response, pointing, beam patterns, or absolute polarization angle. In particular, our analysis shows that the polar angle error can be reduced below 1\,arcmin for each detector. In relation to beam-pattern characterization, the calibration sources could provide noise-floor levels lower than $-70$\,dB in the overall band of interest (from 40 to 400\,GHz) with only one calibration session of about 1 day's duration. This would assure a complete and accurate characterization of the beam and in particular of the far sidelobes, placed at more than $10^\circ$ from the beam center.
We will assess the feasibility as the conceptual study progresses.





\section{Statistical and Systematic Uncertainties in the Tensor-to-scalar Ratio Measurement} 
\label{s:cosmological_forecasts}

This section describes the way that the forecasts on $r$ are derived. Section~\ref{ss:forecasts_imo} gives an overview the model of the instrument that is used for the simulations, with a particular focus on the expected statistical noise derivation. Section~\ref{ss:forecasts_fg_cleaning} summarizes the assumptions made for the foreground modeling, as well as the strategy concerning component separation. In Sect.~\ref{ss:forecasts_syst}, a review of the main systematic studies and their impact on the derivation of the systematic error on $r$ is detailed. Finally in Sects.~\ref{ss:forecasts_delta_r} and \ref{ss:extrasuccess}, a summary of the total expected error on $r$ is given, without and with the use of external data sets. 

\subsection{Instrument Model}
\label{ss:forecasts_imo}

\label{section:sensitivity} 

The \LiteBIRD\ \gls{imo} is a quantitative description of the entire \LiteBIRD\ experiment:
the three telescopes (LFT, MFT, and HFT); the payload; and the observational strategy. 
The IMo is used to store and track changes of the instrument design and is used in and interfaced to 
the forecasts/simulations/data analysis tools. Its use 
is required to ensure consistency between all the instrumental parameters,
and it gives the best current description of the instrumental design. 
This section gives an overview of the main parameters entering the IMo. 

\begin{figure}[htbp!]
\centering
\includegraphics[width = 0.8\textwidth]{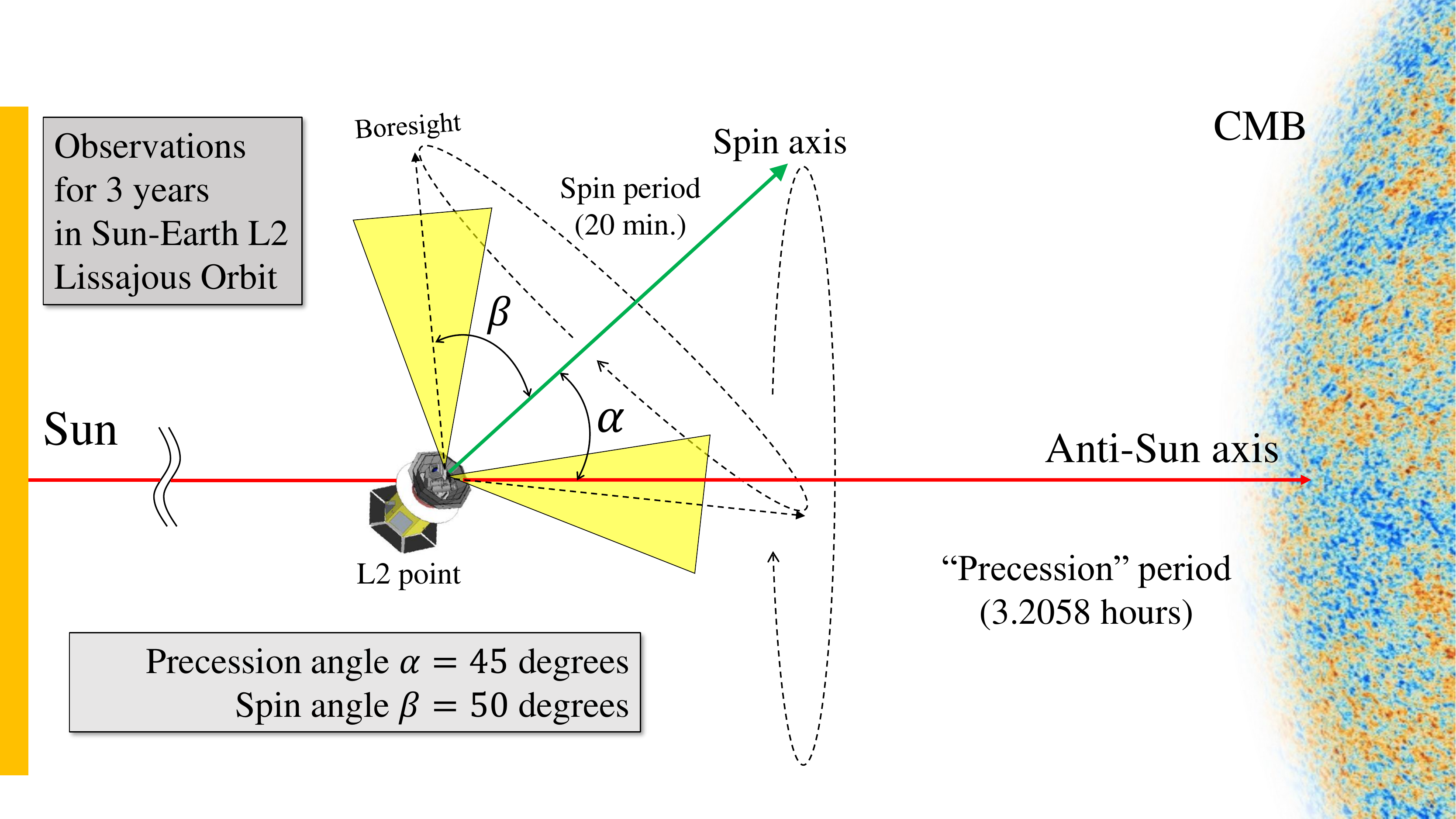}
\caption{Schematic of the observational parameters. The telescope boresight is at an angle $\beta= 50^\circ$ to the spin axis and rotates with a rate of 0.05\,rpm. The spin axis is rotated around the anti-Sun direction through precession with an angle $\alpha = 45^\circ$ in 3.2058\,hours. The anti-Sun axis rotates around the Sun in one year. With a combination of the three motions, the boresight can cover the entire sky in half a year. 
}
\label{fig:PLM-param-Imo}
\end{figure}

Table~\ref{tbl:imo1} 
gives the values of the main baseline parameters common to the three telescopes, LFT, MFT, and HFT.
The parameters related to the observational scan strategy are
further defined by the schematics shown in Fig.~\ref{fig:PLM-param-Imo}. 
The duration of the mission is assumed to be three years throughout this section,
except where explicitly stated otherwise.

We optimize the scan strategy parameters including $\alpha$, $\beta$, and the spin rate
so as to have the distributions as uniform as possible for the hits over the entire sky area and for the scanning directions in individual sky pixels.
A spin rate less than 0.05 rpm is found to give non-uniform distributions for both quantities.
The precession rate is determined to avoid moire patterns using the procedure described in Ref.~\citep{Hoang:2017wwv}.
The HWP spin rates are set so as not to overlap the HWP harmonics with the science observation frequency band determined by the beam FWHM.
The sampling rate is twice the Nyquist frequency of the HWP modulation.

\begin{table}[htbp!]
\setlength{\tabcolsep}{3pt}
\centering
\caption{Baseline observation strategy and parameters related to the sampling rate.}
\label{tbl:imo1}
\begin{tabular}{|c|c|c|c|c|c|c|c|}
\hline
$\alpha$ & $\beta$ & Precession rate & Spin rate & \multicolumn{3}{|c|}{HWP revolution rate [rpm]} & Sampling rate  \\ 
\cline{5-7}
[deg.] & [deg.] & [min.] &  [rpm] & LFT & MFT & HFT & [Hz] \\ \hline\hline
45 & 50 & 192.348 & 0.05 & 46 & 39 & 61 & 19  \\ \hline
\end{tabular}
\end{table}

Table~\ref{tbl:imo:sensitivities} summarizes the expected noise and sensitivities of the
different frequency bands of \LiteBIRD, together with related key parameters such as beam size (calculated using physical optics simulations similar to those described in Ref.~\cite{Gudmundsson:20}), and number
of detectors.
The details of the sensitivity calculations are given in Sect.~\ref{section:sensitivity}. 
By default the bandpasses are assumed to be tophat functions whose bandwidths are given
in the same table. 

\begin{table}[htbp!]
\centering
\caption{\LiteBIRD\ sensitivities. We show values related to the sensitivity in 15 observational bands. From left to right the columns are: the telescope covering the band (or ``comb.'' for combined values); the band identification number; the band center frequency in GHz; the bandwidth in GHz and its ratio to the center frequency; the main beam size FWHM in arcmin; the total number of detectors; the array NET in $\mu$K$\sqrt{\rm s}$; and the sensitivity in $\mu$K-arcmin.
The two values of the beam size and the sensitivity for the LFT 68-, 78-, and 89-GHz bands in  parentheses are for the two detector pixel sizes of 16 and 32\,mm.
See Table~\ref{tbl:plm:sensitivities} for the detector pixel sizes and detector NETs.
}
\label{tbl:imo:sensitivities}
\setlength{\tabcolsep}{3pt}
\begin{tabular}{|c|c|c|c|c|c|c|c|}
\hline
 & ID &  $\nu$   & $\delta\nu$ [GHz] & Beam size &  No. of  & NET$_{\rm arr}$ & Sensitivity   \\ 
 & & [GHz] & ($\delta\nu/\nu$) & [arcmin]  & detectors &
 [$\mu$K$\sqrt{\rm s}$] & [$\mu{\rm K}$-arcmin]\\
 \hline\hline
LFT	      	& 1			& 40 		& 12 (0.30)	 	& 70.5	 	    		& 48			& 18.50				  	& 37.42 \\ 		
\hline
LFT	      	& 2			& 50 		& 15 (0.30)	 	& 58.5					& 24			& 16.54				  	& 33.46 \\ 	
\hline
LFT	      	& 3			& 60 		& 14 (0.23)	 	& 51.1					& 48			& 10.54				  	& 21.31 \\ 	
\hline
LFT	      	& 4			& 68 		& 16 (0.23)	 	& (41.6, 47.1)			& (144, 24)		& (9.84, 15.70)		  	& (19.91, 31.77) \\
comb.	&			&			&				&							&				& 8.34					& 16.87\\ 			
\hline
LFT	      	& 5			& 78 		& 18 (0.23)	 	& (36.9, 43.8)			& (144, 48)		& (7.69, 9.46)			& (15.55, 19.13) \\
comb.	&			&			&				&							&				& 5.97					& 12.07\\ 		
\hline
LFT	      	& 6			& 89 		& 20 (0.23)	 	& (33.0, 41.5)			& (144, 24)		& (6.07, 14.22)			& (12.28, 28.77) \\
comb.	&			&			&				&							&				& 5.58					& 11.30 \\
\hline
LFT/  		& 7			& 100 		& 23 (0.23)		& 30.2/					& 144/			& 5.11/					& 10.34 \\
MFT			&			&			& 				& 37.8 		 			& 366 			& 4.19 					& 8.48 \\
comb.	&			&			&				&							&				& 3.24					& 6.56 \\			
\hline
LFT/		& 8			& 119 		& 36 (0.30)		& 26.3/					& 144/			& 3.8/					& 7.69 \\
MFT			&			&			&				& 33.6					& 488			& 2.82					& 5.70 \\
comb.	&			&			&				&							&				& 2.26					& 4.58 \\
\hline
LFT/		& 9			& 140 		& 42 (0.30)		& 23.7/					& 144/			& 3.58/					& 7.25 \\
MFT			&			&			&				& 30.8					& 366			& 3.16					& 6.38 \\
comb.	&			&			&				&							&				& 2.37					& 4.79 \\		
\hline
MFT	      	& 10		& 166		& 50 (0.30)		& 28.9					& 488			& 2.75					& 5.57 \\
\hline
MFT/   		& 11		& 195		& 59 (0.30)		& 28.0/					& 366/			& 3.48/ 				& 7.05 \\
HFT			&			&			&				& 28.6					& 254			& 5.19					& 10.50 \\	
comb.	&			&			&				&							&				& 2.89					& 5.85 \\
\hline
HFT	      	& 12		& 235		& 71 (0.30)		& 24.7					& 254			& 5.34					& 10.79 \\
\hline
HFT	      	& 13		& 280		& 84 (0.30)		& 22.5					& 254			& 6.82					& 13.80 \\
\hline
HFT	      	& 14		& 337		& 101 (0.30)	& 20.9					& 254			& 10.85					& 21.95 \\
\hline
HFT	      	& 15		& 402		& 92 (0.23)		& 17.9					& 338			& 23.45					& 47.45 \\
\hline
Total   	&			&			&				&						& 4508			&						& 2.16 \\
\hline
\end{tabular}
\end{table}



The \gls{nep} of each detector is predicted from a quadrature sum of the expected photon noise ($\rm{NEP_{\rm ph}}$), thermal carrier noise of the bolometer ($\rm{NEP_{\rm th}}$), readout noise ($\rm{NEP_{\rm read}}$), and extra noise ($\rm{NEP_{\rm ext}}$) sourced from the environment and unknowns. The NEP of a single detector ($\rm{NEP_{\rm det}}$) may therefore be expressed as
\begin{equation}
	\mathrm{NEP_{\rm det}} = \sqrt{\mathrm{NEP_{\rm ph}^{2} + NEP_{\rm th}^{2} + NEP_{\rm read}^{2}+ NEP_{\rm ext}^{2}}},
	\label{eq:nep}
\end{equation}
where all the components are assumed to be mutually uncorrelated. The quantity $\rm{NEP_{\rm ph}}$ is deduced from the expected loading on the detectors (taking account of the various optical elements, their intrinsic characteristics, and the temperature to which they are cooled), while $\rm{NEP_{\rm th}}$ is derived from the temperature of the bath and the expected critical temperature of the bolometers. The readout noise and the external noise are required to increase the internal detector noise (encompassing the detector and the thermal carrier noise) by 
no more than 10\,\% and the total detector noise by 15\,\%, respectively.

We define the \gls{net} of a detector as 
\begin{equation}
	\mathrm{NET_{\rm det}} = \frac{\mathrm{NEP_{\rm det}}}{\sqrt{2} \, (dP/dT_{\rm CMB})},
\end{equation}
where we define the conversion factor from power to CMB thermodynamic temperature through
\begin{equation}
	\frac{dP}{dT_{\rm CMB}} = \int_{\nu_1}^{\nu_{2}} \Big[ \, k_{\rm B}\eta(\nu) \Big(\frac{h \nu}{k_{\rm B}T_{\rm CMB} (e^{h \nu / k_{\rm B} T_{\rm CMB}} - 1)} \Big)^{2} \, e^{h \nu / k_{\rm B} T_{\rm CMB}} \, \Big] \mathrm{d}{\nu}.
	\label{eq:NETcalculation}
\end{equation}
The parameters $\nu_{1}$ and $\nu_{2}$ represent the lower and upper edges of the frequency band, respectively, while $\eta$ is an overall optical efficiency factor. 
Detector NETs for each band are shown in Table~\ref{tbl:plm:sensitivities}.
The NET for an array is calculated as 
\begin{equation}
	\mathrm{NET_{\rm arr}} = \frac{\mathrm{NET_{\rm det}}}{\sqrt{N_{\rm det}\times0.8}},
\end{equation}
where $N_{\rm det}$ is the number of detectors in the given frequency band and
the factor of 0.8 represents a degradation factor for the detector yield. 
The NETs of a frequency band covered by multiple telescopes are combined into the average 
$\rm{NET_{\rm comb}}$ as
\begin{equation}
\mathrm{NET_{\rm comb}} = \sqrt{\frac{1}{ \sum_{i}(1/\mathrm{NET_{i}}^2) }},
\label{eq:ivws}
\end{equation}
where the index $i$ runs over the telescopes of LFT/MFT/HFT.

Finally, the statistical sensitivity ($\sigma_{\rm S}$) in $\mu$K-arcmin is given as
\begin{equation}
	\sigma_{\rm S} = \sqrt{\frac{4 \pi f_{\rm sky} \, 2 \, \mathrm{NET^2_{\rm comb}}}{t_{\rm obs}}} \left( \frac{10800}{\pi} \right),
\end{equation}
where we assume the sky coverage fraction $f_{\rm sky} = 1.0$ 
and $t_{\rm obs} = 3$\,years (94{,}672{,}800 sec) $\times \eta$.  Here
$\eta$ is the observation efficiency, derived by taking into account degradation factors on 
the observation duty cycle
($\eta_{\rm duty}=0.85$),
the data loss due to cosmic rays
($\eta_{\rm cr}=0.95$),
and a margin
$\eta_{\rm margin}=0.95$.  Hence we have 
$\eta = \eta_{\rm duty}\eta_{\rm cr}\eta_{\rm margin}=0.77$.
We note that the factor of $\mathrm{\sqrt{2}}$ here originates from 
the fact that the polarization is measured using two independent pieces of data.
The total sensitivity is obtained by using the NET values combined with all the NET$_{\rm comb}$ values of individual frequency channels using Eq.~(\ref{eq:ivws}).

\subsection{Foreground Cleaning}
\label{ss:forecasts_fg_cleaning}

\begin{figure}[htbp!]
\begin{center}
\includegraphics[width = 1.0\textwidth]{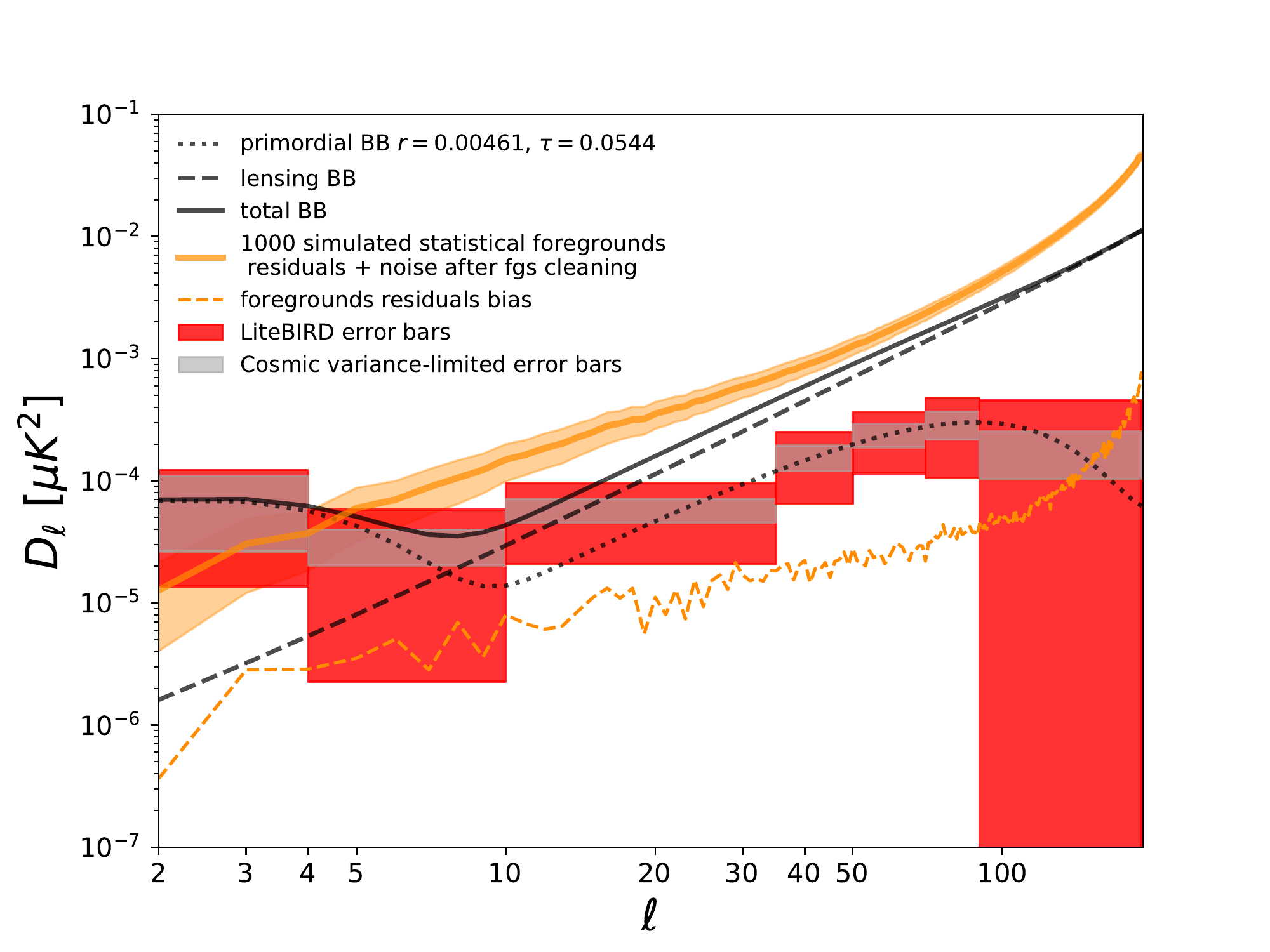}
\end{center}
\caption{\LiteBIRD\ error bars on the primordial $BB$ power spectrum for $r=0.00461$ and $\tau=0.0544$, where $D_\ell = \ell(\ell+1)C_\ell/2\pi$. 
There are two types of error bars: cosmic-variance (including primordial and lensing $B$-mode variance) shown as gray boxes; and total error bars (including cosmic variance and averaged, noisy foreground residuals), shown in red. The input foreground maps are described in Sect.~\ref{sec:reference_skies}. The solid orange curve represents the average total residuals (including statistical and systematic foreground residuals, as well as noise after component separation, accounted as $N_\ell$ in Eq.~(\ref{eq:cl-measured-in-likelihood})), with the light band corresponding to the scatter across noise simulations. The dashed orange curve represents the noiseless foreground residuals, interpreted as the bias on $B$-mode power, eventually leading to a bias on the tensor-to-scalar ratio. Although the contribution of the statistical foreground residuals, noise, and lensing $B$ modes are included in the error bars, the latter are centered on the theoretical primordial $BB$ curve.
}
\label{fig:xforecast-results-pysm}
\end{figure}

In this section, we describe the reference sky we adopt, and which elements from the IMo in Sect.~\ref{ss:forecasts_imo} we exploit in order to generate the \LiteBIRD-simulated multi-frequency input maps for treatment through component separation. We then describe various methodologies, as well as the products of these analyses that are exploited in the following sections. 

\subsubsection{Reference Skies}
\label{sec:reference_skies}
We adopt a native resolution of all maps corresponding to \texttt{HEALPix} $N_{\rm side}=512$~\cite{gorski/etal:2005}. The diffuse emission, corresponding to CMB and Galactic foregrounds, are in thermodynamical CMB temperature units, $\mu$K$_{\rm CMB}$. The CMB angular power spectra correspond to the \Planck\ 2018 best-fit parameters \citep{Planck2018VI}, with the six $\Lambda$CDM parameters corresponding to $\{H_{0}, \Omega_{\rm b}h^{2}, \Omega_{\rm c}h^{2}, \tau, n_{\rm s}, A_{\rm s}\}$ = $\{67.36\ {\rm km}\,{\rm s}^{-1}\,{\rm Mpc}^{-1}, 0.02237, 0.12, 0.0544, 0.965, 2.099\times 10^{-9}\}$, with lensing and with no tensors, i.e., $r=0$.

Diffuse foregrounds are produced with the public package Python Sky Model ({\tt PySM}) with default implementation of the amplitude of foregrounds, and spatially varying spectral indices~\citep{2017MNRAS.469.2821T,2021arXiv210801444Z}, corresponding to the {\tt d1s1} parametrization in {\tt PySM}. The Galactic polarized emission is composed of thermal dust and synchrotron models, dominating the polarization emission at frequencies higher and lower than about 70\,GHz, respectively~\citep{krachmalnicoff/etal:2018}. In this work, the polarized synchrotron emission is modeled as 
\begin{equation}
[Q_{\rm s},U_{\rm s}](\hat n,\nu)=[Q_{\rm s},U_{\rm s}](\hat n,\nu_*)\left(\frac{\nu}{\nu_*}\right)^{\beta_{\rm s}(\hat n)}\,,
\label{eq:synchspectrum}
\end{equation}
where $[Q_{\rm s},U_{\rm s}](\hat n,\nu)$ are the Stokes parameters of synchrotron emission in Rayleigh-Jeans (RJ) temperature units on a given line of sight $\hat n$ at a frequency $\nu$, $\nu_*$ is a reference frequency, and $\beta_{\rm s}(\hat n)$ is the spatially-varying synchrotron spectral index. On the $50\,\%$ sky mask shown in Fig.~\ref{fig:foreground_masks}, the {\tt PySM} template gives $\beta_{\rm s} = -2.993  \pm  0.046$.
The polarized thermal dust emission is modeled as a modified blackbody, corresponding to the expressions 
\begin{equation}
[Q_{\rm d},U_{\rm d}](\hat n,\nu)=[Q_{\rm d},U_{\rm d}](\hat n,\nu_*)\left(\frac{\nu}{\nu_*}\right)^{\beta_{\rm d}(\hat n)-2}\frac{B[\nu,T_{\rm d}(\hat n)]}{B[\nu_*,T_{\rm d}(\hat n)]}\,,
\label{eq:mbbspectrum}
\end{equation}
where $[Q_{\rm d},U_{\rm d}](\hat n,\nu)$ are the Stokes parameters of thermal dust emission in RJ temperature units, $\beta_{\rm d}(\hat n)$ is the dust spectral index (power-law index of dust emissivity),  $T_{\rm d}(\hat n)$ is the effective temperature of dust, and $B(\nu,T_{\rm d})$ is a blackbody spectrum with a temperature $T_{\rm d}$. Similarly to the synchrotron spectral index, {\tt PySM} templates show $\beta_{\rm d} = 1.534 \pm  0.028$ and $T_{\rm d}=22.23 \pm 1.56$K variations across the $f_{\rm sky}=49.5\,\%$ mask shown in Fig.~\ref{fig:foreground_masks}. The total intensity of dust and synchrotron scale in frequency just like polarization. 
According to predictions~\citep{Puglisi_2018,li2021}, obtained using the publicly available Point Source ForeCast package,\footnote{\url{giuspugl.github.io/ps4c/index.html}} polarized compact sources are expected not to play a major role in \LiteBIRD\ measurements; bright radio sources will need to be removed with dedicated filters, and for the infrared populations only upper limits exist in polarization \citep[see e.g.,][]{Planck2015XXVI}. Thus, we do not include their treatment in the present work, focusing on the capabilities of \LiteBIRD\ to deal with diffuse foregrounds.

\subsubsection{Input Maps} 
\label{sec:input_maps}

The templates of the reference sky described above are scaled in frequency using the {\tt PySM} prescriptions and combined with instrumental features from the IMo. These are specifically: 
\begin{itemize}
\item bandpass integration;
\item beam convolution; 
\item addition of white noise. 
\end{itemize} 
The IMo objects used for the above operations are described in Sect.~\ref{ss:forecasts_imo} and we report here their exploitation for achieving the series of multi-frequency maps to be processed by component separation. All templates are convolved with top-hat bandpasses (see Table~\ref{tbl:imo:sensitivities} for the band centers and widths) and co-added to form input sky maps. Concerning beam convolution, two series of maps are produced, for comparison purposes. The first one is a direct convolution with a Gaussian circular beam shape with FWHMs from the IMo. 
The second one is a timeline-based series of simulations, generated with the public package \gls{toast}\footnote{\url{hpc4cmb.github.io/toast}}. 
The convolution of the signal with the beam simulated in {\tt GRASP} (encoding for the whole $4\pi$ beam pattern) is performed with the \texttt{libconviqt}\footnote{\url{https://github.com/hpc4cmb/libconviqt}} library, implementing beam convolution on a sphere based on Ref.~\citep{Pr_zeau_2010}.

For the first set of simulations,
the noise was added to the co-added multi-frequency and multi-component maps for the Gaussian circular beam convolutions. A total of 1000 noise and CMB simulations were made available using the procedures described in the next subsection. 
We use the second set for a study of the beam far sidelobe systematic effects in Sect.~\ref{sec:beam-systematics}. 

\subsubsection{Component Separation}
\label{sec:component_separation}

CMB extraction and diffuse foreground reconstruction may be achieved through broad classes of techniques \citep[see Ref.][and references therein]{Planck2018IV}. Parametric fitting consists of constructing a sky model on the basis of a suitable parametrization of foreground and CMB unknowns and estimating the latter by exploiting a multi-frequency data set. Internal linear combination and template-fitting methods consist of the minimization of a linear mixture of the multi-frequency data and the subtraction of foreground templates that come from external data, or are derived internally, respectively. Baseline results for the present paper were obtained with such a parametric fitting method, as discussed below. 

\begin{figure}[htbp!]
    \centering
    \includegraphics[scale=.4]{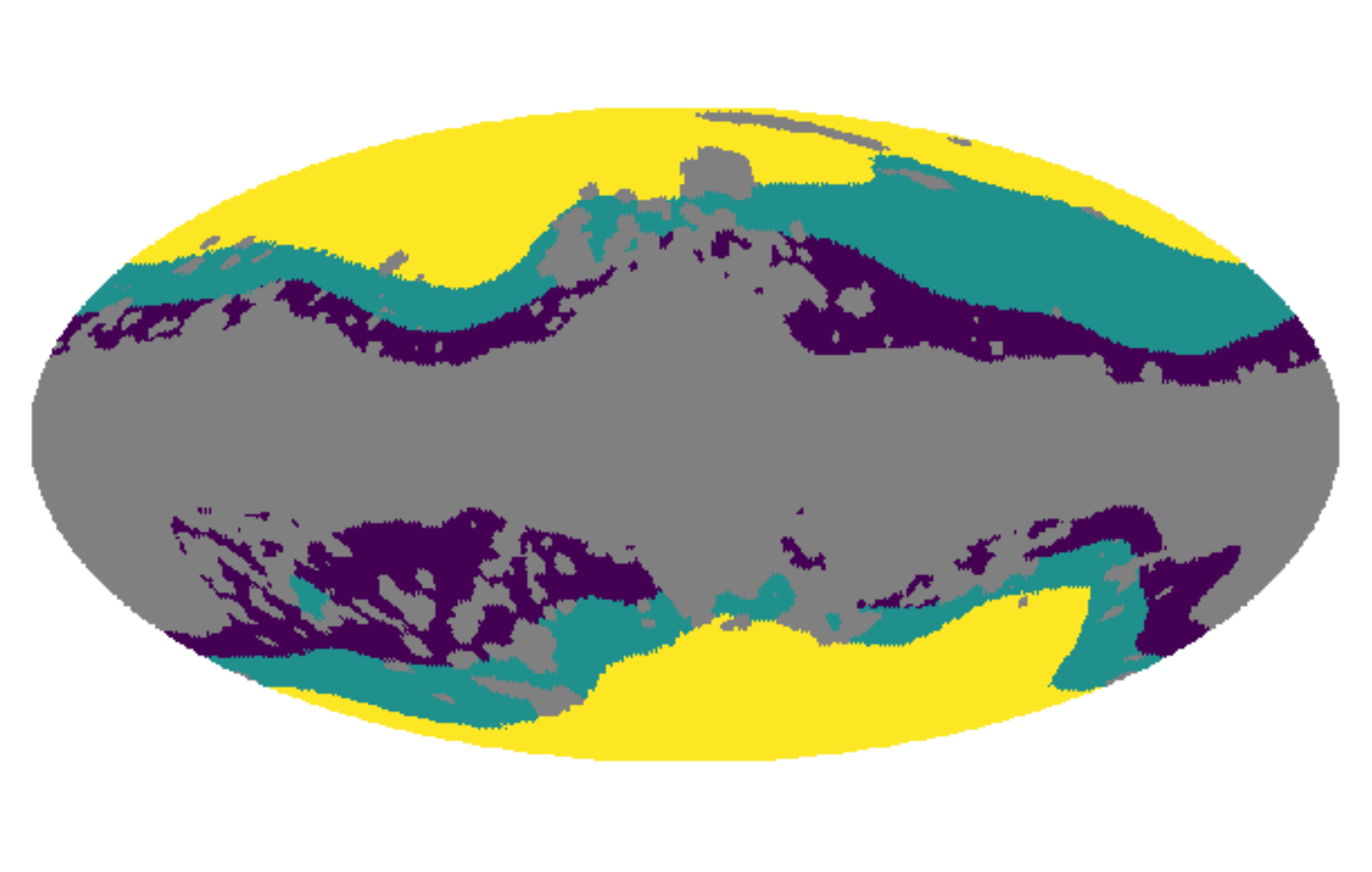}\\
    \vspace{30pt}
    \includegraphics[scale=1]{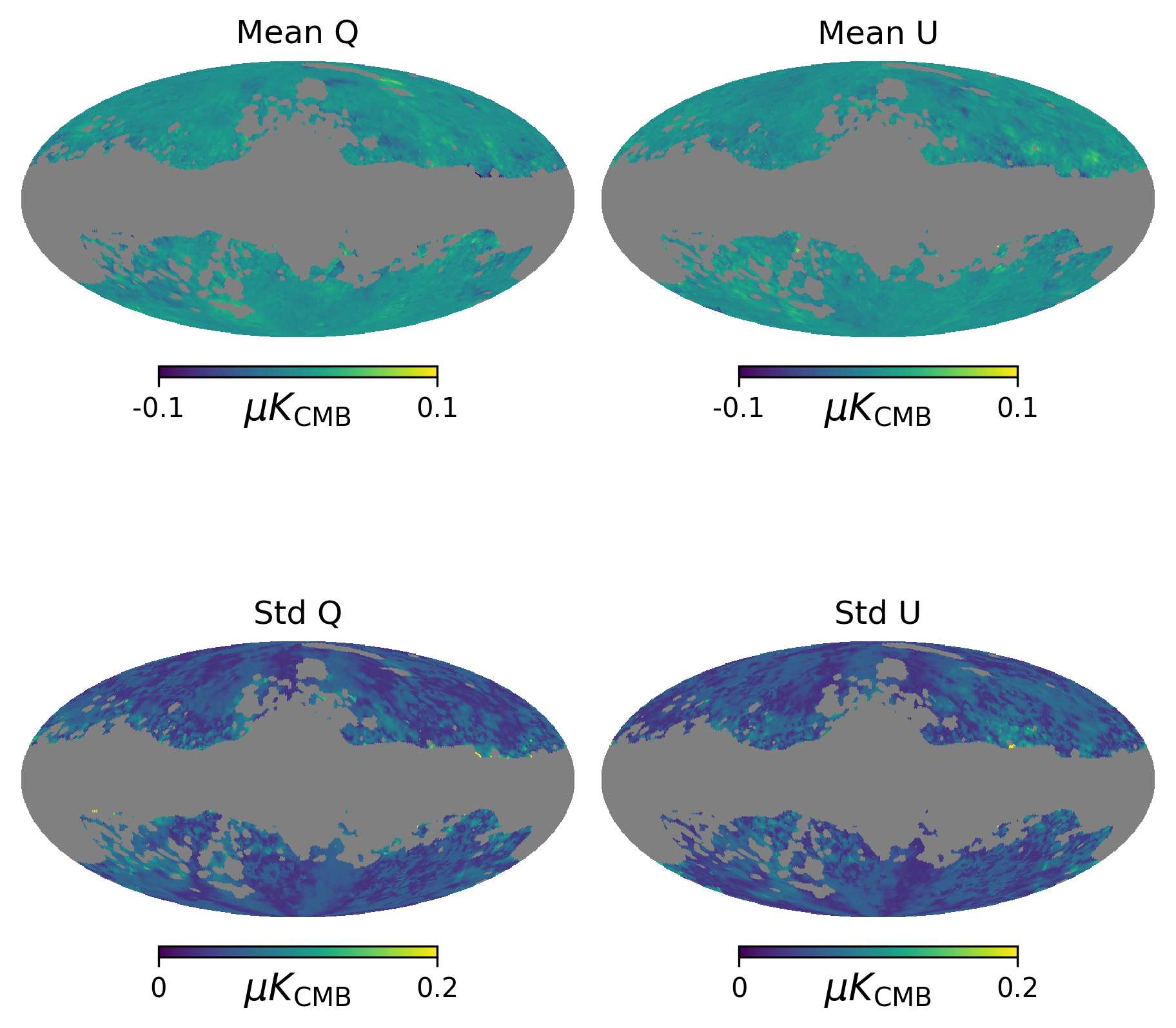}
    \caption{Top panel: Masks used for foreground analyses.  The gray region shows the sky area excluded from the foreground analysis, while the colors show areas with homogeneous resolution for the foreground spectral parameters. Middle (bottom) panel: map of the mean (standard deviation) of the foregrounds residuals.}
    \label{fig:foreground_masks}
\end{figure}

In the parametric map-based component-separation tool ``ForeGround Buster'' (FGBuster\footnote{\url{github.com/fgbuster/fgbuster}} \cite{stompor/etal:2009}), our approach is to assume a foreground model with three spectral parameters determining their \glspl{sed}: a spectral index $\beta_{\rm d}$ and a temperature $T_{\rm d}$ of a modified blackbody for dust; and the spectral index $\beta_{\rm s}$ of a power law for synchrotron. These spectral parameters are assumed to be constant over \texttt{Healpix} pixels at a given ${N_{\rm side}}$. Of course, this represents an approximation of the model, whereas the foreground spectral parameters will in general vary smoothly across the sky. In order to balance between the statistical uncertainties sourced by the number of free parameters, and the need for a flexible enough model capturing the input-sky complexity (especially the spatial variability of the SEDs), we fit the different spectral parameters at different resolutions depending on Galactic latitude. Since the optimal balance depends on the sky region, we divide the sky into (almost) isolatitude areas covering 20\,\% of the sky each; we rank them by their expected level of foreground contamination and consider only the three least contaminated. 
The ${N_{\rm side}}$ parameters for these regions are
given in Table~\ref{tab:ThreeNsides}.
Here the three columns correspond to the three 20\,\%-parts of the sky, from the lowest to highest Galactic latitudes, defined from the \Planck\ post-processing masks adopted for component separation \citep{Planck2018IV}, referred to as \textit{GAL20}, \textit{GAL40}, and \textit{GAL60}, available on the Planck Legacy Archive\footnote{\url{pla.esac.esa.int}} and
shown in Fig.~\ref{fig:foreground_masks}. The frequency coverage and sensitivity of \LiteBIRD\ allow us to estimate foreground parameters and CMB amplitudes in suitably small sky areas or resolution elements. Thus, the number of non-linear parameters, $\{\beta_{\rm d}, T_{\rm d}, \beta_{\rm s}\}$, as driven by the resolutions in Table~\ref{tab:ThreeNsides}, computed in the $f_{\rm sky}=49\%$ observed sky, is precisely $24{,}545$. The number of $Q$/$U$ amplitudes contained in the recovered CMB, dust and synchrotron maps is $144{,}534$. Therefore, in total, the component-separation process deals with $169{,}079$ free parameters. This can be compared to the $10^6$-long data vector as formed by the observed $Q$/$U$ amplitudes in the 22 frequency channels.

\begin{table}[htbp!]
\caption{Values of $N_{\rm side}$ for foreground parameter variations in three different
20\,\% large regions of the sky. The $N_{\rm side}$ values are chosen to capture most of the spatial variability of the spectral indices and to keep the resulting foreground bias under control while not increasing the statistical uncertainty. The angular resolution for $\beta_{\rm d}$ is kept to the maximum allowed ($N_{\rm side}=64$), since \liteBIRD\ turns out to be particularly sensitive to this parameter.}\label{tab:ThreeNsides}
\begin{center}
\begin{tabular}{|c|c|c|c|}
\hline
& \multicolumn{3}{c|}{Galactic latitude} \\
\cline{2-4}
& Low & Medium & High \\
\hline\hline
 $\beta_{\rm d}$ & 64 & 64 & 64 \\ 
 $T_{\rm d}$ & 8 & 4 & 0 \\  
 $\beta_{\rm s}$ & 4 & 2 & 2 \\
 \hline
\end{tabular}
\end{center}
\end{table}

The unmodeled foreground component leaks into the CMB reconstruction and sources the foreground systematic residuals reported in Fig.~\ref{fig:xforecast-results-pysm} and eventually can give a non-zero bias on $r$, as introduced in~\cite{stompor/etal:2016}. The noisy estimation of the spectral parameters, instead, sources the so-called ``statistical foreground residuals'' and eventually the uncertainty $\sigma(r)$. We estimate the latter with a Fisher-matrix approach, where the total covariance ${\sf C}$ contains CMB lensing, noise, and statistical residuals.
Namely, we calculate the second derivative of the likelihood as $F=-\langle \partial^2\log({\mathcal L})/\partial r^2\rangle$ and obtain
\begin{equation}
\sigma(r=0)=1/\sqrt{F}.
\label{eq:sigma-r-definition}
\end{equation}
Here, $-2\log{\cal L}=f_{\rm sky}\left({\rm tr}{\sf C}^{-1}{\bf D}+\log{\rm det}{\sf C}\right)$, where ${\bf D}$ is the simulated data vector~\cite{Errard:2018ctl}, calculated using the noise and residuals spectra averaged over 1000 foregrounds+noise realizations. For each of them, any non-zero recovered CMB $B$ modes are interpreted as foreground leakage. 

\subsubsection{Products}
\label{sec:products}
Once the residuals are estimated on 60\,\% of the sky, their power spectra are evaluated in a sky region corresponding to $f_{\rm sky} = 49.5\,\%$, obtained by intersecting the 60\,\% mask with foreground residuals estimated through component separation (see Fig.~\ref{fig:foreground_masks}). 
The mask is derived by: (1) removing regions with the highest foregrounds residuals; (2) smoothing it with a $2^\circ$ beam; and (3) converting it to a binary mask using a threshold equal to 0.5. The  threshold for step (1) is chosen so that the bias on $r$ is low enough while keeping the sky coverage large enough to lower as much as possible the associated $\sigma(r)$. Although this exact methodology will not be applicable to real data sets, a better understanding of foreground morphology, as well as additional observational/theoretical priors on their SEDs, will help us to better define the Galactic masks by the time of the \LiteBIRD\ launch.
The angular power spectra of the foreground residuals are estimated using pseudo-$C_{\ell}$s corrected for the mask, evaluated using the {\tt Healpy Anafast}\footnote{ \url{https://healpy.readthedocs.io/en/latest/generated/healpy.sphtfunc.anafast.html}} public code. We checked that the power spectra of foreground residuals do not change if they are estimated by the methods such as {\tt NaMaster}\footnote{\url{github.com/LSSTDESC/NaMaster}},
mainly because foreground $E$ and $B$ modes have similar amplitudes.

Figure~\ref{fig:xforecast-results-pysm} shows the result of our foreground-cleaning method applied to the multi-frequency data set specified in Sect.~\ref{sec:input_maps}, in the case of Gaussian beam convolution. We display the expected error bars on the primordial $BB$ spectra for $r=0.00461$, corresponding to the reference Starobinsky model introduced in Sects.~\ref{ss:cmbpol_r}, \ref{ss:cmbpol_beyondpower}~\cite{starobinsky:1980}, 
and $\tau=0.0544$~\cite{Planck2015XIII}. The error bars are shown for pure cosmic variance, including the primordial and lensing $B$-mode variance, as well as for the total uncertainty coming from cosmic variance plus noisy foreground residuals. The total residuals from foregrounds and noise after component separation, averaged over the 1000 simulations, are shown in Fig.~\ref{fig:xforecast-results-pysm} (orange area, with average given by the dark orange line), as well as the noiseless foreground residuals (dashed orange line), otherwise known as ``systematic foreground residuals.''  The first element determines the accuracy on $r$ via power spectrum estimation, namely the statistical error including foreground cleaning and noise, while the second is responsible for the bias on the recovered $BB$ spectrum and thus on $r$. Due to their large amplitudes, the statistical foreground residuals, sourced by statistical errors on the estimated spectral indices, dominate the overall error budget.

From these averaged spectra, the corresponding bias on $r_{\rm FG}$, as well as the accuracy $\sigma(r=0)$ after foreground cleaning, is given by $r_{\rm FG}=\left(3.3 \pm 6.2\right) \times 10^{-4}$. This result is obtained for an input value $r=0.00$, while the same analysis with an input $r=0.01$ leads to $\sigma(r=0.01)= 0.0013$.
We find $\sigma(r=0.01)=0.0036$ from the reionization bump only ($\ell\,{<}\,20$), while $\sigma(r=0.01)=0.0014$ from the recombination bump ($\ell\,{>}\,20$). 
Note that $\sigma(r=0.01)$ is the measurement uncertainty including cosmic variance. Since the posterior probability density of $r$ is asymmetric and falls rapidly towards $r=0$, we expect detections of both bumps at a significance exceeding $5\,\sigma$ (see Sect.~\ref{ss:forecasts_delta_r}).

Relative to the recombination bump, the reionization feature becomes more relevant as one approaches the Starobinsky limit, since the recombination bump would require substantial delensing in order to be measured. We will come back to delensing effects in Sect.~\ref{ss:extrasuccess}, where the averaged total and noiseless foreground residuals mentioned above are the inputs to the estimation of the total uncertainty on the tensor-to-scalar ratio $r$. 

Along with the recovered CMB map, our component-separation approach also provides estimates of the Galactic foregrounds, and in particular of the dust map. This information is used in Sect.~\ref{ss:forecasts_delta_r} to marginalize our estimate of $r$ over the foreground bias.

\subsection{Instrumental Systematic Uncertainties}
\label{ss:forecasts_syst}

We describe here our analysis of instrumental systematic effects. Foreground cleaning and systematic residuals will constitute the total error budget, as described in Sect.~\ref{ss:forecasts_delta_r}. Throughout this section, we describe specific definitions and objects for each of the systematic effects we consider. Common to the whole analysis, simulations of the sky include Galactic foreground components as an input, following the model described in Sect.~\ref{sec:reference_skies}, combined as described in Sect.~\ref{sec:input_maps} to produce simulated frequency-channel data. 

\subsubsection{Introduction to Systematic Uncertainties}
\label{sss:syst_intro}
In this section, we describe potential systematic effects for the \LiteBIRD\ mission and
evaluate their impact on the final results.
Systematic effects originate from imperfect knowledge of the foregrounds, combined with incomplete correction of instrumental or environment effects 
arising from either inaccurate modeling or
limited knowledge of instrumental parameters and environmental contributions.

The main objective of \LiteBIRD\ is to measure the tensor-to-scalar ratio $r$.
We therefore focus our study on the impact of systematic effects
on the measurement of this parameter.
We particularly estimate the systematic contribution on the measurement 
bias $\Delta r$ under the condition of the true value being $r=0$.
We give the detailed definition of $\Delta r$, as well as the total error $\delta r$,
in the next section.

We divide the systematic effects into several categories. The systematic bias produced by the component-separation residual was described in Sect.~\ref{ss:forecasts_fg_cleaning}. 
We describe remaining systematic sources in this section. Table~\ref{tbl:syst} shows a summary of the most relevant systematic effects as the top level category. 
Beam systematic effects are mainly caused by imperfections in the calibrated knowledge of the beam. 
We further divide the beam into three: the main lobe; the near sidelobe; and the far sidelobe. The main lobe is the center of the beam with width shown in Table~\ref{tbl:imo:sensitivities}. 
We define the near sidelobe to be the region out to $3^\circ$ with respect to the main-lobe peak direction. 
The remaining part of the beam is defined as the far sidelobes. The optical system, consisting of the HWP and lenses, is supposed to produce multiple reflections and hence ghosting images for bright sources. The requirement for the suppression of these images is obtained given the required systematic bias in $\Delta r$. 
The systematic effects of the beam variation and polarization (co-polar and cross-polar) in the observation band are also studied. 
For far-sidelobe effects, which give the most sizeable systematic bias, 
we study the effects for all frequency channels of the three telescopes. 
We study the HWP systematic effects in terms of the Mueller matrix, consisting of: the instrumental polarization that leaks from temperature into the polarization signal in the measurement band; the transparency; the polarization efficiency due to the retardance (phase difference) uncertainties of the HWP 
; 
and the polarization angle in multiple bands due to mixing of the two orthogonal polarized signals. 
We study cosmic-ray effects based on a newly developed simulation tool.  Gain-variation systematic effects are also studied. 
Given the systematic error budget, the requirements are obtained on the focal plane temperature stability and the gain variation in detectors. 
Polarization angle systematic effects are sourced by the uncertainties of the absolute angle, the relative angles among all bands, the HWP spinning position, and the time variation due to the attitude and HWP position measurement errors.
Polarization efficiency systematic effects for the detector plane antenna and the optical system are additionally studied. 
The pointing uncertainty is modeled as a constant offset, as well as a random variation. The requirements on the HWP wedge angle and the bandpass calibration uncertainties are also given. 
The transfer-function category includes systematic effects due to the uncertainties of the detector time constant and the crosstalk among detectors. 
Details of each of these items are given in the following sections.
Some of the systematic error budget of individual sources can be further divided into the sub-level requirements. One example is the pointing uncertainty, which will be detailed in Sect.~\ref{sec:other-systematics}.\footnote{We identify about 70 systematic sources in total including the ones in the sub-levels. }

We note that the estimated systematic bias is approximately a linear sum of individual systematic effect biases, 
when those power spectra are much smaller than that of the lensing effect.  
This is because
$\Delta r$ is approximately proportional to the square of the error of 
the systematic effects to first order;
the details are given in Sect.~\ref{ss:forecasts_delta_r}. 
The total error budget assigned for the systematic bias $\Delta r$ is $0.57\times 10^{-3}$~\citep{Hazumi:2019lys}.
We assign an error budget of 1\,\% of the total
systematic bias $5.7\times 10^{-6}$
to individual systematic effects as a nominal error budget
in order to account for dozens of potential systematic sources.
We note that there are some exceptions including the systematic effects 
caused by the far sidelobe uncertainty, however.

To give $\Delta r$, we employ one of two approaches.
One way is to estimate $\Delta r$ values based on the expected uncertainties 
in the calibration in-flight or at the expected precision
of measurement devices.
The other approach is to impose requirements on the calibration accuracy or 
of the measurement devices.
In Table~\ref{tbl:syst}, we divide the systematic sources into two types
by assigning ``E'' (Expectation) and ``R'' (Requirement), respectively.
We show the obtained angular power spectra for systematic effects in Fig.~\ref{fig:cl_sys_all}.

\begin{table}[htbp!]
\setlength{\tabcolsep}{4pt}
\centering
\caption{Summary of sources of systematic effects as the top level category. We categorize the systematic effects
in the first column. The details of individual categories
are shown in the second column. The third column shows the systematic biases on $r$,
except for the cosmic rays, which are supposed to yield an extra noise source.
The fourth column describes the sources of systematic effects. The last column shows the types.
The type ``E'' and ``R'' state the expectation and the requirement, respectively, as described in the text. Note that
we assign an error budget of 1\,\% of the total
systematic bias $5.7\times 10^{-6}$
to individual systematic effects as a nominal error budget
in order to account for dozens of potential systematic sources. 
Details are described in the text.
}
\label{tbl:syst}
\begin{tabular}{|c|c|c|c|c|}
\hline
Category & Systematic effect & $\Delta r$ & Source & Type\\
\hline\hline
Beam & Far sidelobes & $4.4\times 10^{-5}$ & $B\to B$, $E\to B$ & R \\
& Near sidelobes & $5.7\times 10^{-6}$ & $B\to B$, $E\to B$ & R \\
& Main lobe & $<10^{-6}$ & $E\to B$ & E \\
& Ghost & $5.7\times 10^{-6}$ & $E\to B$ & R \\
& Polarization and shape in band & $< 10^{-6}$ & $E\to B$ & R \\
\hline
Cosmic ray & Cosmic-ray glitches & Noise &  Power to B, E & E\\
\hline
HWP & Instrumental polarization & $<10^{-6}$ & $T\to B$ & E \\
& Transparency in band & $5.7\times 10^{-6}$ & $E\to B$ & R \\
& Polarization efficiency in band& $5.7\times 10^{-6}$ & $B\to B$ & R \\
& Polarization angle in band & $5.7\times 10^{-6}$ & $E\to B$ & R \\
\hline
Gain & Relative gain in time & $5.7\times 10^{-6}$ & $E\to B$ & R \\
& Relative gain in detectors & $5.7\times 10^{-6}$ & $E\to B$ & R \\
& Absolute gain & $1.9 \times 10^{-6}$ & $B\to B$ & E \\
\hline
Polarization & Absolute angle & $9.1\times 10^{-6}$ & $E\to B$ & E \\
angle & Relative angle & $5.7\times 10^{-6}$ & $E\to B$ & E \\
& HWP position & $1.0\times 10^{-6}$ & $E\to B$ & E \\
& Time variation & $<10^{-7}$ & $E\to B$ & E \\
\hline
Pol. efficiency & Efficiency & $5.6\times 10^{-6}$ & $B\to B$ & E \\
\hline
Pointing & Offset & $5.7\times 10^{-6}$ & $E\to B$ & R  \\
& Time variation & $<10^{-6}$ & $E\to B$ & E \\
& HWP wedge & $5.7\times 10^{-6}$ & $E\to B$ & R \\
\hline
Bandpass & Bandpass efficiency & $5.3\times 10^{-6}$ & $E\to B$ & R \\
\hline
Transfer & Crosstalk & $5.7\times 10^{-6}$ & $B\to B$ & R \\
function & Detector time constant knowledge & $5.7\times 10^{-6}$ & $E\to B$ & R\\
\hline
\end{tabular}
\end{table}

Our analysis has shown that the employment of the PMU is crucially important to mitigate systematic errors.
Without the PMU, data from a pair of detectors mutually orthogonal in the polarization
orientation are usually combined, causing leakage from temperature to polarization
when there are differences in any features between the two detectors.
The features giving outstanding systematic effects 
include the gain, beam, and bandpass.
The systematic errors for these are evaluated assuming that
the effects could be mitigated by increasing the number of detectors $N$ 
(if the systematic effects on individual detectors are uncorrelated),
with the scaling $\Delta r \propto 1/N$.
Existing studies include: bandpass effects~\citep{Hoang:2017wwv},
yielding $\Delta r$ on the order of $10^{-4}$ with an assumption that
the bandpass boundaries have an uncertainty of 1\%  without any corrections;\footnote{Reference~\cite{Banerji:2019kbc} demonstrates the correction of bandpass systematic effects.}
the instrumental polarization caused by
reflections on the two mirrors in the LFT, giving at most $3\times 10^{-4}$ for $M_{QI}$ components in the Mueller matrix
~\citep{Kashima:2017yub}; and
the Carbon Monoxide (CO) line emission contamination, with $\Delta r\simeq 10^{-4}$
for a bandpass response mismatch of 10\,\% without applying notch filters~\citep{Ghigna2020-dthesis}.
We conduct a simulation study of
the systematic errors caused by differences for a pair of detectors in gain, pointing, beam ellipticity, beam width, cross-polarization, and sidelobes, and find the systematic bias $\Delta r$ to be on the order of $10^{-6}$.
However, these effects would increase significantly when there is some cross-correlation among
the detectors with a correlation coefficient larger than $10^{-3}$.  
A time-correlated noise contribution is expected to contaminate detector
timestream data at long timescales.
If the noise contamination is projected onto the sky map without adequate treatment,
it can cause significant degradation of the sensitivity for the lower multipole range.
We generate $1/f$-noise simulated data, project it onto the sky map and estimate
the degradation to be $\delta r\simeq 3\times 10^{-4}$ for a 20-mHz knee frequency.
All the effects above are found to be significantly reduced when we use the PMU.
For example, Ref.~\citep{Hoang:2017wwv} demonstrates this for the bandpass effects.
In summary, with the PMU we can suppress almost all outstanding systematic sources that give rise to the leakage from temperature to polarization, as well as the $1/f$ noise that significantly deteriorates the sensitivity in the lower multipole region.

In the rest of this section, we focus on systematic effects 
with the PMU.
Among those we will give details of systematic studies of the beam, cosmic rays, HWP, and gain in the following sections.
Other systematic effects are summarized in Sect.~\ref{sec:other-systematics}.
We estimate the systematic effects based on simulated sky maps of the CMB and foregrounds described in Sect.~\ref{sec:input_maps}. The basic strategy is to obtain a residual sky map that is a difference between the reconstructed sky maps with and without a particular systematic effect, which mitigates the model dependence of the sky maps employed. 

\subsubsection{Formalism of the Errors}
\label{sub:ErrorForm}
This section gives the definition of the bias $\Delta r$ and the total error $\delta r$.
We define a likelihood function $L(r)$ expressed in the multipole domain 
due to the large sky coverage of the satellite mission.
Assuming Gaussian stationary and isotropic fields and hence no coupling 
between multipoles (which is an approximation in the presence of a mask of the Galactic plane),
the likelihood function is given by
\begin{equation}
\log L(r) = \sum_{\ell=\ell_{\rm min}}^{\ell_{\rm max}} \log P_\ell(r),
\label{eq:global-likelihood}
\end{equation}
where $\ell_{\rm min}=2$ and $\ell_{\rm max}=191$~\footnote{We use $N_{\rm side}=64$, giving $\ell_{\rm max}=3\times N_{\rm side}-1$.}.
We define
\begin{equation}
\log P_\ell(r) = -f_{\rm sky}\frac{2\ell+1}{2}\left[\frac{\hat{C}_\ell}{C_\ell}+\log C_\ell 
-\frac{2\ell-1}{2\ell+1}\log\hat{C}_\ell\right]
\end{equation}
where $\hat{C}_\ell$ ($C_\ell$) is the measured (modeled) $B$-mode power spectrum~\citep{PhysRevD.77.103013}.
We use a sky mask having $f_{\rm sky}=0.495$.
In order to estimate the potential bias of individual systematic effects on $r$, which we call $\Delta r$, 
we represent the measured $B$-mode spectrum
as a sum of the following contributions (assuming no primordial $B$ modes, i.e.,
$r=0$):
\begin{equation}
\hat{C}_\ell = C_\ell^{\rm sys}+C_\ell^{\rm lens}+N_\ell.
\label{eq:cl-measured-in-likelihood}
\end{equation}
Here $C_\ell^{\rm sys}$ is the estimated systematic effects power spectrum,
$C_\ell^{\rm lens}$ is the lensing $B$-mode power spectrum,
and $N_\ell$ is the expected noise power spectrum,
including the residual noise after the component separation (shown in 
Fig.~\ref{fig:xforecast-results-pysm})
as well as the cosmic-ray contribution.
The modeled power spectrum is given by
\begin{equation}
C_\ell = r C_\ell^{\rm tens}+C_\ell^{\rm lens}+N_\ell,
\label{eq:cl-model-in-likelihood}
\end{equation}
where $C_\ell^{\rm tens}$ is the tensor mode with $r=1$.
We estimate the systematic bias under the condition of 
the existence of the expected noise $N_\ell$ coming from instrumental and foreground residual uncertainties.

The potential systematic bias $\Delta r$ is defined as the value
giving the maximum of the likelihood function:
\begin{equation}
\left. \frac{dL(r)}{dr}\right|_{r=\Delta r}=0.
\label{eq:r-bias-definition}
\end{equation}
We also define the total error on $r$, $\delta r$, as the value covering 68\,\%
of the area of the total likelihood function:
\begin{equation}
\frac{\int_0^{\delta r}L(r)dr}{\int_0^{\infty} L(r)dr} = 0.68.
\label{eq:r-total-error-definition}
\end{equation}

\subsubsection{Beam Systematic Effects}\label{sec:beam-systematics}
Imperfect knowledge of the beam, due to a combination of uncertainties in beam measurements and modeling, are important
sources of systematic effects for the measurement of primordial $B$ modes. 
We model the sky signal $p(t,\nu,\hat{n},\hat{s})$ observed at location $\hat{n}$ at time $t$ with frequency $\nu$, from a beam centered at location $\hat{s}$ in the presence of an ideal HWP, without coupling with other optical elements:
\begin{align}
p(t, \nu, \hat{n}, \hat{s}) &= 
B_{\parallel}(t, \nu, \hat{n}-\hat{s})
[I(\nu, \hat{n})+ Q(\nu, \hat{n})\cos2\varphi(t, \nu)
+ U(\nu, \hat{n})\sin2\varphi(t, \nu)] \nonumber \\
& \qquad +B_{\perp}(t, \nu, \hat{n}-\hat{s})
[I(\nu, \hat{n})- Q(\nu, \hat{n})\cos2\varphi(t, \nu)
- U(\nu, \hat{n})\sin2\varphi(t, \nu)], 
\end{align}
where $\varphi(t,\nu)=4\rho(t,\nu)-2\psi(t,\nu)$,
$\rho(t,\nu)$ is the HWP rotation position,
$\psi(t,\nu)$ is the polarizer measurement angle with respect to the axis fixed in the sky, and $B_{\parallel}(t, \nu, \hat{n}-\hat{s})$ and $B_{\perp}(t, \nu, \hat{n}-\hat{s})$ are the co-polar and cross-polar beams, respectively. The measured signal is the integral of $p(t, \nu, \hat{n}, \hat{s})$ over the direction $\hat{n}$ and over the frequency $\nu$ weighted by the bandpass function.
Different regions of the beam kernel have different effects on the data and are also estimated differently during the calibration process (either in-flight or from the ground). We divide the beam into three regions (as shown in Fig.~\ref{fig:LoK}): the main lobe; the near sidelobes; and the far sidelobes.
Below we describe systematic effects for the far sidelobes, the near sidelobes, and the main lobe in the order that they are sourced by the uncertainties in our knowledge of $B_{\parallel}$.
We also discuss the possible ghosting effect caused by reflections inside the telescope, 
and the polarization and beam shape in the observed frequency band.

\paragraph{Far-sidelobe Systematic Effects}
The main effect of uncertainties in the beam \glspl{fsl} measurement or modeling is that it induces unexpected leakage of the Galactic signal coming from the Galactic plane to higher Galactic latitudes. The mismatch of this excess signal between different frequency bands leads to residuals in the recovered CMB Stokes parameter maps after component separation. With the help of the HWP, allowing quasi-instantaneous measurement of the polarization Stokes parameters ($Q$ and $U$), and of the symmetrization of the effective beam on the maps by the diversity of scanning angle in each pixel of the sky, the dominant systematic effect originates from residual $B$-mode Galactic signal leaking towards the CMB $B$-mode estimate.
On top of this effect, uncorrected effective beam asymmetries might potentially lead to small $E$-to-$B$ leakage effects.

Diffraction by optics elements located between the HWP and the detectors is not expected to induce a significant contribution to the cross-polarization beam $B_{\perp}(t,\nu,\hat{n}-\hat{s})$. This is because the HWP does not modulate potential generated polarization signal and hence negligible levels of $I$-to-$P$ leakage should be induced. 
However potential contributions to $B_{\perp}(t,\nu,\hat{n}-\hat{s})$ are the HWP itself and external elements such as baffles. Those might induce $I$-to-$P$ leakage, which will be further studied in subsequent works.

In this subsection 
we evaluate the impact on the tensor-to-scalar ratio $r$ of the uncertainties on the beam knowledge by perturbing the shape of the beam response at different angles from the beam center. Before describing this procedure we will introduce the simulations that were carried out.

\paragraph{Description of Simulations}
For this study, we employ the beam-convolved sky maps generated by \texttt{TOAST} described in Sect.~\ref{sec:input_maps}.
In order to set the requirements on the far-sidelobe beam calibration, we have considered the beam convolution in the angle ranges $5^\circ<\theta<10^\circ$, $10^\circ<\theta<15^\circ$, and $15^\circ<\theta<90^\circ$, where $\theta$ is the angle with respect to the main lobe peak.
This convolution has been performed for all sets of co-added foreground maps. We do not include noise in the input simulated maps because we are interested in the estimation of the requirements in terms of a bias on $r$ ($\Delta r$) for the beam systematic effects. 
The time-ordered data (TOD) are simulated for the 22 nominal frequency bands of \LiteBIRD. A subset of detectors for each frequency band is considered, instead of the full focal plane. The selected detectors are well spread across the arrays so that we can assume that the results are representative of the full focal plane, and therefore a global rescaling of the results is applied at the end. 

\LiteBIRD's scanning strategy with nominal parameters of the IMo (see Sect.~\ref{ss:forecasts_imo}) is simulated and a full-sky convolution of the beams is performed. Input simulated reference beams are calculated using \texttt{GRASP} software with a simplified optical model, i.e., 
no multiple reflections have been included in the beam model. \texttt{GRASP} beams have been simulated at the central frequency for each band, neglecting the color effects and hence the differential effects on various sky components. 
We will describe the possible systematic effects that are not accounted for in the model above in the last paragraph of this subsection. 

\paragraph{Impact of FSL Systematic Effects on $B$ Modes}
We now evaluate the impact of the lack of knowledge of the beam shape at large angular scales on the measurement of the tensor-to-scalar ratio $r$. With this goal in mind, we process the previously described simulated band-averaged maps produced with non-Gaussian beam convolution through all the steps leading to the measurement of $r$. The effect of uncertainties on the beam shape will then be translated into a bias $\Delta r$ on $r$.

For each perturbed case studied, we proceed with two sets of maps, with and without beam perturbations, which are both processed with component separation to calculate CMB $I$, $Q$, $U$ maps. The first set, including the unperturbed beam, provides reference CMB maps $\mathbf{m}_0$. 
The second set of simulations includes perturbations of the FSL and are run through the same component-separation code with the same assumptions. The perturbations are performed in a specified angle range and for a given frequency band in the following manner:
\begin{equation}
    B_{\rm pert}(\nu,\theta,\phi) = \mu(\alpha) (\left[1 + \alpha W(\theta)\right]\, B_0(\nu,\theta,\phi).
\end{equation}
Here $B_0(\nu,\theta,\phi)$ is the reference \texttt{GRASP} beam for the frequency band centered at $\nu$, at radial and azimuthal angles $\theta,\phi$, $\alpha$ is an amplitude parameter of the perturbation that we are varying to provide the requirements, and $\mu(\alpha)$ is a normalization parameter. The window function $W(\theta)$ is chosen to include
apodization in order to reduce ringing edge effects and is set to
\begin{equation}
    W \left( \theta \right) = A_{\rm inf} - A_{\rm sup}, \quad \text{where} \quad A_{i} \left( \theta \right) = \left\{ \begin{array}{ll}
         0, & \text{if } \theta < \frac{1}{2}\theta_{i} \\
         \frac{1}{2} \left( 1 - \mathrm{cos} \left( \frac{\left( 2\theta - \theta_{i} \right) \pi}{\theta_{i}} \right) \right), & \text{if } \frac{1}{2}\theta_{i} < \theta < \theta_{i} \\
         1, & \text{if } \theta_{i} < \theta,
    \end{array} \right.
\end{equation}
where $i=({\rm inf}, {\rm sup})$, for three different angle ranges $\left[\theta_{\rm inf},\theta_{\rm sup}\right]$, namely $\left[ 5^\circ, 10^\circ \right]$, $\left[ 10^\circ, 15^\circ \right]$, and $\left[ 15^\circ, - \right]$ ($A_{\rm sup}=0$ for the third window). With the previous parameters, the window is actually effective, i.e. $W/W_{\rm max}>0.5$, in the angle ranges $\left[3.75^\circ, 7.5^\circ\right]$, $\left[7.21^\circ,11.69^\circ\right]$, and $\left[11.25^\circ,90^\circ\right]$.

The normalization $\mu$ is calculated for each value of $\alpha$ such that the beam-convolved maps have the same dipole component ($\ell=1$) amplitude as the unperturbed ones.
This procedure allows us to isolate the effect of beam mismatch only from calibration uncertainties that are studied in 
Sect.~\ref{beams-calib-section}.

In our approach, we find the limiting values $\alpha_{\rm lim}$ independently for each frequency band, i.e., we perturb one beam at one frequency at a time, leaving the others constant, such that the induced bias on the tensor-to-scalar ratio is $\Delta r = 1.9\times 10^{-5}$ divided by 66, the number of frequency channels times the number of windows. 
The systematic bias $\Delta r$ is obtained using Eq.~(\ref{eq:r-bias-definition})
given $C_\ell^{\rm sys}$, which is obtained from the residual sky maps $\delta \mathbf{m} = \mathbf{m} - \mathbf{m}_0$ after applying a \Planck-HFI Galactic mask leaving 51\,\% of sky area, where $\mathbf{m}$ ($\mathbf{m}_0$) is the sky map convolved by the far sidelobe with (without) the perturbation of individual $\alpha$ values.

The results for the three angle ranges are shown in Table~\ref{tab:beamFSLrequir}. The $\alpha$s are just intermediate parameters and we indicate the values of the beam perturbation amplitudes in the three windows 
by calculating the following values:
\begin{equation}
    \delta \overline B_{\rm lim} = \,\frac{\int \delta B_{\nu} \left( \theta,\phi \right) W \left( \theta \right) {\rm d}\Omega}{\int W \left( \theta \right) {\rm d}\Omega} ;\, {\rm and}\,\,\,\, \delta R_{\rm lim} = \, \frac{\int \delta B_{\nu} \left( \theta,\phi \right) W \left( \theta \right) {\rm d}\Omega}{\int B_{0} \left(\nu, \theta,\phi \right) {\rm d}\Omega}.
\end{equation}
Here the beam perturbation $\delta B_\nu (\theta,\phi) = \alpha_{\rm lim} B_0(\nu,\theta,\phi)$, including the normalization of the input beam at the peak: $B_0(\nu,\theta = 0) = 1$.
The quoted values $\delta \overline B_{\rm lim}$ represent the precision that is required on the knowledge of the mean amplitude of the beam in the window $W(\theta)$, while $\delta R_{\rm lim}$ represents the required precision on the beam's relative power in the window angle range specified by $W(\theta)$.
Because the amplitude of the beam models used for this analysis drops drastically at angles larger than around $50^\circ$,
we have cut the last window for the perturbation $W(\theta)$ above $70^\circ$ (to be conservative) for the calculation of $ \delta \overline B_{\rm lim}$ and $\delta R_{\rm lim}$, as well as the estimated calibration precision that we will discuss later.

\begin{table}[htbp!]
\setlength{\tabcolsep}{4pt}
\begin{center}
\caption{Beam perturbation requirements for each frequency channel and each of the three angle ranges of the beam perturbations giving $\Delta r = 1.9\times 10^{-5}/66$, fixing the other perturbations for the other frequency channels and angle ranges to 0. Units are dB for $\delta \overline B_{\rm lim}$ and ${\sigma}^{\rm cal}_{0.25^\square}$, and dBi for $\delta R_{\rm lim}$. The indicated values of angles in the first row, providing the typical range in which the beam is perturbed, are such that 95\,\% of the perturbed beam power for a typical CMB channel is within this range (the value varies from channel to channel as the beam shape varies). 
This explains why the upper bound for the last window is only $50^\circ$ while the window extends to $90^\circ$ as the beam drops above 50--$60^\circ$, depending on the channel.
The calculated values of the required precision
${\sigma}^{\rm cal}_{0.25^\square}$ are assuming $0.5^\circ$ separation between noise-limited measurements. An identical calibration precision of $\sigma^{\rm cal}_{0.25^\square}=-56.90$\,dB for all 66 entries leads to a similar bias.}
\label{tab:beamFSLrequir}
\begin{tabular}{|c||>{$}c<{$}|>{$}c<{$}|>{$}c<{$}||>{$}c<{$}|>{$}c<{$}|>{$}c<{$}||>{$}c<{$}|>{$}c<{$}|>{$}c<{$}|}
\hline
& \multicolumn{3}{c||}{\text{$3^\circ<\theta<8^\circ$}} & \multicolumn{3}{c||}{\text{$6^\circ<\theta<13^\circ$}} & \multicolumn{3}{c|}{\text{$10^\circ<\theta<50^\circ$}} \\
\cline{2-10}
\rule[-1.5ex]{0pt}{4.0ex} 
& \delta \overline B_{\rm lim} & \delta R_{\rm lim} & \sigma^{\rm cal}_{0.25^\square} & \delta \overline B_{\rm lim} & \delta R_{\rm lim} & \sigma^{\rm cal}_{0.25^\square} & \delta \overline B_{\rm lim} & \delta R_{\rm lim} & \sigma^{\rm cal}_{0.25^\square} \\
\hline \hline
L1-040 & -42.55 & -23.54 & -28.20 & -46.62 & -25.41 & -30.68 & -66.40 & -27.49 & -42.79 \\ \hline
L2-050 & -34.09 & -13.45 & -19.73 & -38.35 & -15.51 & -22.41 & -57.98 & -17.44 & -34.37 \\ \hline
L1-060 & -39.46 & -17.68 & -25.11 & -43.67 & -19.68 & -27.73 & -63.20 & -21.53 & -39.60 \\ \hline
L3-068 & -36.70 & -13.27 & -22.34 & -41.38 & -15.75 & -25.44 & -61.52 & -18.20 & -37.92 \\ \hline
L2-068 & -30.45 & -8.02 & -16.10 & -35.62 & -10.99 & -19.68 & -56.00 & -13.67 & -32.39 \\ \hline
L4-078 & -42.82 & -18.36 & -28.46 & -45.87 & -19.21 & -29.93 & -61.76 & -17.41 & -38.16 \\ \hline
L1-078 & -39.42 & -16.43 & -25.07 & -42.48 & -17.28 & -26.53 & -59.07 & -16.18 & -35.46 \\ \hline
L3-089 & -48.56 & -23.14 & -34.20 & -51.75 & -24.13 & -35.81 & -68.70 & -23.40 & -45.10 \\ \hline
L2-089 & -38.84 & -15.50 & -24.49 & -42.04 & -16.49 & -26.10 & -59.51 & -16.27 & -35.91 \\ \hline
L4-100 & -51.80 & -25.61 & -37.45 & -54.98 & -26.59 & -39.04 & -72.09 & -26.01 & -48.49 \\ \hline
L3-119 & -54.82 & -27.57 & -40.47 & -57.91 & -28.46 & -41.97 & -75.00 & -27.86 & -51.39 \\ \hline
L4-140 & -51.25 & -23.23 & -36.90 & -54.11 & -23.88 & -38.17 & -74.60 & -26.69 & -51.00 \\ \hline \hline
M1-100 & -50.65 & -26.84 & -36.30 & -53.91 & -27.90 & -37.97 & -69.44 & -25.74 & -45.84 \\ \hline
M2-119 & -54.58 & -29.71 & -40.23 & -57.78 & -30.71 & -41.84 & -73.96 & -29.20 & -50.36 \\ \hline
M1-140 & -49.55 & -23.92 & -35.20 & -52.45 & -24.61 & -36.51 & -75.36 & -29.83 & -51.76 \\ \hline
M2-166 & -60.87 & -34.64 & -46.52 & -64.07 & -35.63 & -48.13 & -80.06 & -33.91 & -56.45 \\ \hline
M1-195 & -63.56 & -37.07 & -49.20 & -66.58 & -37.89 & -50.64 & -80.69 & -34.30 & -57.08 \\ \hline \hline
H1-195 & -59.01 & -33.00 & -44.65 & -62.38 & -34.17 & -46.44 & -78.63 & -32.73 & -55.03 \\ \hline
H2-235 & -62.92 & -35.65 & -48.57 & -66.07 & -36.60 & -50.13 & -81.55 & -34.38 & -57.95 \\ \hline
H1-280 & -60.57 & -32.52 & -46.22 & -63.60 & -33.34 & -47.66 & -75.23 & -27.26 & -51.63 \\ \hline
H2-337 & -70.48 & -41.78 & -56.12 & -73.57 & -42.66 & -57.63 & -86.41 & -37.79 & -62.81 \\ \hline
H3-402 & -67.91 & -38.03 & -53.56 & -70.77 & -38.68 & -54.83 & -88.72 & -38.93 & -65.12 \\ \hline
\end{tabular}
\end{center}
\end{table}

The derived constraints can be 
translated into required accuracy in beam calibration measurements during ground testing.
The quantity effectively measured during the beam calibration can be modeled as
\begin{equation}
    P_{\rm cal}({\vec r}) = \int B_\nu \,\omega({\vec r}' - {\vec r})\, {\rm d}\Omega' \,\frac{1}{\int B_\nu \,\omega({\vec r}')\, {\rm d}\Omega'} + n_{\rm cal},
\end{equation}
with $\omega({\vec r})$ a small integration window of the beam 
and $n_{\rm cal}$ the noise in the beam calibration.
We have estimated the precision required on the measured quantity $P_{\rm cal}({\vec r})$ assuming random uncorrelated errors in each measurement (and hence no systematic effects in the calibration), and a grid of measurements at many angles. In this case the calibration measurement uncertainty $\sigma^{\rm cal}_{\Omega_{\rm pix}} = \sqrt{\langle n^{\sf T}_{\rm cal} n_{\rm cal}\rangle}$ is related to the uncertainty in the beam amplitude averaged over the window area $\delta \overline B_{\rm lim}$ by
\begin{equation}
\label{sigmacalibbeam}
      \sigma^{\rm cal}_{\Omega_{\rm pix}} = \frac{\int W(\theta) {\rm d}\Omega}{\sqrt{\sum_{ij}W^2(\theta_{ij})} \Delta\Omega_{\rm pix}}\,\delta \overline B_{\rm lim},
\end{equation}
with $\Delta\Omega_{\rm pix}$ the solid angle covered by one calibration measurement and $i,j$ the pixel numbers for a pixelized beam calibration map. We assume that the calibration measurements are normalized to unity at the peak (for the beam center). From the previous equation, we can see that the factor ($\int W {\rm d}\Omega/\sqrt{\sum_{ij}W^2(\theta_{ij})} \Delta\Omega_{\rm pix})^2$ is the effective number of pixels in the window area of the beam perturbation. Assuming a pixel width $\sqrt{\Delta\Omega_{\rm pix}}=0.5^{\circ}$, which is the expected beam measurement step size in the ground calibration process,
we have calculated the required precision for the calibration measurements and present the values in Table~\ref{tab:beamFSLrequir}. The effective numbers of measurements for the three windows are 742.0, 1541.9, and 52943.2.

We can see that the requirements on calibration precision ($\sigma^{\rm cal}_{0.25^\square}$) are more stringent at the highest frequencies for the last window, which covers a larger area of the sphere. This shows that it is necessary to accurately calibrate the \glspl{fsl} of the channels, providing a proxy for the Galactic dust for component separation and for which the fraction of power going into the main lobe is lower, since those have the smallest beam FWHM values. 
Assuming that we calibrate the whole area of the beam for all the frequency channels with a constant precision $\sigma^{\rm cal}_{0.25^\square}$, we have evaluated the desired precision is $-56.90$\,dB 
to have requirements equivalent to what is presented in Table~\ref{tab:beamFSLrequir}.
The total contribution of $\Delta r$ has been estimated with hundred realizations by simultaneously varying individual $\alpha$ values with a uniform distribution in the range $[{\rm max}(-1,-\sqrt{3}\alpha_{\rm lim}); \sqrt{3}\alpha_{\rm lim}]$, with $\alpha_{\rm lim}$ calculated from Table~\ref{tab:beamFSLrequir}. The $\sqrt{3}$ factor is to keep the same variance $\alpha_{\rm lim}^2$ as the Gaussian distribution used for Table~\ref{tab:beamFSLrequir} to derive requirements.
We find the systematic bias resulting from the above assumptions to be $\Delta r = 2.2 \times 10^{-5}$.

We want to stress several important limitations of this analysis.  Firstly, one of the key assumptions is that the scanning strategy induces enough symmetrization of the beam so that many calibration measurements can be averaged for large $\theta$ and that the beam systematic effects are then dominated by uncertainties in the transfer function. This explains why the number of effective calibration points is high for each window.  Secondly, in order to be representative of the final results, we have assumed that the bias on $r$ is not too sensitive to the structure of the beam inside each window $W(\theta)$. 
Thirdly, the correlation in the overlaped region in Table~\ref{tab:beamFSLrequir} is assumed to be negligible.
Lastly, we quote the required precision on the calibration measurements assuming that those are limited by calibration noise. This is somewhat unrealistic, since measurements are expected to be limited by systematic effects in the ground measurements. Nevertheless, this analysis still provides a framework for quantifying uncertainties in further studies that might be needed. 

We have checked that the results are not sensitive to the component-separation parameters and the model, in particular the effective beam function assumed as input in Sect.~\ref{ss:forecasts_fg_cleaning}. The variations introduced by the change of the recovered component-separation parameters due to the slight difference of the beam transfer-functions in the recovered CMB maps vanish at first order in the difference $\mathbf{m} - \mathbf{m}_0$.

The beam model used for the simulations contains several uncertainties including the effects of beam asymmetry caused by the multiple reflections and scattering, especially for the detectors placed at the edge of the focal plane. In order to account for the uncertainties, we conservatively multiply by a factor of two the requirements to obtain the FSL total systematic bias of $\Delta r = 4.4\times 10^{-5}$.

\paragraph{Near-Sidelobe Effect}
We pursue two approaches to estimate the systematic effects due to
inaccuracy of the knowledge of the near sidelobe, i.e., the beam structures in the angle range up to about $3^\circ$ with respect to the main-lobe center.
In the first method, we model the beam shape having an axially-symmetric Gaussian main beam
plus a power-law tail. We convolve the sky (including foregrounds) with the beam
and deconvolve the sky map with a beam having a different amplitude of the power-law tail.
Then we take a difference of the sky maps before and after this procedure for the 15 frequency channels,
and apply the component-separation procedure to obtain the residual sky map and power spectrum.
We find $\pm 10$\,\% variations in the amplitude of the near sidelobe, whose nominal magnitude is less than 1\,\% of the main lobe, 
give rise to a systematic bias
of $\Delta r = 5.7\times 10^{-6}$.
In the second method, we model the beam as the sum of the Gaussian main beam 
and a ring shape near the sidelobe, whose cross-sectional shape is a Gaussian with a peak height of 1\,\% of the main-lobe peak. 
The position, width, and height of the modeled near-sidelobes are set to values
close to those in the \texttt{GRASP} simulation.
The typical values are $1.7^\circ$ from the main-lobe center for the position, $0.6^\circ$ for FWHM  and $-20\,$dB with respect to the main lobe for the height.
We repeat a similar analysis by changing the ring position, width, and height, and
find that a $\pm 20$\,\% variation from the nominal values corresponds to $\Delta r = 5.7\times 10^{-6}$.
In summary, we require a calibration accuracy of better than 10\,\% for the near sidelobe in order
to have a systematic bias $\Delta r < 5.7\times 10^{-6}$.

\paragraph{Main-Lobe Effect}
At the simplest level, the beam's main lobe is characterized by two quantities, namely 
the height and the shape.
The height corresponds to the gain that we describe in Sect.~\ref{sec:Gain-systematics}. 
The main-lobe beam has a shape close to elliptical for the detectors
located far from the center of the focal plane.
We quantify this effect using two parameters, namely the width and the ellipticity or flattening.
We conduct a simulation study similar to the one for the near-sidelobe effects.
With an expected calibration accuracy of 
the 1\,\% for both the width and the flattening in-flight (see Sect.~\ref{beams-calib-section}),
we find that the resulting systematic bias is smaller than $10^{-6}$.

\paragraph{Ghosting Effect}
It is known that the multiple reflections between the focal plane and
the optics components, such as lenses and HWP, produce a ghosting image~\citep[e.g., Ref.][]{Bierman_2011}.
To investigate this, we conduct a simulation study to estimate the systematic effects
due to foreground contamination leaking through off-boresight small-scale
structure into the beam. We find that a leakage of 0.05\,\% amplitude through a 30\,arcmin diameter spot
separated from the beam boresight direction by 900\,arcmin produces a systematic bias
equivalent 
to $\Delta r = 5.7\times 10^{-6}$.
The lens reflection effects are included in the discussion regarding the FSL.
We note that the ghosting effect can be partly mitigated by tilting the HWP by an angle larger than half of the \gls{fov}.

\paragraph{Polarization and Beam Shape in Band}
Sinuous antennas are sensitive to polarization at specific angles due to their design. However, the angle of the polarization sensitivity has a dependence on the frequency of the incoming light (commonly referred to as the “wobbling effect”). In our case, this effect is of the order of $3^\circ$ in the observing band~\citep{aritokisuzuki-dthesis}.
It is known that the wobbling effect can be canceled using four detectors, i.e.,
two pairs of detectors with mutually mirrored sinuous patterns~\citep{aritokisuzuki-dthesis}.
We study wobble cancelation effects with the beam patterns obtained from a \texttt{GRASP} simulation
for 100-GHz detectors with 88, 100, and 112\,GHz frequencies, corresponding
to the central and edge frequencies in the band.
We use the \texttt{QuickPol}~\citep{Hivon2017} algorithm, in which the \LiteBIRD\ scanning strategy is taken into account, to estimate the residual systematic effect when the two mutually mirrored patterned pairs
have different polarization efficiencies.
We find that a difference of 20\,dB produces a residual power spectrum
lower by 1\,\% with respect to the one due to the lensing effect,
corresponding to $\Delta r < 10^{-6}$.
Note that the beam calibration will be conducted with a precision of 53\,dB,
as discussed for the FSL systematic effects. Considering this assumption, this systematic effect can be considered to be negligible.

\subsubsection{Cosmic-ray Systematic Effects}\label{sec:CR-systematics}
For the High Frequency Instrument (HFI) of \Planck\, 
cosmic rays produced glitch signals at a rate of 5 hits
${\rm cm}^{-2}\,{\rm s}^{-1}$~\citep{Planck2013X}, giving rise to systematic errors and
requiring a deglitching process.
For \LiteBIRD, we will use a different detector technology
and expect that the outcome of cosmic-ray effects may differ from that of \Planck-HFI.
In order to predict and characterize the potential effects of cosmic-ray glitches,
two studies have been conducted:
one is to use actual laboratory measurements to estimate the response 
of the detectors to energetic particle impacts
by irradiating the detectors with alpha particles, $\gamma$-rays, or cosmic-ray muons; 
and the second
is to realize an end-to-end simulation to evaluate the propagation of 
cosmic-ray effects.
The former is described in Sect.~\ref{sss:detection_fpm}
 (see also Ref.~\cite{beckman2018}).
Here we describe details for the latter approach (see also Refs.~\cite{Tominaga2020SPIE,Stever_2021}).

For \LiteBIRD, we will employ TES arrays on a 10-inch silicon wafer.
We estimate the expected cosmic-ray hit-rate on a single silicon substrate
using PAMELA data fitted with an Usoskin model
at L2 for the cosmic-ray flux determination~\citep{USOSKIN2001571}
and \texttt{GEANT4} for simulating the interaction of
particles (mainly protons) in the substrate and the surrounding materials (mainly
aluminum), including secondaries and electromagnetic showers.
The cosmic-ray rate is expected to be 400\,Hz on a single silicon wafer on which 
the bolometers are placed. Since the time between hits is comparable to the time constant of the 
TES (a few ms), individual cosmic-ray hits may not be identified.
Hence the net effect is an increase in noise, which may be non-Gaussian.

Cosmic rays impacting the silicon wafer deposit part of their 
energy of 1.8\,MeV on average as heat.
The heat propagation in the wafer is simulated using a finite-element thermal
model in the commercial software \texttt{COMSOL},
assuming the wafer heat capacitance and the heat conductance 
to the refrigerator's thermal bath is kept at a temperature of 100\,mK~\citep{Stever_2021}.
The heat produced by the cosmic-ray impact propagates to  the TES through the SiN support structure, producing a transient fluctuation
in the TES current.
The sensitivity of the detector is used to convert the TES current to
an equivalent signal power.
The bolometer analog signal is digitized with a rate of 20\,MHz, 
and down-sampled to about 19\,Hz with the application of a digital low-pass
filters. In this process, the higher frequency components are dropped without
any aliasing effects.
The bolometers' signals are read out in the frequency domain using
cryogenic resonators. The resonant peak overlap in the frequency domain may
cause crosstalk between detectors, since the cosmic-ray glitch signal
is fast and has wide bandwidth. 
In the simulation study, we implement both effects in the \texttt{TOAST} simulator 
that generates the time-ordered data and projects it onto
sky maps~\citep{Tominaga2020SPIE}.

In this study, we do not include the athermal (ballistic) phonon effect, 
since this effect is hard to model at this stage.
The modeling would require detailed comparisons between the measurements
of the athermal phonon signals and the underlying physics models in a simulation.
We plan to develop a ray-tracing technique for producing Monte Carlo
simulations of athermal phonon propagation effects in the \LiteBIRD\ detector wafer,
which are currently ongoing.
It should be noted, however, that with the expected impact rate of 
400 hits s$^{-1}$ in the wafer, the total amount of energy that propagates 
as athermal phonons will contribute to the same amount of average noise across
the wafer, as indicated by modeling the thermal phonons.
Furthermore, we note that athermal phonons would produce short pulses
smoothed by the TES time constant of a few msec, whose
higher frequency components are removed by the digital filters.

In Sect.~\ref{ss:forecasts_delta_r}
we show band-weighted averaged angular power spectrum obtained
with the technique developed in Refs.~\citep{Tominaga2020SPIE} and \citep{Stever_2021}.
We take into account this effect as an additional noise in $N_\ell$
in Eqs.~(\ref{eq:cl-measured-in-likelihood}) and (\ref{eq:cl-model-in-likelihood}).
The magnitude of the cosmic-ray noise is about 1\,\% of the total noise estimated with the component separation in Fig.~\ref{fig:xforecast-results-pysm}.

\subsubsection{HWP Systematic Effects}\label{sec:HWP-systematics}

The fast rotating HWPs are key elements of \LiteBIRD, allowing quasi-instantaneous estimation of the Stokes parameters $I$, $Q$, and $U$, with individual polarized detectors. An ideal HWP rotates the incident polarization by a known angle. However, due to the complexity of the structure of transmissive broadband HWPs, 
requiring multiple stacked sapphire plates with \gls{arc}, dielectrically embedded in a multi-layer structured plate, etc.,
the transformation of Stokes parameters is more complex, leading to potential artifacts in polarization maps (specifically, mixing of Stokes parameters). Those imperfections can be described by the Jones formalism connecting the input and output electric fields, before and after the HWP, respectively~\citep{10.1111/j.1365-2966.2007.11558.x}:
\begin{equation}\label{eq:jones}
\begin{pmatrix}
E_{x;{\rm o}}\\
E_{y;{\rm o}}
\end{pmatrix} = \begin{pmatrix}
J_{xx} & J_{xy}  \\
J_{yx} & J_{yy}
\end{pmatrix}
\begin{pmatrix}
E_{x;{\rm i}}\\
E_{y;{\rm i}}
\end{pmatrix},
\end{equation}
as well as the Mueller matrix formalism relating the input and output Stokes parameters:
\begin{equation}
\begin{pmatrix}
I_{\rm o}\\
Q_{\rm o}\\
U_{\rm o}\\
V_{\rm o}
\end{pmatrix}
 = \begin{pmatrix}
M_{II} & M_{IQ} & M_{IU} & M_{IV} \\
M_{QI} & M_{QQ} & M_{QU} & M_{QV} \\
M_{UI} & M_{UQ} & M_{UU} & M_{UV} \\
M_{VI} & M_{VQ} & M_{VU} & M_{VV}
\end{pmatrix}
\begin{pmatrix}
I_i\\
Q_i\\
U_i\\
V_i
\end{pmatrix}.
\end{equation}
The relation between the Mueller matrix $M$ and the Jones matrix $J$ can be found in Refs.~\citep{Bryan:10}
and \citep{Leahy2010},
for example.

We now introduce different types of perturbations 
in the coefficients describing the HWP
in order to model imperfections. We have carried out two independent studies of the effect of imperfections on the final $B$-mode signal: the first study, based on the Mueller-matrix formalism is focused on the effect of  \gls{ip} for a transmissive HWP; and the second study, based on the Jones formalism, addresses the effects of polarization efficiency and $Q$/$U$ mixing.
Other potential systematic effects include HWP rotation-synchronous signal observed in Refs.~\citep{Johnson_2007,ritacco2017}, which will be studied in future.

We first focus on the effect of instrumental polarization. For oblique incidence angles, a small fraction of the unpolarized incident intensity is transfered into polarization (due to anisotropic transmission for a transmissive HWP), inducing non-zero $M_{QI}$ and $M_{UI}$ elements with azimuthal angle dependence in particular at 4 times the HWP rotation frequency. 
A model of the observations can be written as
\begin{equation}
  s = \Big(1 \,\,\cos 2\psi_0 \,\,\sin 2\psi_0 \,\, 0 \Big)\, M(\Theta, \rho - \psi)\, R(2\psi)
  \label{modelDataHWP}
\begin{pmatrix}
I_i \\ Q_i \\ U_i \\ V_i
\end{pmatrix} + n.
\end{equation}
The various quantities here are: $s$ is the observed signal on a polarized detector at angle $\psi_0$ with respect to a reference axis in the \gls{fp} coordinates; $\psi$ is the rotation angle of the reference axis of the FP with respect to the sky reference frame in which the Stokes parameters $I_{\rm i}$, $Q_{\rm i}$, $U_{\rm i}$ and $V_{\rm i}$ of the incoming wave are calculated; $\rho=\omega_{\rm HWP}\,t + \rho_0$ is the HWP rotation angle with respect to the sky reference frame; $\Theta$ is the incident angle with respect to the optical axis of the HWP; and $R(2\psi)$ is the rotation matrix for Stokes parameters. The HWP Mueller matrix $M$ is written in the frame of the HWP.

The Mueller matrix of the HWP has been calculated using the electromagnetic wave propagation simulation tool \gls{rcwa} assuming a nine-layer HWP with pyramidal \gls{arc} optimized for LFT.\footnote{For the HWPs of MHFT, we give a mathematical model using the Jones Matrix, which will be described in the following section.}
In an electromagnetic simulation, we find that the Mueller matrix can be expanded into three components consisting of the constant one (0f), one modulated with an angular frequency twice that of the HWP rotation (2f), and one with 4 times that frequency (4f).
Each element of the matrix $M_{ij}$ for a given incident angle $\Theta$ is decomposed into sine waves with respect to the angle $\rho-\psi$, and only the first three terms are shown to be significant:
\begin{equation} \label{eq:M}
M_{ij}(\Theta,\rho - \psi) = M^{(0{\rm f})}_{ij}(\Theta)\,+\,M^{(2{\rm f})}_{ij}(\Theta)\cos (2\rho - 2\psi + \phi^{(2f)}_{ij})\,+\,M^{(4{\rm f})}_{ij} (\Theta) \cos(4\rho - 4\psi + \phi^{(4f)}_{ij}).
\end{equation}
Coefficients of the three matrices $M^{(0{\rm f})}$, $M^{(2{\rm f})}$ and $M^{(4{\rm f})}$, as well as the phases $\phi^{(2f)}_{ij}$ and $\phi^{(4f)}_{ij}$\footnote{Those phases are defined for one reference location in the focal plane and are modified for different azimuthal angles around the HWP rotation axis in a fixed frame in the focal plane.} have been estimated for the whole band at 140\,GHz and for many angles $\Theta$. The IP coefficients of the $M^{(4{\rm f})}$ matrix (four times the HWP spinning frequency) are the ones 
inducing a significant effect on the final $B$-mode power
spectrum. The values of those coefficients $\epsilon = M^{(4{\rm f})}_{QI} = M^{(4{\rm f})}_{UI}$ are found to be $\simeq 4\times 10^{-5}$ for $\Theta_0 = 10^{\circ}$. The empirical relation $\epsilon(\Theta) = \epsilon(\Theta_0) \left(\sin\Theta/\sin \Theta_0\right)^2$ is used to extrapolate the coefficients for any $\Theta$ values corresponding to various locations of detectors in the focal plane.

We have generated input TOD simulations implementing Eq.~(\ref{modelDataHWP}), which include polarized CMB anisotropies (with $r=0$), CMB dipole, and polarized Galactic foregrounds from {\tt PySM} described in Sect.~\ref{sec:reference_skies}, with and without noise.\footnote{Since the model and the map-making relations are linear with respect to the input TOD, the effect of HWP imperfection on the recovered signal can be estimated without noise in the simulations.}
We assume the nominal scanning strategy described in Table~\ref{tbl:imo1}. Simulated data are projected into $I$, $Q$, $U$ maps using the optimal GLS map-making method \texttt{SANEPIC}~\cite{Patanchon08} (see also Ref.~\cite{deGasperis16} for a similar derivation and a description of the application for mapping on a sphere) applied to each detector. The residual maps are calculated by differencing the recovered and input maps used for the simulations.

\begin{figure}[htbp!]
\centering
 \includegraphics[width=0.49\textwidth]{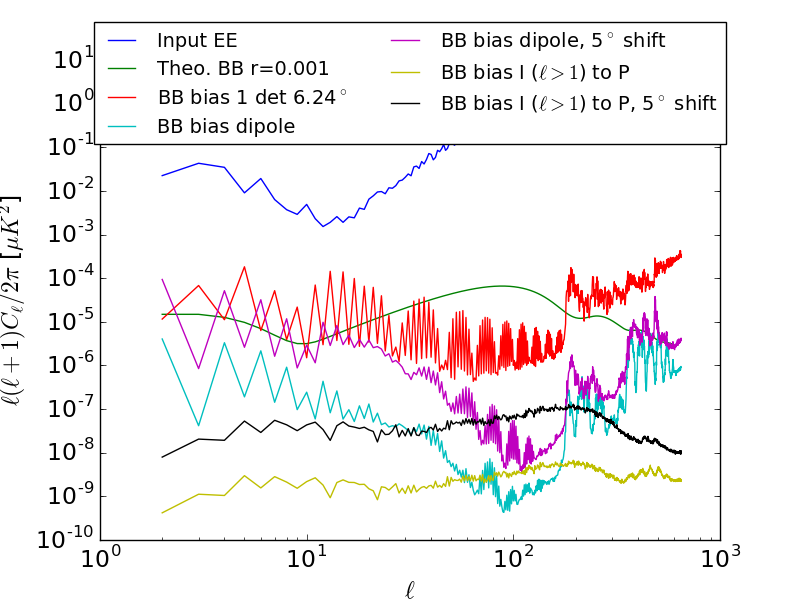}
 \includegraphics[width=0.49\textwidth]{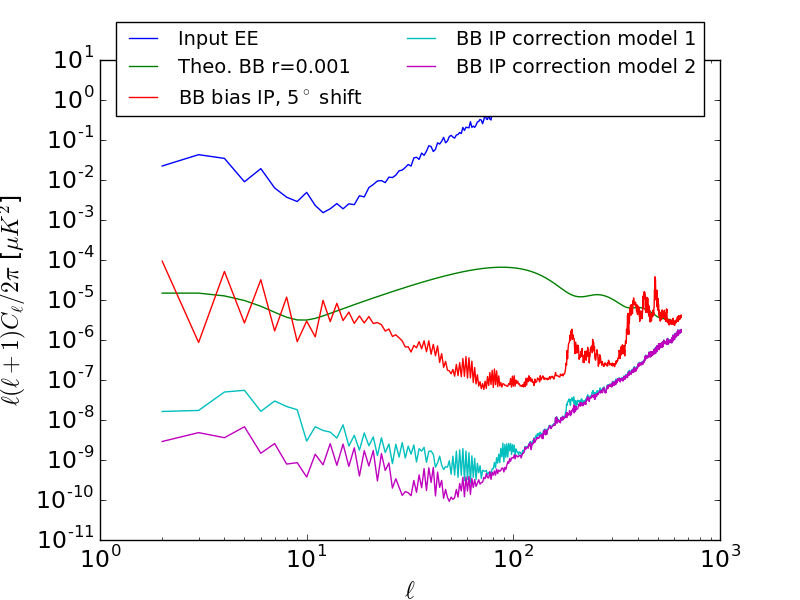}
\caption{\textit{Left}: $B$-mode power spectra of the residual IP leakage. The curves are: the residual power spectrum for one detector at $6.24^\circ$ from the HWP rotation axis (red); the combined effect (all detectors at 140\,GHz LFT) for the dipole only and a centered HWP (cyan); the combined effect of the dipole and $5^\circ$ HWP tilt (violet); and the same as the cyan and violet cases, but for anisotropies without the dipole (yellow and black).
\textit{Right}: $B$-mode power spectra of residual IP after correction with the fit of individual detector IP amplitude in cyan (model~1), and using the $\epsilon(\Theta)$ model in violet (model~2). The power spectrum without correction (effect of the dipole and of the anisotropies combined) is shown in red. We used the less favorable case of a $5^\circ$ tilt of the HWP rotation axis.
}
\label{fig:IPHWP}
\end{figure}

Some cancellation of the leakage contributions is happening when combining several detectors at different location around the HWP rotation axis. This is because for detectors with different azimuthal angles around the HWP rotation axis, the ordinary axis of the HWP is orthogonal to the different detector lines of sight at different times, leading to different phases of the IP contribution for different detectors at a given time. This phase shift is accounted for in our multi-detector simulations.
The resulting contribution of the IP to the $B$-mode spectra after combining all Stokes parameter maps for all detectors at 140\,GHz is shown in Fig.~\ref{fig:IPHWP}. The contributions of the dipole and of the other fluctuations are shown separately. An apodized Galactic mask has been applied, considering 50\,\% of the sky. We have assumed two cases: (i) the optical axis of the HWP is centered in the LFT focal plane; and (ii) the HWP rotation axis is shifted by $5^\circ$. There is less cancellation in the second case. The induced bias $\Delta r$ on the tensor-to-scalar ratio $r$ using the likelihood described in Sect.~\ref{sub:ErrorForm} is indicated in Table~\ref{tab:dr_IP}.

\begin{table}[htbp!]
\captionof{table}{Contribution of the IP to the $\Delta r$ parameter. The most pessimistic case of a shifted HWP is used for evaluation of the effect after correction with two different schemes: for model~1, when the $\epsilon_i$ coefficients are estimated independently for each detector; and for model~2 when we use the scaling relation $\epsilon(\Theta, \phi)$ accounting for the exact relative phase dependence of this parameter with respect to the azimuthal angle of the detector locations $\phi$ around the HWP rotation axis (two detectors at different locations in the FP see the same $4f$ IP effect with a known delay, as described further in the text). 
}
\label{tab:dr_IP}
			\begin{tabular}{|c|c|c|c|c|}
				\hline
				& HWP centered & HWP shifted by 5$^\circ$  & Mitigated, model~1 & Mitigated, model~2\\
				\hline
				\hline
				$\Delta r$ & $1.36 \times 10^{-4}$ & $1.47 \times 10^{-3}$ &  $< 1.20 \times 10^{-6}$ & $< 1.84 \times 10^{-7}$ \\
				\hline
			\end{tabular}

\end{table}

A correction method has been implemented and applied to the simulations including noise. With this method, the Stokes parameter maps, as well as the IP coefficients $\epsilon$, are jointly estimated by maximizing a global likelihood. The recovered polarization maps are then corrected from the effect. Two models used are: (1) when the $\epsilon_i$ coefficients are estimated independently for each detector; and (2) when we use the scaling relation $\epsilon(\Theta, \phi)$ with the relative phase dependence of this parameter with respect to the azimuthal angle of the detector locations $\phi$ around the HWP rotation axis.
The method does not use any external template and exploits from the large intensity signal to recover the $\epsilon$ parameters. Because the monopole is not recovered within the map-making process, it is jointly fitted with the $\epsilon$ parameters, and so is marginalized over, in order to remove its leakage in the polarization maps. The procedure is described in Ref.~\cite{Patanchon21}. The residual $B$-mode spectra of the IP after correction are shown in Fig.~\ref{fig:IPHWP} and the values of $\Delta r$ are indicated in Table~\ref{tab:dr_IP}. Three years of observation is assumed here.
We observe that the bias is greatly reduced after mitigation. We argue that the phase of the residual effect in the map is random after mitigation because the errors on $\epsilon$ are limited by noise, so if more frequency channels are included in the analysis the net effect after component separation should be reduced. The quoted value in Table~\ref{tab:dr_IP} calculated for the 140-GHz LFT channel only is then expected to be pessimistic. We do not consider in this study the coupling with other systematic effects, such as gain variation that might affect the estimation of the parameters and the efficiency of the subtraction. This is postponed to future studies.
In our second study, we only consider orthogonal incidence on the HWP, i.e., $\Theta = 0$ and HWP systematic effects other than the IP leakage.

Even if orthogonal propagation through the HWP is assumed, manufacturing imperfections could still lead to systematic effects in the observed CMB signal. The goal of our second study is to set requirements on the accuracy needed to constrain departures of the HWP from the ideal setup. Requirements are set by imposing a threshold on the maximum bias $\Delta r$ that we could tolerate being due to a combination of HWP systematic effects, including the effect of frequency-dependent HWP parameters~\citep{Duivenvoorden2021, giardiello2021hwp}.

We can make explicit the dependence of the Jones matrix in Eq.~(\ref{eq:jones}) (in the HWP frame) on the HWP non-idealities: 
\begin{equation} \label{eq:realistic}
	J_{\rm HWP} = \begin{pmatrix}
		1+h_{1} & \zeta_{1} e^{i \chi_1}\\
		\zeta_{2} e^{i \chi_2}& -(1+h_{2}) e^{i \beta}
	\end{pmatrix},
\end{equation}
where the frequency dependence of each parameter is understood. In Eq.~(\ref{eq:realistic}), $h_{1,2}<0$ are loss terms, $\beta = \psi - \pi $, where $\psi$ is the phase shift between the orthogonal modes, and $\zeta_{1,2}$ and $\chi_{1,2}$ are amplitudes and phases responsible for $x-y$ polarization mixing. The phases $\chi_{1,2}$ are set to zero in this work, being degenerate at first order with $\zeta_{1,2}$. From the Jones matrix, the equivalent Mueller matrix elements $M^{IX}=M^{IX}(h_{1,2},\zeta_{1,2},\beta)$ with $X=I,Q,U$ can be derived. 

We are considering in the simulation a pair of orthogonal detectors on the boresight with a noise contribution. 
Since we want to propagate to $r$, we make use of CMB bands, specifically four MFT bands and one band for LFT and HFT, respectively (the closest ones to CMB channels). We make use of simulated mesh HWP profiles in frequency for the parameters ($h_1(\nu)$, $h_2(\nu)$, $\beta(\nu)$) in the four MFT bands centered at 100, 119, 140, and 166\,GHz, respectively.  We assume $\zeta_{1,2} = 10^{-2}$, constant in frequency~\citep{Pisano2}. For the LFT/HFT band centered at 100/195\,GHz, we assume only parameters that are constant in frequency, since our study is not very affected by the frequency dependence of the model profiles. The resolution of those profiles is 1\,GHz.

When building the TOD, we perturb the HWP frequency profiles so as to simulate a mismatch with the nominal HWP profile. In particular, we have $M^{IX}_{\rm TOD} \equiv M^{IX}(h_{12}+\Delta h,\zeta_{1,2}+\Delta\zeta,\beta+\Delta\beta)$, where, again, the frequency dependence in each term is understood. 
In each frequency bin within a band, the perturbations $\Delta x$ are drawn from band- and frequency-independent Gaussian distribution with variance $\sigma_{\Delta x}^2$,
where $x$ stands for either of $h$, $\beta$, or $\zeta$.
Instead, in the map-making procedure, we make use of the unperturbed, nominal HWP profile, i.e., $M^{IX}_s\equiv M^{IX}(h_{1,2},\zeta_{1,2},\beta)$.

In order not to be dependent on a specific realization of the systematic perturbations, 
we simulate 10 realizations for each $\sigma_{\Delta x}$, by
always keeping CMB and foreground fixed.
For each band, we compute a template map $\mathbf{m}_{\text{templ}}$ with $M^{IX}_{\rm TOD}(\nu)$ = $M^{IX}_s(\nu)$ (so, no error on the HWP parameters) from the same foreground and CMB maps as above. These templates are used to 
obtain residual maps;
this yields maps of residuals $\mathbf{m}_{\text{res}}$ that are minimally affected by the foreground color effect and mostly due to the mismatch between $M^{IX}_{\rm TOD}(\nu)$ and $M^{IX}_s(\nu)$.

For each map of residuals, masked with a 70\,\% Galactic mask, the corresponding $B$-mode  spectrum is computed and added to the fiducial CMB $B$-mode spectrum. We use the simple \texttt{anafast} pseudo-$C_{\ell}$s for the residuals, since those are not very different from the cases in which corrections for partial sky and $E$-$B$ mixing are applied. 
We use the angular power spectrum obtained from the sum of the sky maps of the fiducial CMB and the residual as the first term in Eq.~(\ref{eq:cl-measured-in-likelihood}) to the likelihood function in Eq.~(\ref{eq:global-likelihood}).
In the exact likelihood approach, we compute the posterior probability distribution of $r_{\Delta x}$ given our data.
The likelihood analysis is performed in the range $2 \leq \ell \leq 200$, considering also the foreground residuals and noise from component separation, which allow us to properly weight each multipole. 
The bias $\Delta r$ due to $\sigma_{\Delta x}$ is finally quantified,
as defined in Eq.~(\ref{eq:r-bias-definition}).

In Table~\ref{tab:dr_dsyst} we report the threshold value of $\Delta x$ for each systematic parameter to have a bias $\Delta r = 5.7 \times 10^{-6}$, when perturbing each systematic one at a time. We have checked that, for small enough values of $\sigma_{\Delta x}$, we have $\Delta r \propto \sigma_{\Delta x}^2$. 
In cases where two systematic effects are introduced simultaneously with uncorrelated errors, we expect the resulting systematic bias to be the sum of the biases resulting from individual systematic errors, i.e., $\Delta r(\sigma_{\Delta x}, \sigma_{\Delta y}) \simeq \Delta r(\sigma_{\Delta x}) + \Delta r(\sigma_{\Delta y})$, where $x$ and $y$ are selected from $h$, $\beta$, and $\zeta$.
This has been checked by taking 200 realizations for each $\sigma_{\Delta x}$, causing approximately the same level of $\Delta r$, and comparing them with the cases of two systematic effects perturbed at the same time with those $\sigma_{\Delta x}$ values.

We checked that if we consider the residual maps for each band generated with errors just smaller than or equal to those shown in Table~\ref{tab:dr_dsyst} and we add them with the corresponding weights from component separation, the $\Delta r$ values associated with that comes out 
less than half of the error budget of
$5.7\times 10^{-6}$. So, we can assume that properly performing component separation might lead to smaller residuals.

The net effcet of $\Delta\zeta$ integrated over the bandwidth
produces the same effect as a shift of the angle of the HWP $\Delta\rho$ and so contributes to the global uncertainty on this parameter. Its impact at first order can be reduced by minimizing the $EB$ correlation of the CMB (as described in Ref.~\cite{minami/komatsu:2020}). Because of the frequency dependence of $\Delta\zeta$, color effects between Galactic dust and the CMB might induce residual systematic contributions. 

\begin{table}[htbp!]
\setlength{\tabcolsep}{4pt}
\captionof{table}{Accuracy level needed for measurements of HWP parameters $h$, $\beta$, and $\zeta$ in order to keep the bias on $r$ below $\Delta r = 5.7 \times 10^{-6}$. Threshold values are given for individual \LiteBIRD\ MFT frequency bands and one band for each of LFT and HFT (quoted with their band centers). The total MFT threshold is set by the lowest threshold in the MFT bands.}  \label{tab:dr_dsyst}
			\begin{tabular}
			{|c|c|c|c|}
				\hline
				\multirow{2}{*}{Band}& $\sigma_{\Delta h}$($\Delta r = 5.7\times 10^{-6}$) & $\sigma_{\Delta \beta}$($\Delta r = 5.7 \times 10^{-6}$) & $\sigma_{\Delta \zeta}$($\Delta r = 5.7 \times 10^{-6}$) \\
				&[$\sqrt{\text{GHz}}$] &  [${\text{deg.}}$-$\sqrt{\text{GHz}}$] & [$\sqrt{\text{GHz}}$]\\
				\hline
				\hline
                100\,GHz (LFT)& $\leq 0.0022$ & $\leq 2.5$  &  $\leq 0.0012$ \\
				\hline
				100\,GHz (MFT)& $\leq 0.0023$  & $\leq 2.0$ & $\leq 0.0013$  \\
				119\,GHz (MFT)& $\leq 0.0031$ & $\leq 1.6$ &      $\leq 0.0011$    \\
				140\,GHz  (MFT)& $\leq 0.0021$ & $\leq 0.8$ &    $\leq 0.0012$   \\
				166\,GHz (MFT)& $\leq 0.0014$ & $\leq 1.1$ & $\leq 0.0010$ \\
				Total (MFT) & $\leq 0.0014$  & $\leq 0.83$  & $\leq 0.0010$  \\
				\hline
				195\,GHz  (HFT)& $\leq 0.0013$ & $\leq 0.83$ & $\leq 0.0008$ \\
				\hline
			\end{tabular}

\end{table}

\subsubsection{Gain Systematic Effects}\label{sec:Gain-systematics}

The gain drift, i.e., the variation of the gain in time, is mainly caused by fluctuations in the focal-plane temperature, which is kept around 100\,mK. Changes in this temperature cause variation of the operational point of
the TES placed on the focal plane and its responsivity. We model the gain for each detector, $i$, as a time-varying gain function, $G_i(t)$, which is injected in a signal timestream as
\begin{equation}
    d_i(t)=G_i(t)\,\left[ I(t) + Q(t) \cos(4\rho - 2\psi) + U(t) \sin(4\rho-2\psi) \right],
\end{equation}
where $\rho$ is the HWP rotation angle and $\psi$ is the detector polarization orientation.
We model the gain function $G_i(t)$ as a product of a term $g_i^0$ that is constant
in time and varying from detector to detector, and another term $g_i(T_{\rm bath}(t))$
that depends on the thermal bath temperature $T_{\rm bath}(t)$, varying in time, i.e., $G_i(t) = g_i^0\, g_i(T_{\rm bath}(t) )$.
In a simulation we generate some time variation of $T_{\rm bath}(t)$ using a model of the power spectrum $P_{\rm bath}$ in the frequency domain:
\begin{equation}
    P_{\rm bath}= A^2_{\rm bath} \frac{1~{\rm Hz}}{f}, \label{eq:gain_datamodel}
\end{equation}
where $A_{\rm bath}$ is the amplitude of $P_{\rm bath}$ at a frequency $f$ of 1\,Hz.
The model is a good match to measurements of the temperature power spectrum using a dilution refrigerator.
We assume the bath temperature $T_{\rm bath} (t)$ to be common for all detectors on the same wafer. 
We assess the systematic effects by simulating realistic \LiteBIRD\ detector response in order to convert the focal-plane bath temperature to the responsivity of the detectors. We assume that the detector constant gain $g_i^0$ varies as a Gaussian distribution with a mean value of unity and a standard deviation of $\sigma_g$.
The time-varying term $g_i(T_{\rm bath}(t))$ starts at a value of unity at $t=0$.
Finally, we simulate 
using the \texttt{TOAST} framework the time samples encoding both astrophysical signals and  thermal bath fluctuations for the full LFT focal plane frequency channels observing for 1~year with \LiteBIRD's scanning strategy.

\begin{figure}[htbp!]
\centering\includegraphics[width=0.48\textwidth]{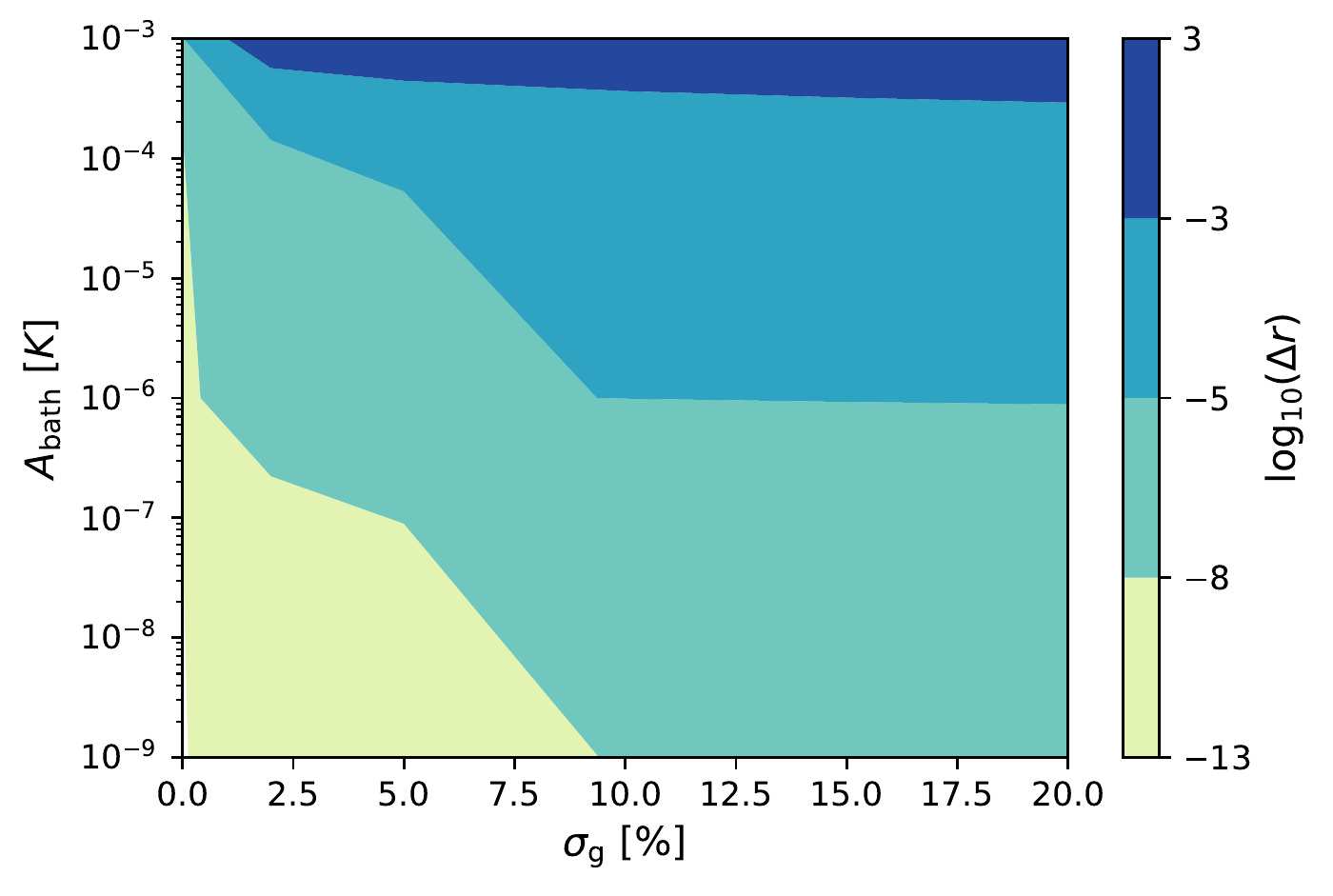}
\centering\includegraphics[width=0.48\textwidth]{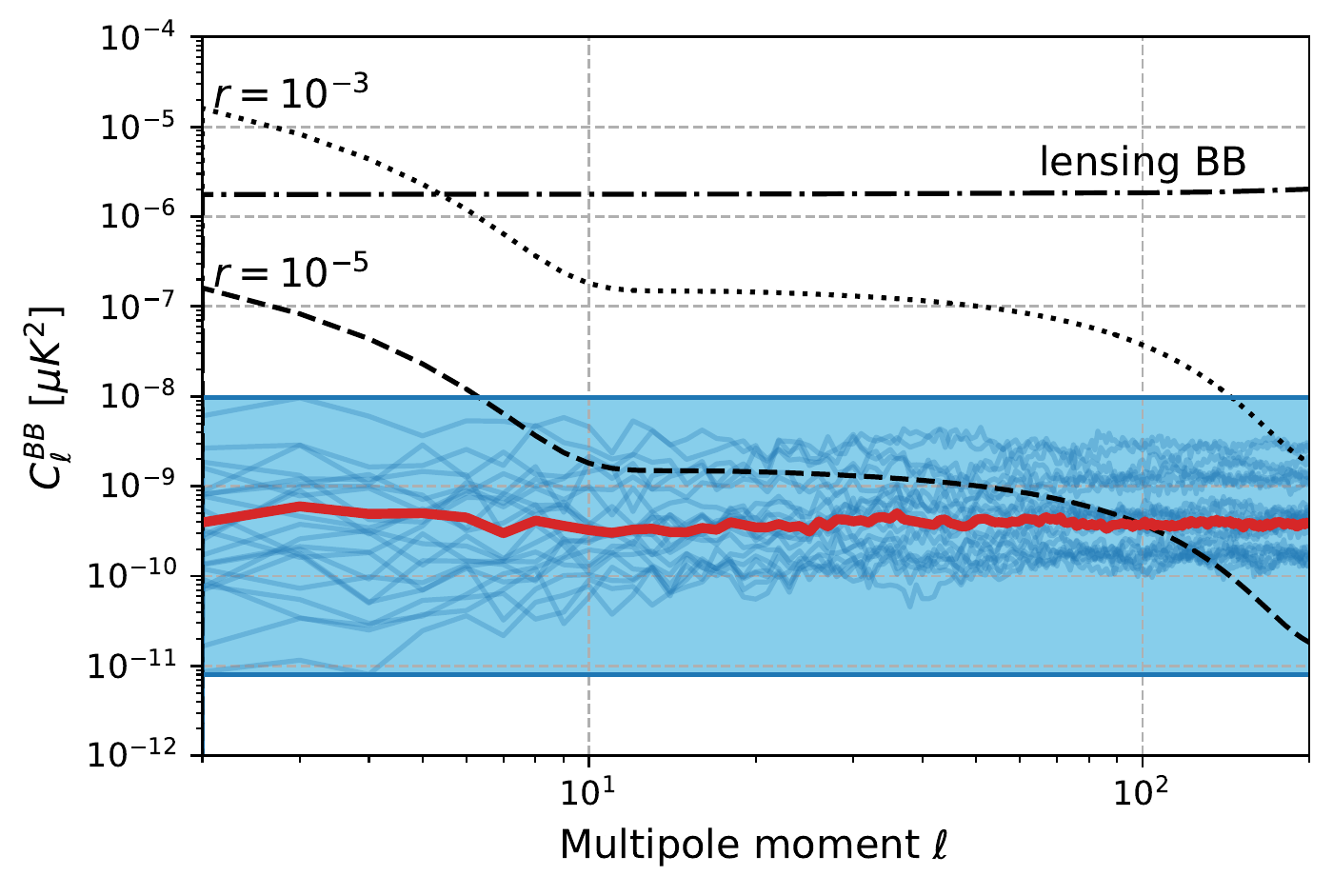}
\caption{\textit{Left}: Systematic bias in the tensor-to-scalar ratio, $r$, equivalent to a bias in the $B$-mode power spectrum between $\ell=2$ and $200$ (the range of the recombination peak), coming from the gain drift $A_{\rm bath}$ and the gain miscalibration $\sigma_g$ introduced in Sect.~\ref{sec:Gain-systematics} for the \LiteBIRD\ LFT 100-GHz channel. 
The quantity $A_{\rm bath}$ is the amplitude of thermal bath fluctuations of a focal-plane wafer at 1\,Hz, while $\sigma_g$ is the relative detector gain miscalibration uncertainty. \textit{Right}: $B$-mode power spectrum showing the bias from unmitigated gain systematic effects in each frequency channel of \LiteBIRD. The $\Lambda$CDM lensing signal is shown as the dash-dotted black line and the tensor signal for $r=10^{-3}$ and $r=10^{-5}$ in the dotted and dashed black lines, respectively. The blue band shows the range of power of the gain systematic biases for each individual frequency channel. The separate spectra for each channel are shown as faint blue lines. The thick red line is the systematic power spectrum from the noise-weighted average over each frequency channel.  
}
\label{fig:systematics_gaindrift_rreq}
\end{figure}

In Fig.~\ref{fig:systematics_gaindrift_rreq} (left) we show the bias introduced in the $B$-mode power spectrum between $\ell=2$ and $200$ 
in terms of the equivalent tensor mode signal and the respective value of $r$ for a simulation of the \LiteBIRD\ 100-GHz channel with 144 detectors. We note that thermal bath fluctuations below $A_{\rm bath}=1\,\mu$K and relative gain miscalibration of $\sigma_g = 10\,\%$ (which is consistent with hardware and calibration requirements), produce a bias of $r<10^{-5}$. Under the assumption that the gain systematic effects are mostly uncorrelated between frequency channels of \LiteBIRD, we can take this result for one of the relevant CMB frequency channels as a conservative upper limit on the systematic bias on $r$ from gain systematic effects.

Figure~\ref{fig:systematics_gaindrift_rreq} (right panel) shows the bias from unmitigated gain systematic effects in the $B$-mode power spectrum. The chosen input parameters for the simulations are $A_{\rm bath}=1\,\mu$K and $\sigma_g=10\,\%$. The bias in $B$-mode power for each frequency channel resembles mostly a white spectrum and lies within the blue band, while the bias from the noise-weighted average over all frequency bands is shown in red. Over the relevant multipole range this bias is well below the requirement of $r<10^{-5}$.

To further lower the residuals several detection and mitigation techniques are now being implemented and tested. They rely on identifying the leakage residuals, estimating these using approximations on the underlying signal, and then mitigating the leakage by subtraction from the data. These mitigation techniques have been shown to reduce the residuals by up to 2 orders of magnitude~\citep{Puglisi2021} and will be presented in Ref.~\citep{puglisi-beck2021}. This is mostly due to the benefit resulting from high sensitivity data and enough cross-linking redundancy provided by the \LiteBIRD\ scanning strategy.
The other subdominant potential sources of gain variation in time include fluctuations of the magnetic field, loading power, bias current, and gains of the cold and warm readout electronics. Given the power spectrum models for those fluctuations, we are in principle able to estimate the systematic effects in a similar manner, but this remains future work.

The absolute gain is a conversion factor from the recorded values, usually in volts, 
to the physical value in units of thermodynamic temperature K$_{\rm CMB}$.
The absolute gain can be calibrated using the solar dipole signal during scans of the sky.\footnote{The orbital dipole can also be used for the absolute gain calibration. We have not yet conducted the study with the existence of possible $1/f$ noise, which remains as future work.}
We estimate the expected accuracy of the absolute gain measurement assuming the existence of
$1/f$ noise with a knee frequency of 20\,mHz for the channels having the lowest white noise for 1~year.
We obtain a value of $5.8\times 10^{-5}$. 
We employ the required value of the gain measurement from Ref.~\citep{Ghigna2020},
on the order of $10^{-4}$, to derive a systematic bias of $5.7\times 10^{-6}$
for the most stringent case in the highest frequency channels.
Therefore the expected bias due to the absolute gain systematic effects is 
$5.7\times 10^{-6} \times (5.8\times 10^{-5} / 10^{-4})^2 = 1.9\times 10^{-6}$.

\subsubsection{Other Systematic Effects}\label{sec:other-systematics}

\paragraph{Polarization Angle}
The error in the polarization angle causes mixing of the Stokes parameters
between $Q$ and $U$, resulting in leakage from $E$ modes to $B$ modes. 
Since the power of the $E$ modes is significantly larger than that of the $B$ modes, 
even a small amount of mixing could cause significant contamination to the $B$ modes. 
The leakage can be described as $C^{EE}_\ell \times K_{\rm pol}$, where $C^{EE}_\ell$ 
is the $E$-mode power spectrum 
and $K_{\rm pol}$ is a factor describing 
the polarization angle homogeneous offset in the entire sky region. 
The effect is sourced by the global offset of the absolute polarization angle 
determined by the roll angle of the spacecraft attitude or the HWP rotation position. 
This effect is modeled as $K_{\rm pol}=\sin^2(2\theta_{\rm g})$, 
where $\theta_{\rm g}$ is the global offset. 
We assume that the expected calibration accuracy of $\theta_{\rm g}$ will be 
2.7\,arcmin, as described in Ref.~\citep{Minami:2020xfg} and in Sect.~\ref{ss:plm_calib}, 
yielding a $\Delta r$ value of $9.1\times 10^{-6}$. 

The systematic effect of relative polarization angle uncertainties between
frequency channels is studied by Ref.~\citep{Vielva_2021}
and described in Sect.~\ref{ss:plm_calib}.
The requirement to give the systematic bias $\Delta r = 5.7\times 10^{-6}$ is shown
in Fig.~\ref{fig:LBangle_requirement}.
The expected precision of measuring the relative angle with the Crab Nebula is shown in Table~\ref{tab:sensitivity_to_crab}.
Since the expectation is comparable to the requirement, we assign a $\Delta r$ value of $5.7\times 10^{-6}$
for the systematic uncertainty of the relative angle.
The frequency dependence of the polarization angle in the observation band includes
the parameter $\zeta$ for the HWP (described in Sect.~\ref{sec:HWP-systematics})
and the sinuous-antenna wobbling effect.
The net effect of those effects with an integration within the bandwidth is calibrated by the global and relative angles
as described above.

The wobble effect in individual detectors
is expected to be smaller than the systematic uncertainties of the HWP $\zeta$,
if we assume that the uncertainty of the wobble effect is uncorrelated among the detectors, resulting in a reduction 
of the bias $\Delta r$ proportional to the inverse of the number of detectors in the band.
We also note that the wobble effect could be canceled further if we use four sinuous antenna patterns~\citep{aritokisuzuki-dthesis},
mutually rotated by $45^\circ$ and inverted. The current focal-plane design employs this technique.

The HWP rotation position is determined by the encoder described in Sect.~\ref{sss:plm_lft_pmu},
with a demonstrated accuracy of less than 1\,arcmin, yielding an expected systematic bias 
of $\Delta r = 1.0\times 10^{-6}$. 
The time variation of the polarization angle determination accuracy of the star trackers
gives negligible systematic effect $\Delta r < 10^{-7}$, 
if we assume that the variation is Gaussian, since the effect cancels out for longer observation times. 
This may not be true when there is a long time correlation in the error;
however, this systematic effect is found to still be much smaller than the constant offset effects shown above. 
The identification of possible error correlations needs further study.

\paragraph{Polarization Efficiency}
The uncertainty of the polarization efficiency is modeled as an uncertainty
of the couplings of the Stokes parameters of $Q$ or $U$, described in 
the Mueller matrix as $M_{QQ}$ and $M_{UU}$. 
The polarization efficiency uncertainties are sourced by the HWP and detectors.
When the polarization efficiency error of the detectors are uncorrelated,
the effect may be scaled proportional to the inverse of the square root of the number of detectors,
yielding effects of negligible amplitude.
On the other hand, an uncertainty in the HWP gives a significant impact, since it is common to all
the focal-plane detectors. 
The polarization efficiency is related to the absolute normalization 
of the $E$- and $B$-mode power spectra. 
The net effect of including the efficiency from the HWP with the frequency band-average 
is calibrated using the $E$-mode power spectrum
for higher multipoles, while its frequency dependence in the band will be calibrated 
using the ground facility. The expected calibration accuracy is found to be 0.2\,\%
using the $E$-mode power spectrum for a higher multipole region, given the detector noise.
If we assume that the $B$-mode lensing effect converts to the tensor signal 
by the amount of the uncertainty, 
then we expect a systematic bias of 
$\Delta r=5.6 \times 10^{-6}$. 
We note that the uncertainties in the frequency dependence of 
the polarization efficiency in the observation band
is taken into account 
through the uncertainties in the parameter $\beta$ in Sect.~\ref{sec:HWP-systematics}.

\paragraph{Pointing}
Pointing errors can arise for several reasons: 
measurement uncertainties in the start tracker; 
mis-alignment between the star tracker and the bore-sight direction of the telescopes; vibration caused by the refrigerators; and deformation of the optical system, which includes the mirrors, lenses, and the support structure. 
The frequency and direction of the pointing deviation depend on the source of the error.
Models of the possible effects on the pointing will be explored in future.
Here we consider two simple cases: a static bias in the pointing; and Gaussian random perturbations at the sampling rate of 19\,Hz. 
The net effect causes polarization leakage from $E$ to $B$ modes, 
resulting in a residual power spectrum similar to that of gravitational lensing. 
This effect is also independent of the existence of the PMU. 
We perform a simulation study for the two cases. 
For the offset case, we find that a pointing bias in the direction orthogonal to the scanning orientation
yields the most stringent requirement of 4.6\,arcmin to give $\Delta r=5.7 \times 10^{-6}$.
On the other hand, the random disturbances expected from the star-tracker pointing accuracy of 0.23\,arcmin yield a negligible bias of $\Delta r < 10^{-6}$.

The wedge shape of the spinning HWP also causes pointing disturbances,
where we define the wedge shape as the lack of parallelism between the plate surfaces. 
We therefore also conduct a simulation study assuming that the wedge produces a pointing disturbance
rotating with the HWP spinning rate around the original detector pointing. 
This gives a requirement on the allowable maximal wedge angle in the HWP wafer fabrication with the relation $\phi=(n-1)\psi$, where $\psi$ is the HWP wedge angle, 
$n$ is the refractive index and $\phi$ is the pointing disturbance angle.
For sapphire, having $n=3.1$ as an example, we set the requirement of $\psi$ to be smaller 
than 4.0\,arcmin to give $\Delta r=5.7 \times 10^{-6}$.

\paragraph{Bandpass}
We study the impact of uncertainties on the frequency bandpass determination 
producing inter-frequency mismatch on the measurement of $r$. 
These uncertainties propagate into uncertainties 
in the amplitude of 
components, including leakage, after applying foreground separation. 
The amplitude offsets are supposed to be an average effect over
the full arrays of detectors, therefore the requirement on individual 
detectors is relaxed by a factor of $\sqrt{N}$, where $N$ is the number of detectors 
in the array, with an assumption that the offset measurements are uncorrelated.
This assumption is not appropriate when the correlation coefficient is larger than $1/N$, however.
We describe the bandpass uncertainty 
using a single parameter $\gamma_{c}$ 
defined in Eq.~(\ref{eq:colorCorrection}), where subscript $c$ is either the dust or synchrotron component. 
We employ the discussions in Ref.~\citep{Ghigna2020}
to give the systematic bias $\Delta r$ for the bandpass uncertainty with the usage of the PMU. 
The systematic effects are evaluated as a function of the uncertainties of 
$\gamma_{c}$ for \LiteBIRD's 15 frequency channels individually 
with the application of the foreground subtraction procedure~\citep{Ghigna2020}. 
With a requirement of $\Delta r$ to be less 
than $5.7 \times 10^{-6}$, 
the requirements on the measurement accuracy 
of $\gamma_{c}$ is in the range of $10^{-4}$ to $2.5 \times 10^{-3}$, 
depending on the frequency channels for the detector arrays. 
The most stringent requirement is given for the two highest frequency channels.
Here we assume that we measure the bandpass with the most stringent requirement
for all the frequency channels, and calculate the systematic residual $B$-mode power spectrum to obtain
a systematic bias of $\Delta r = 5.3\times 10^{-6}$.

\paragraph{Crosstalk}
The readout system makes use of frequency domain multiplexing, 
which is described in Sect.~\ref{ss:we}.
This scheme can reduce the total number of readout wires and the heat load
to the cryogenic detectors, but could also introduce crosstalk effects due to
the interference of the frequency comb.
Such effects are modeled as a single matrix $W_{ij}$ that describes the
leakage of the measured power from detector $j$ to detector $i$.
We study the crosstalk effects with a model similar to the SPT-3G design
\citep{Montgomery2021}. 
We explore ten models of
detector readout orders in the frequency domain and detector physical
positions in the focal plane to give $W_{ij}$, and find that the systematic
effect is almost independent of the models, but does depend on 
the magnitude of the uncertainties in $W_{ij}$.
For the case of the 0.1\,\% uncertainties on that, the systematic error on 
the $B$-mode power (lensing) is 0.07\,\%, corresponding to $\Delta r$ of less than $10^{-7}$ in case of the true $r$ value to be 0 without foregrounds.
We assign a systematic error budget of 
$\Delta r = 5.7\times 10^{-6}$ to this, requiring the 
crosstalk knowledge uncertainty to be less than 1\,\%.

\paragraph{Detector Time Constant}
The TES used for \LiteBIRD\ is known to have a time constant $\tau$ of about 3\,ms,
which can be well modeled by a single exponential function.
The net effect of the convolution in time due to this function is a rotation angle shift $\phi$
of the HWP for the $4f$ modulated signals: $\tan(4\Delta\phi)=4\omega_{\rm HWP}\tau$,
where $\omega_{\rm HWP}$ is the angular speed of the spinning HWP.
This shift remains after applying polarization angle corrections using the $C_l^{EB}$ power spectrum
and has to be further corrected.
The uncertainty of $\tau$, $\delta\tau$, causes a systematic effect, which is 
given as $\delta \tau \simeq \delta\phi/\omega_{\rm HWP}$.
For an HWP spinning rate of 1\,Hz with the 
expectation of $\delta\phi<1$\,arcmin (corresponding to $\Delta r=1.0\times 10^{-6}$) 
we obtain $\delta\tau<47\,\mu$sec.
A measurement of $\tau$ may be conducted in flight using the $2f$-modulated
signals caused by the CMB monopole $I\to P$
leakage through the HWP.
The LFT HWP is expected to give a $2f$-modulated signal with an amplitude
of $p_0=9\simeq 27$\,mK. 
The time duration $T$ required to measure $\tau$ 
with a precision of $\delta\tau$
is given by
\begin{equation}
T =  5.7\times 10^2\,{\rm s}\left(\frac{\rm NET}{50\,\mu{\rm K}\sqrt{\rm s}}\right)^2
\left(\frac{10\,{\rm mK}}{p_0}\right)^2
\left(\frac{1\,{\rm Hz}}{\nu_{\rm HWP}}\right)^2
\left(\frac{47\,\mu{\rm s}}{\delta\tau}\right)^2,
\end{equation}
where $\nu_{\rm HWP}=\omega_{\rm HWP}/2\pi$.
The actual time variation in the time constant is not known yet and requires
study in the future.
We assign $\Delta r$ a value of $5.7\times 10^{-6}$ as the requirement.

\subsection{Total Uncertainties of the Tensor-to-Scalar Ratio}
\label{ss:forecasts_delta_r}

This section provides an evaluation of the total uncertainty on the tensor-to-scalar ratio $r$.
The power spectrum of the systematic effects in Eq.~(\ref{eq:cl-measured-in-likelihood})
is given as
\begin{equation}
C_\ell^{\rm sys} = \sum_i C_\ell^{{\rm sys} (i)} + 
\sum_{i \neq j} \langle B_{\ell m}^{{\rm sys} (i)}B_{\ell m}^{{\rm sys} (j)*}\rangle,
\label{eq:total-systematic-cl}
\end{equation}
where the first term describes the sum of the $i$th systematic effect
power spectrum, and the second term shows the potential correlations between
two systematic effects.  The factors
$B_{\ell m}^{{\rm sys} (i)}$ and
$B_{\ell m}^{{\rm sys} (j)}$ are the $B$-mode
coefficients in the spin-harmonic expansion of the $i$th and $j$th systematic residual sky maps,
where the residual map is the difference between the sky maps
with and without the systematic effects.
Therefore, the first term is given as 
\begin{equation}
C_\ell^{{\rm sys} (i)}=\langle B_{\ell m}^{{\rm sys} (i)}B_{\ell m}^{{\rm sys} (i)*}\rangle.
\end{equation}

\begin{figure}[htbp!]
\centering
 \includegraphics[width=1.0\textwidth]{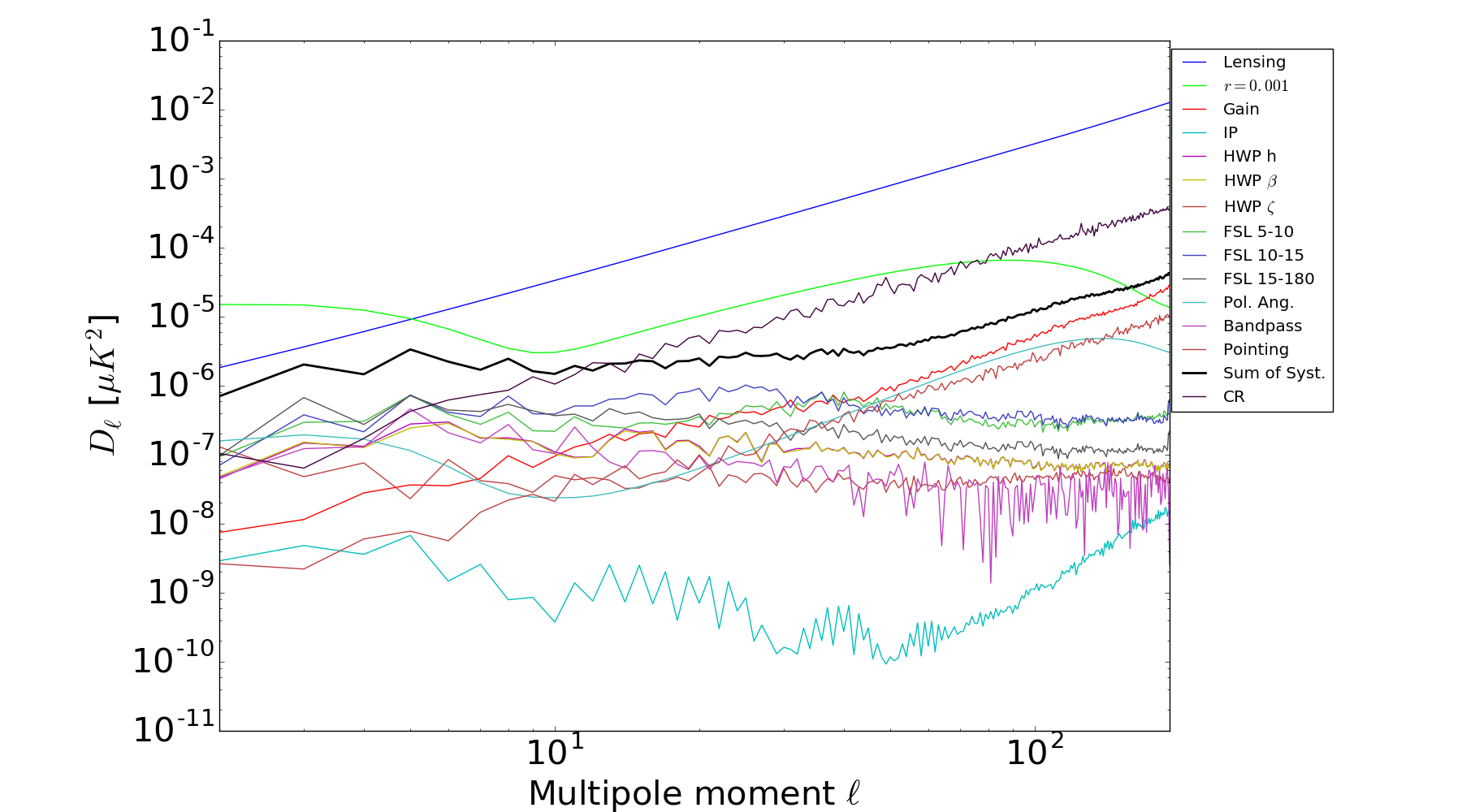}
\caption{Angular power spectra of the sources of systematic effects and cosmological predictions for the lensing and tensor modes, where $D_\ell = \ell(\ell+1)C_\ell/2\pi$.
From the top of the legend downwards, we plot: the lensing $B$ modes; the tensor $B$ modes with $r=0.001$; the systematic effects of the gain; the HWP instrumental polarization (IP); the HWP parameters of $h$, $\beta$, and $\zeta$; the far sidelobes (FSL) in the angle ranges of $5^\circ$--$10^\circ$, $10^\circ$--$15^\circ$, and $15^\circ$--$180^\circ$; the polarization angle (Pol.\ Ang.); the bandpass effects; the pointing effects; and the sum of all the systematic effects. The final line is the cosmic-ray (CR) contribution as an additional noise source.
}
\label{fig:cl_sys_all}
\end{figure}

With the condition of Eq.~(\ref{eq:r-bias-definition}), we may obtain 
an approximation of small systematic biases:
\begin{equation}
\Delta r \simeq \frac{\sum_\ell (2\ell+1)C_\ell^{\rm tens}C_\ell^{\rm sys}/(C_\ell^{\rm n})^2}
{\sum_\ell (2\ell+1)(C_\ell^{\rm tens}/C_\ell^{\rm n})^2},
\label{eq:delta-r-approximation}
\end{equation} 
where $C_\ell^{\rm tens}$ is the tensor mode with $r=1$ and
$C_\ell^{\rm n} = C_\ell^{\rm lens}+N_\ell$ is the sum of the lensing and the noise contributions.
The first term in Eq.~(\ref{eq:total-systematic-cl})
gives the total bias of $r$ as the sum of the individual systematic biases, i.e.,
$\Delta r =\sum_i \Delta r^{(i)}$,
where $\Delta r^{(i)}$ is obtained from Eq.~(\ref{eq:delta-r-approximation}) when $C_\ell^{\rm sys}$ is replaced by $C_\ell^{{\rm sys} (i)}$.
The second term is estimated at the map level by summing up the 
residual sky maps for individual systematic sources. 
We evaluate the second term for the systematic effects including
the beam far sidelobes, the HWP, and the gain.
We compare two $\Delta r$ values: one is the sum 
of individual systematic biases $\Delta r^{(i)}$; and the other is obtained by a single power spectrum
from the superimposed residual sky map of individual systematic effects.
With 20 realizations, we find that the difference is $\pm 20$\,\%.
We therefore assume that the second term in Eq.~(\ref{eq:total-systematic-cl}) cancels.
An evaluation of 
the
correlation between all systematic effects
needs a single combined simulation
tool, accounting for all the effects,
which is beyond the scope of the current work and remains for future study. 
In this paper, we assume that the 
total power spectrum of all systematic effects
is given by the first term of Eq.~(\ref{eq:total-systematic-cl}).
Figure~\ref{fig:cl_sys_all} shows the power spectra of individual systematic effects, as well as their sum.

\begin{figure}[htbp!]
\centering
\includegraphics[width=0.8\textwidth]{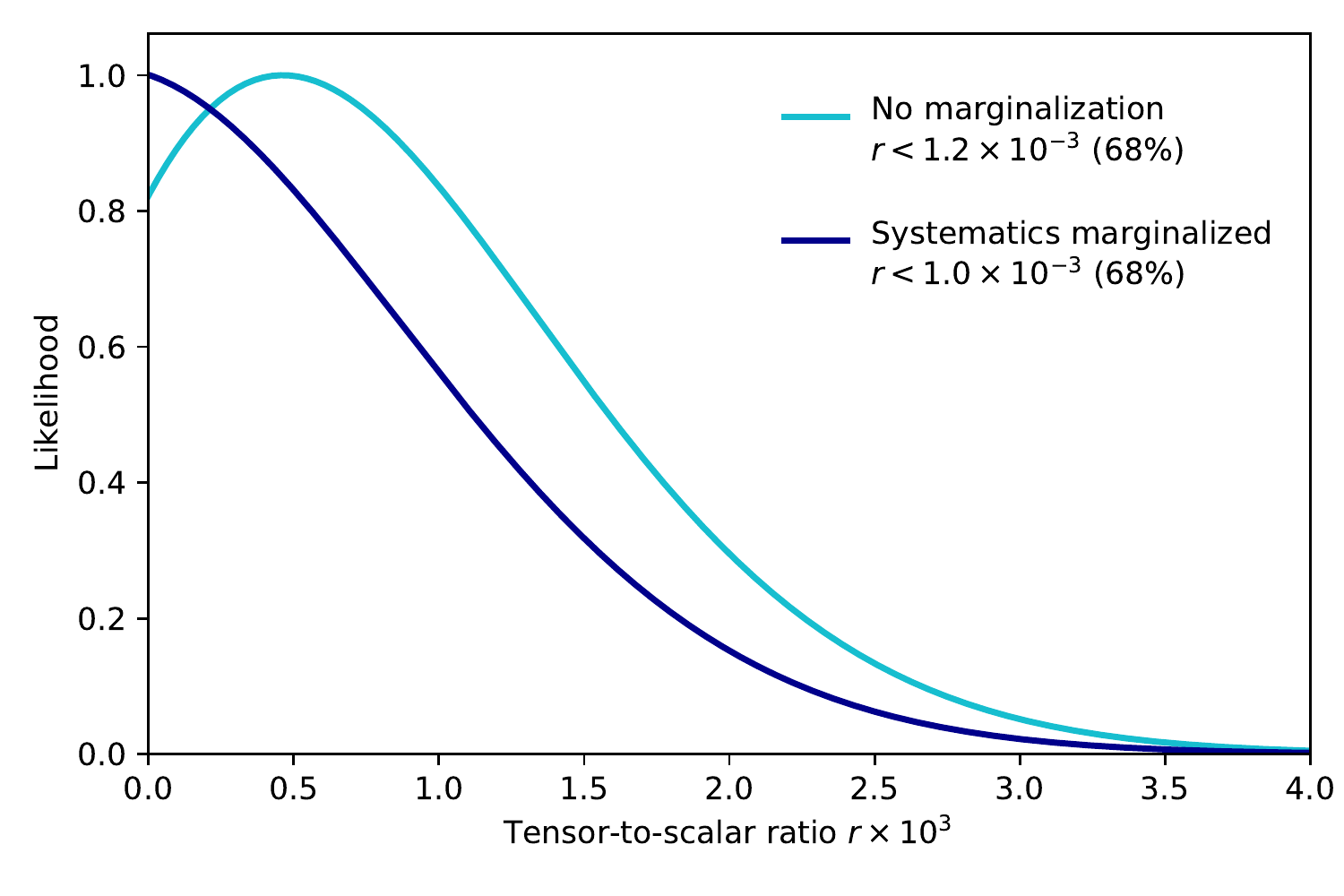}
\caption{Likelihoods as a function of the tensor-to-scalar ratio $r$. The light blue line is that obtained from Eq.~(\ref{eq:global-likelihood}). The dark blue line is obtained after de-bising using marginalization.
}
\label{fig:likelihoods-r}
\end{figure}

Using Eq.~(\ref{eq:total-systematic-cl}) for all sources of systematic effects,
including the gain, the HWP \gls{ip}, the HWP parameters ($h$, $\beta$, and $\zeta$),
the beam far sidelobes, the polarization angle, the bandpass,
pointing, component separation, 
noise contributions $N_\ell$ from
foreground subtraction in Fig.~\ref{fig:xforecast-results-pysm}, and noise arising from cosmic-ray effects in Fig.~\ref{fig:cl_sys_all},
we estimate the total error on $r$ to be $\delta r = 1.2\times 10^{-3}$,
and the systematic bias to be $\Delta r = 0.5\times 10^{-3}$.
Figure~\ref{fig:likelihoods-r} shows the likelihood function obtained, as the right blue curve.
Other systematic effects including the beam near sidelobe, main lobe, ghosting, beam polarization and shape in band, absolute gain, polarization efficiency, detector time constant, and crosstalk give a bias of $0.03\times 10^{-3}$, which is obtained by summing the $\Delta r$ values in Table~\ref{tbl:syst}.

We consider a possible de-biasing of the systematic effects by marginalization. 
We introduce an additional term $\alpha M_\ell$
in Eq.~(\ref{eq:cl-model-in-likelihood}):
\begin{equation}
C_\ell = r C_\ell^{\rm tens}+C_\ell^{\rm lens}+N_\ell + \alpha M_\ell,    
\end{equation}
where $M_\ell$ is a template power spectrum accounting for the systematic bias, and $\alpha$ is a scaling factor to marginalize.
We multiply the likelihood $L(r)$ in Eq.~(\ref{eq:global-likelihood}) by a function $\exp(-\alpha^2/2\sigma_\alpha^2)$
and integrate over $\alpha$ to define
a marginalized likelihood function $\widetilde{L}(r)$:
\begin{equation}
    \widetilde{L}(r) = \int_{\alpha_{\rm l}}^{\alpha_{\rm u}}
    d\alpha L(r)e^{-\alpha^2/2\sigma_\alpha^2}.
    \label{eq:marginalize-integ}
\end{equation}
The net effect of this procedure is to subtract the systematic bias and to inflate the width of the likelihood function by
an increase of the covariance of the power spectrum.
The larger $\sigma_\alpha$ is, the smaller is
the systematic bias,
since the larger $\sigma_\alpha$ gives
almost no constraint in the $\alpha$ value a priori, and is equivalent to fitting $\alpha$.
We employ the template spectrum $M_\ell$ as the dust spectrum, which is evaluated by the component-separation process. 
We justify this by the fact that 
the main systematic effects are caused by polarization leakage from the higher frequency channels where the dust foreground dominates, and the shape of resultant systematic power spectra for those effects is similar to that of the dust component.
For some systematic effects, including the pointing and the gain, the spectra are similar to that of lensing, since those effects are mainly caused by leakage from $E$ modes to $B$ modes,
as shown in Fig.~\ref{fig:cl_sys_all}.
Those effects, however, give rise to power spectra much lower than the others, having a spectral shape similar to that of the dust component for the lower multipole region.
We conduct the marginalization assuming $\sigma_\alpha = 10$,
$\alpha_{\rm l} = 0$,
and $\alpha_{\rm u} = +\infty$
in Eq.~(\ref{eq:marginalize-integ}),
and obtain
$\delta r = 1.0\times 10^{-3}$.
Figure~\ref{fig:likelihoods-r} shows
the likelihood function obtained $\widetilde{L}(r)$ as the dark blue line.

We note that 
the $\delta r$ value is unchanged
when we increase $\sigma_\alpha$ to values larger than 10. This is due to the fact that the large value of $\sigma_\alpha$ allows $\alpha$ to move freely, which completely marginalizes over the systematic biases.
A more sophisticated method is to marginalize individual systematic effects using estimated power spectrum shapes.
However, this requires more precise models of the instruments and the calibration as well as an estimation of $\sigma_\alpha$s that corresponds to uncertainties of the systematic biases.
These issues remain as future work.

\begin{figure}[htbp!]
\centering
\includegraphics[width=0.8\textwidth]{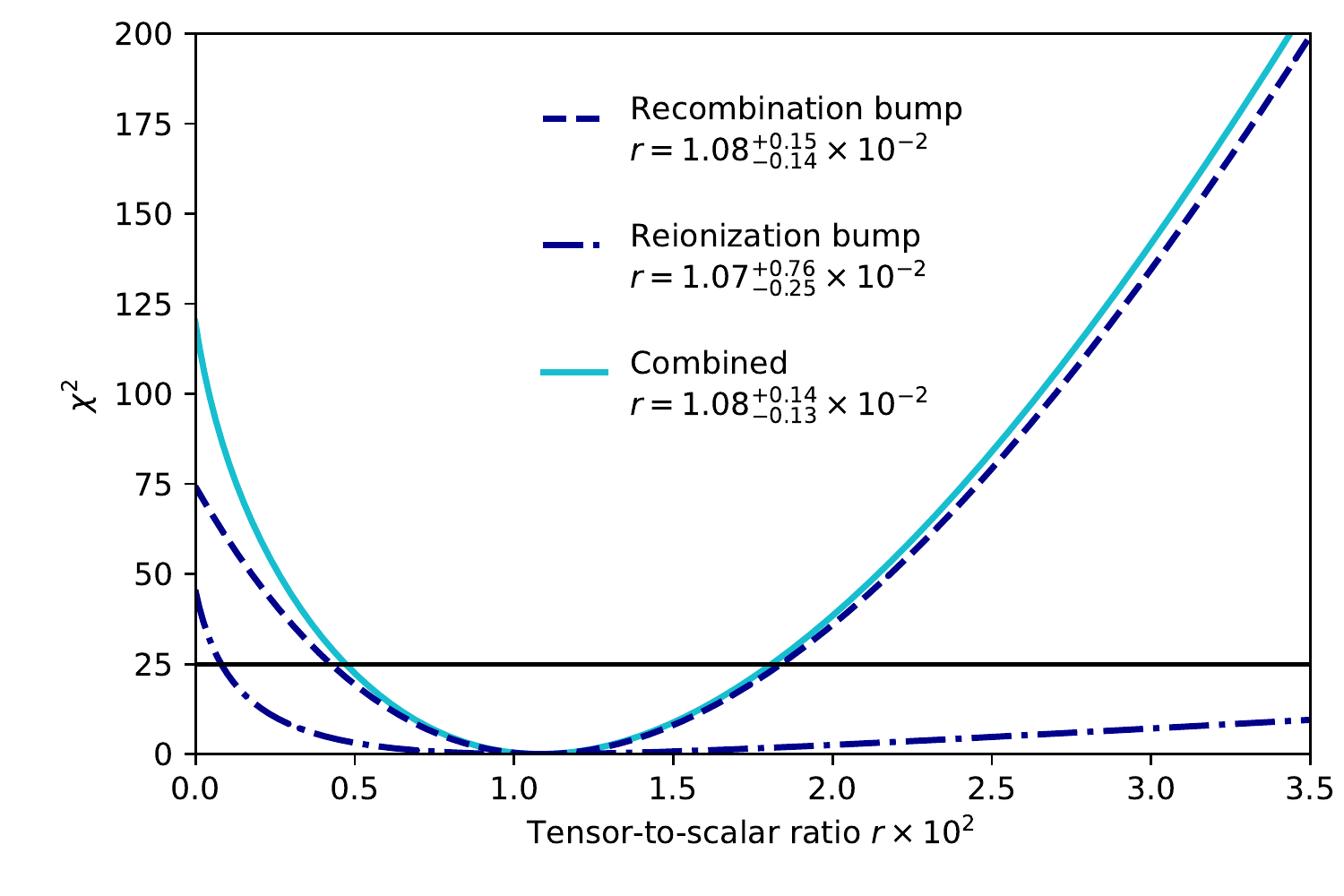}
\caption{$\chi^2$ distribution when $r_{\rm true}=0.01$. The dark blue dashed line is obtained using the multipole range of $11 \leq \ell \leq 191$ containing the recombination bump. 
The dark blue dot-dashed line is obtained using the multipole range of $2 \leq \ell \leq 10$, containing the reionization bump. The light blue curve is the combined $\chi^2$. 
The horizontal line at $\chi^2=25$ indicates 5\,$\sigma$ significance.
}
\label{fig:dchisq-r}
\end{figure}

We estimate the expected significance of the tensor-to-scalar ratio assuming $r_{\rm true}=0.01$ as the true value.
We modify the measured $B$-mode power spectrum in 
Eq.~(\ref{eq:cl-measured-in-likelihood}):
\begin{equation}
\hat{C}_\ell = r_{\rm true}C_\ell^{\rm tens} + C_\ell^{\rm sys}+C_\ell^{\rm lens}+N_\ell,
\label{eq:cl-measured-in-likelihood-r0.01}
\end{equation}
and calculate the likelihood function in Eq.~(\ref{eq:global-likelihood}).
We do not apply the marginalization because the total bias $\Delta r$ is much smaller than $r_{\rm true}$.
We define 
\begin{equation}
    \chi^2(r) = -2 \log \left(\frac{L(r)}{L(\tilde{r})}\right),
    \label{eq:chi2-definition-for-r-signif}
\end{equation}
where $\tilde{r}$ is the $r$ value giving the maximum of $L(r)$, and we
check the requirement shown in Table~\ref{tbl:lv1}, i.e., we calculate the $\chi^2(r)$ value for two multipole ranges of $\ell_{\rm min}=2$ and $\ell_{\rm max}=10$, and
$\ell_{\rm min}=11$ and $\ell_{\rm max}=191$;
the former range includes the reionization bump and the latter includes the recombination bump.
Figure~\ref{fig:dchisq-r} shows the $\chi^2(r)$ distributions for the two cases, as well as the combined one.
We find that the hypothesis of $r=0$ is rejected at a significance more than 5\,$\sigma$ for both cases, implying that the requirement Lv1.02 in Table~\ref{tbl:lv1} is fulfilled.
We note that the asymmetric shape of the
$\chi^2(r)$ distribution for the reionization bump measurement is due to the cosmic variance dependence on $r$. 
The constraint on $r$ of the recombination bump measurement is stronger than that of the reionization bump for $r_{\rm true}=0.01$.
However, the reionization bump becomes significant for lower $r_{\rm true}$ values due to the contribution of the lensing effect. 

\subsection{Enhanced Science Case}\label{ss:extrasuccess}
The criterion for full success of the mission has been defined conservatively and does not rely on any new data sets external to \LiteBIRD. However, we can reduce $\delta r$ further using external data sets, which contribute to the ``enhanced science case.''\footnote{In JAXA, this is also referred to as ``extra success.'' }

In this section, we focus on reducing the statistical part of the total uncertainty, $\sigma (r=0)$, which includes the cosmic variance of the gravitational lensing signal and the noise after the component separation. 
One way to reduce $\sigma (r = 0)$ is to ``delens'' using external data sets \cite{smith/etal:2012,simard/hanson/holder:2015,sherwin/marcel:2015,namikawa/nagata:2015,larsen/etal:2016}. Delensing removes the lensing $B$-modes by subtraction at the map level, thus reducing the lensing $B$-mode cosmic variance contribution described above, rather than simply characterizing its power spectrum~\cite{kesden/cooray/kamionkowski:2002,knox/yong-seon:2002}. Successful delensing using the internal CMB data requires a higher angular resolution than that of \LiteBIRD~\cite{seljak/hirata:2003,smith/etal:2009}, because a low-noise lensing reconstruction requires the imaging of a large number of small-scale modes.

\begin{figure}[htbp!]
\centering
\includegraphics[width=0.8\textwidth]{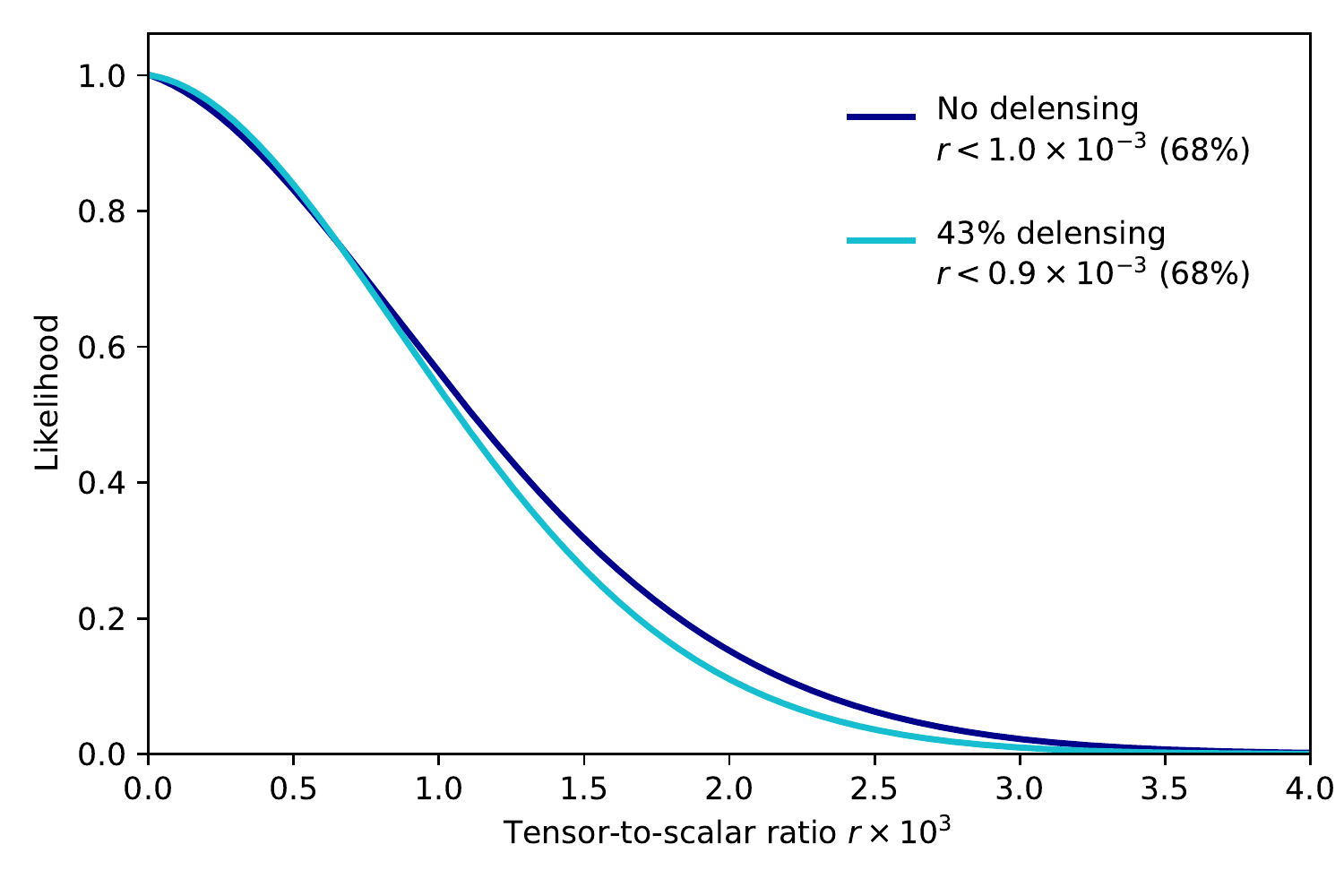}
 \caption{Marginalized posterior distributions of the tensor-to-scalar ratio parameter $r$ with (blue) and without (black) a delensing factor, which could be obtained by a multi-tracer template obtained by \Planck\ CIB and lensing maps reconstructed with the help of \textit{WISE}~\cite{yu/hill/sherwin:2017}. The instrumental systematic uncertainty budget is not included here.}
\label{fig:rdelensed}
\end{figure}

There are several promising ways to delens \LiteBIRD\ with external data sets and thus contribute to the enhanced science case. The most conservative option would be to delens using currently available data sets; for instance, using the multi-tracer template obtained by the \Planck\ \gls{cib} map and the \Planck\ lensing potential map reconstructed with the help 
of \textit{WISE}~\cite{yu/hill/sherwin:2017}. See Fig.~\ref{fig:rdelensed} for the expected improvements on the constraint in $r$ obtained by applying a 43\,\% reduction of the lensing $B$-mode power spectrum as obtained for a large fraction of the sky~\cite{yu/hill/sherwin:2017}. The resulting total uncertainty, including systematic effects, is $\delta r\simeq 0.9\times 10^{-3}$, which enables us to distinguish no primordial gravitational waves from the Starobinsky model~\cite{starobinsky:1980} with a significance  greater than $5\,\sigma$.
\LiteBIRD\ can also help reduce the Galactic dust contamination of the CIB on large angular scales, which may further improve the constraint on $r$.
Further reduction of the lensing $B$-mode power spectrum can be achieved with future lensing measurements from ground-based CMB surveys, such as the CMB-S4 experiment~\citep{Abazajian:2016yjj} and its precursors, but only in the sky region that overlaps with \LiteBIRD.

The other way to reduce $\sigma (r = 0)$ is to use external data sets in frequency bands below the lower edge of \LiteBIRD's lowest frequency band (34\,GHz). Better modeling of the polarized synchrotron emission helps to reduce the foreground residuals. External low-frequency ground-based data sets such as QUIJOTE~\cite{genova-santos/etal:2015,genova-santos/etal:2017}, C-BASS~\cite{irfan/etal:2015}, and S-PASS~\cite{krachmalnicoff/etal:2018} at frequency bands outside those of \LiteBIRD\ ($\nu<34$\,GHz) would be useful for potentially improving foreground cleaning, and hence reducing $\sigma (r = 0)$.

\subsection{Summary of the Statistical and Systematic Uncertainties in the Tensor-to-scalar Ratio Measurement }\label{ss:forcast-summary}
Let us now summarize the results obtained in Sect.~\ref{s:cosmological_forecasts}.
Based on the sky observation strategy with the detector sensitivity shown in Sect.~\ref{ss:forecasts_imo}, we conducted studies on the foreground component separation in Sect.~\ref{ss:forecasts_fg_cleaning},
and the systematic uncertainties in Sect.~\ref{ss:forecasts_syst}.

As explained in Sect.~\ref{sec:component_separation},
we estimate the statistical foreground residual bias $r_{\rm FG}$ and its error $\sigma(r=0)$ is obtained as
\begin{equation}
    r_{\rm FG}=(3.3 \pm 6.2)\times 10^{-4},
\end{equation}
where the definition of $\sigma(r=0)$ is given in Eq.~(\ref{eq:sigma-r-definition}), which is equivalent to the total statistical error $\sigma_{\rm stat}$ in the science requirement Lv2.01 in Sect.~\ref{ss:overview_mission}. The obtained error satisfies the requirement.

In Sect.~\ref{ss:forecasts_syst},
we give details of the studies on the systematic effects which are summarized in Table~\ref{tbl:syst}. The power spectra of the systematic effects are presented in Fig.~\ref{fig:cl_sys_all}. The bias on $r$ is defined in Eq.~(\ref{eq:r-bias-definition}).
The total systematic bias is
estimated to be
\begin{equation}
    \Delta r_{\rm syst}=1.7 \times 10^{-4},
\end{equation}
which is equivalent to the $\sigma_{\rm syst}$
in the science requirement Lv2.02 in Sect.~\ref{ss:overview_mission}.

In Sect.~\ref{ss:forecasts_delta_r}, we estimate the total error $\delta r$ defined  in Eq.~(\ref{eq:r-total-error-definition}) by accounting for the statistical uncertainties with the component separation and the biases produced by the uncertainties of the foregrounds and instrumental systematic effects with the sky observation fraction of $f_{\rm sky}=49.5$\%. We apply the marginalization defined in Eq.~(\ref{eq:marginalize-integ}) to conduct a de-biasing procedure, and obtain the total error
\begin{equation}
    \delta r = 1.0 \times 10^{-3},
\end{equation}
which satisfies the requirement of Lv1.01.

We also estimate the expected significance of $r$ with the assumption that the true value of $r$ is 0.01.
We define $\chi^2(r)$ in Eq.~(\ref{eq:chi2-definition-for-r-signif}),
and compute it for two cases: the multipole range of $\ell_{\rm min}=2$ and $\ell_{\rm max}=10$ covering the reionization bump; and the range of $\ell_{\rm min}=11$ and $\ell_{\rm max}=191$ covering the recombination bump.
We obtain Fig.~\ref{fig:dchisq-r} and find that both cases reject the hypothesis of $r=0$ with a significance more than $5\sigma$, fulfilling the requirement of Lv1.02.

We point out that we could further reduce $\delta r$ using external data sets as described in the enhanced science case shown in Sect.~\ref{ss:extrasuccess}. 
We examine one example by applying delensing, and obtain
\begin{equation}
    \delta r = 0.9 \times 10^{-3},
\end{equation}
with an assumption of 43\% reduction of the lensing effect.



\section{Scientific Outcomes of \lb\ Beyond Primordial Gravitational Waves}\label{s:scientific_outcomes}
The primary science goal of the \lb\ mission is to discover and characterize the signature of the primordial gravitational waves from cosmic inflation in the $B$-mode polarization of the CMB. We have described this goal in detail in Sect.~\ref{s:cmb_polarization} and presented our forecast in Sect.~\ref{ss:forecasts_delta_r}.

In addition to the primary goal, the full sky maps in 15 microwave bands will offer rich new data sets, which will enable exciting breakthroughs in a variety of science areas. We now describe the expected outcomes of the \lb\ mission for some representative science topics including: the reionization of the Universe (Sect.~\ref{ss:tau}); cosmic birefringence (Sect.~\ref{ss:eb}); the hot gas in the Universe probed using the \gls{sz} effect (Sect.~\ref{ss:sz}); spatially varying deviations from a perfect Planckian blackbody CMB spectrum (Sect.~\ref{ss:mu}); primordial magnetic fields (Sect.~\ref{ss:pmf}); tests using polarization of the so-called ``anomalies'' in the temperature data (Sect.~\ref{ss:anomaly}); and Galactic astrophysics (Sect.~\ref{ss:galaxy}). 

\subsection{Optical Depth, Reionization of the Universe and Neutrino Masses}
\label{ss:tau}

The hydrogen atoms in the intergalactic medium are fully ionized in the recent Universe ($z\lesssim 6$). We have multiple lines of evidence for this from the lack of saturated hydrogen absorption lines in the spectra of quasars and afterglows of $\gamma$-ray bursts~\cite{gunn/peterson:1965,fan/etal:2006,bolton/etal:2011,chornock/etal:2013,mcgreer/mesinger/dodorico:2015}. Given that the Universe became almost completely neutral after the epoch of hydrogen recombination (at $z\simeq 1100$), the Universe must have ``reionized'' during some intermediate epoch. However, we still have no clear description of the history of this \gls{eor}. 
In the current picture, early galaxies reionized hydrogen atoms progressively throughout the entire history between $z \simeq 12$ and $z \simeq 6$, while quasars took over to reionize helium atoms from $z \simeq 6$ to 2~\cite{barkana/loeb:2001}.

There are several ways to observe the EoR: the number counts of star-forming galaxies and quasars at $z>6$, which were presumably producing ionizing photons~\cite{robertson/etal:2015}; the shape of the luminosity functions of high-redshift galaxies in the \gls{uv} bands~\cite{Ishigaki_2018}; redshifted 21-cm lines from hydrogen atoms before the completion of reionization~\cite{furlanetto/oh/briggs:2006}; Doppler shifts of CMB photons by the bulk motion of ionized gas~\cite{park/etal:2013} (called the kinetic Sunyaev-Zeldovich effect~\cite{sunyaev/zeldovich:1980}); and finally, the polarization of the CMB produced by electrons scattering quadrupole temperature anisotropies in a reionized Universe~\cite{PlanckIntXLVII}, which will be probed by \lb.

\begin{figure}[htbp!]
 \centering\includegraphics[width=0.8\textwidth]{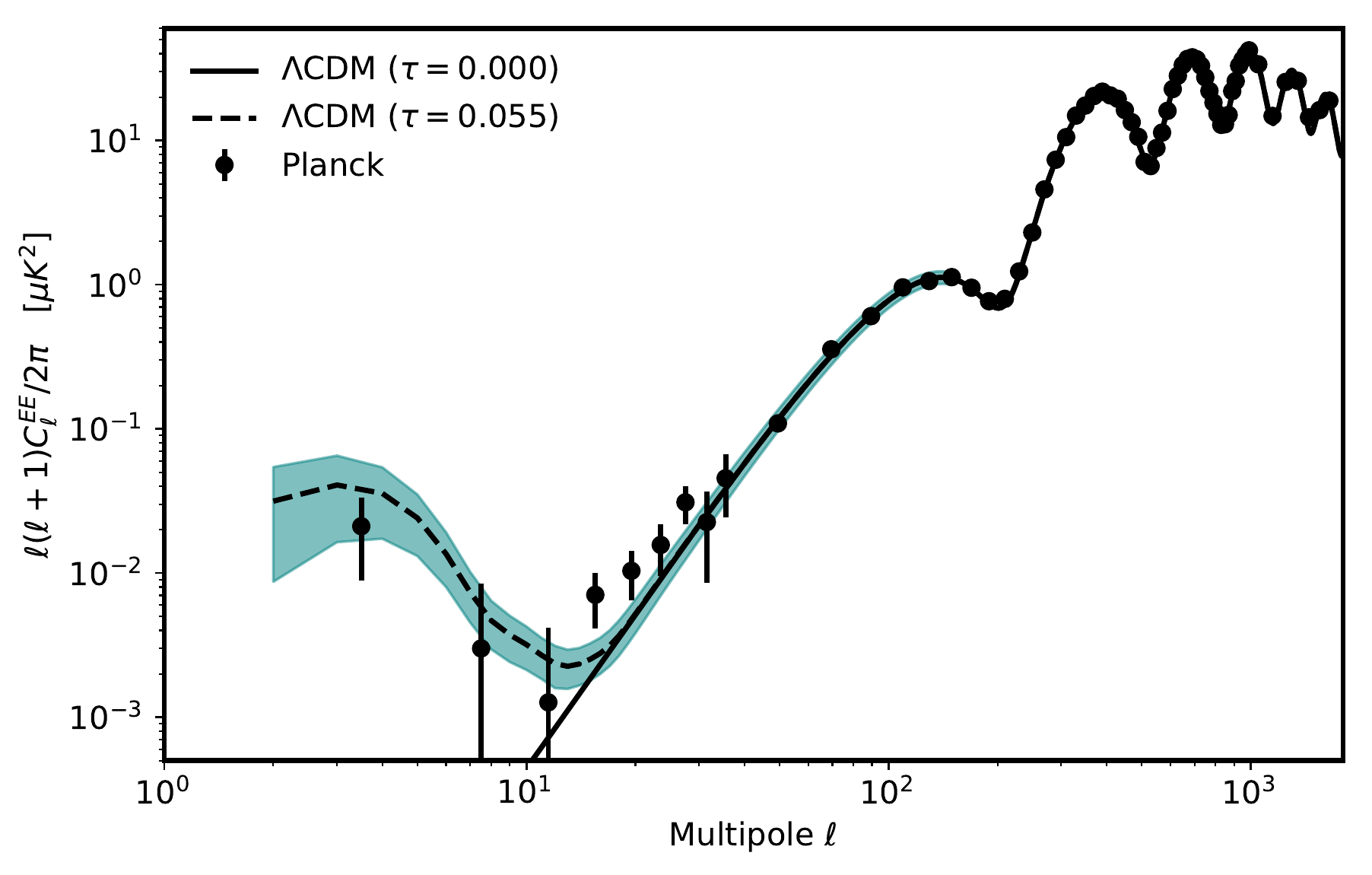}
\caption{\Planck\ $E$-mode power spectrum~\cite{PlanckIntLVII} with optical depths of $\tau=0.055$ and $\tau=0.000$ for constant $\exp(-2\tau)A_{\rm s}$. The cosmological parameters are taken from the best-fitting $\Lambda$CDM \Planck\ 2018 model. The green band shows the expected \lb\ uncertainties for each multipole in the range $\ell=2$--200 (the cosmic-variance-limited uncertainties with $f_{\rm sky}=0.7$), demonstrating significant improvements over the \Planck\ data.}
\label{fig:lowlee}
\end{figure}

Thomson scattering between the CMB photons and free electrons generates linear polarization from the quadrupole moment of the CMB radiation field at the scattering epoch. This occurs at recombination as well as during the EoR. Re-scattering of the CMB photons at reionization generates an additional polarization anisotropy at large angular scales, because the horizon size at this epoch subtends a much larger angular size \cite{zaldarriaga:1997}. The CMB is affected by the total column density of free electrons along each line of sight, parameterized by its Thomson-scattering optical depth, $\tau$.
The wavenumber of the fluctuations contributing to quadrupole temperature anisotropy, as seen by an electron at a redshift $z$, is given by $k\simeq 3/[r_{\rm L}-r(z)]$ where $r_{\rm L}=14$\,Gpc is the comoving distance from Earth to our last-scattering surface, and $r(z)$ is the comoving distance to redshift $z$. For example, a redshift of $z=7.7$ gives $r(7.7)=9.1$\,Gpc. We observe this wavenumber at a multipole of $\ell\lesssim k\,r(7.7)\simeq 6$, which corresponds to the so-called ``reionization bump'' in the polarization power spectra. The effect on the $E$ modes is shown in Fig.~\ref{fig:lowlee}. 

\begin{figure}[htbp!]
 \centering\includegraphics[width=0.8\textwidth]{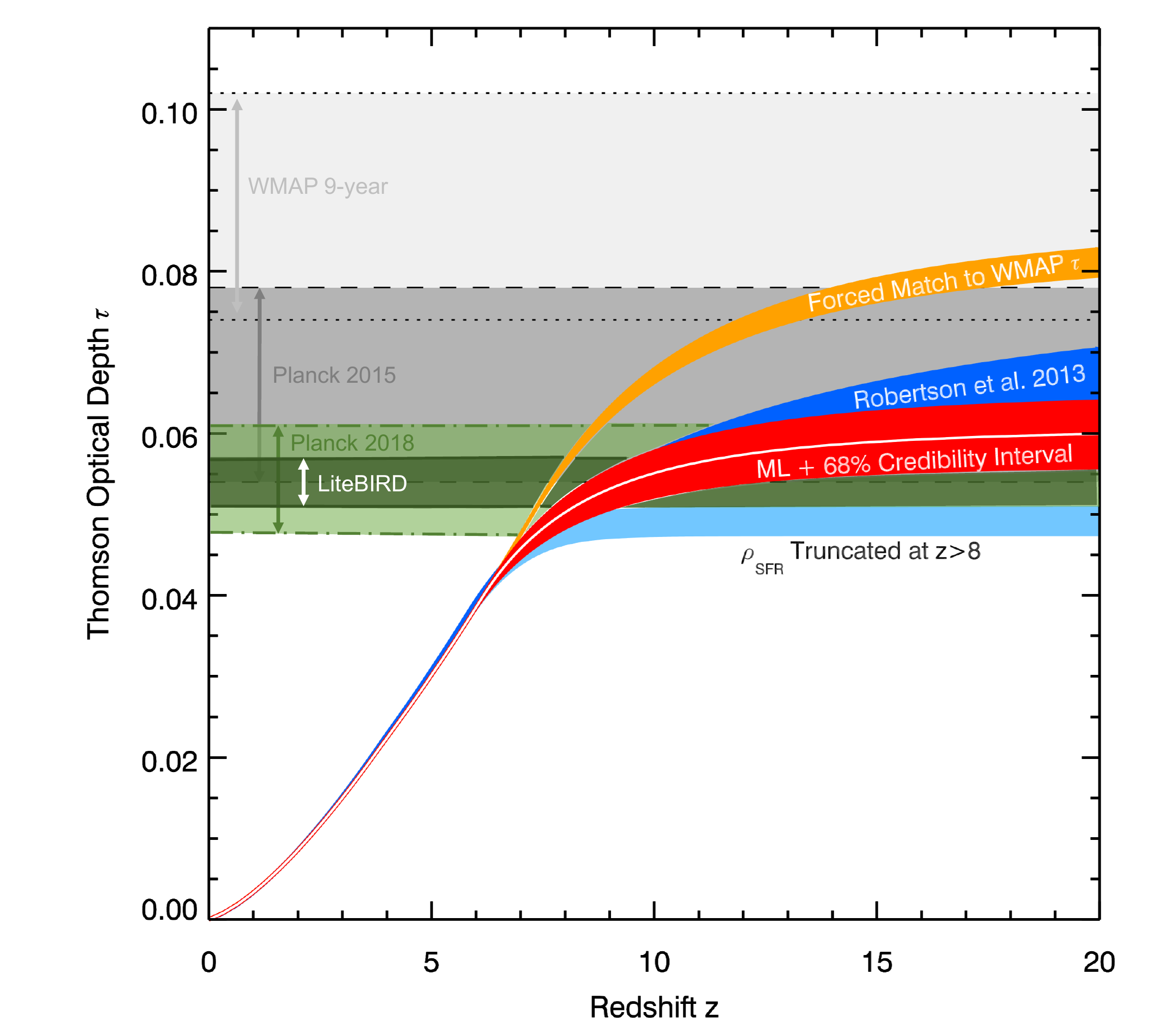}
\caption{Optical depths predicted from various models of the number counts of star-forming galaxies, as a function of the maximum redshift $z$. This figure is adapted from Ref.~\cite{robertson/etal:2015}, showing $\tau$ inferences from different star-formation rates: the best estimates from 2013 (blue); the updated 2015 \gls{ml} constraints (red); a model forced to reproduce the 9-year \WMAP\ $\tau$ constraints (orange); and a model with a truncated star-formation rate (light blue).  The green band shows the expected \LiteBIRD\ 68\,\% and 95\,\% \gls{cl} constraints on $\tau=0.054$. The other bands show the \WMAP\ and \Planck\ constraints.}
\label{fig:sfr}
\end{figure}

The amplitude of the reionization bump is proportional to $\tau^2A_{\rm s}$ where $A_{\rm s}$ is the amplitude of the scalar curvature power spectrum. Because scattering washes out small-scale power, which is proportional to $\exp(-2\tau)A_{\rm s}$, increasing $\tau$ (while fixing $A_{\rm s}$) enhances the reionization bump and suppresses the small-scale power. However, since the small-scale power has been measured precisely, the value of $\exp(-2\tau)A_{\rm s}$ is fixed; thus, for a given measured value of the high-$\ell$ power spectrum, the amplitude of the reionization bump scales as $\tau^2A_{\rm s}\propto \tau^2\exp(2\tau)$. We can use this to determine the value of $\tau$, which in turn provides an integrated constraint on the reionization history of the Universe because $\tau=\sigma_{\rm T}N_{\rm e}$, where the column density of electrons is given by $N_{\rm e}=c\int dt\,n_{\rm e}$ integrated from today to the beginning of reionization. This number can be compared with the expected number of electrons from ionization by star-forming galaxies and quasars (including X-ray emission from accretion disks around black holes); see Fig.~\ref{fig:sfr}.

\begin{figure}[htbp!]
\centering\includegraphics[width=0.8\textwidth]{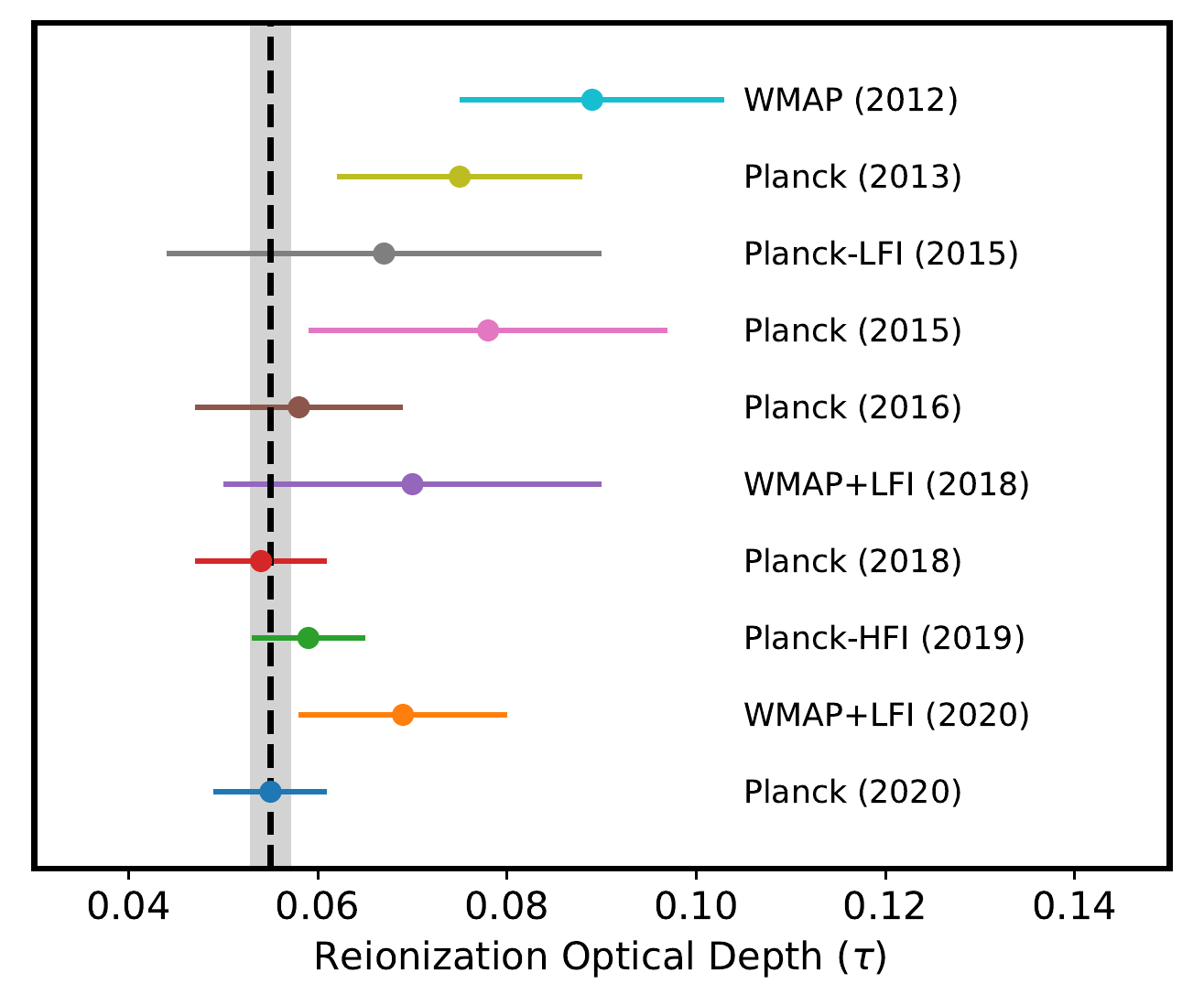}
\caption{
Measurements of $\tau$ based on the CMB power spectra:
\WMAP\ (2012)~\cite{hinshaw/etal:2013};
\Planck\ (2013)~\cite{Planck2013XV};
\Planck\ (2015)~\cite{Planck2015XIII};
\Planck\ (2016)~\cite{PlanckIntXLVII};
\WMAP+LFI (2018)~\cite{Weiland:2018kon};
\Planck\ (2018)~\cite{Planck2018VI};
\Planck-HFI (2019)~\cite{Pagano2020};
\WMAP+LFI (2020)~\cite{Natale:2020owc};
and \Planck\ (2020)~\cite{PlanckIntLVII}.
The gray band shows the \lb\ uncertainty forecast for $\tau$.}
\label{fig:taus}
\end{figure}

Accurate measurements of $\tau$ through the CMB are challenging because of the foreground contamination and  instrumental systematic uncertainties~\cite{Lattanzi:2016dzq}. These are most problematic on large angular scales, where the bulk of the information on $\tau$ is constrained. 
This difficulty is illustrated by the evolution of the constraint over time from the first \WMAP\ release in 2003 ($\tau = 0.17 \pm 0.06$~\cite{Spergel2003}) using the $TE$ cross-correlation, up to the latest \Planck\ collaboration results ($\tau = 0.051 \pm 0.006$~\cite{PlanckIntLVII}, obtained from the ``NPIPE'' reprocessing of the \Planck\ legacy data) using polarized $EE$ measurements (see Fig.~\ref{fig:taus} for a complication of $\tau$ estimates, where time increases from top to bottom).

\lb\ will provide a cosmic-variance-limited determination of $\tau$, i.e., the sampling variance in the limit of zero instrumental noise. This is 
the smallest possible error bar, limited only by the fraction of sky ($f_{\rm sky}$) available for the cosmological analysis.  While ground-based experiments will provide an independent measurement of $\tau$ from gravitational lensing, with an uncertainty similar to the current \Planck\ estimates over the coming years, a new, improved large-scale measurement will come from \lb. Assuming a fiducial value of $\tau = 0.054$ and an available sky fraction of $f_{\rm sky} = 0.7$, the expected $68\,\%$ C.L. uncertainty on $\tau$ is $0.002$~\cite{DiValentino:2016foa}, which is 3 times better than today's tightest bounds from \Planck~\cite{Pagano2020,PlanckIntLVII,DeBelsunce:2021xcp}. Not only is this a significant improvement over current measurements, but it will also be the definitive and most accurate measurement of $\tau$ from the CMB. 

Beyond a cosmic-variance-limited measurement of $\tau$, the $E$-mode measurements carried out by \lb\ will constrain the precise reionization history~\cite{zaldarriaga/etal:2008}. In particular, the ``dip'' in the $E$-mode power spectrum at $\ell\simeq 10$--20 in Fig.~\ref{fig:lowlee} can distinguish between instantaneous reionization at a redshift of $z_{\rm reion}$ and more physical models of reionization~\cite{Barkana:2006ep,Furlanetto:2004nt}, including those with a reionization history extending to longer durations and earlier onsets at $z>z_{\rm reion}$. Although the \Planck\ measurements have reconciled the value of $\tau$ from CMB polarization on large angular scales with the reionization process fueled by star-forming galaxies \cite{Planck2015XIII,robertson/etal:2015,PlanckIntXLVII,Hazra:2019wdn}, there is still considerable uncertainty on its history~\cite{Miranda:2016trf,obied/etal:2018} and degeneracies with other cosmological parameters~\cite{Paoletti:2020ndu}. 
The \lb\ data will provide a significant improvement in constraints on the reionization history models, breaking the remaining degeneracies~\cite{Hazra:2018eib,Watts:2019uvq} and taking the joint CMB constraints with complementary astrophysical probes~\cite{Hazra:2019wdn} to a higher level of precision. For example, in the case of homogeneous reionization models, \lb\ will reduce the uncertainty in the duration of reionization, $\Delta z_{\rm reion}$, for an asymmetric reionization history by approximately 35\,\%~\cite{Hazra:2018eib} with respect to the most recent \Planck\ measurements~\cite{Paoletti:2020ndu}.
The \lb\ data will also allow for testing the inhomogeneity of the reionization process and will contribute to characterizing the patchiness of reionization~\cite{Roy:2018gcv}.

\begin{figure}[htbp!]
\centering
\includegraphics[width=0.8\textwidth]{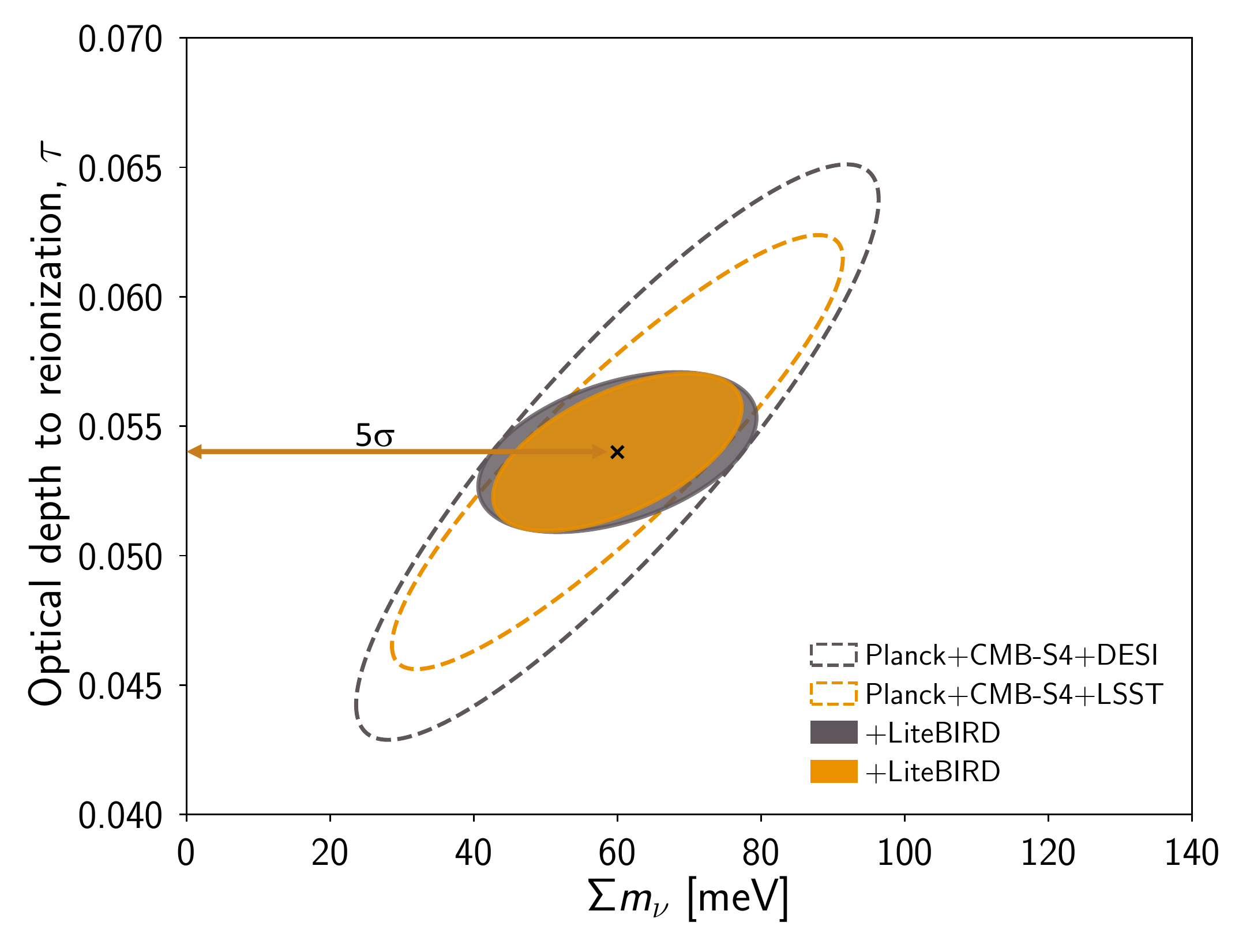}
\caption{Two-dimensional marginalized contour levels at 68\,\% C.L. for $\tau$ and the sum of the neutrino masses as measured by future combinations of CMB and large-scale structure data, for example including \gls{bao} from DESI or galaxy lensing and clustering from LSST, adapted from Ref.~\cite{Calabrese2017}.
The contours are centered on the fiducial values $\tau=0.054$ and $\sum m_\nu = 60$\,meV, as indicated by the cross. A cosmic-variance-limited measurement of $\tau$ is reached with \lb\ (i.e., $\sigma(\tau) = 0.002$). This $\tau$ limit will enable a better neutrino mass measurement, reaching a $5\,\sigma$ detection when combined with DESI or LSST.}
\label{fig:tau_mnu}
\end{figure}

Beyond reionization, $\tau$ impacts the important cosmological and particle physics science topic of determining the sum of the neutrino masses, $\sum m_\nu$~\cite{DiValentino:2015sam,allison/etal:2015,Giusarma:2016phn,Archidiacono:2016lnv,Abazajian:2016yjj,boyle/komatsu:2018}. Such a measurement would allow us to establish the absolute scale of the neutrino masses, and also possibly to distinguish the inverted neutrino mass ordering (i.e., two heavy, one light) from the normal ordering (one heavy, two light)~\cite{Jimenez:2010ev,Hall:2012kg,Hannestad:2016fog,Gerbino:2016ehw,Gerbino:2017ubx}. Massive neutrinos slow down structure formation~\cite{lesgourgues/pastor:2006} and consequently, we can measure the neutrino mass by comparing the amplitude of fluctuations in the low-redshift Universe with that at the last-scattering surface (i.e., $A_{\rm s}$). However, we cannot determine $A_{\rm s}$ unless we know $\tau$. This means that cosmological searches for the neutrino masses can be limited by the inability to measure $\tau$ with sufficient precision.
For example, the exact knowledge of $\tau$ would reduce the uncertainty in the neutrino mass estimates from galaxy redshift surveys by more than a factor of 2 relative to that with \Planck's $\tau$ measurement (see the appendix of Ref.~\cite{boyle/komatsu:2018}). Thus, an improved $\tau$ determination from \lb\ will play a major role in measuring the neutrino mass.

When combined with measurements of the amplitude of density fluctuations at low redshifts, such as those coming from the CMB lensing data and galaxy survey observations of large-scale structure (see below), and possibly with constraints on the expansion history, a cosmic-variance-limited measurement of $\tau$ from \lb\ will enable a statistically significant 
detection of the neutrino mass, even for the minimum value $\sum m_\nu \simeq 60\,{\rm meV}$ allowed by flavor-oscillation experiments.

The information on the amplitude of fluctuations in the low-redshift Universe might be provided by the CMB lensing data from observations of the small-scale anisotropies (e.g., those of the future Simons Array~\cite{westbrook/etal:2018},
SO~\cite{Ade:2018sbj,Abitbol:2019nhf}, and CMB-S4~\cite{Abazajian:2019tiv,Abazajian:2019eic} experiments), and/or by data from large-scale structure surveys tracing the matter distribution (e.g., galaxy surveys with the \gls{desi}~\cite{desi:2016} and the \gls{lsst}~\cite{lsst:2009}). 
Distance measurements, such as those coming from \gls{bao} data, will further improve the constraints by adding information about the cosmic expansion history~\cite{Gerbino:2017ubx}. To give a specific example, combining a cosmic-variance-limited
measurement of $\tau$ ($\sigma(\tau) = 0.002$) with observations of the small-scale CMB anisotropies from CMB-S4 and either 
BAO data from the DESI galaxy survey~\cite{desi:2016} or galaxy lensing/clustering data from the LSST survey of the Vera Rubin Observatory~\cite{lsst:2009}, will in both cases yield $\sigma(\sum m_\nu) =  12\,\mathrm{meV}$~\cite{Calabrese2017}. This will result in a detection of the neutrino
mass at the $5\,\sigma$ level, for the minimum value of $60\,{\rm meV}$, or larger. Fig.~\ref{fig:tau_mnu} shows the constraining power of these data combinations in the $(\tau,\,\sum m_\nu)$ plane, highlighting that a cosmic-variance-limited measurement of $\tau$ from \lb\ will be necessary for reaching a statistically significant detection of the neutrino mass from cosmological data.

To complete the picture on the neutrino sector, the expected uncertainty on the effective number of relativistic species, $N_{\rm eff}$, from \lb\ alone is $\sigma(N_\mathrm{eff}) \simeq 0.15$~\cite{errard/etal:2016}, which is of the same order of magnitude as that obtained by \Planck~\cite{Planck2018VI}. On top of that, it would give an independent measurement, and an important cross-check, since any systematic effects would be different from those relevant for the high-$\ell$ \Planck\ measurements, e.g., the modeling of small-scale foregrounds (see for example Ref.~\cite{henrot-versille/etal:2019} for a discussion of how these might affect the estimate of $N_{\rm eff}$ provided by \Planck). 

The relevance of measuring $N_{\rm eff}$ goes beyond neutrino physics, since this parameter traces the presence of any radiation-like components at the time of CMB decoupling. Such candidates include thermal axions~\cite{Turner:1986tb,Chang:1993gm} and sterile neutrinos~\cite{Boyarsky:2009ix}, as well as other light species that were in thermal equilibrium with the cosmological plasma at some early point in the history of the Universe~\cite{Baumann:2016wac}. Measuring $N_{\rm eff}$ also allows us to probe the thermal history of the Universe with nonstandard particle contents, e.g., the
decay of nonrelativistic massive particles~\cite{Kawasaki:1999na}. A more accurate value of $N_{\rm eff}$ would also help constrain the energy density of the stochastic gravitational wave background~\cite{Smith:2006nka,henrot-versille/etal:2015,Pagano:2015hma} because gravitational waves behave as radiation.

\subsection{Cosmic Birefringence}
\label{ss:eb}

There exists a distinct possibility that either (or both) of dark matter and dark energy is a pseudoscalar field, $\varphi$ that changes sign under parity, i.e., inversion of spatial coordinates \cite{marsh:2016,ferreira:2021}.
If this is the case then this field could leave a unique signature in the polarization of the CMB. More generally,
if the new physics that generated the initial scalar and tensor fluctuations does not violate parity symmetry, and the CMB photons do not experience any parity-violating processes as they propagate for 13.8 billion years from the surface of last scattering to us, then any parity-violating correlation functions such as the temperature-$B$-mode ($TB$) correlation and the $EB$ correlation must vanish. This is because under spatial inversion, the spherical harmonic coefficients transform as
\begin{equation}
a_{\ell m}^T\to +(-1)^\ell a_{\ell m}^T\,,\quad
a_{\ell m}^E\to +(-1)^\ell a_{\ell m}^E\,,\quad
a_{\ell m}^B\to -(-1)^\ell a_{\ell m}^B\,.
\end{equation}
Consequently, the expectation value of any parity-odd observable, such as the $TB$ and $EB$ correlations, must vanish if the underlying physics is parity conserving. On the other hand, if the underlying physics violates parity, these correlation functions can and generically do have non-vanishing expectation values~\cite{lue/wang/kamionkowski:1999,feng/etal:2005,liu/lee/ng:2006,saito/ichiki/taruya:2007,contaldi/magueijo/smolin:2008}.

In this section we describe a physical effect known as ``cosmic birefringence''~\cite{carroll/field/jackiw:1990,Carroll:1991zs,harari/sikivie:1992,carroll:1998}. See Ref.~\cite{Komatsu:2022nvu} for a review. The basic idea is that a new parity-violating coupling of $\varphi$ to the electromagnetic tensor makes the phase velocities of right- and left-handed polarization modes of photons different.  This results in rotation of the direction of linear polarization as the CMB photons propagate through space. In other words, space filled with this $\varphi$ field behaves as if it were a birefringent medium.
A homogeneous $\varphi$ field coupled to the electromagnetic field via the Chern-Simons term (i.e., the Lagrangian contains a term $-\frac14 g_{\varphi\gamma}\varphi F_{\mu\nu}\tilde{F}^{\mu\nu}$), rotates the linear polarization direction uniformly over the sky by an angle $\beta=\frac12g_{\varphi\gamma}\int_{t_{\rm L}}^{t_0} dt~\dot\varphi$ \cite{carroll/field/jackiw:1990,Carroll:1991zs,harari/sikivie:1992,carroll:1998}, converting $E$-mode into $B$-mode polarization. The rotation angle is defined such that $\beta>0$ corresponds to clockwise rotation on the sky in right-handed coordinates with the $z$-axis taken in the direction of observer's lines of sight. Here, $g_{\varphi\gamma}$ is the coupling constant of the interaction of $\varphi$ and photons, $F_{\mu\nu}$ and $\tilde{F}_{\mu\nu}$ are the electromagnetic tensor and its dual, and $t_{\rm L}$ and $t_0$ are the time of last scattering and the present time, respectively. 
We would therefore observe a $B$-mode polarization signal even if there were no $B$-mode polarization initially~\cite{lue/wang/kamionkowski:1999}. The $\varphi$ field can be dark matter, dark energy, or both~\cite{carroll:1998,Finelli:2008jv,marsh:2016,ferreira:2021,fujita/etal:2021}; thus, discovery of such a signal would be a major breakthrough in cosmology and fundamental physics. 

When $\beta$ does not vanish, the observed $E$- and $B$-mode spherical harmonics coefficients of the CMB, $a_{\ell m}^{E,{\rm CMB,o}}$ and $a_{\ell m}^{B,{\rm CMB,o}}$, are related to those at the surface of last scattering as
\begin{eqnarray}
a_{\ell m}^{E,{\rm CMB,o}}&=& a_{\ell m}^{E,{\rm CMB}}\cos(2\beta)-a_{\ell m}^{B,{\rm CMB}}\sin(2\beta)\,,\\
a_{\ell m}^{B,{\rm CMB,o}}&=& a_{\ell m}^{E,{\rm CMB}}\sin(2\beta)+a_{\ell m}^{B,{\rm CMB}}\cos(2\beta)\,.
\end{eqnarray}
However, this effect is degenerate with an instrumental miscalibration of polarization angles~\cite{wu/etal:2009,komatsu/etal:2011}. This means that, in the absence of any other information, we cannot tell whether the polarization angle of the CMB is rotated by the new physics (i.e., $\beta$) or the polarization angle of detectors is rotated with respect to the sky coordinates by a miscalibration angle $\alpha$. As a result, in the absence of any other information, we can only determine the sum of the two angles, $\alpha+\beta$, which explains why the previous determinations of $\beta$ were spread over a wide range beyond the quoted statistical uncertainties (see Refs.~\cite{PlanckIntXLIX,kaufman/keating/johnson:2016,minami/komatsu:2020} for summaries).

Since the miscalibration angle $\alpha$ generates a spurious $B$-mode power spectrum $\sin^2(2\alpha)C_\ell^{EE}$, we must calibrate the angles with a precision sufficient to achieve the full science requirements of \lb. As described in Sect.~\ref{sss:anglecal}, there are no strong polarized astrophysical sources on the sky with precisely known polarization angles; thus, the calibration must rely on measurements on the ground, which currently limit the accuracy of calibration to a bit better than $1^\circ$. This accuracy is much poorer than the requirement, which is of order $0.05^\circ$, demanding substantial improvements in the ground-calibration methodology.
As a result, the option of using the $TB$ and $EB$ correlations (assumed to vanish) to calibrate the instrumental polarization angles~\cite{keating/shimon/yadav:2012} was considered in Sect.~\ref{sss:anglecal}. This method is based on accurate knowledge of the cosmological $TE$ and $EE$ power spectra from the scalar mode, and it is straightforward to fit to the $TB$ and $EB$ power spectra to solve for $\alpha$ with arcminute precision.
However, not only does this ``self-calibration'' procedure eliminate \lb's sensitivity to the new physics of uniform rotation caused by $\varphi$, but it can also bias the angle calibration if non-zero $\beta$ is present in the data.

A new method \cite{Minami:2019ruj,Minami:2020xfg,MinamiKomatsu:2020} has been developed to mitigate this issue. The basic idea is to use the polarized Galactic foreground emission as the angle calibrator. Cosmic birefringence is a cumulative effect, whose magnitude is proportional to the path length of photons. Since the origin of the Galactic emission is much closer to Earth than the CMB, we can ignore $\beta$ in the Galactic emission (labeled `fg' below). On the other hand, the miscalibration angle $\alpha$ affects both the CMB and the Galactic emission. We thus write

\begin{align}
a_{\ell m}^{E,{\rm o}}&=
a_{\ell m}^{E,{\rm fg}}\cos(2\alpha)-a_{\ell m}^{B,{\rm fg}}\sin(2\alpha)+
a_{\ell m}^{E,{\rm CMB}}\cos(2\alpha+2\beta)-a_{\ell m}^{B,{\rm CMB}}\sin(2\alpha+2\beta)\,,\\
a_{\ell m}^{B,{\rm o}}&= a_{\ell m}^{E,{\rm fg}}\sin(2\alpha)+a_{\ell m}^{B,{\rm fg}}\cos(2\alpha)+
a_{\ell m}^{E,{\rm CMB}}\sin(2\alpha+2\beta)+a_{\ell m}^{B,{\rm CMB}}\cos(2\alpha+2\beta)\,,
\end{align}
which gives the ensemble average of the observed $EB$ power spectrum as \cite{Minami:2019ruj}
\begin{eqnarray}
    \langle C_\ell^{EB,{\rm o}}\rangle
    &= &\frac{\tan(4\alpha)}2\left(\langle C_\ell^{EE,{\rm o}}\rangle-\langle C_\ell^{BB,{\rm o}}\rangle\right)
    + \frac{\sin(4\beta)}{2\cos(4\alpha)}\left(\langle C_\ell^{EE,{\rm CMB}}\rangle-\langle C_\ell^{BB,{\rm CMB}}\rangle\right)\cr
    & &\qquad\qquad\qquad+\ \frac1{\cos(4\alpha)}\langle C_\ell^{EB,{\rm fg}}\rangle +\frac{\cos(4\beta)}{\cos(4\alpha)}\langle C_\ell^{EB,{\rm CMB}}\rangle\,.
\end{eqnarray}
This formula allows us to determine $\alpha$ and $\beta$ simultaneously. Note that we do not need to know the intrinsic foreground $EE$ and $BB$ power spectra, but only those of the CMB as well as the intrinsic $EB$ spectra of the CMB and the foreground emission. The last term on the right-hand side is the intrinsic $EB$ power spectrum of the CMB at the surface of last scattering. This can be produced by parity-violating gravitational waves from, e.g., the gravitational Chern-Simons term~\cite{lue/wang/kamionkowski:1999,saito/ichiki/taruya:2007,contaldi/magueijo/smolin:2008} and Abelian~\cite{sorbo:2011} or non-Abelian~\cite{thorne/etal:2018} gauge fields during inflation (see Sect.~\ref{ss:cmbpol_beyondpower}). Because these effects arise from gravitational waves, the shape of the $EB$ power spectrum is completely different from that generated by cosmic birefringence; thus, we can simultaneously fit $\beta$ and  $C_\ell^{EB,{\rm CMB}}$ without losing sensitivity~\cite{Gluscevic:2010vv,thorne/etal:2018}.

The intrinsic $EB$ correlation of the foreground emission, $C_\ell^{EB,{\rm fg}}$, requires a careful treatment. The worst case scenario is that $C_\ell^{EB,{\rm fg}}$ is proportional to $C_\ell^{EE,{\rm fg}}-C_\ell^{BB,{\rm fg}}$. In this case, fitting the observed $EB$ power spectrum gives $\beta-\gamma$ rather than $\beta$, where $\gamma$ is an additional effective angle from the intrinsic foreground $EB$ correlation~\cite{Minami:2019ruj}. Since the \Planck\ data show a positive $TE$ and $TB$ signal for thermal dust emission~\cite{PlanckIntXLIX,Planck2018XI}, we expect a positive $EB$ as well~\cite{huffenberger/rotti/collins:2020}, although this is yet to be found and its strength may depend on the Galactic mask \cite{clark/etal:2021,Diego-Palazuelos:2022dsq}.

\begin{figure}[htbp!]
 \centering\includegraphics[width=0.8\textwidth]{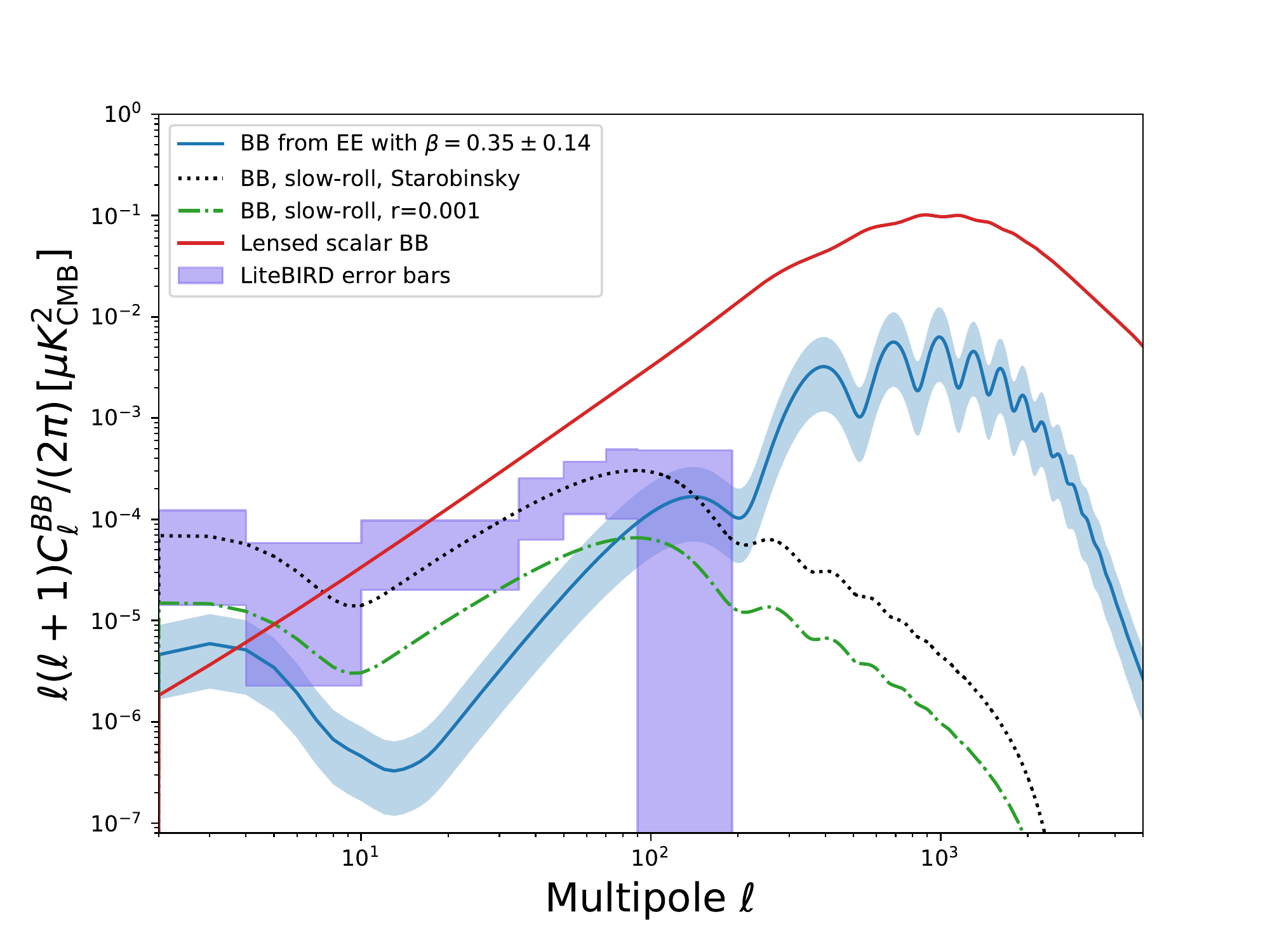}
\caption{The $B$-mode power spectrum from cosmic birefringence with $\beta=0.35^\circ\pm 0.14^\circ$ (the blue shaded area). The dotted and dot-dashed lines show the $B$-mode power spectra from primordial gravitational waves with $r=0.00461$ (Starobinsky's model~\cite{starobinsky:1980}) and 0.001, respectively, both of which are detectable by \lb\ by design. The red line shows the $B$-mode power spectrum from the gravitational lensing effect of the CMB. The purple shaded regions show the expected constraints from \lb, derived in Sect.~\ref{ss:forecasts_fg_cleaning}.}
\label{fig:cleb}
\end{figure}

Recently, the value $\beta-\gamma=0.35^\circ\pm 0.14^\circ$ (68\%~CL) has been obtained from the \Planck\ 2018 data \cite{minami/komatsu:2020}. If $\gamma>0$, as implied by the positive $TE$ and $TB$ measurements of the polarized dust emission, the significance of $\beta$ would increase further. This has been confirmed by the latest analysis using the \Planck\ Public Data Release 4 (PR4) ``\texttt{NPIPE}'' reprocessing~\cite{PlanckIntLVII}.
Thanks to the lower noise and better-characterized systematics, an improved measurement, $\beta-\gamma=0.30^\circ\pm 0.11^\circ$ (68\%~CL), is obtained for nearly full-sky data~\cite{Diego-Palazuelos:2022dsq}. Accounting for $C_\ell^{EB,{\rm fg}}$ using the $TE$ and $TB$ measurements of the polarized dust emission and 
a physical model of Ref.~\cite{clark/etal:2021}, $\beta=0.36^\circ$ is found with the same statistical uncertainty. The impact of the known systematics of the \Planck\ data on $\beta$ is found to be negligible compared to the statistical uncertainty.
It is too early to tell if this tentative hint for $\beta$ is due to cosmic birefringence, some unknown systematics in the \Planck\ data, some unexpected property of the intrinsic foreground $EB$ correlation, or a combination of them. In any case, \lb\ can play a decisive role in understanding the origin of any such signal. With \lb, we can reduce the statistical uncertainty to below $0.1^\circ$~\cite{MinamiKomatsu:2020}, potentially increasing the statistical significance from 3$\,\sigma$ to the level of discovery. 

If cosmic birefringence exists at the level of $\beta=0.35^\circ$, it will produce $B$-mode polarization of $\sin^2(2\beta)C_\ell^{EE}$, which is within reach of \lb's sensitivity (Fig.~\ref{fig:cleb}). Therefore, \lb\ can test for the presence of the cosmic birefringence signal, not only via the $EB$ correlation, but also the $B$-mode power spectrum.

So far we have only discussed birefringence that is the same in every direction, but in principle
we can also look for an {\it anisotropy\/} in $\beta$~\cite{li/zhang:2008,pospelov/ritz/constantinos:2009}. This can arise from inhomogeneity of dark matter (and possibly also dark energy) made of a pseudoscalar field coupled to the electromagnetic tensor via the Chern-Simons term. 
This introduces correlations between $T$ and $B$ and between $E$ and $B$ at different multipoles, (i.e., $\langle a_{\ell m}^Ta_{\ell'm'}^E\rangle$ and $\langle a_{\ell m}^Ea_{\ell'm'}^B\rangle$), in a manner similar to the gravitational lensing effect of the CMB~\cite{kamionkowski:2009,Yadav:2009eb,namikawa:2017}. Such a property makes it possible to create a map of $\beta$ in each \lb\ pixel. This map will be useful not only for probing new parity-violating physics, but also for characterizing instrumental systematics (i.e., anisotropy in $\alpha$). So far there is no evidence for the anisotropic signal of $\beta$~\cite{contreras/etal:2017,namikawa/etal:2020,bianchini/etal:2020,gruppuso/etal:2020}, but \lb\ can certainly improve upon the constraints, particularly on the largest angular scales.

\subsection{Mapping the Hot Gas in the Universe}
\label{ss:sz}

\begin{figure}[htbp!]
 \centering\includegraphics[width=0.9\textwidth]{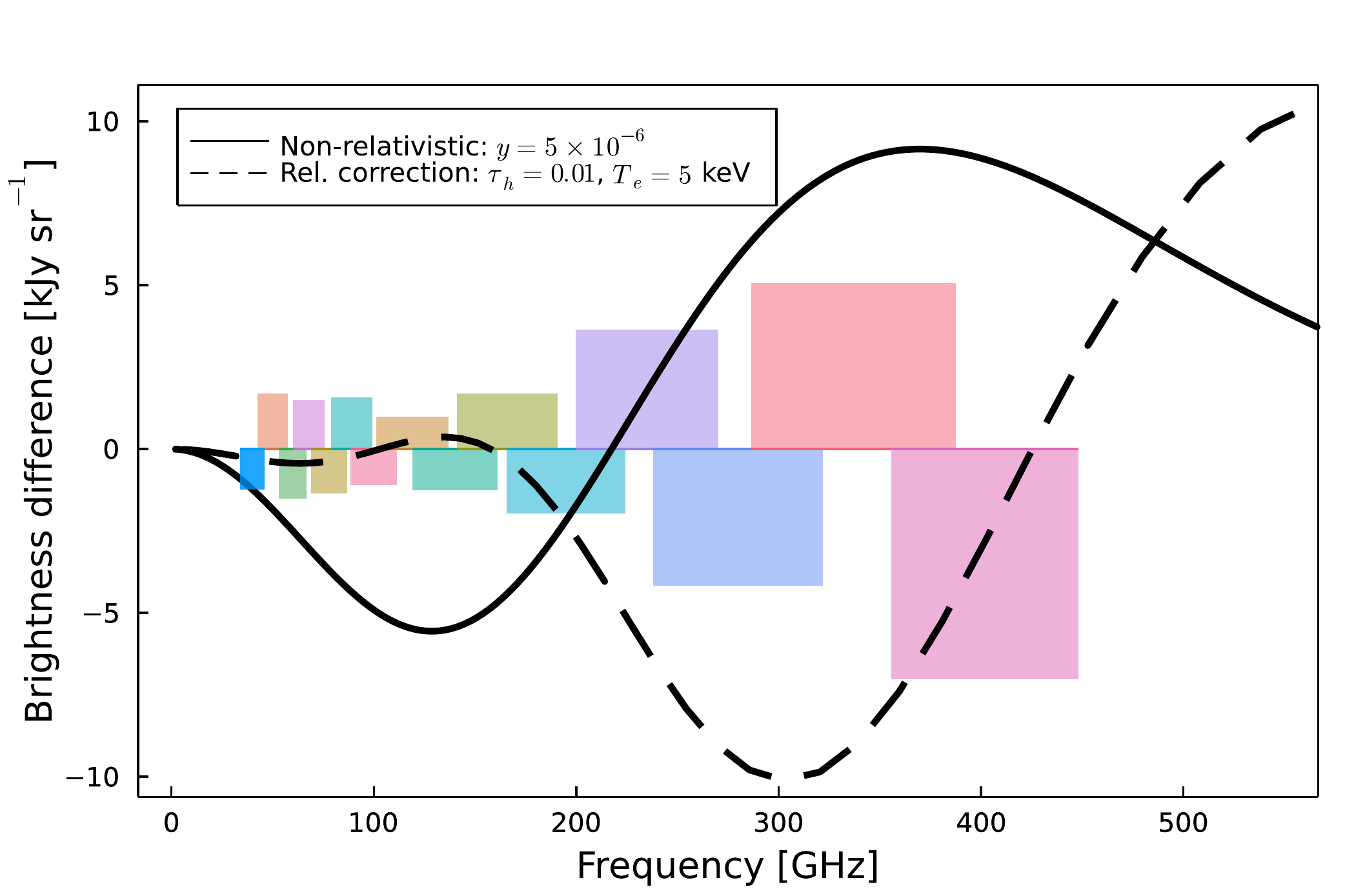}
 \caption{Spectrum of the thermal SZ effect.
 The solid line shows the example spectrum in units of kJy\,sr$^{-1}$ in the non-relativistic limit, $k_{\rm B}T_{\rm e}/m_{\rm e} c^2\ll 1$, with $y=\tau_{\rm h} k_{\rm B}T_{\rm e}/m_{\rm e}c^2= 5\times 10^{-6}$. 
 The dashed line shows the spectrum of the relativistic correction for $\tau_{\rm h}=0.01$ and $k_{\rm B}T_{\rm e}=5\,{\rm keV}$ (which gives $y=9.8\times 10^{-5}$) relative to the non-relativistic SZ effect with $y=9.5\times 10^{-5}$, calculated by the \texttt{SZpack} package~\cite{chluba/etal:2012,chluba/etal:2013}, i.e., the difference between the full and non-relativistic spectra in units of kJy\,sr$^{-1}$. We show the difference between the relativistic and non-relativistic SZ spectra with two different values of $y$ to highlight the genuine effect of the relativistic correction, which cannot be absorbed by changing the value of $y$.
 The colored bars show the sensitivity of the 15 partially overlapping bands of the \lb\ detectors. For clarity we show half of the bands as positive and the other half as negative quantities, but only their absolute values are meaningful.}
\label{fig:sz}
\end{figure}

Electrons in the hot ionized gas transfer their energy to CMB photons by inverse Compton scattering, leading to a characteristic distortion of the blackbody spectrum of the CMB (see Fig.~\ref{fig:sz}). This phenomenon is known as the thermal SZ effect~\cite{zeldovich/sunyaev:1969,sunyaev/zeldovich:1972} and has been routinely detected in the directions of galaxy clusters~\cite{carlstrom/holder/reese:2002,kitayama/etal:2014,bleem/etal:2015,Planck2015XXVII,hilton/etal:2021}. The amplitude of the thermal SZ effect is characterized by the so-called ``Compton $y$ parameter,'' which is given by $y=\tau_{\rm h} k_{\rm B}T_{\rm e}/m_{\rm e} c^2$ where $\tau_{\rm h}$ is the
optical depth of hot gas (which should be distinguished from the optical depth of reionization, $\tau$, discussed in Sect.~\ref{ss:tau}) and $T_{\rm e}$ and $m_{\rm e}$ are the electron temperature and mass, respectively.

\begin{figure}[htbp!]
 \centering\includegraphics[width=\textwidth]{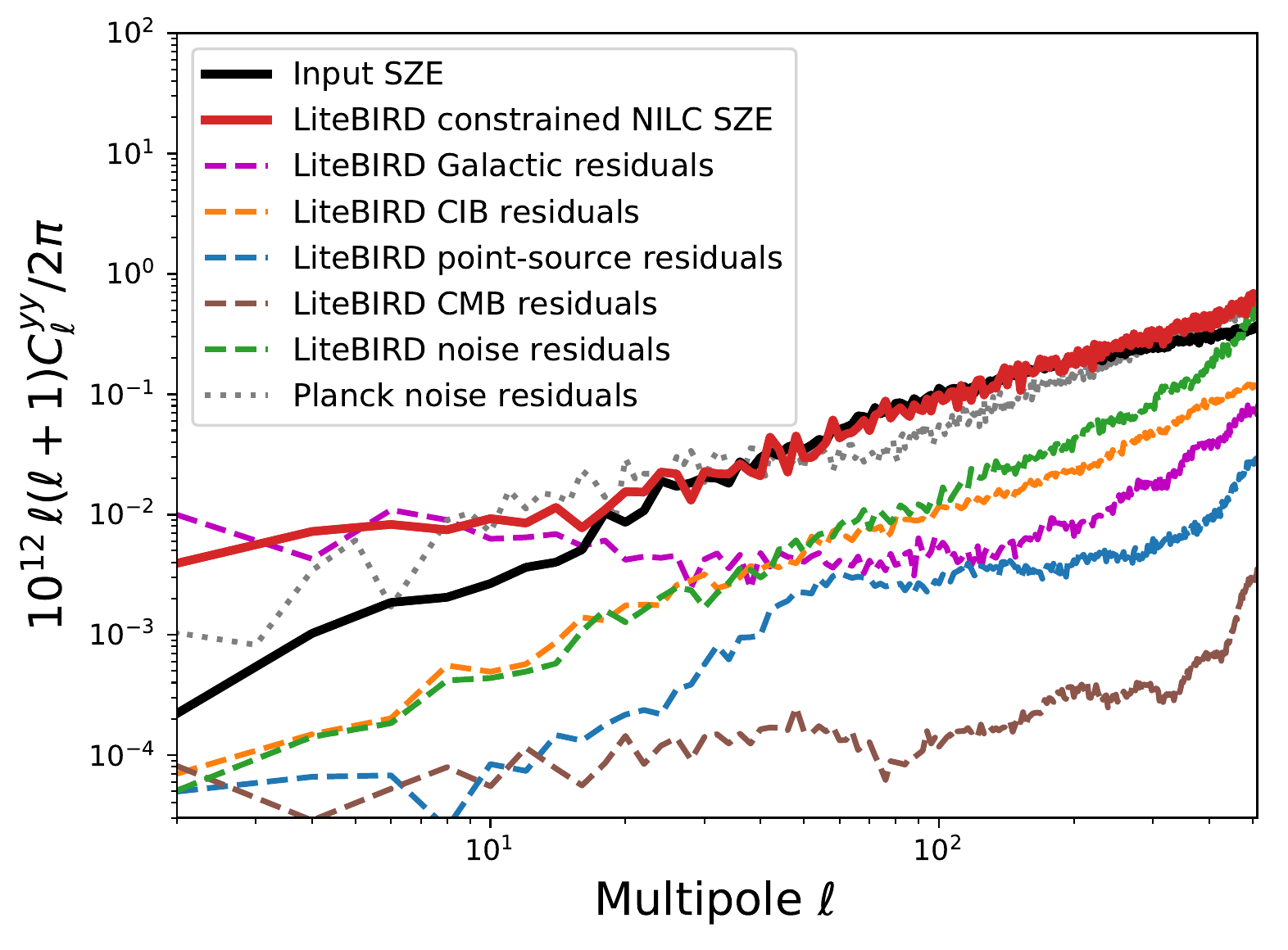}
 \caption{Reconstructed power spectrum of the thermal SZ effect from a simulation of \lb\ (red line), compared with the input one (black line). Both agree well except at $\ell<20$, which still shows residuals from the Galactic emission; however, such low multipoles suffer from large non-Gaussian cosmic-variance error bars in any case. The noise power spectrum of \lb\ (green dashed line) is much lower than that of \Planck\ (gray dotted line), showing the substantially improved sensitivity and fidelity of the thermal SZ map from \lb.}
\label{fig:clsz}
\end{figure}

Using the so-called constrained internal linear combination (cILC) method~\cite{remazeilles/etal:2011,remazeilles/etal:2011b,remazeilles/etal:2013,hurier/etal:2013}, we can reconstruct an all-sky map of the thermal SZ signal and its angular power spectrum with minimal residual foreground contamination~\cite{Planck2015XXII,madhavacheril/etal:2019,bleem/etal:2022}. Applying the same component-separation algorithm that was used on the \Planck\ data to the \lb\ simulations, we find that, while the \Planck\ SZ map still contains contamination of various foreground sources due to the limited number of frequency bands, \lb\ can faithfully reconstruct the SZ map at $\ell>20$. Figure~\ref{fig:clsz} shows the power spectrum of the reconstructed SZ map from this simulation.
We also find that, while the reconstruction noise in the \Planck\ SZ map is comparable to the signal itself, \lb\ will reduce the noise by an order of magnitude relative to the \Planck\ SZ map.

Exploiting the 15 \lb\ frequency bands will yield a much improved, high-fidelity SZ map over the full sky at $\ell\le 200$, essentially free of contamination. This full sky map will show in projection all hot gas in the Universe and will have a lasting impact on astrophysics as legacy data from \lb. An important application of this full sky thermal SZ map will be to cross-correlate it with three-dimensional catalogues of galaxies with the known redshifts, as demonstrated in Refs.~\cite{vikram/lidz/jain:2017,makiya/ando/komatsu:2018,pandey/etal:2019,koukoufilippas/etal:2019,chiang/etal:2020,yan/etal:2021}. This cross-correlation allows us to perform {\it tomography\/} of the hot gas in the Universe as a function of cosmic time, which can test theories of structure formation~\cite{chiang/etal:2021}.
This full-sky thermal SZ map can also be used to search for the \gls{whim} by stacking at the positions of known galaxy pairs~\cite{degraaff/etal:2019}.

The high-fidelity SZ map is also useful for studying an inhomogeneous reionization process via the cross-correlation of the SZ map and fluctuations of the CMB optical depth, $\delta\tau_{\rm h}$~\cite{namikawa/etal:2021}. The SZ effect can be generated by hot electrons during the reionization epoch and the optical depth 
has fluctuations due to the spatial variations of the electron density during the reionization epoch. Thus, an inhomogeneous reionization process leads to a correlation between the SZ map and $\delta\tau_{\rm h}$. This cross-correlation is much less affected by late-time galaxy cluster contributions than the SZ auto-power spectrum. In addition, this cross-correlation can constrain the temperature of ionized bubbles, while the kinetic SZ and $\delta\tau_{\rm h}$ cannot. The cross-correlation signal would be detectable by comparing the \lb\ SZ map with a $\delta\tau_{\rm h}$ map reconstructed from CMB-S4 for an interesting parameter space for the ionized bubbles, e.g., a characteristic size of $5\,$Mpc and a temperature of $5\times 10^4\,$K.

The solid line in Fig.~\ref{fig:sz} shows the SZ spectrum in the non-relativistic limit (where $k_{\rm B}T_{\rm e}/m_{\rm e}c^2\ll 1$) for an example value of $y=5\times 10^{-6}$.
The shape is universal and depends only on the mean CMB temperature; however, small relativistic corrections to this shape exist~\cite{wright:1979,fabbri:1981,rephaeli:1995} and are proportional to $k_{\rm B}T_{\rm e}/m_{\rm e} c^2$ at leading order~\cite{challinor/lasenby:1998,itoh/kohyama/nozawa:1998,sazonov/sunyaev:1998}.
Detecting this relativistic correction averaged over a full sky SZ map~\cite{remazeilles/etal:2019,remazeilles/chluba:2020} can yield the mean gas temperature of the Universe, providing an ``integral constraint'' on physics of the intergalactic medium~\cite{hill/etal:2015} and stringent and robust constraints on the energy feedback from supernovae and \gls{agn}.  This complements information that can be obtained about the relationship between the WHIM and halos from correlating the SZ map with gravitational lensing and other tracers of large-scale structure~\cite{Ma2015,Lim2018}.

The dashed line in Fig.~\ref{fig:sz} shows the relativistic correction for $\tau_{\rm h}=0.01$ and $k_{\rm B}T_{\rm e}=5$\,keV, 
relative to the non-relativistic SZ effect with $y=9.5\times 10^{-5}$. Note that $\tau_{\rm h}=0.01$ and $k_{\rm B}T_{\rm e}=5$\,keV correspond to $y=9.8\times 10^{-5}$. One may wonder why we do not take the difference between the relativistic and non-relativistic SZ spectra with the same $y=9.8\times 10^{-5}$. This would give a larger difference at low frequencies, giving the impression that we can detect the relativistic correction without the high-frequency data; however, this is false, since the difference at low frequencies can be compensated by slightly changing $y$, as we have done here~\cite{chluba/etal:2021}, whereas the distortion at high frequencies is genuine and cannot be compensated by changing $y$.
Observing above 300\,GHz with \lb\ will thus give a great advantage compared to lower frequency ground-based surveys for detecting the relativistic correction, since the electron-temperature dependence of the relativistic correction manifests itself mostly at high frequencies, as shown in Fig.~\ref{fig:sz}.
Finally, the \lb\ sensitivity and high-frequency coverage could be used to apply the method proposed in Ref.~\cite{rephaeli:1980} to improve constraints on the monopole of the $y$-type distortion of the CMB spectrum via the spectrum of the SZ effect.

\subsection{Anisotropic CMB Spectral Distortions}
\label{ss:mu}

Although \lb\ is not sensitive to the spatially uniform (i.e., monopole) component to the distortion of the Planckian spectrum of the CMB, it is very sensitive to any {\it spatially-varying\/} component of the spectral distortion. The thermal SZ effect, described in Sect.~\ref{ss:sz}, is one example of such a spectral distortion and can be used to map the distribution of hot gas in the Universe.

Another example of anisotropies with spectral dependence different from that of the CMB is Rayleigh scattering of the CMB photons~\cite{takahara/sasaki:1991,yu/spergel:2001,lewis:2013}. The CMB decoupled from electrons at a redshift of $z\simeq 1090$, leading to the usual ``surface of last scattering.'' Around this epoch the electrons combined with protons to form neutral hydrogen atoms, which also scatter photons via Rayleigh scattering. The cross-section for Rayleigh scattering has a characteristic frequency dependence of $\sigma_{\rm R}(\nu)\propto \nu^4$~\cite{rayleigh_lord_1881_1431155}, which leaves a frequency-dependent imprint in temperature and polarization anisotropies of the CMB. No monopole spectral distortion is produced by Rayleigh scattering.

The cross-section of Rayleigh scattering is given more explicitly by
\begin{equation}
    \sigma_{\rm R}(\nu)=\sigma_{\rm T}\left(\tilde\nu^4+\frac{638}{243}\tilde\nu^6+\dots\right)\,,
\end{equation}
where $\sigma_{\rm T}$ is the Thomson cross-section, $\tilde\nu\equiv \nu/(\sqrt{8/9}cR_\infty)\simeq \nu/(3.1\times 10^6\,{\rm GHz})$, and $R_\infty$ is the Rydberg constant. This additional scattering produces a frequency-dependent shift of the peak of the visibility function to $z<1090$, which modifies the Silk damping process and shifts the locations of the acoustic peaks in the temperature and $E$-mode polarization power spectra~\cite{yu/spergel:2001,lewis:2013}. 

The Rayleigh-scattering signal can be extracted from the data by cross-correlating the frequency-independent Thomson-scattering component and the frequency-dependent ($\nu^4$) Rayleigh-scattering component~\cite{lewis:2013,beringue/etal:2021}. Since the frequency dependence of these two components is known precisely, we can separate them by expliciting rejecting the other component using the cILC method~\cite{remazeilles/etal:2011,remazeilles/etal:2011b,remazeilles/etal:2013,hurier/etal:2013}. 
The remaining largest contamination is then the \gls{cib}. While the expected Rayleigh scattering signal is small in the frequencies observed by \lb, the superb sensitivity of \lb\ allows for detection of the signal at a statistical significance of 25$\,\sigma$~\cite{beringue/etal:2021}, which is due primarily to the temperature data. The first detection of this signal would not only be a significant achievement in cosmology (because this is a firm prediction of the standard model of cosmology that is yet to be confirmed), but could also improve determination of the cosmological parameters such as $N_{\rm eff}$ and $\sum m_\nu$~\cite{beringue/etal:2021}, as well as primordial non-Gaussianity~\cite{coulton/beringue/meerburg:2021}. 

It may also be possible to probe epochs earlier than CMB last-scattering using different spectral information.
Let us review the well-known physics of the early Universe (see Refs.~\cite{tashiro:2014,chluba/hamann/patil:2015} for reviews). When the temperature of the Universe exceeded $5\times10^6\,$K (or $z>2\times 10^6$ in terms of redshift), double-Compton scattering (which changes the total photon number) was efficient in relaxing the photon spectrum to a Planck spectrum with a vanishing chemical potential, even if some extra energy was injected into the plasma~\cite{burigana/danese/dezotti:1991,hu/silk:1993}. If the energy was injected in the range $5\times 10^4<z<2\times 10^6$, double-Compton scattering would no longer be fast enough to relax the photon spectrum to a blackbody with no chemical potential. Compton scattering is still efficient for redistributing the photon energies so as to maintain an equilibrium distribution (i.e., a Bose-Einstein distribution with {\it non-zero\/} chemical potential, also known as the ``$\mu$ distortion''). If the spectral distortion was caused by an energy injection after $z=5\times 10^4$, it would {\it not\/} relax to an equilibrium distribution because the energy exchange due to Compton scattering would be inefficient, resulting in, for example, a permanent spectral distortion such as the SZ effect described in Sect.~\ref{ss:sz}.

While there exist many theoretical possibilities for energy injection in the early Universe before $z=5\times 10^4$~\cite{tashiro:2014,chluba/hamann/patil:2015}, one mechanism present in the standard model of cosmology is energy injection from the dissipation of sound waves~\cite{silk:1968,sunyaev/zeldovich:1970}. Because this spectral distortion occurs at second order in the perturbation, the energy injection rate due to the dissipation of sound waves is proportional to the sound wave amplitude squared. This property makes it possible to constrain the small-scale power of fluctuations from the chemical potential~\cite{daly:1991,barrow/coles:1991,hu/scott/silk:1994}.

This phenomenon offers the third example of anisotropic spectral distortions of the CMB. While this signal is isotropic in the sky if the fluctuations obey Gaussian statistics, a specific type of non-Gaussian fluctuations, called ``squeezed non-Gaussianity,'' can be produced by certain physical mechanisms during inflation, such as multi-field effects and non-Bunch-Davies vacuum initial conditions, and would generate spectral distortions characterized by a {\it spatially varying\/} chemical potential~\cite{pajer/zaldarriaga:2012,ganc/komatsu:2012}. \lb\ can look for this signal by cross-correlating the measured temperature anisotropies with a map of the chemical potential reconstructed from \lb's multi-frequency data, since this cross-correlation measures a three-point function (temperature fluctuation on large scales correlated with the squared amplitude of sound waves on small scales). Although the multi-field effect of inflation yields only a small signal-to-noise ratio for \lb, given the constraints on this type of non-Gaussianity from the \Planck\ data~\cite{Planck2015XVII,Planck2018IX}, non-vacuum effects can yield a large signal-to-noise ratio, offering a powerful test of the physics of inflation at its onset~\cite{ganc/komatsu:2012}. Additional sources of anisotropic spectral distortion arise when the background spacetime is itself anisotropic~\cite{shiraishi/liguori/bartolo:2015,shiraishi/bartolo/liguori:2016,ota:2019}.

There may be additional spectral-spatial variations in the CMB that can be probed by \lb, including those that affect polarization. As an example,
light axion-like particles are converted into photons in the presence of magnetic fields, generating anisotropic distortions in the CMB spectrum~\cite{mukherjee/khatri/wandelt:2018}. In particular, ``resonant conversion'' of axions into photons by the Galactic magnetic field yields {\it polarized\/} spectral distortions of the CMB with the spatial distribution of the signal tracking the Galactic magnetic field. \lb\ can search for signals of this type.

\subsection{Primordial Magnetic Fields}
\label{ss:pmf}

Magnetic fields are ubiquitous in the Universe at all scales from planetary systems to clusters of galaxies, with hints of their presence also in filaments of the large-scale structure~\cite{govoni/etal:2019,vernstrom/etal:2021} as well as in intergalactic voids~\cite{ando/kusenko:2010,essey/etal:2011,dolag/etal:2011}. The dynamo is a popular explanation for the origin of the cosmic magnetic fields, but it requires an initial field. In other words, a dynamo does not generate the field from nothing, but amplifies the existing ``seed'' field (see Ref.~\cite{kulsrud/zweibel:2008} for a review). Even the simplest adiabatic compression of magnetic fields frozen to plasma could explain the observed strength of $\mu$G in galaxy clusters if the pre-compression strength were 0.1\,nG ($=10^{-10}$\,G) in intergalactic space~\cite{grasso/rubinstein:2001}. Lower bounds on the intergalactic magnetic fields of $\gtrsim 10^{-16}$--$10^{-18}$\,G have been inferred from the lack of extended $\gamma$-ray halos towards blazars in the GeV energy bands (see Ref.~\cite{fermi-lat:2018} and references therein). If they exist in intergalactic space and within filaments of the large-scale structure, where do they come from (see Refs.~\cite{widrow:2002,kandus/kunze/tsagas:2011,widrow/etal:2012,durrer/neronov:2013,subramanian:2016} for reviews)?

Cosmic inflation may provide this origin, i.e., the \glspl{pmf}. In the standard model of elementary particles and fields, massless gauge fields (such as electromagnetism) are conformally coupled to gravity; thus, no metric excitation of spin-1 fields is possible in an expanding Universe, unlike for scalar~\cite{mukhanov/chibisov:1981} and tensor~\cite{grishchuk:1975,starobinsky:1979} perturbations. In other words, we must break conformal invariance of the massless gauge fields to generate the PMFs during inflation~\cite{turner/widrow:1988,ratra:1992}. Although theoretical challenges remain~\cite{demozzi/mukhanov/rubinstein:2009,barnaby/namba/peloso:2012}, searching for the PMFs has profound implications for our understanding of physics of inflation beyond the origin of the scalar and tensor perturbations that we have already discussed extensively in this paper. 

The power spectrum of the stochastic PMFs, $P_B(k)$, is defined by
\begin{equation}
    \langle B_i({\bf k})B^*_j({\bf q})\rangle = (2\pi)^3\delta_{\rm D}^{(3)}({\bf k}-{\bf q})P_B(k)\left(\delta_{ij}-\frac{k_ik_j}{k^2}\right)\,,
\end{equation}
where $\delta^{(3)}_{\rm D}({\bf k})$ is the Dirac delta function and $\delta_{ij}$ is the Kronecker delta. A power-law power spectrum, $P_B(k)=A_Bk^{n_B}$, is characterized by its spectral index, $n_B$, and amplitude, $A_B$. Spectral indices generated during inflation are usually negative~\cite{turner/widrow:1988}, e.g., a nearly scale-invariant spectrum of $n_B=-2.9$ (the exact scale invariance $n_B=-3$ gives a diverging energy density of the magnetic field), whereas they are positive if generated by causal processes such as the electroweak phase transition~\cite{vachaspati:1991,joyce/shaposhnikov:1997,ahonen/enqvist:1998}, i.e., an even integer of $n_B\ge 2$~\cite{durrer/caprini:2003}. 

In the literature, the magnetic field strength is usually quoted as the value extrapolated to the present epoch assuming that the fields are frozen to the plasma. Moreover, we often quote the field value smoothed over a length scale $\lambda$, calculated from the smoothed energy density of the magnetic field:
\begin{equation}
    \rho_{B,\lambda} = \int_0^\infty \frac{d^3k}{(2\pi)^3} P_B(k)e^{-k^2\lambda^2}
    = A_B\frac{\Gamma[(n_B+3)/2]}{4\pi^2\lambda^{n_B+3}}\,.
\end{equation}
Here we choose $\lambda=1\,{\rm Mpc}$ and compute the field strength, $B_\lambda$, from the energy density $\rho_{B,\lambda}=B_\lambda^2/2$, in natural units. We then convert the units and quote the value of $B_\lambda$ in units of nG. 
The convenient quantity is the ratio of $\rho_{B,\lambda}$ to the CMB photon energy density, $\rho_\gamma$:
\begin{equation}
    \frac{\rho_{B,\lambda}}{\rho_\gamma}=9.53\times 10^{-8}\left(\frac{B_\lambda}{1~{\rm nG}}\right)^2\,,
\end{equation}
for the present-day CMB temperature of 2.7255~K.

The sensitive CMB polarization measurements provided by \lb\ can constrain the PMFs with unprecedented precision, using three of the distinct ways that the PMFs affect the CMB.

\begin{enumerate}
\item A stochastic background of PMFs provides the stress-energy source, which gravitationally induces scalar, vector, and tensor perturbations in the CMB, hence the $TT$, $EE$, and $BB$ power spectra (see Refs.~\cite{grasso/rubinstein:2001,yamazaki/kajino/ichiki:2012} for reviews). In addition, helical magnetic fields can produce parity-violating correlations such as the $TB$ and $EB$ power spectra~\cite{pogosian/vachaspati/winitzki:2002,caprini/durrer/kahniashvili:2004,ballardini/finelli/paoletti:2015}; Ref.~\cite{Planck2015XIX} describes the constraints on helical fields from the \Planck\ data.
Because the temperature and polarization anisotropies from the magnetic-field stress-energy tensor are highly non-Gaussian, higher-order correlations are a powerful probe of the PMF~\cite{shiraishi:2013,shiraishi/sekiguchi:2014,trivedi/subramanian/seshadri:2014}.
\item Dissipation of the fields heats gas in intergalactic space, altering the thermal history of the Universe~\cite{sethi/subramanian:2005}. This affects the CMB primarily via the optical depth of electron scattering (Sect.~\ref{ss:tau})~\cite{kunze/komatsu:2015} and the isotropic (monopole) distortion of the blackbody spectrum of the CMB~\cite{jedamzik/katalinic/olinto:2000,kunze/komatsu:2014}. 
\item Faraday rotation induces frequency-dependent ($\propto\nu^{-2}$) {\it inhomogeneous\/} rotation of polarization angles of the CMB over the sky~\cite{kosowsky/loeb:1996,kosowsky/etal:2005,yadav/pogosian/vachaspati:2012}, which has been constrained by \Planck~\cite{Planck2015XIX,gruppuso/etal:2020} and also by ground-based experiments~\cite{birefringence-polarbear:2015,birefringence-bicep:2017}. 
\end{enumerate}
\lb\ is sensitive to all of the above effects, except for the monopole spectral distortion of the CMB.

\begin{figure}
\centering\includegraphics[width=0.8\textwidth]{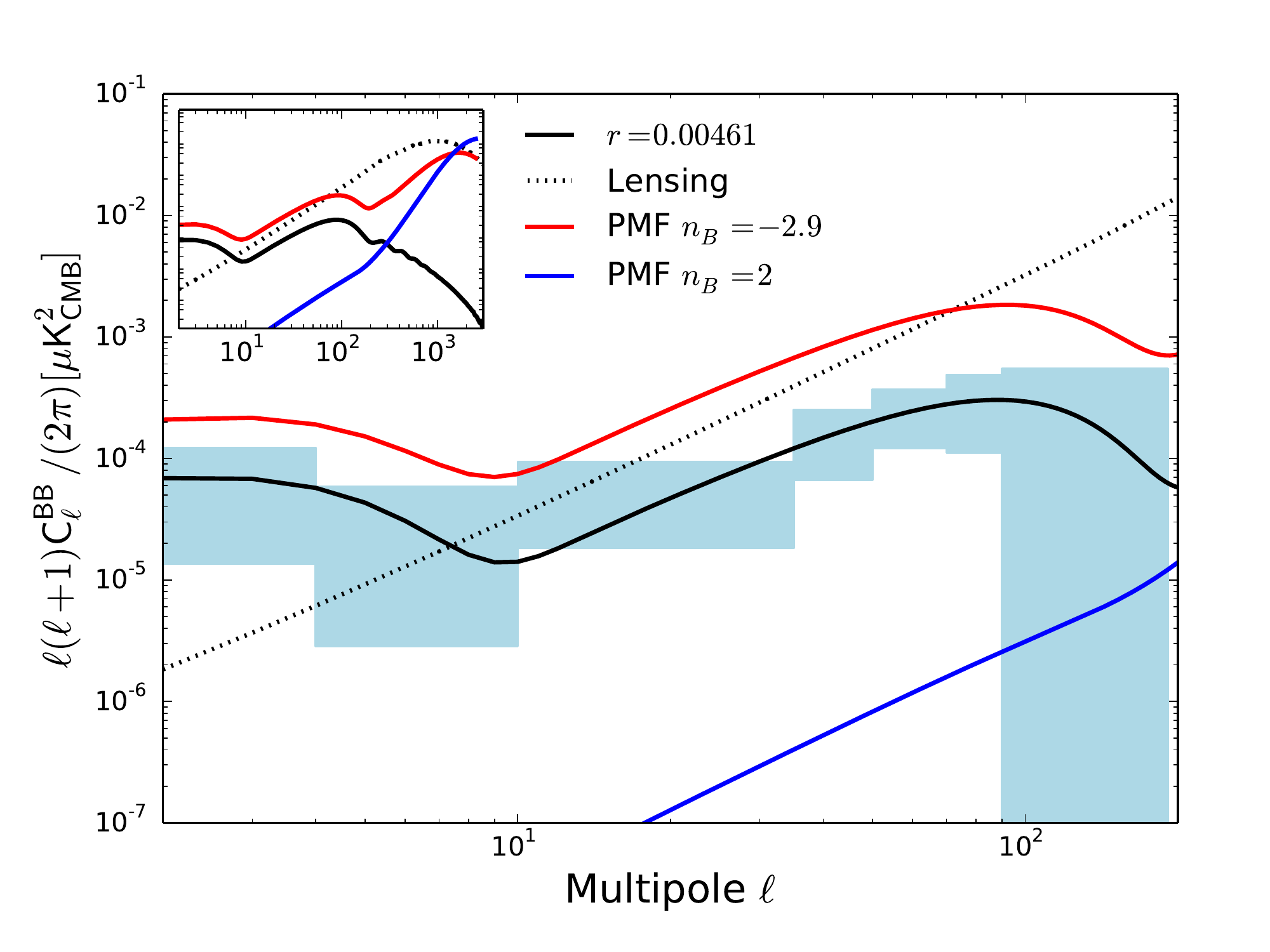}
\caption{$B$-mode power spectra from the gravitational effects of an inflationary 
PMF with $n_B=-2.9$ (red) and 
a causally generated PMF with $n_B=2$ (blue), in comparison to those of the primordial tensor perturbation of the Starobinsky model of inflation with $r=0.0461$ (solid black)~\cite{starobinsky:1980} and gravitational lensing (dotted). The amplitudes of the PMF curves are chosen to match the current limits,
$B(1\,{\rm Mpc})=2$\,nG and 0.003\,nG for $n_B=-2.9$ and 2, respectively. The blue shaded regions show the expected constraints from \lb, derived in Sect.~\ref{ss:forecasts_fg_cleaning}. The inset shows the power spectra up to higher multipoles.}
\label{fig:pmf}
\end{figure}

For the first effect, the study in Ref.~\cite{paoletti/finelli:2019} shows how \lb's  $B$-mode power spectrum can improve upon the constraints on the amplitude of a stochastic background of PMFs, and how in combination with future ground-based experiments it can reach the nG-level field strength for a nearly scale-invariant spectrum ($n_B=-2.9$), and sub-pG level for causally generated fields ($n_B=2$). However, since the gravitational contribution of PMFs to the $B$-mode power spectrum can be degenerate with those of primordial gravitational waves and lensing in the multipole range probed by \lb~\cite[][see Fig.~\ref{fig:pmf}]{renzi/etal:2018,paoletti/finelli:2019}, we can use higher-order correlations (i.e., non-Gaussianity) to further constrain PMFs.
Using \lb's $B$-mode information, the sensitivity to the magnetized tensor bispectrum is improved by more than two orders of magnitude compared to the \Planck\ results, yielding a 1-nG-level constraint~\cite{shiraishi:2019}.
For the second effect, \lb's cosmic-variance-limited measurement of $E$ modes at low and intermediate multipoles (Sect.~\ref{ss:tau}) will significantly tighten the constraints on the post-recombination heating due to PMFs~\cite{kunze/komatsu:2015,chluba/etal:2015,paoletti/etal:2019}, going beyond the nG-level constraints. 
For the third effect, \lb\ can target Faraday rotation with its unique frequency dependence, $\nu^{-2}$, which can be used to cross-check the PMF constraints from the $B$- and $E$-mode spectra. In particular, using the Faraday rotation angle power spectrum, \lb\ can improve upon the limit on the amplitude of a nearly scale-invariant PMF by more than an order of magnitude, reaching the nG threshold, provided that the contamination of Faraday rotation from our Galaxy is properly modeled~\cite{pogosian/etal:2019}. 

While simultaneous analyses of all of these effects have not been performed in the literature, it is possible that the combination of all information can push the limit down to the 0.1-nG level robustly for a power spectrum with power-law index $n_B=-2.9$. A comprehensive analysis of the forecasts for \lb\ including all these effects is left for future work.

\subsection{Elucidating Spatial Anomalies with Polarization}
\label{ss:anomaly}

A 6-parameter cosmological model seems to explain the observed
structure of the CMB in remarkable detail~\cite{Planck2013XVI,Planck2015XIII,Planck2018VI}. Nevertheless,
there exist some features in the data that exert a mild tension
against such a model, potentially providing hints of new physics to be
explored.
Below, we summarize the most important of these spatial ``anomalies,'' found at
modest levels of statistical significance (i.e., $2$--$3\,\sigma$) in
the \WMAP\ and \planck\ temperature data~\cite{Bennett2010,Planck2013XXII,Planck2015XX,Planck2018X}.

\begin{enumerate}
\item \textit{Low-$\ell $ power deficit (low variance)}.
A simple statistical measure of the data is afforded by the variance of the
signal as a function of angular scale. Analysis of low
resolution maps reveals a lack of variance when compared to simulations
based on the best-fit cosmological model. The map-based variance is
dominated by contributions from large-angular scales on the sky,
whereas the cosmological parameter fits are insensitive to these modes
and mostly determined by scales corresponding to $\ell>50$. The dearth
of large-angular-scale power in the \planck\ power spectrum for
$\ell<30$ results in an apparently anomalous variance, with  a
$p$-value (the probability that simulations yield a lower value than
the data) of order 1\,\%.

\item \textit{Lack of correlation on large angular scales.}
A lack of structure in the angular two-point correlation function is
observed for angular separations
larger than $60^{\circ}$, with the observed values lying close to zero
between $\simeq\, 60^{\circ}$ and $\simeq\, 170^{\circ}$. This is captured
by the use of a posteriori statistic, $S_{1/2}$, as proposed by the
\WMAP\ team and given by the integral of the squared correlation
function for angular separations larger than $60^{\circ}$. A
corresponding $p$-value of less than 1\,\% was determined from the 2018
\planck\ data.

\item \textit{Alignments of low multipole moments.}
Detections of the alignment of the quadrupole and octupole
moments of the CMB temperature distribution have been found since
\WMAP's first release. This is somewhat surprising, given that the temperature
anisotropies are expected to have random phases in the standard
cosmological model, implying that the multipole moments should be
uncorrelated. The analysis in Ref.~\cite{Planck2013XXII} finds that the
quadrupole and octupole orientations are aligned to
within about $10^{\circ}$, with a $p$-value lower than 1\,\%.

\item \textit{Hemispherical asymmetry.}
The standard cosmological model predicts that the same power spectrum
should be measured in different patches of the sky, except for
variations connected with sample variance. However, \WMAP\ and \planck\
data show evidence for a hemispherical asymmetry (or dipolar
modulation) of power in a particular direction. For the \planck\ data in particular, several
methods to test for such asymmetry have been applied and compared
\cite{Planck2015XX,Planck2018X}. These are sensitive to either
amplitude, directionality, or both, although they do differ in terms
of their weighting of power on different scales. Nevertheless, the results
are all consistent with a modulation of power of around 7\,\% between
two hemispheres defined by the preferred direction $(l,b) =
(209^{\circ},-15^{\circ})$, extending over scales to
$\ell_\mathrm{max}\simeq 60$ with a significance approaching
$3\,\sigma$.  Interestingly, one such test, based on the anomalous clustering of
directions within bands of multipoles, suggests that this asymmetry holds even to relatively small
scales. 

\item \textit{Parity asymmetry.}
To test whether the CMB is symmetric with respect to reflections about
the origin, ${\hat{n}} \rightarrow -{\hat{n}}$, the CMB
anisotropy field can be divided into symmetric and anti-symmetric
functions with even and odd parity, corresponding to spherical
harmonic modes with even and odd $\ell$ values, respectively. On the
largest angular scales, the Universe should be parity neutral, yet an
odd-parity preference has been
established from analysis of the \WMAP\ and \planck\ data sets.
By computing the ratio between the sum of even
and sum of odd modes up to a given $\ell_\mathrm{max}$, it was found
that the significance varies with the maximum multipole chosen, with
$p$-values of about 1\,\% for $\ell_\mathrm{max}= 20$--30 \cite{Planck2018X}.

\item \textit{Cold spot.}
A particularly large cold region was originally discovered in the \WMAP\ first-year data
from the study of the kurtosis of \gls{smhw}
coefficients over a range of angular scales. It corresponds to an
anomalous temperature feature in these coefficients on angular scales
of $\simeq 10^\circ$, with the structure centered at Galactic
coordinates $(l, b) = (210^\circ, -57^\circ)$.
Less than 1\,\% of simulations based on the standard $\Lambda$CDM
cosmological model yield a kurtosis at least as large as that seen in
the data~\cite{Planck2015XX}, with some dependence on a~posteriori choices.
Novel theories have been invoked to explain the cold spot,
including the gravitational effect produced by a collapsing cosmic
texture.
\end{enumerate}

A conservative explanation for these anomalies, given their
claimed levels of significance, is that they are mild statistical excursions whose significance is overestimated due to the application of a posteriori
statistics. However, whether this is the case, or,
alternatively, that they reflect real physical properties of the
Universe cannot be elucidated further using the temperature
anisotropies, which are already cosmic-variance limited. Instead, new
observations are needed that independently probe the fluctuations that
source the temperature field. Maps of the CMB polarization provide
exactly such information.

The obvious premise is that much of the progress expected from
\litebird\ in this area will stem from working with $E$-mode data
(though statistical tests of the $B$-mode maps should also prove
very valuable near the cosmic-variance limit).  Indeed, \litebird\ should
almost double the statistical information concerning these anomalies.
Of course, given the modest significance of the temperature anomalies,
high significance detections in polarization will still prove challenging; however, only an experiment like \litebird\ has the possibility of providing statistically independent information on the largest scales. 

The first comprehensive search for anomalies in polarization on large
angular scales was presented in Ref.~\cite{Planck2018X}. No definitive
evidence was found in the polarization data for anomalous features
corresponding to those observed in the temperature
data. Nevertheless, several tests related to dipolar modulation showed
hints of asymmetry on scales up to $\ell_{\rm
  max}\,{\simeq}\,250$. More specifically, a variance asymmetry
estimator found an alignment between the preferred directions of the
temperature and $E$-mode dipolar modulation at a modest significance.
However, although residual systematics did not dominate the signal, as
was the case for the 2015 polarization data set \cite{Planck2015IX}, it was
apparent that the various tests of isotropy continued to be limited by
their presence, as well as the dominance of noise over signal in the large-scale \Planck\ polarization data. Indeed, a notable feature of the analyses was the
variation in results with the four component-separated maps studied,
presumably related to their different responses to noise and
systematic residuals, and an incomplete understanding of the noise
properties of the data.  Such effects should not be significant in the
case of \litebird\, since the $E$-mode signal in particular is
expected to be measured at close to the cosmic variance level, with
the actual sensitivity defined by the sky fraction available for
analysis.

A simple measure of whether \litebird\ $E$-mode data are approaching
this level of sensitivity can be provided by the inferred error on
$\tau$, since in the sample-variance limit, this should be of order
$0.002$. Estimates from parameter fits to simulated \litebird\ data
for 70\,\% sky coverage and including foreground residuals related to
component separation, indicate an error consistent with this
expectation (see Sect.~\ref{ss:tau}).
Nevertheless, making inferences about the amplitude of anomalous
features that might be observed in the \litebird\ polarization data
based on what is seen in temperature is non-trivial. Specifically,
predictions must be based on models constructed in three-dimensional position space
then propagated to spherical harmonic space. This mapping is different
for temperature and polarization~\cite{2017PhRvD..96l3522C}.

Despite extensive work, it remains the case that no theoretical model of primordial
perturbations has been constructed that can explain all of the
temperature anomalies.  Hence we need to consider general approaches for testing the hints of anomalies in the temperature data.  This can be achieved by comparing the
distribution of a specific statistic in polarization built from
constrained simulations (where the part of the $E$-mode anisotropy
correlated with temperature is fixed by observations of the
latter) with that constructed from unconstrained realizations.
For example, Ref.~\cite{copi/etal:2013} used constraints from the \WMAP\
7-year temperature power spectrum to compute such distributions
for the $S$-statistic due to the cross-correlation between temperature
and polarization. They determined that a value of the measured $S^{TQ}$
statistic over the angular separation range $[48^\circ, 120^\circ]$
exceeding the value 1.403\,$\mu{\rm K}^4$ would allow the hypothesis
to be ruled out at the 99\,\% confidence level.
Similarly, Ref.~\cite{odwyer_Variance_2017} considered the variance of the
polarization amplitude when the $E$-mode signal is constrained by the
\planck\ 2015 \texttt{SMICA} reconstruction of the CMB temperature
anisotropy. In this case, the temperature data reveal an anomalously
low variance in the northern Ecliptic hemisphere, but 
the constrained realizations show no evidence of a low amplitude
variance in polarization. The measurement of such a signal
would argue against the temperature anomaly being merely a statistical
excursion. 

Irrespective of our expectations, it remains important to search for
characteristic signatures of spatial anomalies in the \litebird\ data, whether or not they are
related to interesting features of the temperature field.
Any detection of anomalies in the polarized sky signal will inevitably
hint at physics beyond that captured by the standard model of
cosmology.

\subsection{Galactic Astrophysics}
\label{ss:galaxy}

Observations of Galactic polarization were among the main outcomes of the \Planck\ space mission. Spectacular images combining the intensity of dust emission with the magnetic field orientation derived from polarization data (so called ``drapery'' patterns, also called ``line integral convolution''~\cite{CabralLeedom}) have received worldwide attention and have become part of the general scientific culture~\citep{Planck2015I}. Beyond this popular impact, the \Planck\ polarization maps have represented 
a big step forward for Galactic astrophysics~\citep{Planck2018XII}. We anticipate a comparable breakthrough with \litebird, which will provide full-sky maps of Galactic polarized emission with a sensitivity many times better than that of \Planck, both for dust and synchrotron polarization. The data will complement the rich array of other polarization observations including: (1) stellar polarization surveys to be combined with {\it Gaia\/} astrometry~\citep{Tassis18} to map Galactic dust polarization in 3D; (2) synchrotron observations, together with Faraday-rotation measurements at radio wavelengths with the \gls{ska} and its precursors~\citep{Haverkorn15}; and (3) ground-based CMB experiments that probe smaller angular scales over restricted parts of the sky with more modest frequency coverage~\cite{Hensley21,CCAT-Prime:2021}.  We now sketch out the expected contributions from \litebird\ to two main directions of Galactic research.

\subsubsection{Magnetic Fields}

Dust polarization probes the magnetic field orientation in dusty regions, mostly in the cold and warm neutral phases of the \gls{ism} that account for the bulk of the gas mass and turbulent energy~\citep{Hennebelle12}. 
Among the various means available to map the structure of interstellar magnetic fields, dust polarization is particularly good for tracing the dynamical interplay between magnetic fields, turbulence, and gravity in the ISM. This interplay helps define the structure of interstellar matter and star
formation. 
The multiphase magnetized ISM is too complex to be described by an analytic theory. Our understanding in this research field progresses through observations, \gls{mhd} simulations, and phenomenological models. Spectroscopic observations  obtained from ground-based telescopes give access to the gas column density and its kinematics. \litebird\ will provide complementary data about magnetic fields. 

\begin{figure}[htbp!]
\hspace{-1cm}
\includegraphics[width=1.1\textwidth]{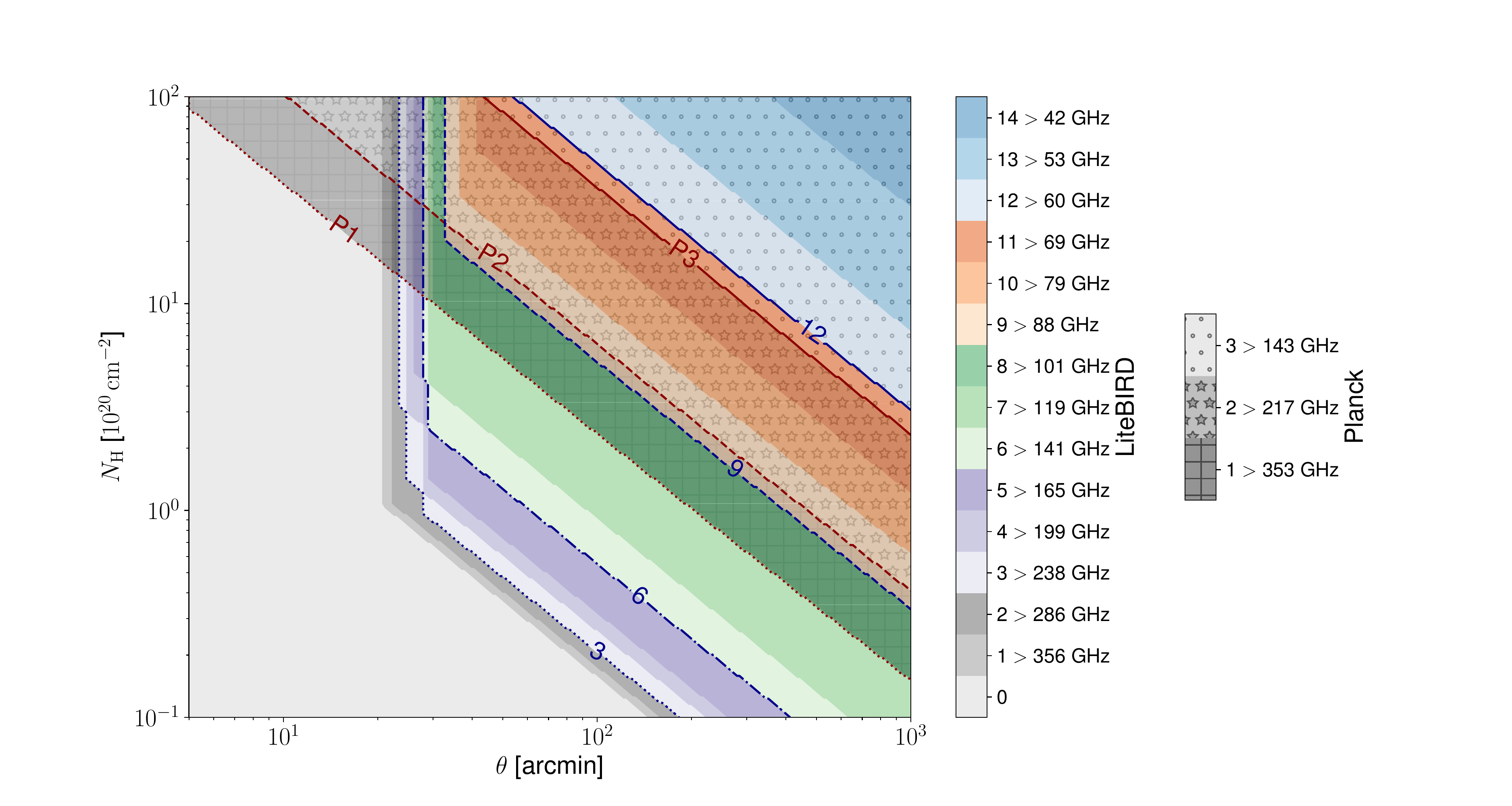}
\caption{\litebird's spectral coverage and sensitivity to polarized dust emission at different angular scales. This plot shows how many \litebird\ bands will yield ${\rm S/N}>1$ as a function of gas column density and angular scale. For comparison, the same information is provided for \Planck, with dots, stars, and hatches delineating regions where dust polarization is detected only at $353\,$GHz, at both 217 and $353\,$GHz, and at the three frequencies 143, 217 and $353\,$GHz, respectively. }
\label{fig:lb_vs_plck_Nbands}
\end{figure}

\begin{figure}[htbp!]
 \hspace{-3cm}\includegraphics[width=1.3\textwidth]{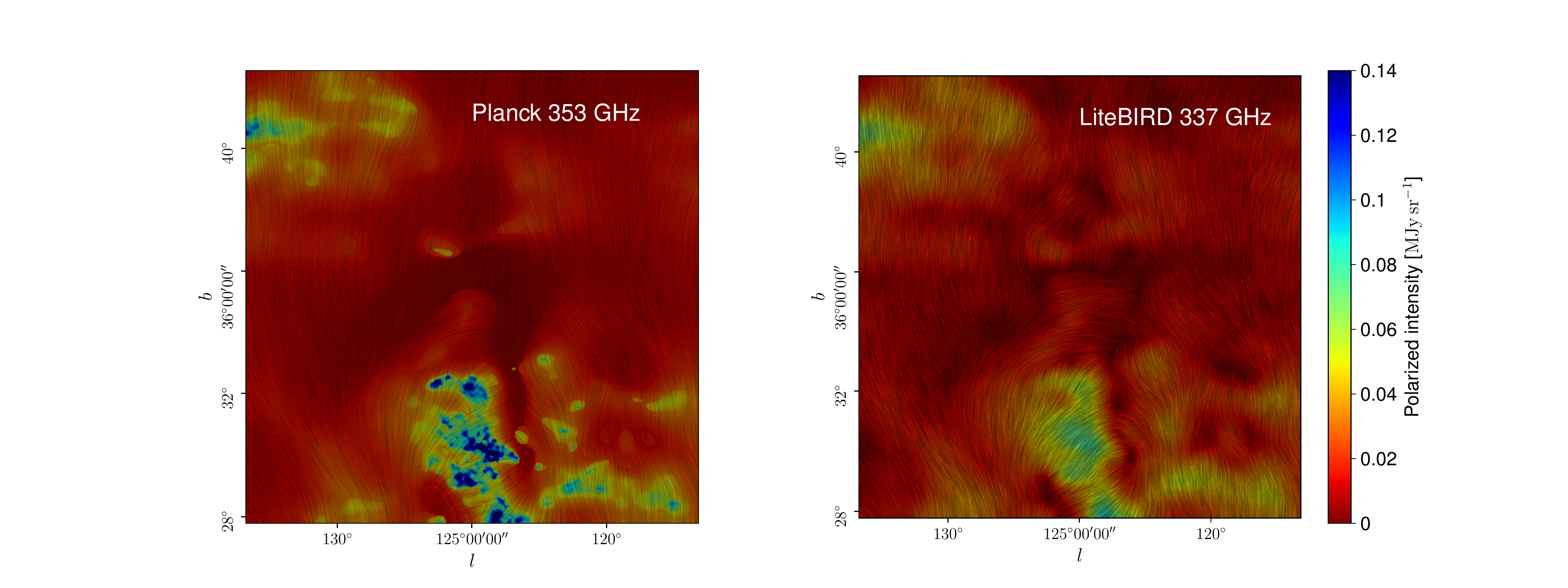}
\caption{Dust polarization images. \textit{Left}: Simulated 
\Planck\ 353-GHz polarization intensity map of the diffuse ISM at high Galactic latitude computed with the model of Ref.~\cite{Vansyngel17}, and centered close to the Polaris Flare~\cite{Levrier18}. The local resolution is adjusted to guarantee ${\rm S/N}>3$ in polarization from pixel to pixel, spanning from 5\,arcmin (in the brightest parts) to 140\,arcmin (in the more diffuse regions). The lines represent the magnetic field orientation derived from the dust polarization angle and the colors show the dust polarized intensity. \textit{Right}: Simulation of the 337-GHz \litebird\ polarization intensity map computed with the same model, at the native resolution of 20.9\,arcmin, which guarantees ${\rm S/N}>3$ in the whole region.  
}
\label{fig:lb_vs_plck_LIC}
\end{figure}

\litebird\ observations at its five highest frequencies, from 166 to 448\,GHz, will improve on the \Planck\ 353-GHz sensitivity to dust polarization by an order of magnitude (Fig.~\ref{fig:lb_vs_plck_Nbands}), increasing the dynamic range of observations by a comparable factor. 
While the analysis of dust polarization from the diffuse ISM at high Galactic latitudes with \Planck\  has been limited to an effective angular resolution of 80\,arcmin, \litebird\ has the required sensitivity to map almost the whole sky down to the 17.9\,arcmin beam size at 402\,GHz (Fig.~\ref{fig:lb_vs_plck_LIC}). The interplay between magnetized turbulence and gas-phase transitions builds the structure of the ISM and seeds the formation of molecular gas; however, these processes are not well understood. \litebird\ will contribute unique polarization maps that will become the new gold standard for analyses of magnetic fields in the diffuse ISM and in the outskirts of nearby molecular clouds. In particular, we expect \litebird\ to reveal coherent magnetic structures that result from the nonlinear interplay between turbulent gas motions and magnetic fields, and the dissipation of turbulence~\citep{Wilkin07,Momferratos14}. Much of our current understanding of the turbulent energy cascade in the ISM derives from MHD simulations that are very far from reproducing its high magnetic Reynolds number~\citep{Federrath16}. The \litebird\ data will be crucial to test how well these simulations match the observations. 

Ongoing developments~\citep{Allys19,Regaldo20,Regaldo21} promise to yield powerful statistical tools for characterizing the \litebird\ dust data and its comparison with simulations, ultimately leading to improved simulations that fully describe the physical processes in our Galaxy. A specific scientific objective will be to elucidate the origin of parity violation (the TB correlation~\cite{Planck2018XI}) of polarized dust emission. Today with the \Planck\ data, we cannot decide whether this is a generic feature of interstellar turbulence rather than a random statistical fluctuation \citep{clark/etal:2021}. 

\litebird\ will improve on \Planck's sensitivity to polarized synchrotron emission by a factor of 5 at 40\,GHz. This gain will extend the range of scales over which the correlation between dust and synchrotron polarization is characterized \citep{Choi15,PlanckIntXXIII}. More generally, the analysis of \litebird\ dust and synchrotron data will contribute to a community effort directed towards modelling of the 3D structure of the Galactic magnetic field \citep{Imagine18}, in particular within the Solar neighbourhood.    

\subsubsection{Interstellar Dust}

The analysis of the \litebird\ data will also yield the spectral characterization of Galactic polarization. The current description of the Galactic contribution to the \Planck\ and \WMAP\ polarization data as arising from two components, namely  thermal dust emission and synchrotron~\citep{Planck2018IV,Planck2018XI}, is likely to prove inadequate for \litebird. By providing data at 15 frequency bands between 34 and 448\,GHz \litebird\ will challenge our current understanding of Galactic emission. The gain is most significant for studying the nature of dust grains and the origin of the so-called \gls{ame} \cite{Banday2003,PlanckIntXV}.

Interstellar dust is often modeled as a mixture of silicate and carbon grains, but polarization observations in emission provide a stringent test challenging existing models~\citep{Guillet18,Hensley19}. The analysis of dust polarization at far-IR wavelengths obtained with the balloon-borne experiment BLASTPol~\citep{Gandilo16} and at microwave frequencies with \Planck~\citep{Planck2018XI} suggests that dust emission at long wavelengths is dominated by one single type of grain, the same for polarization as for total intensity~\citep{Draine2021}.  Thanks to its sensitivity in many spectral bands (Fig.~\ref{fig:lb_vs_plck_Nbands}), \litebird\ observations may unravel additional emission components. In particular, if silicates contain magnetic inclusions, or if free-flying magnetic grains are present, Galactic polarization may include a significant contribution from magnetic dipole emission~\citep{Draine13}.

Dipole emission from spinning dust grains is thought to account for the AME; however, the nature of the carriers remains uncertain~\citep{Dickinson18}. The competing hypotheses, namely polycyclic aromatic hydrocarbons, small silicates, and magnetic nanoparticles, differ in their predictions for the AME polarization~\citep{Draine13,Hoang2013, Draine2016, Hoang2016}. A detection of AME polarization will thus constrain the nature of its carriers. The current limit on AME polarization from the diffuse ISM is set by the cross-correlation between dust and synchrotron polarization~\citep{PlanckIntXXII}. The \litebird\ data, combined with observations from the ground at lower frequencies from e.g., the C-BASS, QUIJOTE and S-PASS surveys \citep{Jones18,Quijote21,S-PASS19}, promise to establish new constraints.

The emission properties of dust at long wavelengths have been shown by \Planck\ to vary throughout the ISM~\citep{Planck2013XI,PlanckIntXVII}. Likewise, the alignment efficiency depends on both the local physical conditions and the dust composition~\citep{Hoang16}. Variations in dust emission properties and alignment efficiency are likely to be correlated with the density structure of the ISM, which in turn is known to be correlated with the magnetic field structure. These couplings break the simple assumption where the spectral frequency dependence of the Galactic polarization and its angular structure on the sky are separable. If a line of sight  intercepts multiple dust clouds with different spectral energy distributions and magnetic field orientations, the frequency scaling of each of the Stokes $Q$ and $U$ parameters of the thermal dust emission may be different. Evidence for this effect has been reported using \Planck\ data \citep{Pelgrims21}. In this context, the interpretation of \litebird\ data in terms of dust properties in polarization will need to be coupled with the modeling of the 3D structure of the Galactic magnetic field.



\section{Potential Design Extensions and Synergies} 
\label{s:discussion}

\subsection{Possible Extension to the Baseline Mission Design -- Shifting the Highest Frequency Channel}
\label{ss:extensions}

\begin{figure}[htbp!]
    \centering
    \includegraphics[width=0.8\textwidth]{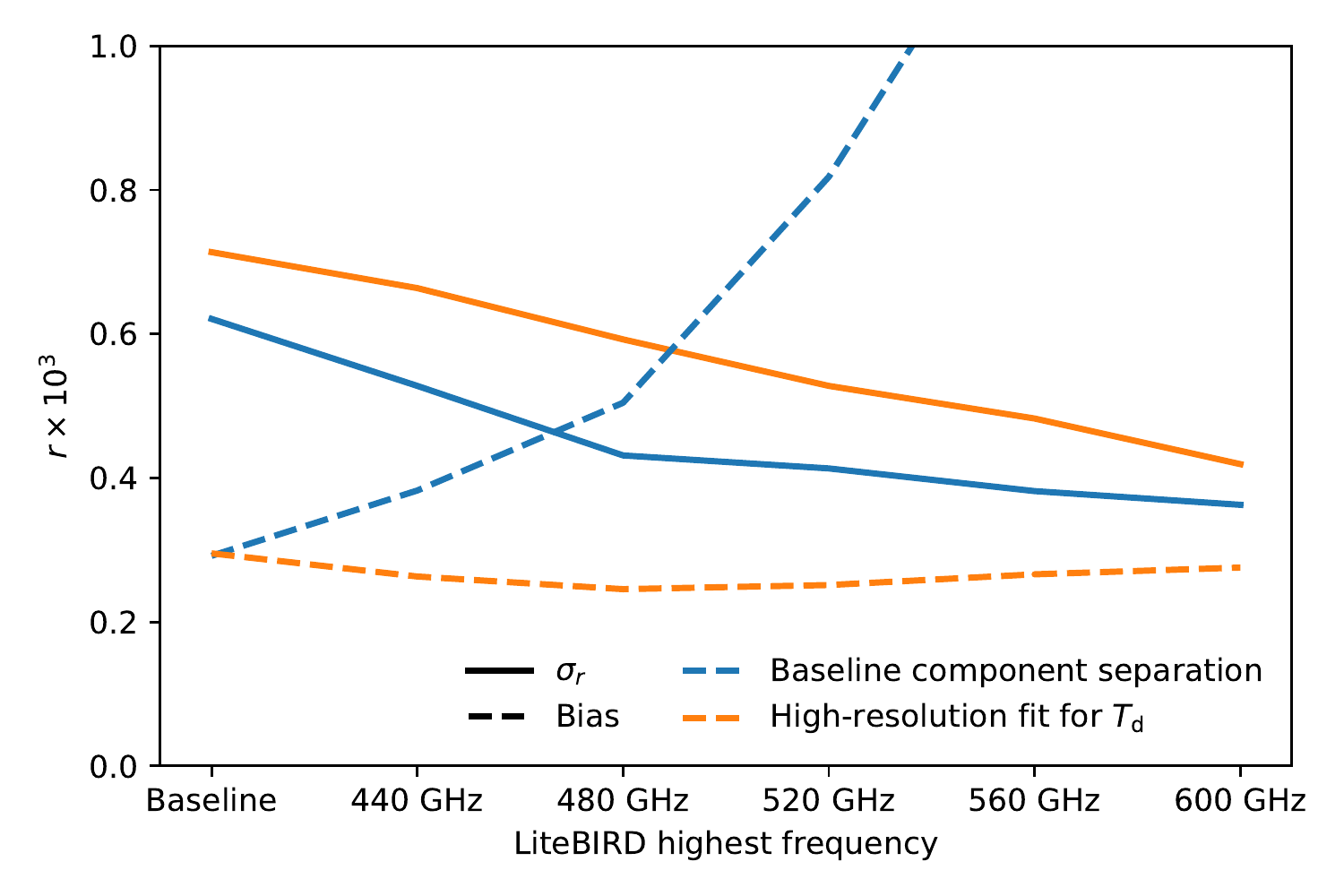}
    \caption{Tensor-to-scalar ratio uncertainty (solid lines) and bias (dashed lines) as a function of the maximum LiteBIRD frequency, ranging between 402 and 600\,GHz. The blue line shows the performance of the component separation configuration discussed in Sect.~\ref{sec:component_separation} or a similar one that fits the dust temerature at high resolution.
  }
    \label{fig:sigmar_vs_freq}
\end{figure}

As discussed in Sect.~\ref{ss:overview_mission}, the current \litebird\ focal-plane configuration results in a conservative mean Fisher uncertainty of $\sigma(r=0) =\sigma_r = 6.2\times 10^{-4}$ after component separation, which conditionally satisfies the full-success criterion of $\delta r<10^{-3}$, as defined in Sect.~\ref{ss:overview_project}. A critical design feature of this configuration is a range for the central frequency bands between 40 and 402\,GHz (34 and 448\,GHz from the edge to edge of the bands), which is a compromise between maximizing our ability to separate polarized foreground emission from CMB emission while still maintaining a lightweight and thermally efficient instrument. In addition, a relatively compact frequency range allows for good overlap between the three individual instruments, which is useful for internal cross-checks and systematics control.

Nevertheless, it is of great interest for the mission as a whole to reduce $\sigma_r$ further in order to increase the margin for other types of systematic errors, both known and unknown. In this respect, we note that one of the dominant sources of uncertainty in the current configuration is a significant degeneracy between CMB and thermal dust emission. With the current focal-plane configuration, we find that statistical uncertainties in the thermal dust temperature (or, equivalently, the second derivative of the thermal dust SED) account for as much as 30--40\,\% of the total error bar on $r$ after component separation. One straightforward approach to decrease this uncertainty is to extend the overall frequency range to higher frequencies. 
The CMB spectral energy density (as measured in brightness density units) falls faster than exponentially above 300\,GHz, while the thermal dust spectral energy density scales as $\nu^{1.5}$. Even a very modest increase in the maximum frequency can therefore dramatically reduce degeneracies between these two components. 

To explore this issue quantitatively, we are currently undertaking a detailed study of alternative extended \litebird\ focal planes, shifting the HFT range to higher frequencies, while leaving the LFT and MFT unchanged. For each configuration, the baseline HFT frequency range is multiplied by some constant factor, such that the ratio between the maximum and minimum frequencies is unchanged, which is essential for the HWP. We consider factors between 1.0 and 1.5 in steps of 0.1. A preliminary example of the outcomes from this study is shown in Fig.~\ref{fig:sigmar_vs_freq}, where we plot the bias (dashed lines) and uncertainty (solid lines) on the tensor-to-scalar ratio for the six different focal plane configurations as derived by \texttt{FGBuster}. While the bias is obtained maximizing the likelihood Eq.~(\ref{eq:global-likelihood}) (with no systematic marginalized), the $\sigma_r$ is a Fisher estimate. This choice allows a decoupling of the statistical uncertainty from the bias, providing two independent figures of merit to compare the instrumental configurations. The colors refer to two component separation setups: the one discussed in Sect.~\ref{ss:overview_project}; and a similar one that differs only for the resolution at which the thermal dust temperature is reconstructed ($N_\text{side}$ equal to 32, 16, and 16 instead of 8, 4, 0, see Table~\ref{tab:ThreeNsides}). The sky model is the same as that employed in the analysis presented in Sect.~\ref{sec:component_separation}, only evaluated at different frequencies.

We see that the statistical uncertainty drops rapidly when extending the frequency range from 400 to 480\,GHz. The flip side is the rise in the bias due to the increased sensitivity to the spatial variations of $T_\text{d}$. However, increasing the degrees of freedom associated to $T_\text{d}$ in the fit keeps the bias under control while still improving the statistical constraints in most of the configurations analyzed. The outcomes of this study will be discussed in detail in a dedicated future paper. For now, we note that the statistical uncertainty is for some \litebird\ configurations 20--30\,\% lower than for the nominal case. Going beyond 500\,GHz, the incremental decrease slows down, and the reason for this is simply the exponential fall-off of the CMB spectrum; at 400\,GHz there is still significant sensitivity to CMB fluctuations given the \litebird\ noise level, while at 500\,GHz it has effectively vanished. Extending the frequency range further does give some additional sensitivity through better constraints on the thermal dust SED, but this is much slower due to the weak power-law dependence of thermal dust, as compared to the super-exponential CMB spectrum. These results are independent of analysis pipelines, and we find that they apply equally to both blind and parametric component-separation methods (e.g., \texttt{Commander} and \texttt{GNILC}).

At the moment, the implications of these issues are being carefully considered by the full \litebird\ collaboration, from hardware through to parameter estimation. 
The main advantages of a slightly shifted focal plane are obvious, namely that the tensor-to-scalar uncertainty could be reduced by as much as 30\,\%, and our ability to reject a potentially false detection due to thermal dust mismodeling is greatly improved. A wider frequency range would also significantly increase the legacy value of the \litebird\ mission with respect to ground-based $B$-mode and Galactic science experiments. However, there are also notable disadvantages associated with shifting the HFT to higher frequencies, the most important of which is a lack of frequency overlap between the MFT and HFT; this could turn out to be important for discovering unknown systematics in either of two instruments when the data actually arrive. A second important drawback of modifying the focal plane at this stage is purely programmatic, in that many current instrument designs would have to be revised if the focal plane is modified, and this could result in a lower overall \gls{trl}. Thirdly, a higher maximum frequency would improve our sensitivity with respect to detailed thermal dust modeling. In the absolutely worst-case scenario in which it turned out to be impossible to model thermal dust emission at 500\,GHz or above at the required level of precision, so that we would have to exclude the highest channel from the CMB analysis stage, the final uncertainty on the tensor-to-scalar ratio would \emph{increase} by $\lesssim$\,5\,\% compared to the current baseline configuration, due to slightly lower sensitivity in the overlap region between the MFT and HFT. 

\subsection{Synergy with Other Projects}

The \lb\ space mission has strong synergy with ground-based CMB polarization anisotropy experiments including SO, SPO, and CMB-S4.  Given the small amplitude of the inflationary $B$-mode signal and the obscuring effects of foregrounds, lensing, and instrumental systematic uncertainties, having highly sensitive data from both \lb\ and ground-based 
experiments will contribute to building confidence in the robustness of an inflationary signal detection.   Furthermore, \lb\ provides data on foregrounds at frequencies above the highest frequency that will be observed by above-mentioned ground-based experiments (280\,GHz), and the \lb\ high-frequency data can be combined with $B$-mode data from the ground improving foreground separation.  As well as \lb\ augmenting ground-based experiments, CMB lensing data from the ground can be combined with \lb\ data to improve the sensitivity of \lb\ to primordial $B$ modes at the angular scales of the recombination peak. 

Ground-based experiments generally focus on a relatively small region of sky, which has the benefit of giving high S/N per spatial mode but on a relatively low number of spatial modes.  \lb, in contrast, will measure the entire sky and measure all the spatial modes that are available at moderate S/N.  Both deep and wide observations can contribute to our understanding of foreground emission, which is essential in the search for inflationary $B$ modes.  Deep measurements give a high S/N characterization of foreground emission, but only on a relatively small region of the sky.  \lb's measurement of the entire sky will probe the variability of the emission in different directions and test the fidelity of the model.  There is a similar complementarity in terms of instrumental systematic uncertainties, where the high S/N per mode of ground-based measurements provides a deep probe for discovering weak systematic errors, whereas \lb\ has the statistical power to measure many spatial modes, in addition to the advantage of making the observations in the benign and stable environment of space.  Finally, \lb\ also has the unique ability to detect an inflationary signal at both the reionization peak (where gravitational lensing does not interfere) and at the recombination peak.  A detection of a signal at both peaks would greatly increase confidence in these challenging measurements.



\section{Conclusions} 
\label{s:conclusions}

\LiteBIRD\ will provide full-sky CMB polarization maps with unprecedented precision in 15 frequency bands between 34 and 448\,GHz.  These capabilities will enable \LiteBIRD\ to satisfy the basic science requirement on the tensor-to-scalar ratio, $\delta r\,{<}\,0.001$, by probing both the reionization and recombination bumps in the primordial $B$-mode power spectrum expected from inflationary models.  This sensitivity level will lead to either detection of primordial gravitational waves or will rule out a large class of popular
inflationary models, yielding insights into physics at the very highest energies.

The \LiteBIRD\ Collaboration has more than 300 researchers from Japan, Europe, and North America, and
has successfully completed the Pre-Phase-A2 concept development studies of \LiteBIRD.
Table~\ref{tbl:specifications} shows baseline specifications of the mission as the result of these studies.

\begin{table}[htbp!]
\centering
\caption{Main specifications of \LiteBIRD. Parameters are from the \LiteBIRD\ pre-phase-A2 concept development studies
and additional studies in 2020, as preparation for the system-requirements review.}
\label{tbl:specifications}
\begin{tabular}{|l|l|} 
\hline
Item & Specification\\
\hline \hline
Science requirement & $\delta r < 0.001$ for $2 \leq \ell \leq 200$\\
\hline
Target launch year & 2029\\
\hline
Launch vehicle & JAXA H3\\
\hline
Observation type & All-sky CMB surveys\\
\hline
Observation time & 3 years\\
\hline
Orbit & L2 Lissajous orbit\\
\hline
Scan and  		& $\cdot$ Spin and precession (prec.\ angle $\alpha = 45^\circ$, spin angle $\beta = 50^\circ$)\\
data recording	& $\cdot$ Spin period = 20\,minutes, precession period = 3.2058\,hours\\
				& $\cdot$ PMU revolution rate = 46/39/61\,rpm for LFT/MFT/HFT\\
				& $\cdot$ Sampling rate = 19.1\,Hz\\
\hline
Observing frequencies  & 34--448\,GHz\\
\hline
Number of bands & 15 \\
\hline
Polarization sensitivity & $2.2\,\mu$K-arcmin (after 3 years) \\
\hline
Angular resolution & 0.5$^\circ$ at 100\,GHz (FWHM for LFT) \\
\hline
Mission instruments & $\cdot$ Superconducting detector arrays \\
                    & $\cdot$ Crossed-Dragone mirrors (LFT)\\
                    & ~~+ two refractive telescopes (MFT and HFT) \\
                    & $\cdot$ PMU with continuously-rotating HWP on each telescope\\
                    & $\cdot$ 0.1-K cooling chain (ST/JT/ADR) \\
\hline
Data size & $17.9\,{\rm GB}\,{\rm day}^{-1}$ \\
\hline
Mass & 2.6\,t \\
\hline
Power & 3.0\,kW \\
\hline
\end{tabular}
\end{table}

\LiteBIRD\ is a mission that will set the course for the future of cosmology.
There are many different inflationary models under active discussion, which predict different values of $r$.
Among them, there are well-motivated inflationary models that predict $r > 0.01$~\cite{kamionkowski/kovetz:2016}.
Since our requirement is $\delta r < 0.001$, we can provide more than $10\,\sigma$ detection significance for these models. On the other hand,
if \LiteBIRD\ finds no primordial $B$ modes and obtains an upper limit on $r$,
this limit should be stringent enough to set severe constraints on the physics of inflation.
As discussed in Ref.~\citep{Linde:2016hbb},
if we obtain an upper limit at $r\simeq 0.003$, we can completely rule out
one important category of models, namely any single-field model in which
the characteristic field-variation scale of the inflaton potential is greater than the reduced Planck mass.

Once we carry out the observations successfully, we can use \LiteBIRD\ data to study many additional topics in cosmology, particle physics, and astronomy.
Examples include:
(1) statistical characterization of large-scale $B$ modes and $E$ modes, including tests of scale-invariance and non-Gaussianity;
(2) investigation of possible power-spectrum features in polarization;
(3) sample-variance-limited measurements of large-scale (low $\ell$) $E$ modes, with implications for the reionization history and the sum of neutrino masses;
(4) searches for cosmic birefringence and parity violation;
(5) maps of cosmological hot gas through the Sunyaev-Zeldovich effects and relativistic corrections;
(6) elucidation of large-angle anomalies in polarization; and
(7) Galactic astrophysics.
\LiteBIRD\ has very focused mission requirements, and at the same time will provide rich scientific outcomes.

	

\section*{Acknowledgment}
\label{acknowledgment}
This work is supported in Japan by ISAS/JAXA for Pre-Phase A2 studies, by the acceleration program of JAXA research and development directorate, by the World Premier International Research Center Initiative (WPI) of MEXT, by the JSPS Core-to-Core Program, and by JSPS KAKENHI. The Italian \LiteBIRD\ phase A contribution is supported by the Italian Space Agency (ASI Grants No. 2020-9-HH.0 and 2016-24-H.1-2018), the National Institute for Nuclear Physics (INFN), the National Institute for Astrophysics (INAF), and a PGR grant from the Italian Ministry of Foreign Affairs and
International Cooperation. The French \LiteBIRD\ phase A contribution is supported by the Centre National d’Etudes Spatiale (CNES), by the Centre National de la Recherche Scientifique (CNRS), and by the Commissariat à l’Energie Atomique (CEA). The Canadian contribution is supported by the Canadian Space Agency. The US contribution is supported by NASA grant no. 80NSSC18K0132. 
Norwegian participation in \LiteBIRD\ is supported by the Research Council of Norway (Grant No. 263011). The Spanish \LiteBIRD\ phase A contribution is supported by the Spanish Agencia Estatal de Investigación (AEI), project refs. PID2019-110610RB-C21 and AYA2017-84185-P. Funds that support the Swedish contributions come from the Swedish National Space Agency (SNSA/Rymdstyrelsen) and the Swedish Research Council (Reg. no. 2019-03959). The German participation in \LiteBIRD\ is supported in part by the Excellence Cluster ORIGINS, which is funded by the Deutsche Forschungsgemeinschaft (DFG, German Research Foundation) under Germany’s Excellence Strategy (Grant No. EXC-2094 - 390783311). \LiteBIRD\ work has received funding from the European Research Council (ERC) under the Horizon 2020 Research and Innovation Programme. This research used resources of the Central Computing System owned and operated by the Computing Research Center at KEK, as well as resources of the National Energy Research Scientific Computing Center, a DOE Office of Science User Facility supported by the Office of Science of the U.S. Department of Energy.




\clearpage


\printnoidxglossary[style=long]


\bibliographystyle{ptephy}
\bibliography{LB,litebird_ptep}

\end{document}

%% file: acronyms.tex
\newacronym{fpu}{FPU}{Focal-Plane Unit} 

\newacronym{fpm}{FPM}{focal-plane module} 
\newacronym{fps}{FPS}{focal-plane structure}
\newacronym{fph}{FPH}{focal-plane hood}
\newacronym{cru}{CRU}{cryogenic readout units}
\newacronym{crh}{CRH}{cryogenic readout harness} 
\newacronym{cr}{CR}{cryogenic readout}
\newacronym{wr}{WR}{warm readout}
\newacronym{hf}{HF}{High-Frequency}
\newacronym{lf}{LF}{Low-Frequency}
\newacronym{mf}{MF}{Mid-Frequency}
\newacronym{hft}{HFT}{High-Frequency Telescope}
\newacronym{mft}{MFT}{Mid-Frequency Telescope}
\newacronym{lft}{LFT}{Low-Frequency Telescope}
\newacronym{mhft}{MHFT}{Mid- and High-Frequency Telescopes}

\newacronym{lffpu}{LF-FPU}{Low-Frequency Focal-Plane Unit}
\newacronym{mffpu}{MF-FPU}{Mid-Frequency Focal-Plane Unit}
\newacronym{hffpu}{HF-FPU}{High-Frequency Focal-Plane Unit}
\newacronym{lffpm}{LF-FPM}{Low-Frequency Focal-Plane Module}
\newacronym{lffps}{LF-FPS}{Low-Frequeny Focal-Plane Structure}
\newacronym{lffph}{LF-FPH}{Low-Frequency Focal-Plane Hood}
\newacronym{lfcrh}{LF-CRH}{Low-Frequency Cryogenic Readout Harness}
\newacronym{2kadrfs}{2K-ADRCC}{2-K adiabatic demagnetization refrigerator cryocooler and controller}
\newacronym{2kadr}{2K-ADR}{adiabatic demagnetization refrigerator that cools the 2-K stage}
\newacronym{skadr}{Sub-K ADR}{adiabatic demagnetization refrigerator that cools the sub-kelvin stages}
\newacronym{adr}{ADR}{adiabatic demagnetization refrigerator}
\newacronym{adrs}{ADRs}{adiabatic demagnetization refrigerators}
\newacronym{adrc}{ADRC}{adiabatic demagnetization refrigerator controller}
\newacronym{JT}{JT}{Joule-Thomson}
\newacronym{jt2}{2-K JT}{Joule-Thomson 1.75-K cooled stage}
\newacronym{jt4}{4-K JT}{Joule-Thomson 4.8-K cooled stage}
\newacronym{em}{EM}{engineering model}
\newacronym{fm}{FM}{flight model}
\newacronym{dm}{DM}{development model}
\newacronym{bbm}{BBM}{breadboard model}
\newacronym{aocs}{AOCS}{attitude orbit-control system}
\newacronym{mw}{MW}{momentum wheel}
\newacronym{rcs}{RCS}{reaction control system}
\newacronym{stt}{STT}{star tracker}
\newacronym{sap}{SAP}{solar array panel}
\newacronym{pmu}{PMU}{polarization modulator unit}
\newacronym{smb}{SMB}{superconducting magnetic bearing}
\newacronym{pp}{PP}{polypropylene}
\newacronym{spa}{SPA}{signal-processing assembly}
\newacronym{spu}{SPU}{signal-processing unit}
\newacronym{catr}{CATR}{compact antenna test range}
\newacronym{vna}{VNA}{vector network analyzer}
\newacronym{fsl}{FSL}{far sidelobe}

\newacronym{fte}{FTE}{full-time equivalent}
\newacronym{csr}{CSR}{Concept Study Report}
\newacronym{am}{AM}{Assurance Manager}
\newacronym{pdam}{PDAM}{Project Deputy Assurance Manager}
\newacronym{srr}{SRR}{System Requirement Review}
\newacronym{mdr}{MDR}{Mission Definition Review}
\newacronym{pdr}{PDR}{Preliminary Design Review}
\newacronym{cdr}{CDR}{Critical Design Review}
\newacronym{jcdr}{JCDR}{JAXA Concept Design Report}
\newacronym{mel}{MEL}{Master Equipment List}
\newacronym{empfm}{EM/PFM}{Engineering Model / Pre-Flight Model} 
\newacronym{icd}{ICD}{interface control documents}
\newacronym{sir}{SIR}{System Integration Review}
\newacronym{trlpr}{TRLPR}{Technology Readiness Level Path Review}
\newacronym{plar}{PLAR}{Post-Launch Assessment Review}
\newacronym{pi}{PI}{Principal Investigator}
\newacronym{we}{WE}{warm electronics}
\newacronym{wbs}{WBS}{work breakdown structure}
\newacronym{evm}{EVM}{earned value management}
\newacronym{rtm}{RTM}{requirements and traceability matrix}
\newacronym{stm}{STM}{science traceability matrix}
\newacronym{nte}{NTE}{not-to-exceed}
\newacronym{mam}{MAM}{Mission Assurance Manager}
\newacronym{dmam}{DMAM}{Deputy Mission Assurance Manager}
\newacronym{cmp}{CMP}{Configuration Management Plan}
\newacronym{jset}{JSET}{Joint Systems Engineering Team}
\newacronym{it}{I\&T}{Integration \& Test}
\newacronym{fmea}{FMEA}{failure modes and effects analysis}
\newacronym{fta}{FTA}{fault tree analysis}
\newacronym{sma}{SMA}{safety mission assurance}
\newacronym{semp}{SEMP}{Systems Engineering Management Plan}
\newacronym{ccb}{CCB}{Configuration Control Board}
\newacronym{srb}{SRB}{Standing Review Board}
\newacronym{tbd}{TBD}{to be done}
\newacronym{tbr}{TBR}{to be resolved}
\newacronym{cbe}{CBE}{current best estimate}
\newacronym{mev}{MEV}{maximum expected value}
\newacronym{plm}{PLM}{Payload Module}
\newacronym{svm}{SVM}{Service Module}
\newacronym{paip}{PAIP}{Performance Assurance Implementation Plan}
\newacronym{orr}{ORR}{Operational Readiness Review}
\newacronym{l1}{L1}{Level 1}
\newacronym{l2}{L2}{Level 2}
\newacronym{l3}{L3}{Level 3}
\newacronym{l4}{L4}{Level 4}
\newacronym{l5}{L5}{Level 5}
\newacronym{pmo}{PMO}{Partner Mission of Opportunity}
\newacronym{mo}{MO}{Mission of Opportunity}
\newacronym{ac}{AC}{Advisory Committee}
\newacronym{trl}{TRL}{Technology Readiness Level}
\newacronym{dcr}{DCR}{Data Completeness Review}
\newacronym{moc}{MOC}{Mission Operations Center}
\newacronym{soc}{SOC}{Science Operations Center}
\newacronym{boe}{BoE}{basis of estimate}
\newacronym{mdra}{MDRA}{Mission Definition Requirements Agreement}
\newacronym{poc}{POC}{point of contact}
\newacronym{too}{TOO}{target-of-opportunity}
\newacronym{ait}{AIT}{assembly, integration, and testing}
\newacronym{aiv}{AIV}{assembly, integration, and verification}
\newacronym{tmr}{TMR}{triple modular redundancy}
\newacronym{ecc}{ECC}{error-correcting codes}
\newacronym{da}{DA}{digitizer assembly}
\newacronym{sca}{SCA}{SQUID controller assembly}
\newacronym{mdpu}{MDPU}{mission data processing unit}
\newacronym{mdpul}{MDPU-L}{mission data processing unit for LFT}
\newacronym{mdpuhf}{MDPU-MH}{mission data processing unit for MFT and HFT}
\newacronym{sdpu}{SDPU}{science data processing unit}
\newacronym{cre}{CRE}{cold readout electronics}
\newacronym{hau}{HAU}{housekeeping acquisition unit }
\newacronym{jtd}{JTD}{Joule-Thomson cooler driver}
\newacronym{pcd}{PCD}{pre-cooler driver}
\newacronym{imo}{IMo}{instrument model}

\newacronym{lcdm}{$\Lambda$CDM}{$\Lambda$ cold dark matter}
\newacronym{fts}{FTS}{Fourier-transform spectrometer}
\newacronym{tod}{TOD}{time-ordered data}
\newacronym{cib}{CIB}{cosmic infrared background}
\newacronym{sts}{STS}{star-tracker system}
\newacronym{drie}{DRIE}{deep reactive ion etching}
\newacronym{css}{CSS}{coarse Sun sensor}
\newacronym{iru}{IRU}{inertial reference unit}
\newacronym{gse}{GSE}{ground-system equipment}
\newacronym{sq}{SQUID}{superconducting quantum interference device}
\newacronym{scu}{SCU}{SQUID Controller Unit}
\newacronym{dlfov}{DLFOV}{diffraction-limited field of view}
\newacronym{hwp}{HWP}{half-wave plate}
\newacronym{ahwp}{AHWP}{achromatic half-wave plate}
\newacronym{hwpss}{HWPSS}{HWP synchronous signal}
\newacronym{tes}{TES}{transition-edge sensor}
\newacronym{fov}{FoV}{field of view}
\newacronym{jsg}{JSG}{joint study group}
\newacronym{ncr}{NCR}{noise-to-carrier ratio}
\newacronym{ar}{AR}{anti-reflection}
\newacronym{rwa}{RWA}{reaction-wheel assembly}
\newacronym{omt}{OMT}{orthomode transducer}
\newacronym{dac}{DAC}{digital-to-analog converter}
\newacronym{adc}{ADC}{analog-to-digital converter}
\newacronym{cmb}{CMB}{cosmic microwave background}
\newacronym{did}{DID}{data item description}
\newacronym{saa}{SAA}{SQUID array amplifier}
\newacronym{dpu}{DPU}{digital processing unit}
\newacronym{pcb}{PCB}{printed circuit board}
\newacronym{co}{CO}{carbon monoxide}
\newacronym{rf}{RF}{radio-frequency}
\newacronym{lc}{LC}{inductor/capacitor}
\newacronym{dfmux}{DfMux}{digital frequency-domain multiplexing}
\newacronym{net}{NET}{noise-equivalent temperature}
\newacronym{fpga}{FPGA}{field-programmable gate array}
\newacronym{fll}{FLL}{flux-locked loop}
\newacronym{dan}{DAN}{digital active nulling}
\newacronym{vhdl}{VHDL}{very high-speed integrated-circuit hardware description language}
\newacronym{asic}{ASIC}{application-specific integrated circuit}
\newacronym{nep}{NEP}{noise-equivalent power}
\newacronym{cpw}{CPW}{coplanar-waveguide}
\newacronym{cvd}{CVD}{chemical vapor deposition}
\newacronym{afm}{AFM}{atomic force microscopy}
\newacronym{sem}{SEM}{scanning electron microscopy}
\newacronym{mems}{MEMS}{micro-electro-mechanical systems}
\newacronym{ms}{MS}{microstrip}
\newacronym{mkid}{MKID}{microwave kinetic-inductance detector}
\newacronym{ame}{AME}{anomalous microwave emission}
\newacronym{toast}{TOAST}{Time-Ordered Astrophysics Scalable Tools}
\newacronym{flrw}{FLRW}{Friedmann-Lema\^itre-Robertson-Walker} 
\newacronym{fwhm}{FWHM}{full width at half maximum}
\newacronym{psf}{PSF}{point spread function}
\newacronym{tml}{TML}{total mass loss}
\newacronym{cvcm}{CVCM}{collected volatile condensable material}
\newacronym{arc}{ARC}{anti-reflection coating}
\newacronym{ptfe}{PTFE}{poly-tetrafluoroethane}
\newacronym{cfrp}{CFRP}{carbon fiber reinforced plastic}
\newacronym{sed}{SED}{spectral energy distribution}
\newacronym{ip}{IP}{instrumental polarization}
\newacronym{rcwa}{RCWA}{rigorous coupled-wave analysis}
\newacronym{sz}{SZ}{Sunyaev-Zeldovich}
\newacronym{eor}{EoR}{epoch of reionization}
\newacronym{uv}{UV}{ultra-violet}
\newacronym{ml}{ML}{maximum-likelihood}
\newacronym{cl}{CL}{confidence level}
\newacronym{bao}{BAO}{baryonic accoustic oscillation}
\newacronym{whim}{WHIM}{warm-hot intergalactic medium}
\newacronym{agn}{AGN}{active galactic nuclei}
\newacronym{pmf}{PMF}{primordial magnetic field}
\newacronym{smhw}{SMHW}{spherical Mexican-hat wavelet}
\newacronym{ism}{ISM}{interstellar  medium}
\newacronym{mhd}{MHD}{magnetohydrodynamic}

\newacronym{carma}{CARMA}{Combined Array for Research in Millimeter-wave Astronomy}
\newacronym{alma}{ALMA}{Atacama Large Millimeter/submillimeter Array}
\newacronym{cmbs4}{CMB-S4}{CMB Stage-4}
\newacronym{so}{SO}{Simons Observatory}
\newacronym{sa}{SA}{Simons Array}
\newacronym{spo}{SPO}{South Pole Observatory}
\newacronym{cobe}{\textit{COBE}}{Cosmic Background Explorer}
\newacronym{wmap}{\textit{WMAP}}{Wilkinson Microwave Anisotropy Probe}
\newacronym{hfi}{HFI}{High-Frequency Instrument}
\newacronym{xrism}{XRISM}{X-ray Imaging and Spectroscopy Mission}
\newacronym{planck}{Planck}{Planck}
\newacronym{gpb}{GPB}{Gravity Probe B}
\newacronym{spica}{SPICA}{Space Infrared Telescope for Cosmology and Astrophysics}
\newacronym{cdms}{CDMS}{Cryogenic Dark Matter Search}
\newacronym{safari}{SAFARI}{SpicA FAR-infrared Instrument}
\newacronym{htt}{HTT}{Huan Tran Telescope}
\newacronym{ebex}{EBEX}{E and B EXperiment}
\newacronym{desi}{DESI}{Dark Energy Spectroscopic Instrument}
\newacronym{lsst}{LSST}{Legacy Survey of Space and Time}
\newacronym{ska}{SKA}{Square Kilometer Array}

\newacronym{cu}{CU}{University of Colorado, Boulder}
\newacronym{lasp}{CU/LASP}{CU-Boulder Laboratory for Atmospheric and Space Physics}
\newacronym{ucsd}{UC San Diego}{University of California, San Diego}
\newacronym{nist}{NIST}{the National Institute of Standards and Technologies}
\newacronym{uw}{UC San Diego}{the University of California, San Diego}
\newacronym{ucb}{UC Berkeley}{the University of California, Berkeley}
\newacronym{ccc}{$C^3$}{Computational Cosmology Center}
\newacronym{lbnl}{LBNL}{Lawrence Berkeley National Laboratory}
\newacronym{hpc}{HPC}{High-Performance Computing}
\newacronym{ssl}{UCB/SSL}{UC Berkeley Space Sciences Laboratory}
\newacronym{cnes}{CNES}{National Centre for Space Studies}
\newacronym{esa}{ESA}{European Space Agency}
\newacronym{great}{GREAT}{GRound station for deep space Exploration And Tele-communication}
\newacronym{csa}{CSA}{Canadian Space Agency}
\newacronym{kek}{KEK}{High Energy Accelerator Research Organization in Tsukuba, Japan}
\newacronym{nersc}{NERSC}{National Energy Research Scientific Computing center}
\newacronym{nasa}{NASA}{National Aeronautics and Space Administration}
\newacronym{nsf}{NSF}{National Science Foundation}
\newacronym{eu}{EU}{Euopean Union}
\newacronym{jaxa}{JAXA}{Japan Aerospace Exploration Agency}
\newacronym{gsfc}{GSFC}{Goddard Space Flight Center}
\newacronym{ipmu}{IPMU}{Kavli Institute for the Physics and Mathematics of the Universe}
\newacronym{nec}{NEC}{Nippon Electric Corporation}
\newacronym{lambda}{LAMBDA}{Legacy Archive for Microwave Background Data Analysis}
\newacronym{shi}{SHI}{Sumitomo Heavy Industries}
\newacronym{stanford}{Stanford}{Stanford University}
\newacronym{cea}{CEA}{French Alternative Energies and Atomic Energy Commission}
\newacronym{apc}{APC}{Laboratoire Astroparticule et Cosmologie}
\newacronym{isas}{ISAS}{Institute of Space and Astronautical Science}
\newacronym{usc}{USC}{Uchinoura Space Center}

\newacronym{pse}{PSE}{Project System Engineer}
\newacronym{pset}{PSET}{Project System Engineer Team}
\newacronym{pm}{PM}{Project Manager}
\newacronym{pfm}{PFM}{Project Financial Manager}

\newacronym{rpm}{RPM}{revolutions per minute}
\newacronym{rms}{RMS}{root mean square}

\newacronym{fp}{FP}{focal plane}
\newacronym{ci}{CI}{cold intermediate}
\newacronym{wi}{WI}{warm intermediate}